\message{**  PAPERMAC.TEX: DJD 06/10/89  **}
\font\twelvebf=cmbx12
%
%
%
%
%
%
%
%
%
%
\def\today{\ifcase\month\or
  January\or February\or March\or April\or May\or June\or
  July\or August\or September\or October\or November\or December\fi
  \space\number\day, \number\year}
\def\prtoday{\number\day\space\ifcase\month\or
  January\or February\or March\or April\or May\or June\or
  July\or August\or September\or October\or November\or December\fi
  \space\number\year}
\def\draftid{\relax}

\def\equtag{\relax}
\def\figtag{\relax}

\def\hideunder#1#2{
    \oalign{{#2}\crcr\hidewidth{#1}\hidewidth}}
%
%
%
\normalbaselineskip= 10pt
\def\singlespace{\baselineskip=\normalbaselineskip}

\def\doublespace{\baselineskip=\normalbaselineskip
                 \multiply \baselineskip by 2}
\def\skipline{\vskip\baselineskip}
%
        \magnification=\magstep1
	\parskip=0pt plus6pt
	\doublespace
   	\hsize   = 5.750truein
	\vsize   = 8.750truein
	\hoffset = 0.500truein
	\voffset = 0.125truein
	\widowpenalty=10000  
	\clubpenalty=10000   
	\pretolerance=800    %
	\tolerance=10000     
	\overfullrule=0pt    
%
	\scriptscriptfont0 = \scriptfont0
	\scriptscriptfont1 = \scriptfont1
	\scriptscriptfont2 = \scriptfont2
	\scriptscriptfont3 = \scriptfont3
%
%
%
%
%
%
\newif\ifnewfoot
\newif\ifnewhead
\def\newchapter#1{%
	\def\secname{#1}
	\global\newfoottrue
	\global\newheadtrue
	\def\refkeys{}
	\def\equkeys{}
	\def\figkeys{}
	\def\tabkeys{}
	\def\equlist{}
	\def\figlist{}
	\def\tablist{}
	\headline={\ifnewhead\newheader\global\newheadfalse \else\oldheader\fi}
	\footline={\ifnewfoot\newfooter\global\newfootfalse \else\oldfooter\fi}
        }
\def\newheader{\hfil}
\def\newfooter{\centerpagenum}
\def\oldheader{\centerpagenum}
\def\oldfooter{\hfil}
\def\centerpagenum{\hss\tenrm\folio\hss\draftid}
%
%
\newchapter{}
\def\refkeys{}
\def\biblist{}
\newtoks\ta
\newtoks\tb
\newif\ifmember
\newcount\citenum
\newcount\refnum
\newcount\seqnum
\newcount\seqfirst
\newcount\seqlast
%
%
%
\def\cite#1{\buildlist{#1}\updaterefs{\listed}$^{\citerefs{\listed}}$}
\def\pcite#1{\cite{#1}\hskip 4.4443pt plus 4.99997pt minus 0.37036pt}
\def\refnumber#1{\buildlist{#1}\updaterefs{\listed}\citerefs{\listed}}

\def\bibitem#1#2{\add{#1}\to{\biblist}%
                 \add{#2}\to{\biblist}}
%
%
%
\long\def\add#1\to#2{\ta={\\{#1}}%
                     \tb=\expandafter{#2}%
                     \global\edef#2{\the\tb\the\ta}}%
%
%
\def\buildlist#1{\def\listed{}\makelistmacro#1,@}
\def\makelistmacro#1#2,#3@{\add{#1#2}\to{\listed}%
                         \def\restoflist{#3}%
                         \ifx\restoflist\empty\relax%
                         \else\makelistmacro#3@%
                         \fi}
%
%
\def\updaterefs#1{\citenum=0%
                 \def\\##1{\updatedef{##1}}%
                 #1}%
\def\updatedef#1{\advance\citenum by 1%
                 \ifx\refkeys\empty
                    \add{#1}\to{\refkeys}%
                 \else%
                    \memberfalse%
                    \ismember{#1}\of{\refkeys}
                    \def\\##1{\updatedef{##1}}
                    \ifmember
                    \else%
                       \add{#1}\to{\refkeys}
                    \fi%
                 \fi}%
%
%
\def\ismember#1\of#2{\def\given{#1}%
                     \def\\##1{\def\next{##1}%
                               \ifmember%
                               \else%
                                  \ifx\next\given%
                                      \membertrue%
                                  \fi%
                               \fi}%
                     #2}%
%
%
\def\citerefs#1{\refnum=0%
                \seqnum=0%
                \def\\##1{\citedef{#1}{##1}}%
                \refkeys%
                \def\\##1{}} 
\def\citedef#1#2{\ifnum\citenum>0
                     \advance\refnum by 1%
                     \memberfalse%
                     \ismember{#2}\of{#1}
                     \def\\##1{\citedef{#1}{##1}}
                     \ifmember
                        \advance\citenum by-1 %
                        \ifnum\seqnum=0 %
                           \ifnum\citenum=0 %
                              \number\refnum %
                           \else %
                              \advance\seqnum by 1 %
                              \seqfirst=\refnum %
                           \fi %
                        \else %
                           \advance\seqnum by 1 %
                           \ifnum\citenum=0 %
                              \ifnum\seqnum=2 %
                                 \number\seqfirst ,\number\refnum %
                              \else %
                                 \number\seqfirst \hbox{--}\number\refnum %
                              \fi %
                           \else %
                              \seqlast=\refnum %
                           \fi %
                        \fi %
                     \else %
                        \ifnum\seqnum>2 %
                           \number\seqfirst \hbox{--}\number\seqlast ,%
                        \else %
                           \ifnum\seqnum=2 %
                              \number\seqfirst ,\number\seqlast ,%
                           \fi %
                           \ifnum\seqnum=1 %
                              \number\seqfirst ,%
                           \fi%
                        \fi%
                        \seqnum=0 %
                     \fi %
                  \fi}
%
%
%
%
\long\def\listrefs{\refnum=0%
                   \def\\##1{\advance\refnum by 1%
                             \writeref{\number\refnum}{##1}}%
                   \refkeys}
%
%
\long\def\writeref#1#2{\memberfalse%
                       \def\nextkey{#2}%
                       \long\def\\##1\\##2{\def\nextbibkey{##1}%
                                           \ifmember
                                           \else
                                              \ifx\nextkey\nextbibkey
                                                 \membertrue
                                                 \printref{#1}{##2}
                                              \fi
                                           \fi}%
                       \biblist%
                       \def\\##1{\advance\refnum by 1%
                                 \writeref{\number\refnum}{##1}}}
%
%
\long\def\printref#1#2{%
               \begingroup
               \noindent
               \singlespace
               \advance\leftskip by 0.5truein
               \llap{#1.\ \ }{#2}\par
               \skipline
               \endgroup}
%
%
%
%
%

\def\APL#1#2#3{Appl.\ Phys.\ Lett.\ {\bf #1}, #2 (#3)}

\def\JAPSOC#1#2#3{J.\ Phys.\ Soc.\ Jpn.\ {\bf #1}, #2 (#3)}

\def\JVST#1#2#3{J.\ Vac.\ Sci.\ Technol.\ {\bf #1}, #2 (#3)}

\def\JVSTB#1#2#3{J.\ Vac.\ Sci.\ Technol.\ B {\bf #1}, #2 (#3)}

\def\PR#1#2#3{Phys.\ Rev.\ {\bf #1}, #2 (#3)}

\def\PRB#1#2#3{Phys.\ Rev.\ B {\bf #1}, #2 (#3)}
\def\PRL#1#2#3{Phys.\ Rev.\ Lett.\ {\bf #1}, #2 (#3)}
\def\RMP#1#2#3{Rev.\ Mod.\ Phys.\ {\bf #1}, #2 (#3)}

\def\SS#1#2#3{Surf.\ Sci.\ {\bf #1}, #2 (#3)}

%
%
%
\def\equkeys{}
\def\equlist{}
%
%
\def\equn#1{\buildlist{#1}\updateequs{\listed}\citeequs{\listed}}
%
%
%
\def\updateequs#1{\citenum=0%
                 \def\\##1{\updateequ{##1}}%
                 #1}%
\def\updateequ#1{\advance\citenum by 1%
                 \ifx\equkeys\empty
                    \add{#1}\to{\equkeys}%
                    \add{#1}\to{\equlist}%
                    \add{#1}\to{\equlist}%
                 \else%
                    \memberfalse%
                    \ismember{#1}\of{\equkeys}
                    \def\\##1{\updateequ{##1}}
                    \ifmember
                    \else%
                       \add{#1}\to{\equkeys}
                       \add{#1}\to{\equlist}%
                       \add{#1}\to{\equlist}%
                    \fi%
                 \fi}%
\def\citeequs#1{\refnum=0%
                \seqnum=0%
                \hideunder{\equtag}{\secname
                \def\\##1{\citedef{#1}{##1}}%
                \equkeys%
                \def\\##1{}}}
%
%
\long\def\listequs{{\bf EQUATION NAMES:}
                   \refnum=0%
                   \def\\##1{\advance\refnum by 1%
                             \writeequ{\number\refnum}{##1}}%
                   \equkeys}
\long\def\writeequ#1#2{\memberfalse%
                       \def\nextkey{#2}%
                       \long\def\\##1\\##2{\def\nextbibkey{##1}%
                                           \ifmember
                                           \else
                                              \ifx\nextkey\nextbibkey
                                                 \membertrue
                                                 \printequ{#1}{##2}
                                              \fi
                                           \fi}%
                       \equlist%
                       \def\\##1{\advance\refnum by 1%
                                 \writeequ{\number\refnum}{##1}}}
%
%
\long\def\printequ#1#2{%
               \begingroup
               \par
               \noindent
               \advance\leftskip by 1truein
               \llap{Eq.\ (\secname#1)\ =\ }#2
               \endgroup}

%
%
%
\def\figkeys{}
\def\figlist{}
%
%
\def\fign#1{\buildlist{#1}\updatefigs{\listed}\citefigs{\listed}}
\def\figitem#1#2{\add{#1}\to{\figlist}%
                 \add{#2}\to{\figlist}}
%
%
\def\updatefigs#1{\citenum=0%
                 \def\\##1{\updatefig{##1}}%
                 #1}%
\def\updatefig#1{\advance\citenum by 1%
                 \ifx\figkeys\empty
                    \add{#1}\to{\figkeys}%
                 \else%
                    \memberfalse%
                    \ismember{#1}\of{\figkeys}
                    \def\\##1{\updatefig{##1}}
                    \ifmember
                    \else%
                       \add{#1}\to{\figkeys}
                    \fi%
                 \fi}%
\def\citefigs#1{\refnum=0%
                \seqnum=0%
                \hideunder{\figtag}{\secname
                \def\\##1{\citedef{#1}{##1}}%
                \figkeys%
                \def\\##1{}}}
%
%
\long\def\listfigs{\refnum=0%
                   \def\\##1{\advance\refnum by 1%
                             \writefig{\number\refnum}{##1}}%
                   \figkeys}
\long\def\writefig#1#2{\memberfalse%
                       \def\nextkey{#2}%
                       \long\def\\##1\\##2{\def\nextbibkey{##1}%
                                           \ifmember
                                           \else
                                              \ifx\nextkey\nextbibkey
                                                 \membertrue
                                                 \printfig{#1}{##2}
                                              \fi
                                           \fi}%
                       \figlist%
                       \def\\##1{\advance\refnum by 1%
                                 \writefig{\number\refnum}{##1}}}
%
%
\long\def\printfig#1#2{%
               \begingroup
               \bigskip
               \noindent
               \singlespace
               \advance\leftskip by 1truein
               \llap{\bf Fig.\ #1\ \ }#2\par
               \endgroup}
%
%
%
%
\def\figinsert#1#2{\pageinsert
                   \vbox{#1}
                   \figcap{#2}
                   \vfil\vfil
                   \endinsert}
\def\figcap#1{\buildlist{#1}\updatefigs{\listed}\listonefig}
\long\def\listonefig{\def\\##1{%
                           \writefig{\citefigs{\listed}}{##1}}%
                     \listed}
%
%
%
\def\tabkeys{}
\def\tablist{}
%
%

%
%
%
\def\updatetab#1{\advance\citenum by 1%
                 \ifx\tabkeys\empty
                    \add{#1}\to{\tabkeys}%
                 \else%
                    \memberfalse%
                    \ismember{#1}\of{\tabkeys}
                    \def\\##1{\updatetab{##1}}
                    \ifmember
                    \else%
                       \add{#1}\to{\tabkeys}
                    \fi%
                 \fi}%
%

%
%
\long\def\listtabs{\refnum=0%
                   \def\\##1{\advance\refnum by 1%
                             \writetab{\number\refnum}{##1}}%
                   \tabkeys}
\long\def\writetab#1#2{\memberfalse%
                       \def\nextkey{#2}%
                       \long\def\\##1\\##2{\def\nextbibkey{##1}%
                                           \ifmember
                                           \else
                                              \ifx\nextkey\nextbibkey
                                                 \membertrue
                                                 \printtab{#1}{##2}
                                              \fi
                                           \fi}%
                       \tablist%
                       \def\\##1{\advance\refnum by 1%
                                 \writetab{\number\refnum}{##1}}}
%
%
\long\def\printtab#1#2{%
               \begingroup
               \bigskip
               \noindent
               \singlespace
               \advance\leftskip by 1truein
               \llap{\bf Table\ #1\ \ }#2\par
               \endgroup}
%
%
%

%

%
%
\def\v{\vskip .21in}
\def\frac#1/#2{\leavevmode\kern.1em
  \raise.5ex\hbox{\the\scriptfont0 #1}\kern-.1em
  /\kern-.15em\lower.25ex\hbox{\the\scriptfont0 #2}}

\def\boxit#1{\vbox{\hrule\hbox{\vrule\kern3pt
  \vbox{\kern3pt#1\kern3pt}\kern3pt\vrule}\hrule}}

\def\today{\ifcase\month\or
  January\or February\or March\or April\or May\or June\or
  July\or August\or September\or October\or November\or December\fi
  \space\number\day, \number\year}
\def\prtoday{\number\day\space\ifcase\month\or
  January\or February\or March\or April\or May\or June\or
  July\or August\or September\or October\or November\or December\fi
  \space\number\year}

\nopagenumbers
\def\rom#1{\uppercase\expandafter{\romannumeral#1}}
{\nopagenumbers
\headline={\hfil}
\vskip0.5truein
\vbox to \vsize {
\centerline{\bf THE DENSITY OF STATES IN}
\centerline{\bf THE TWO-DIMENSIONAL ELECTRON GAS}
\centerline{\bf AND QUANTUM DOTS}
\vfill
\centerline{A Dissertation}
\centerline{Presented to the Faculty of the Graduate School}
\centerline{of Cornell University}
\centerline{in Partial Fulfillment of the Requirements for the Degree of}
\centerline{Doctor of Philosophy}
\vfill
\centerline{by}
\centerline{Raymond Cameron Ashoori}
\centerline{January 1991}}
\vfill\eject}
{\nopagenumbers
\headline={\hfil}
\vbox to \vsize {
\vfil
\vskip 0.1 true in
\centerline{\copyright\ Raymond Cameron Ashoori\quad 1990}
\centerline{ALL RIGHTS RESERVED}
\vfil\eject}}
{\nopagenumbers
\headline={\hfil}
\null\vskip 0.2 true in
\centerline{\bf THE DENSITY OF STATES IN}
\centerline{\bf THE TWO-DIMENSIONAL ELECTRON GAS}
\centerline{\bf AND QUANTUM DOTS}
\bigskip
\centerline{Raymond Cameron Ashoori, Ph.D.}
\centerline{Cornell University 1991}
\vskip 1 true in
The density of states (DOS) in both the two-dimensional (2d) electron gas and
arrays of ``quantum dots'' is studied using capacitive and tunneling
techniques. A capacitance bridge described
in this thesis, is used to make
high sensitivity capacitance measurements on GaAs samples produced using
molecular beam epitaxy.

We have made quantitative determinations of the
``thermodynamic'' DOS of Landau levels
in a 2d system whose electronic density can be varied
by means of a gate bias. A novel technique
which, by taking advantage of two normalization conditions based on
knowledge of the Landau level degeneracy and level spacing,
allows extraction of the DOS from capacitance data using no
sample parameters. The method yields the DOS as a function of Fermi
energy in the 2d electron gas. We find that Lorentzian
lineshapes give an excellent fit to the Landau level structure.
Further, the
widths of these lines are independent of the strength of the
magnetic field. In high fields, the exchange enhanced spin splitting
is observed and the exchange energy is determined.

The ``single-particle'' DOS is measured in the same samples.
Zero bias tunneling of electrons between a
quantum well and an $n^+$ substrate is studied with excitation
voltages smaller than $k_BT$. At low temperatures and only with
magnetic field applied perpendicular to the plane of the electron gas
in the well, the tunneling rate develops a novel temperature dependent
suppression. The data are interpreted in terms of a magnetic field
induced energy gap, at the Fermi level, in the single-particle
spectrum of electrons in the well.

We laterally confine electrons in a quantum well into
arrays of quantum dots
using a technique which requires only slight (300$\AA$)
surface corrugation of the sample.
Electron beam lithography and reactive ion etching
techniques are used to produce this corrugation.
These samples are studied using a capacitance method which has
allowed us to determine the lateral area of dots.
Further, on the femto-Farad
scale, quasi-periodic capacitance fluctuations are observed. We show,
through comparisons with simulated results,
that these fluctuations arise as due to an almost regular
spectrum for adding single electrons to the dots
which develops as a consequence of the Coulomb blockade.

\vfill\eject}
\pageno=-3
\headline={\newheader}
\footline={\newfooter}
\baselineskip=20 true pt
\null\vskip 12pt\centerline{\bf Table of Contents}
\vskip 0.7 true in
\noindent\line{{\bf I Introduction}\dotfill 1}
\indent1.1 The Two-Dimensional Density of States\dotfill2\break
\indent1.2 Quantum Dot Systems\dotfill7\break
References\dotfill13\break
\medskip
\noindent\line{{\bf II Computerized Capacitance Bridge}\dotfill14}
\indent2.1 Introduction\dotfill14\break
\indent2.2 Rough Schematic\dotfill15\break
\indent2.3 Theory of Operation\dotfill17\break
\indent2.4 Evaluation of Sample Impedance\dotfill20\break
\indent2.5 Implementation of Bridge\dotfill21\break
\indent2.6 Special Features\dotfill24\break
References\dotfill27\break
\medskip
\noindent{\bf III Quantitative Measurement of the Two-Dimensional}\hfil\break
\indent{\bf Thermodynamic Density of States}\dotfill28\break
\indent 3.1 Introduction\dotfill28\break
\indent 3.2 Sample Design\dotfill33\break
\indent 3.3 Measurements\dotfill36\break
\indent 3.4 Analysis Procedure\dotfill42\break
\indent 3.5 Low Field Landau Level Fitting\dotfill60\break
\indent 3.6 High Fields\dotfill75\break
\indent 3.7 Observations Regarding the Spatial Extent of States\dotfill85\break
\indent 3.8 Summary\dotfill101\break
Appendix A) Explanation of $\eta$ term\dotfill102\break
Appendix B) Details of the DOS Calculation\dotfill105\break
Appendix C) Comparison of Abscissas on DOS Plots\dotfill114\break
References\dotfill116\break
\medskip
\noindent{\bf IV Equilibrium Tunneling from the Two-Dimensional}\hfil\break
\indent{\bf Electron Gas}\dotfill119\break
\indent 4.1 Introduction\dotfill119\break
\indent 4.2 Samples and Method\dotfill120\break
\indent 4.3 Extraction of Tunneling Conductance from Capacitance
Data\dotfill123\break
\indent 4.4 Sample Characterization\dotfill130\break
\indent 4.5 Tunneling in the Presence of Magnetic
Field\dotfill131\break
\indent 4.6 Average Conductance\dotfill138\break
{\indent 4.7 Low Density Tunneling Suppression in the}\hfill\break
{\parindent=40pt \indent Absence of Magnetic Field\parindent=20pt}\dotfill143\break
\indent 4.8 Low Density Region with Magnetic Field
Applied\dotfill153\break
{\indent 4.9 Speculations Regarding Possible Causes for the}\hfill\break
{\parindent=40pt \indent Suppression Effects\parindent=20pt}\dotfill155\break
\indent 4.10 Summary\dotfill161\break
References\dotfill162\break
\medskip
\noindent\line{{\bf V Creation of Quantum Dots}\dotfill164}
\indent 5.1 Introduction\dotfill164\break
\indent 5.2 Physics of the Confinement Technique\dotfill168\break
\indent 5.3 Device Fabrication Sequence\dotfill172\break
\indent 5.4 Measurement of Dot Area\dotfill179\break
References\dotfill191\break
\medskip
\noindent{\bf VI Observation of the Single Electron Addition}\hfil\break
{\indent\bf Spectrum of Quantum Dots}\dotfill193\break
\indent 6.1 Introduction\dotfill193\break
\indent 6.2 Equilibrium State of Quantum Dots\dotfill195\break
\indent 6.3 Spectrum of One Electron Additions to a Quantum
Dot\dotfill198\break
{\indent 6.4 Capacitance Fluctuations Expected in an Array of}\hfill\break
{\parindent=40pt \indent Quantum Dots\parindent=20pt}\dotfill203\break
\indent 6.5 Comparison of Simulation to Observed Capacitance
Spectra\dotfill221\break
\indent 6.6 Capacitance Spectra in Magnetic Field\dotfill230\break
\indent 6.7 Tunneling Rate Measurements\dotfill236\break
\indent 6.8 Shape of the Relaxation Peak\dotfill242\break
\indent 6.9 Summary\dotfill248\break
References\dotfill251\break
\medskip
\noindent\line{{\bf VII Computer Simulation of Quantum Dot
Capacitance}\dotfill252\break}
References\dotfill256\break
\vfill\eject
\headline={\newheader}
\footline={\newfooter}
\null\vskip 12pt\centerline{\bf List of Figures}
\vskip 0.7 true in
{\noindent2.1 Rough schematic of automated bridge\dotfill16\break}
2.2 Illustration of inner workings of bridge\dotfill22\break
3.1 Layer structure of samples used in DOS determination\dotfill32\break
3.2 Band structure of samples\dotfill34\break
3.3 Capacitance and loss tangent of sample {\bf B} vs.\ frequency\dotfill38\break
3.4 Low and high frequency capacitances in zero field\dotfill40\break
3.5 Low and high frequency capacitances in field\dotfill41\break
3.6 DOS in sample {\bf B} at 2 Tesla\dotfill52\break
3.7 Lever arm from sample {\bf B} at 2 Tesla\dotfill55\break
3.8 Zero field DOS in sample {\bf B}\dotfill57\break
3.9 Thermal broadening of Landau levels\dotfill59\break
3.10 Fits to DOS in sample {\bf A} at 4 Tesla\dotfill63\break
3.11 Fits to DOS in sample {\bf A} at 3 Tesla\dotfill66\break
3.12 Fits to DOS in sample {\bf B} at 2 Tesla\dotfill67\break
3.13 Fits to DOS in sample {\bf A} at 2 Tesla\dotfill69\break
3.14 Fits to DOS in sample {\bf C} at 4 Tesla\dotfill70\break
3.15 Widths of Landau levels as a function of density\dotfill72\break
3.16 Illustration DOS determination method at 8.5 Tesla\dotfill78\break
3.17 DOS at 8.5 Tesla fit with Lorentzian lineshapes\dotfill80\break
3.18 DOS at 8.5 Tesla fit with Gaussian lineshapes\dotfill81\break
3.19 Sample model for low in-plane conductivity\dotfill88\break
3.20 Capacitance and loss tangent with broadening\dotfill90\break
3.21 Broadening parameter as a function of temperature\dotfill93\break
3.22 Broadening parameter at 8.5 Tesla as a function of Fermi
Energy\dotfill96\break
3.23 Broadening parameter at 4 Tesla as a function of Fermi
Energy\dotfill98\break
3.24 Pictorial description of $\eta$ term\dotfill103\break
3.25 Values of $C_{geom}$ vs.\ $\eta$\dotfill108\break
4.1 Essential structure of samples\dotfill121\break
4.2 Figure illustrating equilibration of well and substrate charge
densities\dotfill124\break
4.3 Tunneling suppression in sample {\bf A} at 4 Tesla\dotfill132\break
4.4 Tunneling suppression in sample {\bf A} at 2 Tesla\dotfill135\break
4.5 Tunneling suppression in sample {\bf A} at 8.5 Tesla\dotfill136\break
4.6 Tunneling suppression in sample {\bf B} at 4 Tesla\dotfill137\break
4.7 Summary plot of tunneling suppression with fits\dotfill141\break
4.8 Tunneling conductivity in zero field\dotfill145\break
4.9 Sample model for low densities\dotfill146\break
4.10 Fraction of well area occupied with electrons\dotfill150\break
4.11 Low density tunneling suppression\dotfill152\break
4.12 Low density tunneling suppression as a function of
temperature\dotfill154\break
4.13 Low density tunneling conductivity at 290~mK for different
fields\dotfill156\break
5.1 Layer structure of samples used to make quantum dots\dotfill169\break
5.2 Basic schematic of quantum dot sample\dotfill170\break
5.3 Band structure under a pillar\dotfill171\break
5.4 Band structure between pillars\dotfill173\break
5.5 Electron micrograph of patterns used to produce dots\dotfill176\break
5.6 Electron micrograph of pillars after etching\dotfill178\break
5.7 Electron micrograph of completed sample\dotfill180\break
5.8 Low and high frequency capacitances for a broad area
sample\dotfill182\break
5.9 Low and high frequency capacitances for dot sample\dotfill183\break
5.10 Illustration of model used to determine dot area\dotfill185\break
5.11 Dot area determined from data for different samples\dotfill187\break
5.12 Dot area from data compared to simulation\dotfill188\break
6.1 Illustration of Coulomb effects in quantum dots\dotfill197\break
6.2 Single electron addition spectrum for quantum dots\dotfill201\break
6.3 Capacitance from model compared with measured
capacitance\dotfill209\break
6.4 Illustration of the convolution function for lock-in
measurements\dotfill213\break
6.5 Fluctuations due to $n^{\rm{th}}$ electron addition to dot
array\dotfill216\break
6.6 Simulated capacitance spectrum at low gate biases\dotfill217\break
6.7 Simulated capacitance spectrum at high gate biases\dotfill218\break
6.8 Simulated capacitance spectrum for random electron
additions\dotfill222\break
6.9 Capacitance fluctuation data at 615 Hz\dotfill224\break
6.10 Fluctuation data from sample conductance at 1.7 kHz\dotfill228\break
6.11 Fluctuation data from sample capacitance at 1.7 kHz\dotfill229\break
6.12 Simulated capacitance spectrum for magnetic field
applied\dotfill232\break
6.13 Conductance fluctuation data in magnetic field at high
bias\dotfill234\break 
6.14 Conductance fluctuation data in magnetic field at low
bias\dotfill235\break
6.15 Relaxation peak frequency as a function of
amplitude\dotfill241\break
6.16 Capacitance and loss tangent vs.\ frequency for quantum
dots\dotfill244\break
6.17 Broadening parameter as a function of amplitude\dotfill245\break
6.18 High frequency capacitance as a function of
amplitude\dotfill247\break
6.19 Capacitance vs.\ frequency curves for different
amplitudes\dotfill249\break
7.1 DOS from computer model and Hansen's data\dotfill254\break
\vfill\eject
\pageno=1
\magnification=1200
\baselineskip=20pt

%
%
\def\v{\vskip .21in}
\def\frac#1/#2{\leavevmode\kern.1em
  \raise.5ex\hbox{\the\scriptfont0 #1}\kern-.1em
  /\kern-.15em\lower.25ex\hbox{\the\scriptfont0 #2}}

\def\boxit#1{\vbox{\hrule\hbox{\vrule\kern3pt
  \vbox{\kern3pt#1\kern3pt}\kern3pt\vrule}\hrule}}

\def\today{\ifcase\month\or
  January\or February\or March\or April\or May\or June\or
  July\or August\or September\or October\or November\or December\fi
  \space\number\day, \number\year}
\def\prtoday{\number\day\space\ifcase\month\or
  January\or February\or March\or April\or May\or June\or
  July\or August\or September\or October\or November\or December\fi
  \space\number\year}

%
%
%
%
\bibitem{LSWa}{J.A. Lebens, R.H. Silsbee, and S.L. Wright, 
        \APL{51}{840}{1987} }
\bibitem{LSWp}{J.A. Lebens, R.H. Silsbee, and S.L. Wright, 
	\PRB{37}{10308}{1988} }
%
\bibitem{DICK-dyn}{R.E. Cavicchi and R.H. Silsbee, \PRB{37}{706}{1988} }
\bibitem{DICK-bridge}{R.E. Cavicchi and R.H. Silsbee, Rev.\ Sci.\
Instrum.\ {\bf 59}, 176, (1988)}
\bibitem{DICK-thes}{R.E. Cavicchi, Ph.D.\ Thesis, Chapter~4, Cornell University (1987)}
\bibitem{CHAP6}{{\it See Chapter 6}}
\bibitem{CHAP4}{{\it See Chapter 4}}
%
\bibitem{SILS-coul}{R.E. Cavicchi and R.H. Silsbee, \PRL{52}{1453}{1984} }
%
%
%
\bibitem{JOHN}{J.A. Lebens, Ph.D. thesis, Cornell University (1988) }
%
%
\bibitem{NORMAN}{N. Ramsey, Master's thesis, Cornell University (1986) }
%
%
\bibitem{YO}{R.C. Ashoori et al. (to be published)}
\bibitem{THESIS}{R.C.\ Ashoori, Ph.D.\ thesis, Cornell University
(1991) }
\bibitem{YOPRL}{R.C. Ashoori, J.A. Lebens, N.P. Bigelow, and R.H. Silsbee,
	\PRL{64}{681}{1990} }
\bibitem{YOCOM}{R.H. Silsbee and R.C. Ashoori, \PRL{64}{1991}{1990} }
%
%
%
\bibitem{HIG-cap}{R.K. Goodall, R.J. Higgins, and J.P. Harrang,
	\PRB{31}{6597}{1985} }
%
%
\bibitem{LAMB-coul}{John Lambe and R.C. Jaklevic, \PRL{22}{1371}{1969} }
%
%
\bibitem{TREY-dos}{T.P. Smith, B.B. Goldberg, P.J. Stiles, and M. Heiblum,
	\PRB{32}{2696}{1985} }
%
\bibitem{TREY-ext}{T.P. Smith III, W.I. Wang, and P.J. Stiles,
	\PRB{34}{2995}{1986} }
\bibitem{TREYext}{T.P. Smith III, W.I. Wang, and P.J. Stiles,
	\PRB{34}{2995}{1986} }
\bibitem{exa-TREYext}{{\it see for example:} T.P. Smith III, W.I. Wang,
and P.J. Stiles, \PRB{34}{2995}{1986} }
%
\bibitem{STIL-mos}{L.C. Zhao, D.A. Syphers, B.B. Goldberg, and P.J. Stiles,
	Solid State Commun.{\bf 49}{859}{1984} }
%
\bibitem{STIL-tilt}{F.F. Fang and P.J. Stiles, \PR {174}{823}{1968} }
%
\bibitem{TREY-1dfqhe}{T.P. Smith III, K.Y. Lee, J.M. Hong, C.M Knoedler,
	H. Arnot, and D.P. Kern, \PRB{38}{1558}{1988} }
%
\bibitem{TREY-aniso}{T.P. Smith III, J.A. Brum, J.M. Hong, C.M. Knoedler,
	H.Arnot, and L. Esaki, \PRL{61}{585}{1988} }
%
\bibitem{TREY-zeem}{W. Hansen, T.P. Smith, III, K.Y. Lee, J.A. Brum,
	C.M. Knoedler, J.M. Hong, and D.P. Kern,
	\PRL{62}{2168}{1989} }
%
\bibitem{TREY-fab}{K.Y. Lee, T.P. Smith, III, H. Arnot, C.M. Knoedler,
J.M. Hong, D.P. Kern, and S.E. Laux, \JVSTB{6}{1856}{1988} }
%
\bibitem{TREY-wire}{T.P. Smith, III, H. Arnot, J.M. Hong, C.M.
Knoedler, S.E. Laux, and H. Schmid, \PRL{59}{2802}{1987} }
%
%
%
%
\bibitem{ALL-si-ext}{B.A. Wilson, S.J. Allen, Jr., and D.C. Tsui,
	\PRL{44}{479}{1980} }
%
\bibitem{ALLEN-freqdep}{S.J. Allen, Jr., D.C. Tsui, and F. DeRosa,
\PRL{35}{1359}{1975}}
%
%
\bibitem{BRIG-qhelow}{A. Briggs, Y. Guldner, J.P. Vieren, M. Voos,
J.P. Hirtz, and M. Razeghi, \PRB{27}{6549}{1983} }
%
%
\bibitem{DYNE-1d-gap}{Alice E. White, R.C. Dynes, and J.P. Garno,
	\PRL{56}{532}{1986} }
%
\bibitem{DYNE-2d-dos}{Alice E. White, R.C. Dynes, and J.P. Garno,
	\PRB{31}{1174}{1985} }
%
%
%
\bibitem{EAVE-par}{B.R. Snell,K.S. Chan, F.W. Sheard, L. Eaves,
	G.A. Toombs, D.K. Maude, J.C. Portal, S.J.Bass, P. Claxton,
	G. Hill, and M.A. Pate, \PRL{59}{2806}{1987} }
%
%
\bibitem{EIS-DvA}{J.P. Eisenstein, H.L. St{\"o}rmer, 
	V. Narayanmurti, A.Y. Cho, A.C. Gossard,
	and C.W. Tu, \PRL{55}{875}{1985} }
%
\bibitem{EIS-tc}{J.P. Eisenstein, H.L, A.C. Gossard, and V. Narayanmurti, 
        \PRL{59}{1341}{1987} }
%
%
\bibitem{FOWL-magcond}{A.B. Fowler, F.F. Fang, W.E. Howard, and P.J.
Stiles, \PRL{16}{901}{1966}}
%
%
\bibitem{FULT-charge}{T.A. Fulton and G.J. Dolan, \PRL{59}{109}{1987}}
%
%
\bibitem{GEER-turnstile}{L.J. Geerligs, V.F. Anderegg, P.A.M. Holweg,
J.E. Mooij, H. Pothier, D. Esteve, C. Urbina, and M.H. Devoret,
\PRL{64}{2691}{1990} }
%
%
\bibitem{GER-cumulant}{R.R. Gerhardts, \SS{58}{227}{1976} }
%
%
%
\bibitem{GIA-tun}{I. Giaever and H.R. Zeller, \PRL{20}{1504}{1968} }
%
\bibitem{GOLD-exch}{B.B. Goldberg, D. Heiman, and A. Pinczuk,
	\PRL{63}{1102}{1989} }
%
%
%
\bibitem{GORD-stab}{Gordon T. Saville, John M. Goodkind, and P.M. Platzman
	\PRL{61}{1237}{1988} }
%
%
\bibitem{GORN-doubwell}{J. Smoliner, W. Demmerle, G. Berthold, E.
Gornik, G. Weimann, and W. Schlapp, \PRL{63}{2116}{1989}}
%
%
%
\bibitem{GRI-wig}{C.C. Grimes and G. Adams, \PRL{42}{795}{1979} }
%
\bibitem{MOTY-high}{U. Meirav, M. Heiblum, and Frank Stern, 
\APL{52}{1268}{1988} }
%
%
%
\bibitem{HEIT-res-width}{D.Heitmann, M. Ziesmann, and L.L. Chang,
\PRB{34}{7463}{1986} }
%
%
\bibitem{HICK-fqhe}{T.W. Hickmott, \PRL{57}{751}{1986} }
%
\bibitem{HICK-magtun}{T.W. Hickmott, \PRB{32}{6531}{1985} }
%
%
\bibitem{IMRY-tun}{Yoseph Imry and Zvi Ovadyahu, \PRL{49}{841}{1982} }
%
%
\bibitem{KAIS-beem}{L.D. Bell and W.J. Kaiser, \PRL{61}{2368}{1988} }
%
\bibitem{KAIS-schottky}{M.H. Hecht, L.D. Bell, W.J. Kaiser, and F.J.
Grunthaner, \APL{55}{780}{1989}}
%
%
\bibitem{KASH-small}{K. Kash, A. Scherer, J.M. Worlock, H.G.
Craighead, and M.C. Tamargo, \APL{49}{1043}{1986} }
%
%
\bibitem{KAST-dot-mag}{P.L. McEuen, E.B. Foxman, U. Meirav, M.A.
Kastner, Yigal Meir, Ned S. Wingreen, and S.J. Wind
\PRL{66}{1926}{1991} }
%
%
\bibitem{KOTT-vdots}{J. Alsmeier, E. Batke, and J.P. Kotthaus,
\PRB{41}{1699}{1990} }
%
\bibitem{KOTT-saw}{A. Wixforth, J.P. Kotthaus, and G. Weinmann,
\PRL{56}{2104}{1986} }
%
%
\bibitem{KUCH-mic}{F. Kuchar, R. Meisels, G. Weimann, and W. Schlapp,
\PRB{33}{2965}{1986}}
%
%
\bibitem{LIK-both}{P. Delsing, K.K. Likharev, L.S. Kusmin, and T.
Claeson, \PRL{63}{1180}{1989} and \PRL{63}{1861}{1989}}
%
%
\bibitem{NICO-broad}{E.H. Nicollian and A. Goetzberger, \APL{10}{60}{1967} }
%
%
\bibitem{PAAL-qhelow}{M.A. Paalanen, D.C. Tsui, and A.C. Gossard,
\PRB{25}{5566}{1982}}
%
%
\bibitem{PEPP-wire}{T.J. Thornton, M. Pepper, H. Ahmed, D. Andrews,
and G.J. Davies, \PRL{56}{1198}{1986} }
%
\bibitem{PEPP-magwire}{K.F. Berggren, T.J. Thornton, D.J. Newson, and
M. Pepper, \PRL{57}{1769}{1986} }
%
%
\bibitem{REED-dot}{M.A. Reed, J.N. Randall, R.J. Aggarwal, R.J. Matyi,
T.M. Moore, and A.E. Wetsel, \PRL{60}{535}{1988} }
%
%
\bibitem{ROUK-resol}{A. Scherer and M.L. Roukes, \APL{55}{377}{1989}}
%
\bibitem{ROUK-quench}{M.L. Roukes, A. Scherer, S.J. Allen, Jr., H.G.
Craighead, R.M. Ruthen, E.D. Beebe, and J.P. Harbison,
\PRL{59}{3011}{1987} }
%
\bibitem{ROUK-unpub}{M.L. Roukes, A. Scherer, and B.P. Van der Gaag,
(unpublished)}
%
%
\bibitem{SAK-mob}{Kazuhiko Hirakawa and Hiroyuki Sakaki, \PRB{33}{8291}{1986}}
%
%
\bibitem{STM-spc-ht}{E. Gornik, R. Lassnig, G. Strasser, H.L. St{\"o}rmer,
	A.C. Gossard, and W. Wiegmann, \PRL{54}{1820}{1985} }
%
\bibitem{STM-fqhe}{H.L. St\"ormer, A. Chang, D.C. Tsui,
	J.C.M. Hwang, A.C. Gossard, and W. Wiegmann,
	\PRL {50}{1953}{1983} }
\bibitem{STM-pri}{H.L. St\"ormer, private communication}
%
%
%
\bibitem{TSUI-heat}{J.K. Wang, J.H. Campbell, D.C. Tsui, and A.Y. Cho,
\PRB{38}{6174}{1988} }
%
\bibitem{TSUI-temp}{H.P. Wei, A.M. Chang, D.C. Tsui, and M. Razeghi,
\PRB{32}{7016}{1985}}
%
\bibitem{STM-zero}{D.C. Tsui, H.L. St\"ormer, and A.C. Gossard
\PRB{25}{1405}{1982} }
%
\bibitem{TSUI-deloc}{H.P. Wei, D.C. Tsui, M.A. Paalanen, and A.M.M.
Pruisken, \PRL{61}{1294}{1988}}
%
%
\bibitem{HOUT-pnt}{B.J. van Wees, H. van Houten, C.W.J. Beenakker, J.G.
	Williamson, and C.T. Foxon, \PRL{60}{848}{1988} }
%
\bibitem{HOUT-magres}{H. van Houten, C.W.J. Beenakker, P.H.M. van
	Loosdrecht, T.J. Thornton, H. Ahmed, M. Pepper, C.T. Foxon,
	and J.J. Harris, \PRB{37}{8534}{1988} }
%
\bibitem{HOUT-qhe-pnt}{B.J. van Wees, E.M.M. Willems, C.J.P.M. Harmans,
	C.W.J. Beenakker, H. ban Houten, J.G. Williamson, C.T. Foxon,
	and J.J. Harris, \PRL{62}{1181}{1989} }
%
%
\bibitem{KVK-mcap}{V. Mosser, D. Weiss, K. von Klitzing, K. Ploog, and
G. Weimann, Solid State Comm. {\bf 58}, 5 (1986) }
%
\bibitem{KVK-photo}{D. Stein, K. v. Klitzing, and G. Weimann,
\PRL{51}{130}{1983}}
%
\bibitem{KVK-bar}{R.J. Haug, A.H. MacDonald, P. Streda, and K. von Klitzing,
	\PRL{61}{2797}{1988} }
%
\bibitem{KVK-qhe}{K. von Klitzing, G. Dorda, and M. Pepper,
	\PRL{45}{494}{1980} }
%
\bibitem{KVK-magtun}{E. B{\"o}ckenhoff, K. v. Klitzing, and K. Ploog,
	\PRB{38}{10120}{1988} }
%
\bibitem{KVK-exch}{R.J. Nicholas, R.J. Haug, K.v. Klitzing, and
G. Weimann \PRB{37}{1294}{1988} }
%
%
%
\bibitem{WASH-back}{S. Washburn, A.B. Fowler, H. Schmid, and D. Kern,\
	\PRL{61}{2801}{1988} }
%
%
\bibitem{ZEMEL-cap}{M. Kaplit and J.N. Zemel, \PRL{21}{212}{1968}}
%
%
%
%
%
\bibitem{ANDER-scale}{E. Abrahams, P.W. Anderson, D.C. Licciardello,
and T.V. Ramakrishnan, \PRL{42}{673}{1979}}
%
%
\bibitem{ANDO-broad}{T. Ando and Y. Murayama, \JAPSOC{54}{1519}{1985} }
%
\bibitem{ANDO-mob}{T. Ando, \JAPSOC{51}{3900}{1982} }
%
\bibitem{ANDO-osc}{T. Ando and Y. Uemura, \JAPSOC{37}{1044}{1974} }
\bibitem{ANDO-ellip}{T. Ando and Y. Uemura, \JAPSOC{36}{959}{1974} }
\bibitem{ANDO-F-S}{T. Ando, A.B. Fowler, and F. Stern,
\RMP{54}{437}{1982} and references therein}
%
\bibitem{ANDO-loc}{Hideo Aoki and Tsuneya Ando, \PRL{54}{831}{1985}}
%
%
%
\bibitem{APP-accum}{G.A. Baraff and Joel A. Appelbaum,
\PRB{5}{475}{1972} }
%
%
\bibitem{ALT-zero-bias}{B.L. Altshuler and A.G. Aronov,
	Solid State Comm. {\bf 30}, 115 (1979) }
%
\bibitem{ALTSH-gap}{B.L. Altshuler, A.G. Aronov, and P.A. Lee,
\PRL{44}{1288}{1980}}
%
\bibitem{BUTT-time}{M. B\"uttiker and R. Landauer, \PRL{49}{1739}{1982} }
%
%
%
\bibitem{CRAN-wig}{R.S. Crandall and R. Williams,
	Physics Lett.{\bf (34A)}, 404 (1971) }
%
%
%
\bibitem{DSAR-dos}{S. Das Sarma and X.C. Xie, \PRL{61}{738}{1988} }
%
\bibitem{DSAR-dos}{Qiang Li, X.C. Xie, and S. Das Sarma,
\PRB{40}{1381}{1990} } 
%
\bibitem{DSAR-1dmob}{S. Das Sarma, Song He, and X.C. Xie,
\PRL{61}{2144}{1988} }
%
%
\bibitem{DAVIES-pud}{John A. Nixon and John H. Davies, \PRB{41}{7929}{1990} }
%
%
\bibitem{EFROS-Coul}{A.L. Efros and B.I. Shklovskii, J.\ Phys.\ C {\bf
8}, L49, (1975)}
%
%
\bibitem{GERH-between}{Vidar Gudmundsson and Rolf R. Gerhardts,
\PRB{35}{8005}{1987} }
%
%
\bibitem{GLYDE-dos}{V. Sa-yakanit, N. Choosiri, and H. R. Glyde,
\PRB{38}{1340}{1988} }
%
%
\bibitem{GOLD-freqdep}{A. Gold, S.J. Allen, B.A. Wilson, and D.C.
Tsui, \PRB{25}{3519}{1982}}
%
%
\bibitem{HERM-g}{Claudine Hermann and Claude Weisbuch,
\PRB{15}{823}{1977}}
%
%
\bibitem{HALP-bet-lev}{C. Kallin and B.I. Halperin, \PRB{30}{5655}{1984} }
%
\bibitem{HALP-edge}{B.I. Halperin, \PRB{25}{2185}{1982} }
%
%
%
\bibitem{KIRC-plas-sup}{Wei-ming Que and George Kirczenow,
	\PRL{62}{1687}{1989} }
%
%
%
\bibitem{KRAMER-scale}{Bodo Huckestein and Bernhard Kramer,
\PRL{64}{1437}{1990}}
%
%
\bibitem{LAND-1d}{R. Landauer, IBM J. Res. Dev {\bf 1}, 223 (1957) }
\bibitem{LAND-1d-2}{R. Landauer, Philos. Mag. {\bf 21}, 863 (1970) }
%
%
\bibitem{LAUGH-quant}{R.B. Laughlin, \PRB{23}{5632}{1981} }
%
%
\bibitem{LEE-mob-edge}{P.A. Lee, \PRB{26}{5882}{1982} }
%
\bibitem{LEE-2d-cdw}{D. Yoshioka and P.A. Lee, \PRB{27}{4986}{1983} }
%
%
\bibitem{LIK-tunneltheory}{K.K. Likharev, IBM J. Res. Develop. {\bf
32}, 144, (1988)}
%
\bibitem{ONO-diverge}{Y. Ono, \JAPSOC{51}{2055}{1982} }
%
%
\bibitem{PLAT-cdw}{H. Fukuyama, P.M. Platzman, and P.W. Anderson,
	\PRB{19}{5211}{1979} }
%
\bibitem{PLAT-wig}{P.M. Platzman and H. Fukuyama, \PRB{10}{3150}{1974} }
%
%
%
\bibitem{PRUISK-scale}{A.M.M. Pruisken, \PRL{61}{1297}{1988}}
%
%
%
\bibitem{RAS-sup}{Mark Rasolt, \PRL{58}{1482}{1987} }
%
%
\bibitem{SHARV}{Yu. V. Sharvin, Zh. Eksp. Teor. Fiz. {\bf 48}, 984 (1965)
	[Sov. Phys. JETP {\bf 21}, 655 (1965)]}
%
%
\bibitem{STER-mob}{S. Das Sarma and Frank Stern, \PRB{32}{8442}{1985} }
\bibitem{STER-pri}{F. Stern and S.E. Laux, private communication}
%
%
\bibitem{STRED-edge-exa}{{\it see for example} P. Streda, J. Kucera,
and A. H. MacDonald, \PRL{59}{1973}{1987}}
%
\bibitem{WEGNER-dos}{F. Wegner, Z. Phys. B {\bf 51}, 279 (1985)}
%
%
%
\bibitem{QHE}{R.E. Prange and S.M. Girvin, {\it The Quantum Hall
Effect} (Springer-Verlag, New York, 1987)}
%
\bibitem{HORO&HILL}{Paul Horowitz and Winfield Hill, {\it The Art of
Electronics} (Cambridge University Press, 1989)}
%
%
%
\bibitem{NR}{W.H. Press, B.P. Flannery, S.A. Teukolsky, and W.T. Vetterling,
        {\it Numerical Recipes} (Cambridge University Press, Cambridge, 1986)}
\bibitem{AS}{M. Abramowitz and I.A. Stegun, {\it Handbook of Mathematical
        Functions} (Dover, New York, 1970)}
%
%
\bibitem{WOLF-tunchap2}{E.L. Wolf, {\it Principles of Electron
Tunneling Spectroscopy,} Chap.\ 2 (Oxford Univ.\ Press, 1985)}
%
%
%
\bibitem{LOREN}{L. Pfeiffer, private communication}
%
%
%
\bibitem{OHMIC}{N. Blanc, P. Gu\'eret, P. Buchmann, K. D\"atwyler, and
P. Vettiger, \APL{56}{2216}{1990} }
%
%
\bibitem{RHEED}{J.M. Van Hove, C.S. Lent, P.R. Pukite, and P.I. Cohen,
\JVSTB{1}{741}{1983}}
%
\bibitem{BRILLSON-schottky}{L.J. Brillson, Surf.\ Sci.\ Rep.\ {\bf
2}, {123} (1982) }
%
\bibitem{SPICER-schottky}{W.E. Spicer, I. Lindau, P. Skeath, and C.Y.
Su, \JVST{13}{831}{1976}}
%
\bibitem{Cr-schottky}{Z. Lilental-Weber, N. Newman, J. Washburn, E.R.
Weber, and W.E. Spicer, \APL{54}{356}{1989} }
%
%
\bibitem{JERRY}{J.\ Coumeau, private communication}
%
%
\bibitem{JEOL}{JEOL JB5DX11 users manual, unpublished}
%
\bibitem{WOODARD}{D. Woodard, private communication}
%
%
\bibitem{CCL2F2}{K. Hikosaka, T. Mimura, and K. Joshin, Jpn. J. Appl.
Phys. {\bf 20}, L847 (1981)}
\bibitem{CCL2F2-stop}{K.L. Seaward, N.J. Moll, and W.F. Stickle,
\JVSTB{6}{1645}{1988} }
%
%
\bibitem{D/Aconv}{D. Sheingold, J. Wilson, and G. Whitmore,
Application Guide to CMOS Multiplying D/A Converters, Analog Devices,
Norwood, MA, 1978}
%
%
%
\bibitem{CHAP3}{{\it See Chapter 3}}
\bibitem{CHAP2}{{\it See Chapter 2}}
\bibitem{CHAP1}{{\it See Chapter 1}}
\bibitem{CHAP4}{{\it See Chapter 4}}
\bibitem{CHAP5}{{\it See Chapter 5}}
\bibitem{CHAP6}{{\it See Chapter 6}}
\bibitem{CHAP7}{{\it See Chapter 7}}

\newchapter{1.}
\centerline{\twelvebf Chapter I}
\v
\v
\centerline{\twelvebf Introduction}
\v
\v
\v
\v
\v
\v

This thesis deals with the behavior of electrons in dimensions less
than three. Specifically, it uses measures of the electronic density
of states (DOS) to study these systems.
The thesis is roughly divided into two parts.
The first part deals with the two-dimensional (2d)
electron gas in GaAs. The 2d electron gas forming
at an inversion layer at the silicon-silicon dioxide interface has
been studied heavily since the 1960's.
It is a statement of the wealth of physics contained by semiconductor
2d systems that new and
exciting results, such as the quantum Hall effect\cite{KVK-qhe}
(circa 1980), are still being discovered in them. Most present
experiments are being done in much higher mobility materials such as
in the GaAs/AlGaAs system.
The second part of the thesis deals with a newer, less studied
mesoscopic system, an array of very small electron packets. Only very
recently has it become technologically feasible to produce feature
sizes small enough in semiconductor systems to do these experiments.
Such packets have become known as ``quantum dots''. One
salient feature in the study of these ``dots'' is the production of a
system so small that the energies associated with the addition of one
electron to the system are so large, either due to a quantum level
spacing or electrostatic effects in the system, as to be
experimentally observable.

\v\par\penalty-1000
{\noindent{\bf 1.1 The Two-Dimensional Density of States}
\nobreak}\v\nobreak
Since the early magnetoconductance work of Fowler et.\
al.,\cite{FOWL-magcond} which first demonstrated that the electronic
density of states in a deeply inverted silicon surface was
two-dimensional, there has been much interest in the effects of
a perpendicularly applied
magnetic field on the 2d electron gas. The reason for this is
clear: because of the confinement, there is no dispersion in the
direction of the magnetic field so that a magnetic field applied
perpendicular to the plane of the electron gas creates highly
degenerate electron energy levels. In a single particle picture and
with no disorder in the 2d electron gas, the magnetic field
essentially places electrons in identical harmonic oscillator
potential wells of number equal to the number of magnetic flux quanta
passing through the 2d system. The quantum energy levels, equally
spaced in energy, in these harmonic oscillator potential wells are
known as ``Landau levels''. Depending on the magnetic field strength,
these levels can have enormous degeneracy. For example, the highest
magnetic field strength used in the data presented in this thesis is
8.5~T. At this field strength, a spin degenerate Landau level
contains $4.1\times10^{11}$ states per square centimeter.
Both the energy
spacing of these levels and their degeneracy increase linearly with
the applied field. Thus for a given density of electrons in the 2d
gas, the position of the Fermi energy moves between different Landau
levels as the magnetic field strength is varied.

In the absence of a magnetic field, the DOS in two dimensions is
a constant, dependent only on fundamental constants and the effective
mass of the confined electrons. In the single particle picture with no
scattering of electrons, the
addition of a magnetic field perpendicular to the plane of the 2d
electron gas produces a series of Dirac delta functions in the DOS
spectrum.

Of course, the DOS spectrum does not involve true delta
functions: instead, one expects that these levels should 
contain some broadening.
For one thing, the electrons in the 2d gas have
a finite scattering time for elastic collisions with impurities.
Moreover, changes in the DOS
spectrum created by the magnetic field modify scattering
times as the density of
final states into which electrons can scatter and lengths for
screening impurities are changed.
This suggests that even in a single particle picture
a self consistent approach is needed to understand
the DOS theoretically.
Further, particle exchange and many body effects may also change the
character of these levels.

One intriguing consequence of confinement of the electron gas into two
dimensions is the quantum Hall effect (QHE).
Ordinarily, in most bulk metals and semiconductors, the Hall
resistance is linear in the magnetic field. For the 2d electron gas,
though, plateaus in the Hall resistance appear,
quantized to values of ${\nu}h/e^2$. The integer,
$\nu$, is equal to the number of filled Landau levels. The plateaus
occur when the Fermi energy is located in the region between Landau
levels.
A key question is: How does DOS spectrum relate
to the formation of these plateaus?

Basic theories\cite{LAUGH-quant,HALP-edge}
of the quantum Hall effect rely on the existence
of a nonzero DOS between Landau levels in the DOS spectrum. Given
that the Hall plateaus occur with the Fermi level between Landau levels,
these states are required to ``pin'' the Fermi level in the
region between Landau levels
over some range in magnetic field to give the plateaus the observed width.
Further, crucial to theories is the idea that these states existing
between
Landau levels are localized; electrical conduction cannot occur
through them.
One may expect that information about
the shape of the DOS peaks will lead to clues
about the nature of the electronic states and what allows them to
produce such bizarre phenomena as quantum Hall plateaus.

\v\par\penalty-1000
{\noindent\sl Density of States Measurement}
\nobreak\v\nobreak
In the last few years there have been attempts, both experimentally as
well as theoretically, to determine the shapes of Landau level DOS
peaks. Most of the experimental studies have been
either largely qualitative in nature or restricted to low magnetic fields,
interpreting peaks in the specific heat, capacitance, or
magnetization of the the 2d electron gas
as the magnetic field strength is varied
as reflecting an underlying energy spectrum of the Landau level DOS.
The early capacitance work of Kaplit and Zemel\cite{ZEMEL-cap}
on a MOSFET structure demonstrated the sensitivity of capacitance
measurements to the DOS in an inversion layer.

In chapter 3, we present a technique which allows us
to do precision quantitative spectroscopy of these levels now {\it
as a function of Fermi energy}
of the 2d electron gas. This technique has permitted
a systematic study of the DOS at various magnetic
field strengths and over a
wide range of electron densities in three samples.
Through characterizing Landau levels
this way we have determined that both their shapes and the dependence
of their
level widths on magnetic field strength are different from that
previously thought. Specifically, we observe that Landau level DOS
peaks are well characterized by Lorentzian lineshapes whose level
widths, although dependent on the electronic density of the 2d
electron gas, are independent of magnetic field strength.

A measure, other than the DOS,
of the ``nature'' of electronic states is their spatial extent or
``localization length''. In a pioneering theoretical work,
Abrahams et.\ al.\cite{ANDER-scale} predicted, using scaling theory,
that in two dimensions electrons are localized
by an arbitrarily small amount of
disorder, and at zero temperature the electronic conduction thus drops
to zero. When a magnetic field is applied, the localization length of
electrons is thought\cite{PRUISK-scale}
to oscillate in electronic energy as this energy is moved
from Landau level to Landau level. The localization length is thought
to become very large in a region near the DOS peak, usually called the
band of ``extended states'', and much shorter in the region between
Landau levels, called the band of ``localized states''.
This variation in localization length is thought to be
ultimately responsible for the Hall resistance 
plateaus of the quantum Hall effect.

Chapter~3 also develops a method, integrated with the
capacitance technique used to determine the DOS, which probes
the localization length of states in the region between Landau levels.
The results show evidence of the localization length smoothly decreasing
and achieving a minimum length at a position nearly half way between
Landau levels, supporting some of these basic ideas on the
localization length.

\v\par\penalty-1000
{\noindent\sl Two different Densities of States\nobreak}\v\nobreak
In a many body system of electrons it is typically not
simple or even
possible to describe the system in terms of an independent
electron model.
The behavior of the system is discussed
in terms of an antisymmetrized many body wavefunction for
the system of identical particles.
Often one speaks of ``quasi-particle'' excitations that
have the appearance of a single particle, but contain some amount of
``dressing'' due to Coulomb, phonon, and other interactions.

An interesting feature of tunneling is that an exponential decrease in
probability for several particles tunneling together compared to that
of a single particle makes it very
unlikely that any electrons dressing a particular electron can
accompany it during tunneling. Electrons must shed their dressing
before they can tunnel.
The electronic system typically gains or loses energy during this process.
The tunneling of an electron into or out of the system may force the
system into a nonequilibrium state.
This state of affairs requires the definition of two different densities of
states, one for a system which has been allowed to relax back to
equilibrium after a tunneling event, and the other which describes the
nonequilibrium situation.

The thermodynamic DOS,\cite{LEE-mob-edge} $\partial{n}/\partial\mu$,
is the change in the particle density of a
system for a change in chemical potential after a time period
longer than the system's internal equilibration time.
The means by which the system moves between the initial state with one
chemical potential to the final state with another affects only the
time scale for changing states and is irrelevant to the value of the
thermodynamic DOS.
In the audio frequency capacitance measurements described above
potentials are changed so slowly that the 2d electron gas system
evolves reversibly during the course of a measurement; those
experiments determine thermodynamic DOS.

The single-particle or tunneling DOS is instead measured
when the system of particles does not have
sufficient time to rearrange during the sudden
subtraction or addition of a single electron, as is the case with 
tunneling measurements. The time an electron spends underneath a
heterostructure tunnel
barrier such as those used in the experiments described in this
thesis is thought to be around $10^{-13}$ seconds.\cite{BUTT-time}
Tunneling currents in our experiments depend on the single particle
DOS. If energy is required to ``undress'' an electron in the 2d
electron gas so that it can tunnel out, we expect the tunneling DOS
may be smaller than the thermodynamic DOS.

The tunneling conductance thus reflects the single-particle
DOS.\cite{ALTSH-gap} In chapter~4, we report results from a novel type
of tunneling measurement which we call ``Equilibrium Tunneling''. We
measure the tunneling of particles between a two dimensional electron
gas and a three dimensional electron gas. These two electron gases are
kept in quasi-equilibrium during the experiments, with voltages no
larger than $k_BT$ applied across the tunnel barrier. This allows for
measurement of the single-particle DOS at the Fermi energy in the
quantum well to an energy resolution of order $k_BT$.
Capacitive coupling allows us to do these small signal measurements
which contrast with earlier, more standard I-V measurements, in which
difficulties associated with making electrical contact to a 2d layer
did not allow an examination of tunneling at voltages across
the tunnel barrier smaller
than the bandwidth of states in the 2d gas. These prior experiments
thus had little energy resolution of features in the single particle DOS.
Our tunneling experiments
are done in the same samples in which the thermodynamic DOS
measurements of chapter~3 are made, integrating into one experiment
measures of both the single-particle and thermodynamic DOS. 

The tunneling 
experiments of chapter~4 have led to some striking results. The most
dramatic feature that we have seen is a strong (up to a factor
of 10) temperature dependent suppression of the tunneling conductance
when a magnetic field is applied perpendicular to the plane of the
2d electron gas. These data are fit well by a model which presumes the
formation of an energy gap at the Fermi energy, induced by the
magnetic field, in the
single-particle spectrum of electrons in 2d electron gas of width
equal to about 5\% of the cyclotron energy.

\v\par\penalty-1000
{\noindent\bf 1.2 Quantum Dot Systems}
\nobreak\v\nobreak
In recent years, the subject of quantum confinement to dimensions
fewer than the two dimensions discussed above has become a
popular topic in the literature.
The simplest structures, ``quantum wires'' and ``quantum dots'', have
produced captivating results. Although single mode quantum wires
appear only now on the threshold of being produced,\cite{STM-pri}
results have appeared from multi-mode wires ($\approx10$ modes), and
these have produced peculiar and unexpected results, particularly in Hall
effect measurements.\pcite{ROUK-quench} Experiments have been carried
out on both the photoresponse and transport properties of quantum dot
systems. Some evidence now exists for resolution of lateral states in
a quantum dot system.\pcite{REED-dot} In this thesis, we present
measurements on arrays of quantum dots. We pursue not
only resolution of lateral states but electrostatic ``energy levels''
having to do with the very small size of electron packets.

The main difficulties in quantum dot experiments have been in finding
appropriate techniques for lateral confinement (confinement other than
the initial 2d confinement built into the heterostructure devices) of
electrons. In chapter~5, we present the fabrication process that we
have used to produce lateral confinement of electrons. The size and
number of electrons in these dots is varied by means of a gate
voltage. Chapter~5 also presents a measure of the size of these
electron packets.

\v\par\penalty-1000
{\noindent\sl Important Energy Scales in Quantum Dots}
\nobreak\v\nobreak
As alluded to above, there are two energy scales that are important in
quantum dot systems. One is the mean quantum level spacing of the one
electron energy states
$\delta_{mean}$ (the inverse of the density of states in the dot) and the
other is the energy $\Delta$ of charging a dot associated with the addition of
a single electron. The latter can be described in terms of the
capacitance of the dot to its surroundings $C$ and is given by
$\Delta=e^2/C$. In the configuration of our experiments, $\Delta$ is
of order 10 times larger than $\delta_{mean}$. Hence the effects of the
charging energy
on an experiment in which the population of electrons in the dot is
changed should be more pronounced than those of the quantum level
spacing.

Effects of the single electron
charging energy have been explored in capacitance experiments
before, the earliest being the ``tunnel capacitor'' work of Lambe and
Jacklevic.\pcite{LAMB-coul} Later work by Cavicchi and
Silsbee\cite{DICK-dyn} in our
lab demonstrated the consequences of the charging energy in features in the
tunneling conductance of electrons between small metal particles and
an electrode. Both of these experiments were done in oxide capacitors
with the packets of electrons being
contained in a large number of nonuniformly sized small metal
particles close enough to one electrode so that tunneling of electrons
between the particles and the electrode could occur. The electron
occupancy of the small metal particles is controlled by varying the
voltage between the capacitor electrodes.
More recent experiments by Fulton and
Dolan\cite{FULT-charge} have demonstrated one electron charging
effects in tunneling in a single metal particle coupled to electrodes by two
tunnel junctions.

In the oxide capacitors of Cavicchi and Silsbee, for a single small
metal particle, one expects a transfer, periodic in voltage applied
across the capacitor, of single electrons between the small particle
and the nearby electrode. This periodic spectrum of electron additions
to the particle
arises from the Coulomb blockade which demands that the electrostatic
potential at the particle be changed by an amount $\Delta$ between
electron additions (or subtractions) to the particle. For a particular
small metal particle, there are well defined voltages at which
electron transfer between the particle and the electrode is allowed.
At these voltages, there is a peak in the device capacitance, $dQ/dV$,
as a small change in bias voltage leads to a sudden change in the
charge configuration of the device.

At first glance, one
expects a capacitor containing an ensemble of particles all with
the same charging energy, $\Delta$, to have sharp periodic peaks in
its capacitance as the voltage across the capacitor is varied.
The problem with this idea is that even assuming that all particles
have the same periodicity of electron additions,
inhomogeneities in real devices will offset these individual particle
spectra from one another by fixed device voltages. For instance,
particles may allow electronic additions to occur at every 5~mV change
in gate bias; some particles will add electrons at 0,5,10,...~mV
device bias while
others will add them at 2,7,12,...~mV. There is a distribution in
these offset voltages for the spectra of different particles.
Typically, it is so wide as to indicate a nearly uniform distribution of
offset voltages over a device bias range for one electron addition to
a particle. One might then think that capacitance peaks due to
individual electron additions would be hopelessly smeared out. The
distribution though does contain fluctuations from uniformity,
and they will be reflected as features in the device capacitance
repeating with the periodicity in device voltage of the single
electron additions to the particles. 

There have been attempts in our lab,\cite{NORMAN} in tunnel capacitors
similar to those used by Cavicchi and Silsbee,
to observe capacitance fluctuations from single electron
additions to small metal particles. These however were largely 
thwarted by conduction between the small metal particles in the sample
configuration used and
polarization effects in the oxide of the type described by
Lambe and Jaklevic.\pcite{LAMB-coul} Further, the nonuniformity
of the sizes of the small metal particles and their capacitance to the
surroundings may cause difficulties in the interpretation of data
revealing capacitance fluctuations. The oxide capacitors are
expected to have a wide distribution of charging energies among the
small metal particles. The polarization
effects are not expected to occur in GaAs/AlGaAs heterostructures, and
packets of electrons can be constructed with much greater uniformity
of charging energy than is the case in oxide capacitors. The other
feature of the GaAs system is the small effective mass which enhances
the contribution of the quantum level spacing to the capacitance
spectrum and increases the likelihood that quantum size effects can be
observed.

In chapter~6 we describe both computer simulations and observations of
capacitance fluctuations in quantum dot arrays in GaAs associated with
the addition of single electrons to small packets of electrons. Using
the area of the quantum dots as a function of gate bias obtained from
capacitance measurements in chapter~5, we determine a spectrum of
single electron additions to the quantum dots as a function of gate
bias. These are only ``quasi-periodic'' electron additions. For one
thing the charging energy decreases as the size of the dots grows
leading to an increase in the number of allowed electron
additions to a dot per unit gate voltage as the dot size grows.
Further, the fact that $\delta$, the average energy spacing between
quantum levels not including the zero energy spacing of degenerate levels
(different from $\delta_{mean}$ which is the average splitting
including the zero energy spacings), is equal to about a third of the
charging energy also adds irregularity to the spectra.

Again, spectra for different dots are expected to be offset in gate
bias from one another due to device inhomogeneities. 
The width of the distribution
of these offsets can be estimated from device capacitance
features. This distribution contains deviations from a smooth shape
due to statistical fluctuations. We
have carried out computer simulations incorporating these statistical
fluctuations which indicate that, despite the deviations from perfect
regularity of the electron addition spectra, peaks of order a few
femto-Farads occur in the capacitance spectrum of the quantum dot
array, and these peaks occur with the same frequency in gate bias as
the underlying electron additions to the quantum dots.

In Chapter~6,
we explore the capacitance of quantum dot arrays to a resolution of
0.1~fF, and have seen peaks in the capacitance spectrum.  These peaks
occur quasi-periodically, in a fashion similar to those predicted by
our model, and are also of the same amplitude as indicated by the
model. In essence, comparison to computer simulation indicates that
through capacitance measurements, we have observed, albeit with less
than perfect resolution, the spectrum of single electron additions to
quantum dots.

Hansen et.\ al.\cite{TREY-zeem} have made capacitance measurements
to much lower resolution on quantum dot arrays very similar to ours. They
observed fluctuations in the device capacitance as the gate bias on
their device was varied. They considered these fluctuations to be
a reflection of the underlying quantum level structure in a quantum
dot. We believe that this interpretation neglects charging effects in
the dot.
In chapter~7, we present a computer simulation indicating that, if the
charging energy is taken into account,
these capacitance features can be the result of
statistical fluctuation in the spectrum of capacitance
peaks due to one electron addition to the the quantum dots
and not clearly associated with quantum levels.

First, we start with chapter~2, which describes the apparatus used to
take the data presented in this thesis. It is a bridge which
measures the capacitance and loss tangent of samples as a function of
frequency over the entire audio range.
Its special facilities for the use of small measuring signals,
for making precision measurements, and most importantly
for highly automated data taking, have allowed us to acquire the data
necessary for the DOS, tunneling, and fluctuation spectra results
discussed above.

\vfill\supereject
\centerline{\bf References}
\v
\listrefs
\vfill\eject

\newchapter{2.}
\figitem{barebones}{The figure displays a basic schematic of the
automated bridge. The voltages $V_c$ and $V_r$ summed and applied to
the standard capacitor (of capacitance $C_{stand}$ and impedance
$Z_{stand}$) have phases $180^\circ$ and $+90^\circ$ respectively
compared with the drive voltage applied to the sample, $V_d$. The
basic idea behind the bridge is to null the voltage $V_x$ at the
balance point so that the sample impedance is given by
Eqs.~2.5~\&~2.6.}

\figitem{detail}{The figure shows the basic inner workings of the
bridge circuitry. Signal levels are varied using two computer
controlled multiplying D/A
converters (MDACs). The digital inputs to these MDACs are typically cycled
through the algorithm of section 2.3. 
The output of one MDAC is fed into an op-amp integrator
which provides a $90^\circ$ phase shift to balance resistive
components of the sample impedance. The signals from this integrator
circuit and the other MDAC are summed and inverted and applied to the
arm of the bridge containing the standard capacitor.}
{\centerline{\twelvebf Chapter II}}
\v
\v
{\centerline{\twelvebf Computerized Capacitance Bridge}}
\v
\v
\v
\v
\v
\v

{\noindent\bf 2.1 Introduction}
\v
All of the data presented in this thesis was derived from
capacitance-loss tangent measurements on our samples. We have
experimented with many samples
and taken enormous number of capacitance measurements.
Initially, these measurements were all taken by hand balancing a
capacitance bridge. This is an extraordinarily tedious process that
requires at least 20~sec for each capacitance measurement. Further,
for A.C.\ signal levels less than a few mV, noise would make the
determination of balance subjective at best. Often we have needed to
take measurements at very low signal levels,\pcite{YOPRL,CHAP4,CHAP6}
a feat which would have
been impossible in measurements taken by hand.
We needed an automated
system capable of signal averaging over very long periods
to take these measurements.

This chapter describes an automated capacitance-loss tangent bridge
that we have developed which
has made measurements of samples of capacitance to below 10~pF over
the range of frequencies from 2~Hz to 30~kHz. This
is done with signal levels as low as 30~$\mu$V~rms. The bridge can
accurately measure loss tangents as low as $10^{-4}$.
Depending on the background noise, measuring signal level, and signal
averaging time, the bridge can measure absolute capacitance to a
precision of .01\% and make relative capacitance measurements good to
1 part in $10^6$. Typically with a 1~mV~rms measuring signal at 1~kHz, samples
of capacitance of around 50~pF can be measured to 0.1\% precision in 15~sec.
The bridge has taken over 1,000,000 capacitance measurements
including all of the data presented in this thesis.

The bridge is in some ways similar to the earlier generation bridge of Cavicchi
and Silsbee\cite{DICK-bridge}. Analogously to that bridge, it adjusts
voltages on the arms of the bridge instead of adjusting the value of
circuit elements. However unlike the Cavicchi bridge, it uses a
computerized algorithm to find balance instead of an analog feedback
network. The operation of the bridge described here is continuously
monitored by a computer whose software has built-in solutions for
most problems that might cause the bridge to ``hang up'' on a measurement.
This allows the bridge to run unsupervised for days at a time. The
other main difference between the bridge described here and the
Cavicchi bridge is the elimination of a resistance standard in the
balancing arm of the bridge. In the new bridge, only a capacitance
standard is
used, with the resistive part of the impedance of samples balance by
adjusting the phase on the voltage applied to this standard. This
eliminates the Johnson noise produced by a resistance standard.

\figinsert{\vfil\vskip6.5truein\includegraphics{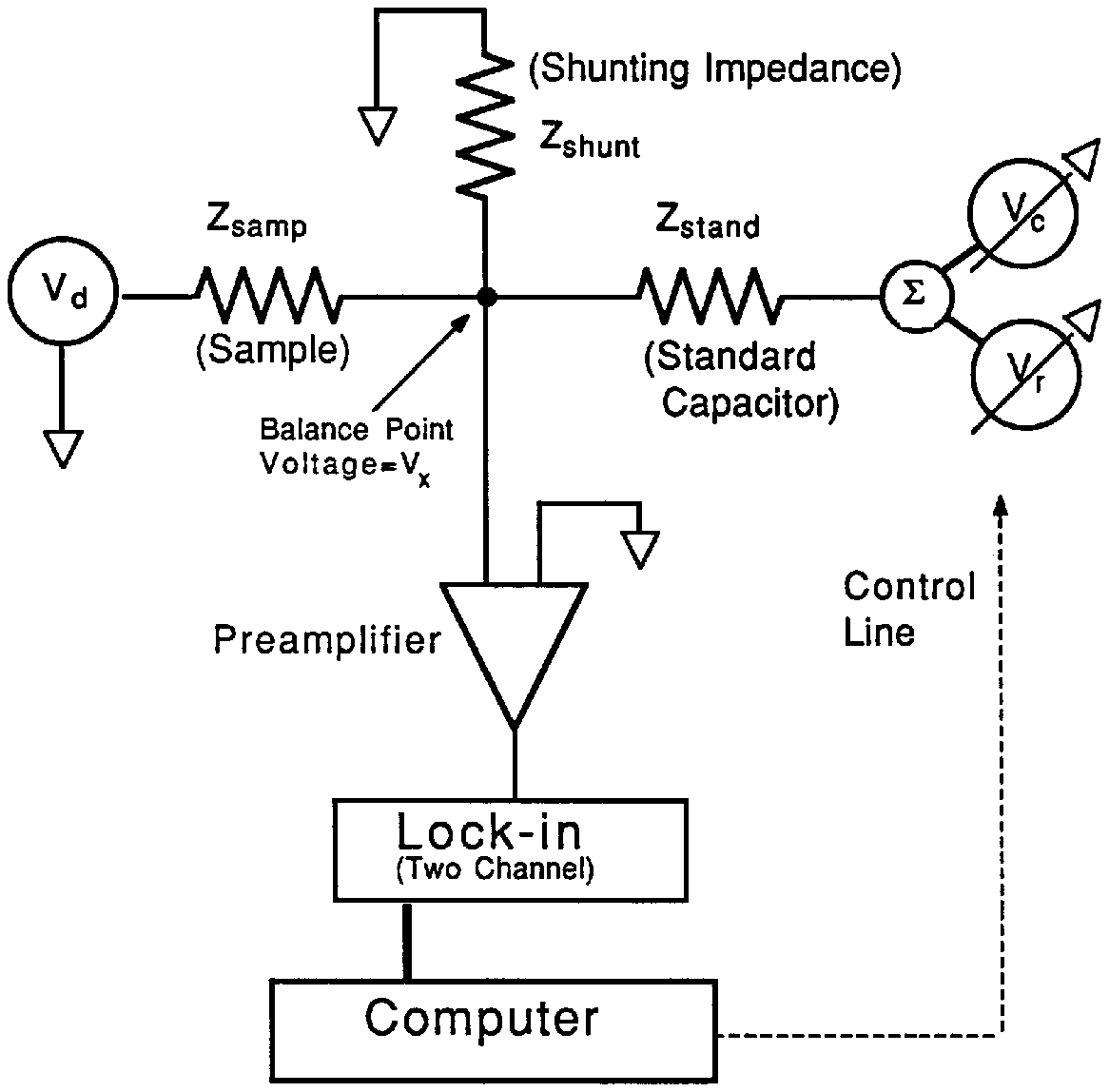}}{barebones}

\v\par\penalty-1000
{\noindent\bf 2.2 Rough Schematic}
\nobreak\v\nobreak
Fig.~\fign{barebones} displays the essential elements of the balancing
circuit. An A.C.\ voltage $V_{d}$ (drive voltage),
whose amplitude remains fixed in the bridge balancing, is applied to one
side of the sample of unknown impedance, $Z_{samp}$. Two
voltages, each of variable amplitude, are are summed and
applied to a precision air dielectric standard
capacitor whose impedance is $Z_{stand}$. One of the voltages, $V_{c}$
180$^{\circ}$ out of phase with the drive, and the other
$V_{r}$, leading the drive voltage by approximately
90$^{\circ}$ are applied to the
standard capacitor whose impedance is $Z_{stand}$. The basic idea
behind the bridge is to null the voltage, $V_{x}$ at the ``balance
point'' where the standard capacitor and the sample meet. In the simplest
configuration (as shown in Fig.~\fign{barebones}), knowledge of the values of
$|V_c|/|V_d|$ and $|V_r|/|V_d|$ needed to achieve balance, the phase
relations of $V_r$, $V_c$, and $V_d$,
and $Z_{stand}$ is sufficient to determine $Z_{samp}$. Because
measurement is made when the
voltage $V_x$ is nulled, no current flows across the impedance (mostly
from shunt capacitance)
shunting the balance point to ground, $Z_{shunt}$, at balance, and the
value of $Z_{shunt}$ does not enter into the determination of $Z_{samp}$.

A preamplifier feeds the voltage signal at the balance point into a
lock-in amplifier. A computer (IBM, PC-AT)
reads the signals $L_1$ and $L_2$ from
the two channels of the lock-in. These measure the in phase and out of
phase signals of the voltage $V_x$ with respect to the reference
signal entering the lock-in. The phase of the lock-in with respect to
$V_d$ is irrelevant to the balancing procedure and to the
determination of the sample impedance and need not considered in this
discussion.
The computer controls the amplitudes $|V_r|$ and $|V_c|$ and uses the
procedure described in the next section to adjust these to null the
voltage $V_x$.

\v\par\penalty-1000
{\noindent\bf 2.3 Theory of Operation}
\nobreak\v\nobreak
{\noindent\sl Determination of Balancing Voltages}
\nobreak\v\nobreak
This section describes mathematically how the balance condition is
determined by the computer. Because most of our samples are, to an excellent
approximation for the measuring voltage used here, linear devices, we
can immediately write two linear equations describing the signals at
the lock-in output. The multi-step balancing procedure described here
can still achieve an accurate balance even with samples which contain
some nonlinearity in their current-voltage relation.
$L_1$ and $L_2$ can be written in terms of the
voltage amplitudes $|V_r|$ and $|V_c|$ by:
$$\eqalign{L_1&=K_{r,1}|V_r|+K_{c,1}|V_c|+M_1\cr
L_2&=K_{r,2}|V_r|+K_{c,2}|V_c|+M_2.}\eqno(\equn{lvals})$$
The variables $K$ are independent of the voltage amplitudes,
and $M_1$ and $M_2$ are the lock-in output
voltages which are
measured on channels 1 and 2 when the voltages $V_c$ and $V_r$
are set to zero. The phases of $V_r$ and $V_c$ are constant with
respect to the phase of $V_d$ throughout the balancing procedure.

First, consider the constants, $K$, as known variables. Their values
depend on the values of $Z_{shunt}$, $Z_{stand}$, and $Z_{samp}$.
We describe later the algorithm which determines them. With the $K$'s
known,
reading the lock-in output with the voltage amplitudes $|V_r|$ and
$|V_c|$ set to
arbitrary values $|V_r^{\alpha}|$ and $|V_c^{\alpha}|$ determines the
values of $M_1$ and $M_2$ and thus specifies the values to which $|V_r|$ and
$|V_c|$ should be set to null the lock-in outputs $L_1$ and $L_2$.
Denoting the lock-in output values as $L_1^{\alpha}$ and $L_2^{\alpha}$
when $|V_r|$ and $|V_c|$ are set to $|V_r^{\alpha}|$ and $|V_c^{\alpha}|$,
it is easy to show that these values, $|V_r^0|$ and $|V_c^0|$, are given by:
$$|V_r^0|=|V_r^{\alpha}|+{P\over K_{r,1}}{\bigl(}-L_1^{\alpha}+{K_{c,1}\over
K_{c,2}}L_2^{\alpha}{\bigr)},\eqno(\equn{rbal})$$
and
$$|V_c^0|=|V_c^{\alpha}|+{P\over K_{c,2}}{\bigl(}-L_2^{\alpha}+{K_{r,2}\over
K_{r,1}}L_1^{\alpha}{\bigr)},\eqno(\equn{cbal})$$
where
$$P={\biggl(}1-{K_{c,1}K_{r,2}\over K_{r,1}K_{c,2}}{\biggr)}^{-1}.$$

Equations~\equn{rbal}~\&~\equn{cbal} make it clear how balance is
found once the values of the variables $K$ are known. Determination of
the values of these variables follows very simply by monitoring
changes in the lock-in output values $L_1$ and $L_2$ as the voltage
amplitudes $|V_r|$ and $|V_c|$ are varied. This can be done in three
successive stages: {\bf 1)} With $|V_r|$ and $|V_c|$ at fixed starting values,
$L_1^{\bf1}$ and $L_2^{\bf1}$ are measured; {\bf 2)} Only $|V_r|$
is changed to a
value $|V_r|+{\Delta}|V_r|$ and $|V_c|$
kept fixed at the same value as in step (1). $L_1{\bf^2}$ and $L_2{\bf^2}$ are
measured; {\bf 3)} $|V_c|$ is changed to a value $|V_c|+{\Delta}|V_c|$
and $|V_r|$ is changed back to its starting value
in step (1). Finally, $L_1{\bf^3}$ and $L_2{\bf^3}$ are measured.
It is then trivial to show that:
$$\eqalign{K_{r,1}&={L_1^{\bf2}-L_1^{\bf1}\over{\Delta}|V_r|},\cr
K_{c,1}&={L_1^{\bf3}-L_1^{\bf1}\over{\Delta}|V_c|},}
\qquad\eqalign{K_{r,2}&={L_2^{\bf2}-L_2^{\bf1}\over{\Delta}|V_r|},\cr
K_{c,2}&={L_2^{\bf3}-L_2^{\bf1}\over{\Delta}|V_c|}.\cr}\eqno(\equn{kval})$$
Quite obviously, for fixed noise errors $\delta L_1$ and $\delta L_2$,
the greatest precision in the values of the various $K$ variables is
obtained when $\Delta|V_r|$ and $\Delta|V_c|$ are made as large as
possible.

\v\par\penalty-1000
{\noindent\sl Two Cycles to Achieve Balance}
\nobreak\v\nobreak
The bridge thus operates in two cycles. During the first cycle, the
procedure of the last paragraph is carried out and each of the $K$
variables is determined using Eqs.~\equn{kval}. During the second
cycle, balance is achieved by adjusting the
amplitudes $|V_r|$ and $|V_c|$ according to
Eqs.~\equn{rbal}~\&~\equn{cbal}.

A certain amount of (digital) signal averaging takes place 
place during each measurement of $L_1$ and $L_2$ during each part of
both cycles. An
attempt is made to adjust the time spent in each part of both cycles
to minimize the errors $\delta{V_r^0}$ and $\delta{V_c^0}$ in the
final ``balancing'' values of $|V_r|$ and $|V_c|$ for the given total
amount of time desired to be spent on signal averaging. Further, we 
find that these errors are smaller if the balancing (second) cycle
proceeds in two steps rather than one step.
The first step achieves approximate
balance, setting the amplitudes $|V_r|$ and $|V_c|$ to the values
indicated by Eqs.~\equn{rbal}~\&~\equn{cbal}, starting with amplitudes
$|V_r|$ and $|V_c|$ arbitrarily far away from their balancing values,
$|V_r^0|$ and $|V_c^0|$. The second step in the balancing cycle starts
from this approximate balance. Because the signal is near balance,
digitization error of the lock-in output values $L_1$ and $L_2$ can
become significant. For this reason, during this step
the gain on the preamplifier feeding into the lock-in is increased by a
factor of 10. Then the second balancing step using
Eqs.~\equn{rbal}~\&~\equn{cbal} is undertaken to achieve better
experimental values of $|V_r^0|$ and $|V_c^0|$.

\v\par\penalty-1000
{\noindent\bf 2.4 Evaluation of Sample Impedance}
\nobreak\v\nobreak
We point out that
the treatment of the operation of the bridge has up to this point made
no use of the phase relation between the voltages
$V_r$ and $V_c$, applied to the standard capacitor,
with each other or with the drive voltage, $V_d$.
However, knowledge of these phase relations is needed to
determine the value of $Z_{samp}$. Consider $V_r$ to lead $V_d$ by
exactly $90^{\circ}$ and $V_c$ to be $180^{\circ}$ out of phase with
$V_d$. In the bridge that we constructed, $V_r$ is actually a known
phase shift (between $0$ and $0.3^{\circ}$ of phase lag) away from
being exactly $90^{\circ}$ ahead of $V_d$. This complicates the
$Z_{samp}$ determination only slightly; but
for simplicity here, we consider $V_r$
as leading $V_d$ by exactly $90^{\circ}$.

We consider $Z_{samp}$ to consist of a parallel combination of a
capacitance, $C_{samp}$, and a conductance, $G_{samp}$.
In this case one finds, summing all of the currents into
the balance point when the voltage at the balance point is nulled, the
following relations:
$$\eqalignno{G_{samp}&={\omega}C_{stand}{|V_r^0|\over|V_d|},&(2.5)\cr\noalign{\hbox{and}}
C_{samp}&=C_{stand}{|V_c^0|\over|V_d|}.&(2.6)\cr}$$
The solution for arbitrary phase of $V_r$ is of course obtained simply
by splitting $V_r$ into components in and out of phase with the drive
voltage, adding the in phase part to $|V_c|$ in Eq.~2.6 and replacing
$|V_r|$ in Eq.~2.5 with the out of phase part.

\v\par\penalty-1000
{\noindent\bf 2.5 Implementation of Bridge}
\nobreak\v\nobreak
The hardware in the bridge was purposely kept very simple, leaving
the control of almost all functions to the computer. Fig.~\fign{detail}
displays a more detailed view of the function of the bridge.
The voltages $V_c$, $V_r$, and $V_d$ are all derived from the same
signal generator. This signal, aside from going
directly to the sample (actually by means of a voltage divider and a
circuit which applies a bias voltage), goes to the inputs of two
12 bit multiplying D/A converters (MDACs).
These MDACs control the voltage amplitudes $|V_c|$ and $|V_r|$.

\figinsert{\vfil\vskip6.5truein\includegraphics{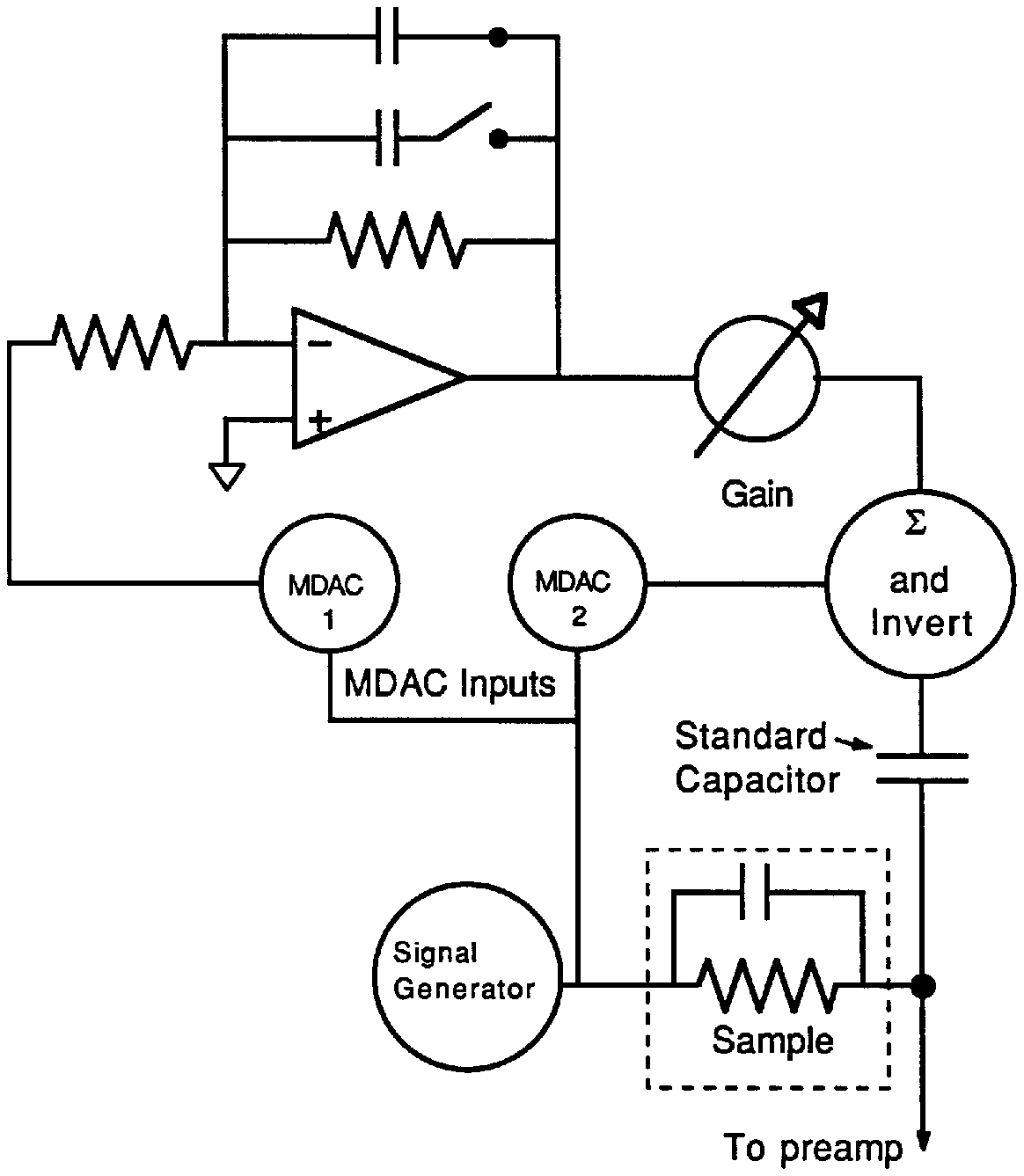}}{detail}

The output of the MDAC (MDAC\#1) in Fig.~\fign{detail}
which controls $|V_r|$ is fed into 
an op-amp integrator\cite{HORO&HILL} which
is used to produce the $+90^\circ$ phase shift in the voltage $V_r$.
A feedback resistor (40~k$\Omega$) shunts the
capacitor in integrator to keep it from
charging to a large D.C.\ voltage. Because we need the bridge to
operate over a wide range of frequencies, and because 
the gain of the integrator
is inversely proportional to
the applied frequency and the integrating capacitance,
relays are used to switch to different integrating capacitances (from
5000~pF to $5~\mu{F}$ in powers of 10)
depending on the frequency and considerations described below.
We note that the feedback
resistance causes the final signal phase to lag slightly from a
$90^\circ$ degree phase advance relative to
the signal from the signal generator.

Although this is not a problem because we know the value of this phase
lag, we prefer to keep it small. The reason for this is that the elements
used are off the shelf capacitors and resistors whose value
may drift with time. Typically we need for our experiments greater
accuracy in capacitance measurements than for loss measurements; hence
we would like any signal coming through the integrator circuit
that deviates from $90^\circ$ to be small so that the errors due to
imprecise knowledge of component values in the integrator circuit
are small in the capacitance measurement.
For this reason, we keep the integrating
capacitance as large as possible without making it so large that the
output of the integrator stage is so small as to give poor signal to
noise in the input of the next stage. The next stage is a variable
gain amplifier (gain variable discretely in powers of 10) which adjusts
the strength of the signal leaving the integrator before it is
summed with the signal leaving the MDAC which controls $|V_c|$ (MDAC\#2
in Fig.~\fign{detail}). The summed signal is then inverted and divided
by a factor of 1000 using a resistive voltage divider (not shown in
Fig.~\fign{detail}) before it is applied to the sample. The large
voltage dividers allow us to process signals at large signal levels
inside the bridge to improve signal to noise while permitting sample
measurement at low signal levels and at low output impedance
($5\Omega$) from the bridge signal sources.

The bridge is typically run through the algorithm described in section
2.3. The idea is to adjust MDAC\#1 and MDAC\#2 so that the
signal at the balance point is nulled.
The bridge software adjusts the gain on the stage after the integrator
so that if
MDAC\#1 and MDAC\#2 are the same ``distance'' (in MDAC input bit counts)
off balance, similar sized off balance signals appear in the lock-in
in both channels.
This ensures that resolution is not lost in either the capacitance or
the loss tangent measurement due to the signals from one MDAC being
much larger than from the other.

The lock-in, preamplifier, and signal generator are also under computer
control. The bridge software adjusts the lock-in and the preamplifier
gain to be appropriate for the signal levels used in the measurement.
The bridge also contains circuitry to apply a bias voltage on samples.
This bias voltage, the amplitude of signals used to measure samples,
and the frequency of these signals, are all under the control of the
bridge software.

We mention briefly a few features used in the bridge to minimize 60~Hz
and other noise. Most of the bridge circuitry is run by batteries to
eliminate potential ground loops in power lines. All computer lines to
the bridge circuitry and the preamplifier
are optically isolated. Also, the reference signal from
the bridge circuits to the lock-in is optically isolated as well.
The preamplifier, which
is also run off of batteries, has its output coupled into the lock-in
via an audio frequency transformer. This allows the bridge circuits,
which are grounded to the signal generator, to attach to power line
ground at only one point, eliminating ground loops. All of the bridge
circuitry (which is in shielding metal boxes)
and batteries are kept in a cage surrounded by copper screening.

\v\par\penalty-1000
{\noindent\bf 2.6 Special Features}
\nobreak\v\nobreak
We mention briefly a few features implemented on the bridge
which have facilitated data taking and data analysis. One important
feature is the calculation of errors in the final values obtained for
the real and imaginary parts of the sample impedance. All data
received from the lock-in is digitally signal averaged, and the
standard deviation in each set of averaged values is calculated. These
errors on each measurement are propagated as indicated by the
algorithm of section 2.3 to determine error bars on the sample
impedance measurement.
Knowledge of these errors is helpful because in curve fitting to our
measurements (described in later chapters) it is important to know how
much ``weight'' to give to various measurements.

In some of our work\cite{CHAP4} it has been very important to ensure
that stray high frequency signals did not appear across samples. For
this reason, we have occasionally added low pass filters (cutoff
$\approx$30~kHz) on both sample leads. Provisions are built into the
bridge software to compensate sample impedance results for these
filters. Another correction that the bridge software makes in obtained
values for the sample impedance is needed
because phase shifts result in signals passing through the MDACs due
to the fact that the MDACs contain some small output
capacitance ($\approx$30~pF)
and the signals from the MDACs are fed into op-amps with
finite open loop gain.\pcite{D/Aconv} These phase shifts
grow larger with frequency. Because these phase shifts depend somewhat
on the settings of the MDACs it becomes difficult to correct bridge
measurements for them at high frequencies. These phase shifts are
responsible for the high frequency limit of operation of the bridge at
30~kHz.

Finally for very high resolution measurements, the bridge is run using
a different procedure than that described in section 2.3. The MDACs
contain a gain error of about 1 part in $10^4$. For higher precision
than this in impedance measurements, the MDACs must be kept at a fixed
value (their gain appears to be stable to better than 1 part in $10^6$
over one hour), and the amount to which signals at the balance point
are different from zero are interpreted to determine the sample
impedance. The bridge software has built in it the ability to measure
shunt impedance and then interpret an off balance signal in terms of
capacitance and loss tangent of the sample.
Typically, we wish to measure how much the sample
capacitance and loss tangent change as the applied bias across the
sample is varied. The protocol now becomes as follows. At a particular
applied gate bias, the bridge is balanced. The gate bias is now varied
and measurements are made with the MDAC digital inputs fixed at these
values. This high resolution mode is especially important in the work
of Ch.6, where we measure samples of capacitance 35~pF to a resolution
of .1~fF.

\vfil\supereject
\centerline{\bf References}
\v
\listrefs
\vfill\eject

\newchapter{3.}
\figitem{sampdos}{The essential layer structure of the samples. In
samples {\bf A} and {\bf C} the blocking barrier contains Si doping
commencing at distances of $100~\AA$ and $150~\AA$ respectively
away from the edge of the quantum well.}

\figitem{samp}{(a) shows the essential structure of the samples.
Electron transfer through the tunnel barrier brings the electron
gases in the substrate and the quantum well into equilibrium.
The density of electrons in the quantum well may be varied through
the application of a gate bias ($V_{gate}$ in the figure). Capacitive
coupling to the well allows us to measure both the thermodynamic
density of states (DOS) in the well and tunneling from the well to
the substrate. (b)~displays the
model of the sample which is used in curve fitting to extract low and
high frequency capacitances used to determine
density of states in the quantum well.}

\figitem{freq}{Capacitance and loss tangent of a 200~$\mu$m mesa
etched on sample {\bf B}. The solid and dotted curves are fits from
the circuit model presented in Fig.~\fign{samp}b. At low frequencies,
electron transfer from the substrate to the well can take place, and
the capacitance $C_{low}$ is measured. At frequencies high compared to
the RC time of the tunnel barrier, no electron transfer to the well
takes place, and the measured capacitance, $C_{high}$
drops to a value consistent
with the distance from the top gate to the substrate layer.}

\figitem{nofield}{Capacitances $C_{low}$ and $C_{high}$ (described in
the text) determined from fits to capacitance vs.\ frequency data on
sample {\bf B}. The sharp increase in $C_{low}$ occurs as the well
begins to fill with electrons.  Here, there is no magnetic field
applied and the temperature is 4.2~K.}

\figitem{wfield}{Capacitances $C_{low}$ and $C_{high}$ for sample {\bf
B} at 2.0~T and 2.1~K. Note the oscillations in the low frequency
capacitance. These arise from the Landau level DOS in the well. The
high frequency capacitance is insensitive to the DOS in the well and
does not display any such oscillations. The two dashed vertical lines
are limits of integration described in section 3.4.3 of the text.}

\figitem{2.0TH}{Density of states ($\partial n/\partial\mu$)
vs.\ Fermi energy in sample {\bf B} at 2.0~T and 2.1~K,
extracted using the analysis procedure described in the paper, from
the data presented the data of Fig.\ \fign{wfield}. The vertical
lines drawn are $\hbar\omega_c$
apart in energy.
The deviation of the
spacing between the first and second peaks arises from inhomogeneous
``puddling'' of electrons in the well as the well is emptied. The
first peak is thus outside of the region of validity of our model.}

\figitem{intlev}{The integrated ``lever-arm' used to convert DOS
results from gate bias to Fermi energy in the well. The lever-arm along
with the DOS is determined directly from the data. The flatter
regions of the curve occur where the DOS in the well is high, and
addition of electrons to the well changes the Fermi energy little. The
steeper regions occur where the DOS is low.}

\figitem{zerofield}{Zero field DOS results from sample {\bf B} with and
without correction to $C_{geom}$ described in the text. Note the long
flat region of the corrected curve (circles) at which the DOS is set
equal to the expected $2.8\times10^{13}$eV$^{-1}$cm$^{-2}$ by
the appropriate choice of $C_{geom}$ and $\eta$ (see Appendix~B).}

\figitem{fold}{$\partial n/\partial\mu$ from the data of Fig.~\fign{fit4ta}
(for our purposes here, assumed to be the same as the zero
temperature DOS) convolved with the
derivative of the Fermi Distribution function at 7.0~K (crosses), compared to
$\partial n/\partial\mu$ determined from the data at 7.0~K (squares). The
excellent agreement between the 7.0~K and the results of the
convolution serves as another consistency check of the model.
The slight deviation of the peaks is believed to arise largely from
the fact that the data used in the convolution is not the zero
temperature DOS but instead at 1.9~K. Note that the horizontal
(energy) scale is different from that plotted in Fig.\ \fign{2.0TH}
to allow the full density range of sample {\bf A} to be plotted.}

\figitem{fit4ta}{$\partial n/\partial\mu$ of sample {\bf A} at 4.0~T and
1.9~K. Circles are data.
The solid curve is a fit to Lorentzian lineshapes, and the dotted
curve is a fit to Gaussian shapes plus an added energy independent
background of $\beta=1.52\times10^{13}$ eV$^{-1}$cm$^{-2}$. The other
fitting parameters from Gaussian plots give: $\Gamma_1=$1.08~meV,
$\Gamma_2=$.689~meV, and $\Gamma_3=$.597~meV. Note that the large
background causes the fitted widths to be much narrower than fits
that do not include a background. The fitting parameters for the
Lorentzian fits are $\Gamma_1=$1.34~meV, $\Gamma_2=$.969~meV, and
$\Gamma_3=$.847~meV.}

\figitem{fit3ta}{$\partial n/\partial\mu$ of sample {\bf A} at 3.0~T and
1.9~K. Circles are data.
The solid curve is a fit to Lorentzian lineshapes, and the dotted
curve is a fit to Gaussian shapes. The fitting parameters for the
Lorentzians are: $\Gamma_1=$1.98~meV, $\Gamma_2=$1.18~meV,
$\Gamma_3=$.976~meV, and $\Gamma_4=$.885~meV. The fitting parameters for the
Gaussian fits are: $\beta=1.0\times10^{9}$ eV$^{-1}$cm$^{-2}$
(negligible background), $\Gamma_1=$2.11~meV, $\Gamma_2=$1.43~meV,
$\Gamma_3=$1.21~meV, and $\Gamma_4=$1.14~meV.}

\figitem{fit2tb}{$\partial n/\partial\mu$ of sample {\bf B} at 2.0~T
and 2.1~K. Circles are data. 
The solid curve is a fit to Lorentzian lineshapes. The fitting
parameters are: $\Gamma_1=$1.47~meV, $\Gamma_2=$.810~meV,
$\Gamma_3=$.737~meV, and $\Gamma_4=$.678~meV}

\figitem{fit2ta}{$\partial n/\partial\mu$ of sample {\bf A} at 2.0~T
and 1.9~K. The solid curve is a fit to Lorentzian lineshapes with
fitting parameters given by: $\Gamma_1=$2.52~meV, $\Gamma_2=$1.21~meV,
$\Gamma_3=$1.10~meV, $\Gamma_4=$.987~meV, $\Gamma_5=$.909~meV, and
$\Gamma_6=$.892~meV}

\figitem{fit4tc}{$\partial n/\partial\mu$ of sample {\bf C} at 4.0~T
and 4.2~K. The solid curve is a fit to Lorentzian lineshapes with
parameter values of: $\Gamma_4=$1.70~meV, and $\Gamma_2=$1.14~meV.}

\figitem{widths}{Widths of Landau levels plotted as a function of the
density at which the Landau level peak occurs for magnetic fields of
2, 3 and 4~T. The solid curve is a power law fit described in the
text. The figure suggests that the Landau level widths are
independent of magnetic field strength.}

\figitem{8.5all}{Plotted is $\partial n/\partial\mu$ at 8.5~T and
1.85~K. This figure illustrates the main difficulty in the DOS
determination at high fields. Because the second level is at the high
density limit of operation of the device, it is difficult to do a
precise adjustment of the
parameters of the DOS determination ($C_{geom}$ and $\eta$).}

\figitem{8.5lor}{Plotted is $\partial n/\partial\mu$ of the lowest
Landau level at 8.5~T and 1.85~K. The solid curve is a fit using the
theory of Ando and Uemura and assuming underlying Lorentzian
lineshapes. The exchange energy in the fit is 3.8~meV. The initial
parameters describing the Lorentzians (see Eq.~\equn{lors}) for the
up and down spin bands are $\Gamma_{+}=$1.0~meV and
$\Gamma_{-}=$0.9~meV.}

\figitem{8.5gauss}{Plotted is $\partial n/\partial\mu$ of the lowest
Landau level at 8.5~T and 1.85~K. The solid curve is a fit using
Gaussian lineshapes and the theory of Ando and Uemura. The exchange
energy from the fit is 3.3~meV. The parameters describing the
Gaussians are $\Gamma_{+}$=1.2~meV and
$\Gamma_{-}$=1.05~meV.}

\figitem{samp2}{The figure shows a more realistic model of the sample
than that of Fig.~\fign{samp} when the in-plane conductance of the
2d gas vanishes. The sample can be thought of as broken up into
separate domains connected by the resistances $R_S$. At zero
temperature, these resistances diverge.}

\figitem{broad}{Plotted are capacitance and loss tangent curves at a
particular gate bias for sample {\bf A} at 8.5~T and 875~mK. Notice
that the curves are broader in frequency than those
of Fig.~\fign{freq}. In this case, the broadening parameter
(described in the text), $\chi$, has a value of 3.46. The dashed line
is a theoretical loss tangent
curve with the same value of $f_{peak}$ but no broadening ($\chi=1$).}

\figitem{broadtemp}{The figure plots the broadening parameter, $\chi$,
as a function of temperature at a fixed Fermi energy. The Fermi
energy is fixed at a value of 4~meV on Fig.~\fign{8.5broad}. Note
the saturation of $\chi$ below about 0.5~K. The solid curve is a guide
to the eye.}

\figitem{8.5broad}{Plotted is $\chi$ as a function of Fermi energy in
sample {\bf A} at 8.5~T for temperatures ranging from 0.2~K to 3~K
along with $\partial n/\partial\mu$ at 8.5~T and 1.85~K.
Note that the temperature dependence of $\chi$ is roughly independent
of Fermi energy; $\chi$ saturates everywhere along the curve at about
0.5~K. The fact that the
region between the spin split Landau levels shows a
different temperature dependence does not reflect the physics of the
broadening; instead it arises because the exchange enhancement of the
spin splitting continues to increase as the temperature is lowered.}

\figitem{4broad}{The figure plots $\chi$ as a function of Fermi
energy at 4~T and 140~mK in sample {\bf A}. Also plotted is $\partial
n/\partial\mu$ at 4~T and 1.9~K.}

\figitem{eta}{This figure pictorially describes the origin of the
term $\eta$ in our DOS calculation. 
In the sheet charge model,
$U_w$ is the energy of the band
edge at the position of electrons and $U_w+E_0$ is the energy of the
bound state. A distributed charge model
spreads out the electrons, and the energy of the band edge at the
mean position of electrons is reduced to $U_{w (dist)}$. This, and the
fact that electrons sense the potential energy at positions other
than $x_w$ reduces the bound state energy to $U_w+E_0-\eta\sigma_w$.}

\figitem{imag}{The curve plotted is the locus of points in parameter
space for which the Landau level DOS, deduced using the analysis of
this chapter, contains peaks spaced $\hbar\omega_c$ apart. Different
points on the curve correspond give different values of the zero field
DOS; labels next to the box symbols give the zero field DOS obtained
using these values of $C_{geom}$ and $\eta$.
The error bars on the points illustrate the error in $C_{geom}$ after
using the techniques involving the zero field DOS discuss in the text.
Because of the steepness of the slope of the curve, the error in
$C_{geom}$ is translated into a much larger uncertainty in the value
of $\eta$.}
\centerline{\twelvebf Chapter III}
\v
\v
\centerline{\twelvebf Quantitative Measurement of the Two-Dimensional}
\centerline{\twelvebf Thermodynamic Density of States}
\v
\v
\v
\v
\v
\v
{\noindent{\bf 3.1 Introduction}}
\v
The shape and size of Landau level density of states (DOS) peaks in a two
dimensional (2d) electron gas in the presence of a magnetic field
applied perpendicular to the plane of the electron gas has generated
considerable interest over the past two decades.\pcite{ANDO-F-S}
The thermodynamic DOS is an equilibrium property of the 2d electron gas.
It makes no distinction between localized or extended states.
However, ideas involving the processes resulting in
localization of states\cite{QHE} can be
investigated through observation of the shape density of states peaks
associated with Landau levels. The notion of a nonzero DOS between Landau
levels has been crucial in formulating ideas of the Quantum Hall
Effect.

A variety of
experiments, including specific heat,\pcite{STM-spc-ht,TSUI-heat}
magnetization,\pcite{EIS-DvA}
and capacitance\cite{HIG-cap,TREY-dos,KVK-mcap} 
studies have sought to probe the 2D DOS.
In most cases, measurements have determined the DOS of a system with a
fixed electron density in the 2D layer, with variation of parameters
such as applied field and temperature. 
Typically, models of the 2d electron gas refer instead to the density
of states at the Fermi energy as the Fermi energy is varied.
Information from such a model (as for example in
Ref.~\refnumber{STM-spc-ht}) must be transformed in an appropriate form in
order to compare with the data.  This procedure may introduce
ambiguity as to the shape of the original DOS peak as a function of
energy. Many of these experiments have, in fact, given only qualitative
results on the DOS shape, and not a quantitative description of the
DOS. 

\v\par\penalty-1000
{\noindent{\sl {\bf 3.1.1} Extraction of the DOS from capacitance measurements}
\nobreak\v\nobreak
The technique presented in this chapter uses capacitance measurements to
determine the 2d DOS.
Previous experiments have suffered from large errors in
the 2d DOS determined through capacitance measurements\cite{TREY-ext} due to
imprecise knowledge off sample parameters (barrier thicknesses,
dielectric constants, positions of electronic charges, etc.).
In the experiments described here, two normalization conditions,
derived from the known parameters of Landau level degeneracy and
Landau level energy spacing, remove uncertainties normally present
in other techniques and provide us a means for accurate measurement
of the DOS at the Fermi energy versus energy.

Unlike many experiments, ours are carried out a constant
magnetic field strength; we instead vary the electronic density in a
quantum well by means of a gate bias.
The technique used here relates
this gate bias to the Fermi energy in the quantum well and yields the DOS
from the capacitance data, which can then be plotted as a function of
Fermi energy.

Results from Goodall, Higgins, and
Harrang\cite{HIG-cap} have revealed some of the difficulties involved
in the DOS determinations using capacitance spectroscopy
due to the large conductivity variations in the plane of the 2d
electron gas as the Landau index is varied. 
Attempts at circumventing these problems have been made
by restricting the experiments to
low measuring frequencies\cite{TREY-dos,KVK-mcap}
and low magnetic fields ($<$2T).\pcite{TREY-dos}
Our experiments are done on
samples where the 2d electron gas is coupled by tunneling to a
conducting substrate. This allows charge to be transferred to all
regions of the of the 2d electron gas even when the ``in-plane
conductance'' is very low, allowing us to make capacitance
measurements which are largely insensitive to the magnetoresistance of
the 2d layer. Similarly designed samples have been used previously to
qualitatively determine the DOS at the Fermi energy as a function of
magnetic field up to very large field strengths.\pcite{TREY-ext}

\v\par\penalty-1000
{\noindent{\sl {\bf 3.1.2} Essential DOS Results}
\nobreak\v\nobreak
A detailed outline of this new technique is presented here, as well
as DOS results obtained using this technique
on samples from three different wafers grown using
molecular beam epitaxy (MBE).
The bulk of the chapter focuses on the DOS in magnetic
fields 4.0 T and below, with some results from higher fields (8.5 T)
presented near the end of the chapter.
In low magnetic fields ($<$4T),
where the effect of the spin splitting is too small to
be observed in the level shape, 
we observe a striking difference
between lineshapes determined from our experiment,
displayed here as a function of Fermi energy, and
lineshapes inferred by others from data taken at
constant electron density in the 2d gas. The lines are well fit by
Lorentzian lineshapes in contrast with most existing literature
in which Gaussian shapes with perhaps the addition of
a constant background DOS\cite{STM-spc-ht,GERH-between} are used.
Lorentzians give good fits to DOS results from all three
samples even though the three samples are very different in terms of
the positioning of dopants in an AlGaAs barrier near the quantum well
which contains the 2d electron gas.
Also remarkable in our data
is the fact that the widths of DOS peaks observed in
our experiment, while dependent on the Landau level index, show almost
no dependence on magnetic field strength. In essence, the picture of
the Landau level DOS developed here is a significant departure from
that of previous work.

We also obtain the DOS
in the high field region (up to 8.5~T), where there is strong exchange
enhancement of the Zeeman energy of electrons in the 2d gas. We use
the exchange model of Ando and Uemura\cite{ANDO-osc} using both
Gaussian and Lorentzian fits for determination of the
spin exchange energy, $E_{exch}$, to provide comparison with other recent
determinations of this parameter.\pcite{GOLD-exch,KVK-exch}

\v\par\penalty-1000
{\noindent{\sl {\bf 3.1.3} Measurements probing the localization
length of states}
\nobreak\v\nobreak
Another important question in the physics of the 2d electron gas asks:
As the Landau level filling factor, $\nu$, in the 2d electron gas is
varied how does the localization length of states vary? A great
deal of theoretical work,\cite{ONO-diverge,ANDO-loc} and more recently
experimental work,\cite{TSUI-deloc} has been 
devoted to understanding the behavior of the localization length.
Much of this work has been concerned with the region around a Landau
DOS maximum, where it is thought that the localization length
diverges. Effort has been focused on the nature of the
divergence; much less work has been done on the absolute length scales of the
localized levels or their behavior well away from the divergence.

In a later section of this chapter, we introduce a new technique
for the determination of the localization length as a
function of Fermi energy. This technique is particularly sensitive to
the localization length when the Fermi energy is between Landau
levels, a region where the localization length is not easily probed
by transport experiments.
We present qualitative results from this technique.

\figinsert{\vfil\vskip6.5truein\includegraphics{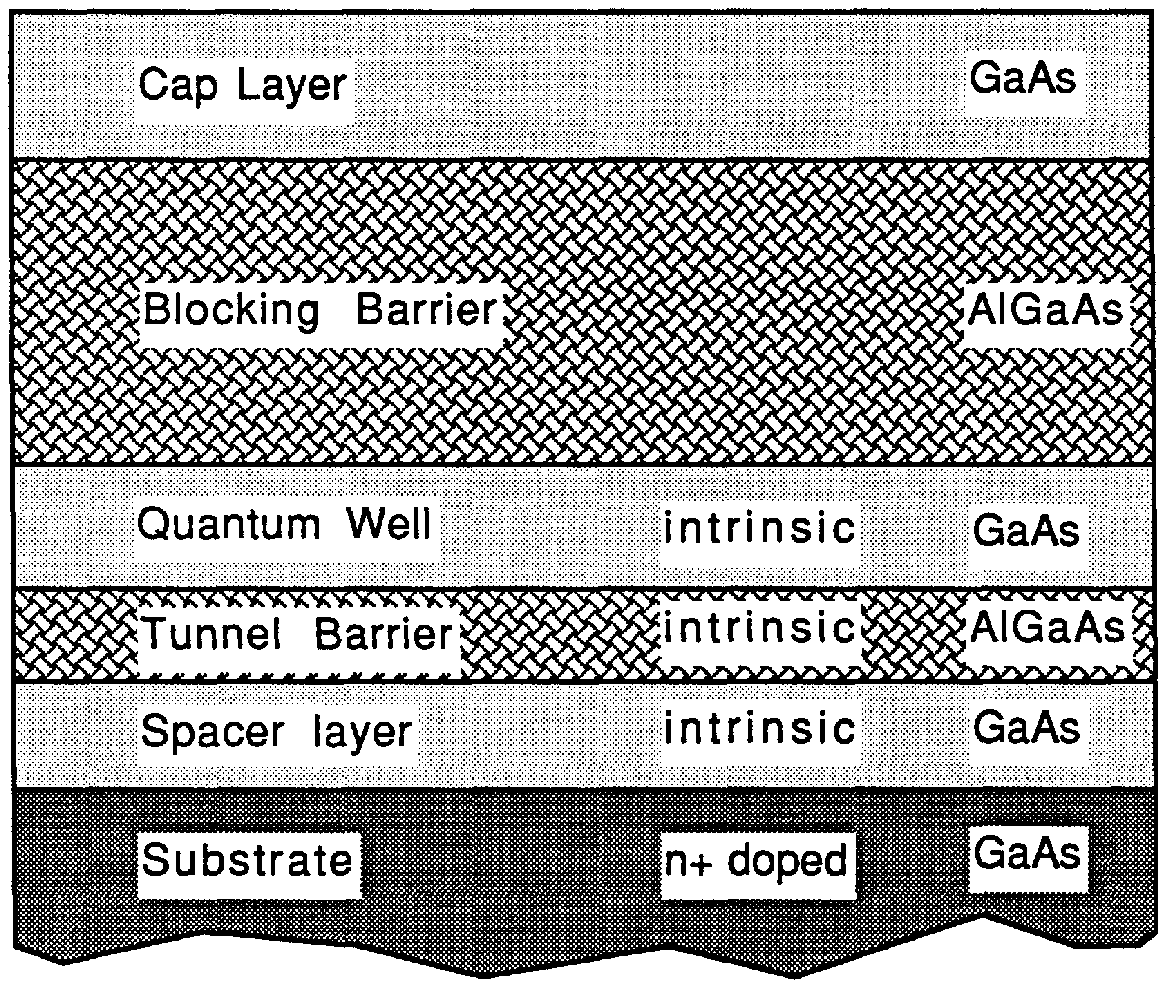}}{sampdos}

\v\par\penalty-1000
{\noindent{\bf 3.2 Sample Design}}
\nobreak\v\nobreak
We will refer to the three different samples used in this study as
{\bf A}, {\bf B}, and {\bf C}. Figure~\fign{sampdos} displays the
essential layer by layer construction
of our MBE grown wafers, and Fig.~\fign{samp}a shows
the conduction band edge structure in
sample {\bf B} which is very similar to that in the other samples.
All wafers are grown on
$n^+$ GaAs substrates which remain conducting at 4~K and lower
temperatures. On top of the substrate, an AlGaAs
tunnel barrier is grown, and then a GaAs quantum well. Beyond
this, there is an AlGaAs ``blocking barrier'' which allows no
electrical conduction over the range of gate biases applied in the
measurements described here. Samples {\bf A} and {\bf C} each have a
dopant layer in the blocking barrier; sample {\bf B} has no dopants in
the blocking barrier. A GaAs cap layer is grown above the blocking barrier
in all of the samples. In samples {\bf A} and {\bf B}, this layer is
heavily doped, and ohmic contact is made to this layer, which
then serves as a ``gate'' for our device. In sample {\bf C}, the cap
layer is undoped; a Cr Schottky contact is made to the surface
and the Cr metal acts as the gate.

\figinsert{\vfil\vskip6.5truein\includegraphics{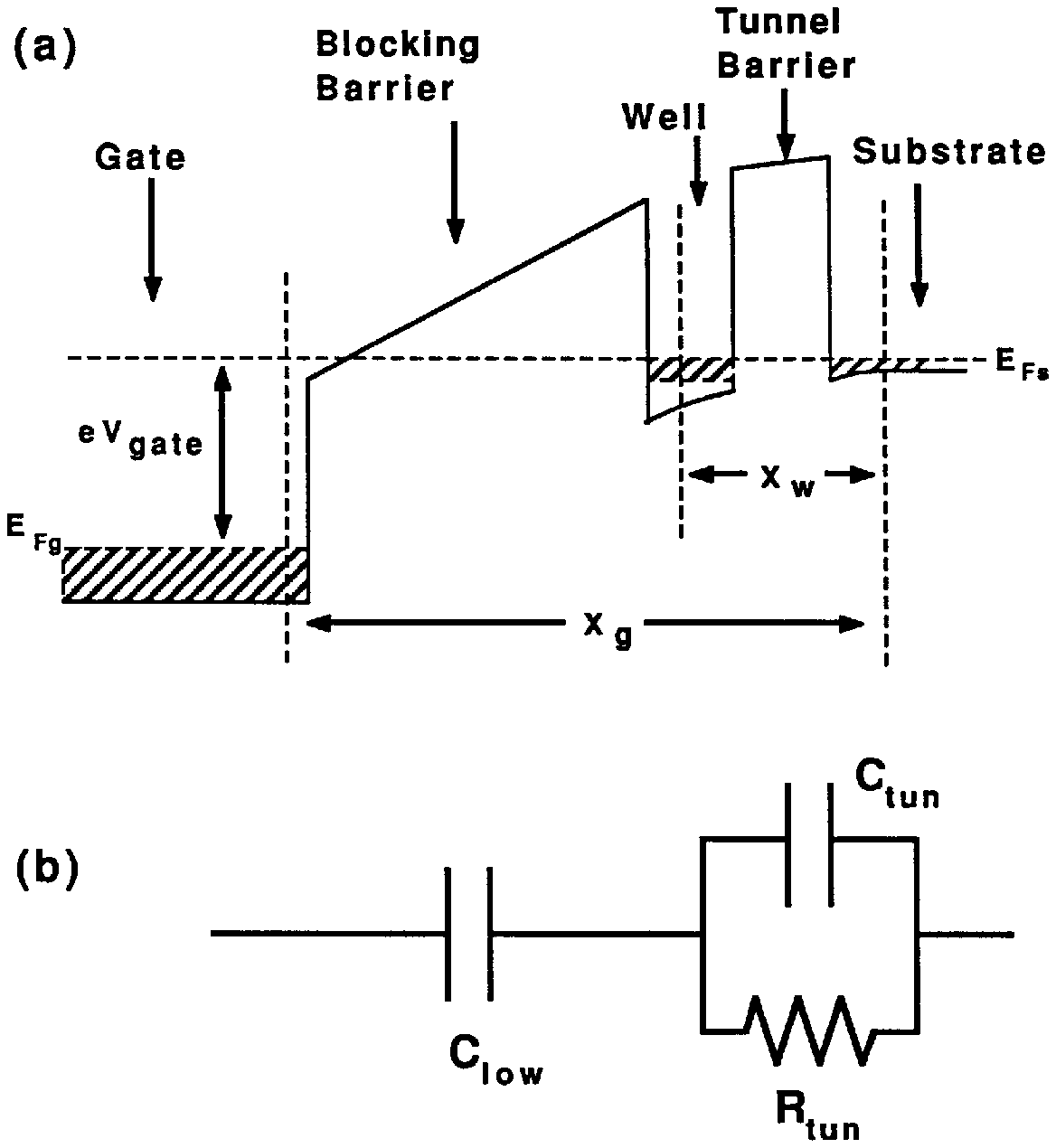}}{samp}

Sample {\bf A}
has been described extensively.\pcite{LSWa,JOHN}
The wafer used to produce this sample consists of a degenerately n
(Si-$\rm1\times10^{17}cm^{-3}$) doped substrate in GaAs, a 30 $\AA$
GaAs undoped spacer layer, an AlGaAs
tunnel barrier (160 $\AA$ wide), a GaAs quantum well (150 $\AA$ wide),
a thick AlGaAs blocking barrier (1550 $\AA$ wide), and a degenerately n
doped GaAs surface contact region.  A doped region
($\rm5\times10^{17}cm^{-3}$) exists from  $100\AA$ to $200\AA$ away
from the edge of the quantum well in the blocking barrier.
These donors are fully ionized for the range of electron filling of the
quantum well from $\rm0-5\times10^{11}cm^{-2}$.
All AlGaAs in sample {\bf A} has an Al concentration of 30\%.

Sample {\bf B} has a GaAs substrate doped at
Si-$\rm4\times10^{17}cm^{-3}$, a 30 $\AA$
undoped GaAs spacer layer, an AlGaAs
tunnel barrier (133 $\AA$ wide), a GaAs quantum well (150 $\AA$ wide),
an AlGaAs blocking barrier (800 $\AA$ wide), and a degenerately n
($\rm3\times10^{18}cm^{-3}$) doped GaAs surface contact region.
To help achieve high mobility, this sample contains no dopants in the
blocking barrier. Moreover, the sample temperature was lowered during
the growth of the GaAs spacer layer to reduce Si donor impurity
migration and the tunnel barrier was grown with periodic
growth interruptions (every 23 $\AA$ and with one monolayer of GaAs
grown at each interruption) to increase interface
smoothness. These techniques are thought to enhance the mobility of
systems with AlGaAs grown below the 2d layer.\pcite{MOTY-high}
All AlGaAs in sample {\bf B} has an aluminum concentration of 41\%.

Sample {\bf C} has a degenerately n
(Si-$\rm4\times10^{17}cm^{-3}$) doped substrate in GaAs, a 150 $\AA$
GaAs undoped spacer layer, an AlGaAs
tunnel barrier (150 $\AA$ wide), a GaAs quantum well (150 $\AA$ wide),
an AlGaAs blocking barrier (500 $\AA$ wide) and an undoped 300 $\AA$
wide GaAs cap layer. The cap layer never contains electrons over the
range of gate biases applied to the sample for the measurements in
this chapter. The 300 $\AA$ cap layer can thus be considered as part of
the blocking barrier. Cr Schottky contact is made to the surface of the sample.
A heavily doped ($\rm6\times10^{17}cm^{-3}$) region exists in the AlGaAs
from $150\AA$ to $500\AA$ away
from the edge of the quantum well and the AlGaAs blocking barrier.
These donors are fully ionized for the range of measurements presented
here. All AlGaAs in sample {\bf C} has an aluminum concentration of 30\%.

Metallic discs (AuGe ohmic contacts for samples {\bf A} and {\bf B}
and Cr for {\bf C}) ranging in size from 200 $\mu$m to 400 $\mu$m
in diameter act as gate contacts and also served as etch masks for a
mesa etch defining devices upon which capacitance measurements are made.
In each of the samples, the electron concentration in the quantum
well can be varied 
by application of a gate bias.
The 2d density in {\bf A} can be varied from
$\rm0-6\times10^{11}cm^{-2}$ before there is any measurable current
through the blocking barrier.
Samples {\bf B} and {\bf C} each have a range of
$\rm0-4\times10^{11}cm^{-2}$.

\v\par\penalty-1000
{\noindent{\bf 3.3 Measurements}}
\nobreak\v\nobreak
Our DOS determination derives from
measurement of two capacitances of our devices,
one measured at low and the other at
high frequencies.
The samples have been designed so that the frequency,
$$f_{peak}\approx{1\over2\pi R_{tun}C_{tun}},$$
lies within the range audio frequencies at which we make capacitance
measurements. Here $R_{tun}$ and $C_{tun}$ are the resistance and
capacitance of the tunnel barrier respectively, and $f_{peak}$ is the
frequency at which the loss tangent for the device reaches a maximum.
A frequency dependence of the sample capacitance arises near the
frequency $f_{peak}$. This can be intuitively understood by considering
that at low frequencies,
where the measured capacitance is $C_{low}$, there is enough time
during one half cycle of the measuring frequency for charge to move into
the quantum well from the substrate and bring the two into
equilibrium. At high frequencies, during one half
cycle of the measuring frequency little charge can be transferred
between the substrate and the well,
and the measured capacitance $C_{high}$
is lower, of value appropriate to the distance from the gate contact
region to the substrate charge.
An essential issue our DOS determination is that $C_{low}$
depends on the DOS in the 2d electron gas.
If the quantum well had an infinite DOS, then it would fully shield
the substrate from the gate field, and $C_{low}$ would be the
capacitance simply deduced from the distance from the gate to the 2d
gas. For a finite DOS, however, this shielding is not complete, and our
analysis provides a means to deduce the DOS from the measured
capacitances.

\figinsert{\vfil\vskip6.5truein\includegraphics{freq.ps}}{freq}

In practice, at each value of temperature, magnetic field, and gate
bias for which we wish to measure the DOS, we measure the capacitance
as a function of frequency over the range from 15 Hz to 30 kHz and
obtain $C_{low}$ and $C_{high}$ from fits to this data.
Figure~\fign{freq} displays the measured capacitance and loss tangent
from a 200 $\mu$m diameter mesa from sample {\bf B}.
The capacitance remains constant, of value $C_{low}$, at the lowest
frequencies, decreases over a range of about a decade in frequency, at
frequencies around $f_{peak}$,
then levels off to $C_{high}$ at the highest frequencies.
The loss tangent moves
through a peak at the same frequencies where the measured capacitance is
decreasing most sharply with frequency.

The capacitance $C_{low}$ is the capacitance of the nonshunted
capacitor in Fig.~\fign{samp}b. $C_{tun}$ is the capacitance of the capacitor
shunted by the tunneling resistance $R_{tun}$ in the model, and
$C_{high}$ is the series combination of $C_{low}$ and $C_{tun}$.
We fit the data to forms indicated by
the simple circuit model given in Fig.~\fign{samp}b
and extract the parameters $C_{low}$,
$C_{high}$, and $f_{peak}$ (and hence $C_{tun}$ and $R_{tun}$).
The form for the capacitance fit as a function of frequency is:
$$C(f)={{C_{high}C_{low}[1+(f/f_{peak})^2]\over
C_{high}+C_{low}(f/f_{peak})^2}}.\eqno{(\equn{cfit})}$$
In other work,\pcite{LSWa,LSWp,YOPRL,CHAP4} we have focused on results
for the tunneling conductance.
The model of Fig.~\fign{samp}b is only appropriate
for extracting the tunneling conductance in the limit of infinite DOS
in the 2d electron gas; for finite DOS, the interpretation of the
tunneling conductance in terms of $f_{peak}$ and the other fitting
parameters must be modified.
However, the functional forms indicated by the model
for the loss tangent and the capacitance vs.\ frequency are valid
for all values of the DOS.\pcite{CHAP4}
For the present work only the
values of $C_{low}$ and $C_{high}$, extracted correctly
from fitting data using Eq.~\equn{cfit}, are important in determining
the DOS.

\figinsert{\vfil\vskip6.5truein\includegraphics{nofield.ps}}{nofield}

Figure~\fign{nofield} displays $C_{low}$ and $C_{high}$ obtained from
fits as a function of gate bias for sample {\bf B} at 4.2~K and with
no applied magnetic field. At the lowest gate biases (below -80~mV)
the quantum well is completely devoid of electrons. At these gate
biases, the capacitance of the sample measured as a function of
frequency is a constant.
As the gate bias is increased, electrons enter the well, and there is
a sharp rise in the low frequency capacitance. At this step, the bound
state energy in the well is dropping below the Fermi energy in the
substrate. Beyond this step, the low frequency capacitance is almost
constant aside from a slight slope due to the shift of the mean
position of charges in the well and the substrate with gate bias.
This constancy reflects the fact
the 2d DOS in the absence of magnetic field is a constant in energy.
The high frequency capacitance continues to measure the capacitance
with no charge transfer to the well and no significant
change is observed in this capacitance as the well fills. The slight
slope in $C_{high}$ seen in Fig.~\fign{nofield} is due to a variation
with gate biases in the position of charges in the substrate.

\figinsert{\vfil\vskip6.5truein\includegraphics{wfield.ps}}{wfield}

Figure~\fign{wfield} shows $C_{low}$ and $C_{high}$, again for sample
{\bf B}, at 2.0~T and 2.1~K. $C_{low}$ develops some obvious
oscillations in the presence of magnetic field in regions of gate bias
where the well contains electrons. These are due to the Landau level
DOS now reflecting the variations of the 2d DOS in the well in the
presence of magnetic field. Again,
$C_{high}$ does not probe the DOS in the well, and consequently does
not undergo oscillations as the gate bias is varied.

In previous
capacitance DOS determinations from samples with the same essential
structure as ours, where charge can be
transferred from a 2d gas to a conducting substrate,
only capacitances
analogous to $C_{low}$ have been measured.\pcite{TREY-ext}
The next section of this
chapter will show that in measuring both $C_{low}$ and $C_{high}$
a much more complete determination of the DOS can be made
with no sample parameters other than sample area
entering the calculation.

\v\par\penalty-1000
{\noindent{\bf 3.4 Analysis Procedure}}
\nobreak\v\nobreak
We present an analysis procedure which allows us to
extract the density of states in the quantum well. The desired result
is to extract the DOS $g$ from the measurements of $C_{low}$ and
$C_{high}$, and to plot this DOS as a function of Fermi energy in the
well, measured with respect to the bound state energy in the well.
It should be understood that references to the DOS or $g$ in this
section describe the thermodynamic DOS, $\partial{n}/\partial\mu$. (In
a single particle picture it may be thought of as the zero temperature
DOS $g_0$ convolved with the derivative of the Fermi distribution function.)
This section describes how the DOS is extracted from the data
and how a conversion is made from gate bias to energy in the well.

We start with
a model which places all of the electronic charge in the well at
a plane in the well and treats charges in the substrate and contact as
planar charges near, but not necessarily at, the physical interfaces
between insulating and metallic regions.
Later, we make a correction for the
sheet charge approximation of the distribution of charge in the well.
As we will discuss, the shape of the
electron charge distribution in the $x$ (vertical) direction in the
well is important.
There is no change of the DOS
determination in going to a more precise model which considers
charges in the substrate and contact regions to be distributed in the
$x$ dimension.
For simplicity, we neglect the effect of the doping
spike here. This has no effect on the DOS determination.
Also, we will neglect the effects of the nonzero polarizability of
electrons in the well. The
contribution of this polarizability to the measured capacitance is
less than 1\% of the sample capacitance, and a simple correction can
be made for its effects on the DOS results.

Referring to the sample geometry outlined in Fig.~\fign{samp},
Poisson's equation determines the following set of equations:
$${U_g={e^{2}\over\epsilon}[\sigma_gx_g+\sigma_wx_w]}\eqno{(\equn{U_{gate}})}$$
$$U_w=-{e^{2}\over\epsilon}\sigma_sx_w\eqno{(\equn{U_w1})}$$
$$eV_{gate}=U_g+E_{Fg}-E_{Fs}.\eqno{(\equn{vgate})}$$
All definitions given in this paragraph are understood to be in the
sheet charge model.
$x_w$ and $x_g$ are the distances
between the mean position of the excess charge density
in the substrate and the mean position of the electronic
charge distribution in the well and gate (top contact) respectively.
$U_w$ is the potential energy at the conduction band edge,
again compared to the band edge in the substrate,
in the well at position $x_w$.
$U_g$ is the energy of electrons at the band edge
in the gate region compared to the
band edge in the substrate.
$V_{gate}$ is the gate voltage applied from gate to substrate,
and the electronic charge is taken
as positive in this discussion in order to make comparison with to
figures. $E_{Fg}$ is the Fermi
energy in the contact region (gate) measured with respect to the conduction
band edge in the contact region, and $E_{Fs}$ is the Fermi energy in
the substrate measured with respect to the conduction band edge in the
substrate.
$e$ is the magnitude of the electronic
charge, and $\epsilon$ is the dielectric constant of the medium
(differences in dielectric constant between GaAs and AlGaAs do not
influence our DOS determination).
$\sigma_g$, $\sigma_w$, and $\sigma_s$ are the number densities of
excess electrons in the gate region, the well, and the substrate
region respectively.

Electron transfer through
the tunnel barrier brings the electron gases in the substrate and in
the well into equilibrium. If the density of states in the well is
constant, the number density of electrons in the well is given by,
$$\sigma_w=\int\limits_{U_{bound}}^{E_{Fs}}g(E)dE
=g[E_{Fs}-U_{bound}].\eqno{(\equn{fill1})}$$
$U_{bound}$ is the energy of the ground state in the well, which is
the only well electronic state ever occupied in our measurements,
again measured with respect to the band edge in the substrate. If, in the
sheet charge model, the bound state energy is considered to be at a
fixed energy $E_0$ with respect to the potential energy in the well at
position $x_w$ (the position of the sheet), then we could write
$$U_{bound}=U_w+E_0.\eqno{(\equn{usheet})}$$
The more realistic model of charge distributed in the well and a bound
state energy which depends on the potential at positions other than $x_w$
gives
$$U_{bound}=U_w+E_0-\eta\sigma_w.\eqno{(\equn{eta2})}$$
The additional term,
linear in the well charge density,
corrects the bound state energy for the nonzero width of the
electronic charge distribution in the well.
The bound state energy difference between the sheet and distributed
charge models,
$\eta\sigma_w$, serves as a correction
in the sheet charge model for both
the band edge energy difference given by the models at position $x_w$
and the quantum mechanical
variation in the bound state energy
in the well with respect to $U_w$ due to changes in the shape of the
potential along the well bottom as the well is filled.
Modeling the electrons as
being in a plane at the center of the well overestimates the
electrostatic energy at position $x_w$. Also, the quantum
mechanical energy of the bound state compared to the conduction band
edge at position $x_w$ decreases due to increased curvature of the
well bottom as electronic charge is added to the well.
These facts indicate that $\eta>0$.
A detailed discussion of the term, $\eta$, is given in Appendix A.

Replacing $U_{bound}$ in Eq.~\equn{fill1} with the expression from
Eq.~\equn{eta2}, we obtain
$$\sigma_w={g\over 1-g\eta}[(E_{Fs})-(U_w+E_o)].\eqno{(\equn{fill})}$$
Our analysis requires the differential form
$$\delta\sigma_w={-g\over 1-g\eta}\delta U_w,\eqno{(\equn{delsig})}$$
since our interest is in a nonconstant DOS function $g$.
Finally, charge neutrality dictates that
$$\sigma_g+\sigma_w+\sigma_s=0.\eqno{(\equn{sumsig})}$$
Solving differential forms of
Eqs.\ \equn{U_{gate}}, \equn{U_w1}, \equn{vgate}, \& \equn{sumsig}
and Eq.~\equn{delsig}
we obtain the following differential
changes in charge density per change in the voltage
applied to the gate:
$$\delta\sigma_w={-\epsilon gx_w\over
{e^2g(x_g-x_w)x_w+\epsilon(1-g\eta)x_g}}\delta(eV_{gate}),$$
and
$$\delta\sigma_s={\epsilon(1-g\eta)\over e^2gx_w}\delta\sigma_w.$$
Further, we know by definition that $d\sigma_w/dU_{bound}=g$.
Thus the ``lever-arm'', or change in energy of the bound state
per change in gate energy, is
$${dU_{bound}\over d(eV_{gate})}={x_w\epsilon\over{e^2g}(x_g-x_w)x_w+\epsilon(1-g\eta)x_g}.\eqno{(\equn{levx})}$$
In the case $g=0$ this expression reduces to the simple
``geometric lever-arm'' or\break
$dU_{bound}/d(eV_{gate})=x_w/x_g$. Knowledge of
the lever-arm will allow us to
convert the DOS known as a function of gate bias into DOS as a
function of $E_{Fs}-U_{bound}.$

The capacitance $C_{low}$, measured at frequencies small compared with
$1/2{\pi}RC$, where $RC$ is the characteristic relaxation time of the
tunnel barrier, is then
$$C_{low}=-Ae{\delta\sigma_w+\delta\sigma_s\over\delta V_{gate}}=
{A\epsilon}{e^2gx_w+\epsilon(1-g\eta)\over
e^2g(x_g-x_w)x_w+\epsilon(1-g\eta)x_g},\eqno{(\equn{clow})}$$
where $A$ is the area of the sample.
Pausing briefly to examine this equation, we see that the value of
$C_{low}$ can vary between $C_{low}=Ae/(x_g-x_w)$ for infinite DOS to
$C_{low}=Ae/x_g$ for zero DOS in the well. Physically, this can be
explained as follows. In an infinite DOS capacitor model, electrons
added to the well completely screens the region between the well and
the substrate where there can thus be no electric field.
However, in the finite $g$ model, the energy of the bound
state, $U_{bound}$ must decrease at a rate inversely proportional to
$g$, thus leaving an electric field in the region between the well and
the substrate. A change in the gate voltage induces charge in the
substrate along with the quantum well; the
applied voltage is dropped over a longer distance,
and the capacitance is less than $Ae/(x_g-x_w)$.

The other measured quantity in our experiment is the high frequency
capacitance, $C_{high}$, measured at frequencies large compared to
$1/2{\pi}RC$, such that no charge transfer can occur between substrate
and well. In our model, the value of $C_{high}$ is
$$C_{high}={A\epsilon\over x_g}.\eqno{(\equn{chigh})}$$
It is clear that for zero DOS, $C_{low}=C_{high}$, and for $g>0$,
$C_{low}>C_{high}$. These ideas can be compared to the data shown in
Fig.~\fign{wfield}. At gate biases below $-$50 mV, the bound state in
the well is at a higher energy than the Fermi energy in the substrate.
In this case, the DOS in the well can be considered to be zero. For
this range of voltages, the figure shows that $C_{low}=C_{high}$ as
indicated by the ideas given here. At higher gate voltages, the bound
state in the well begins to fill, and $C_{low}$ increases to a value
larger than $C_{high}$. The value of $C_{low}$ however, varies due to
the variation in the DOS at the Fermi energy as the gate bias is
varied. The maxima and minima in $C_{low}$ correspond to maxima and
minima in the Landau level DOS respectively.

\v\par\penalty-1000
{\noindent{\sl {\bf 3.4.1} Lever-Arm and DOS
Determined from Measured Quantities}}
\nobreak\v\nobreak
Now we consider the lever-arm once again. We introduce the quantity
$$C_{geom}={A\epsilon\over x_w},$$
the ``geometric capacitance'' between the well and the substrate.
In terms of this quantity and $C_{low}$, the lever-arm becomes
$${dU_{bound}\over d(eV_{gate})}={C_{low}\over
Ae^2g+C_{geom}(1-g\eta)}.\eqno{(\equn{lev1})}$$
Eqs.\ \equn{clow}, \equn{chigh}, \& \equn{lev1} may be solved to give
$${dU_{bound}\over d(eV_{gate})}={C_{low}\over
C_{geom}}-({C_{low}\over C_{high}}-1)(1-{\eta
C_{geom}\over Ae^2}),\eqno{(\equn{lever})}$$
and
$$Ae^2g={C_{geom}({C_{low}\over
C_{high}}-1)}{d(eV_{gate})\over dU_{bound}}.\eqno{(\equn{dos})}$$

Equations \equn{lever} \& \equn{dos} are the core of our analysis.
Both the lever-arm and the DOS are obtained through them.
Equation \equn{dos}, can be stated another way. The differential
change in charge density in the well for a differential change in gate
voltage is
$$gdU_{bound}=d\sigma_w={C_{geom}\over Ae}({C_{low}\over
C_{high}}-1)dV_{gate}.$$
For infinite $g$, $C_{high}$ becomes the series combination of
$C_{low}$ and $C_{geom}$, and this equation reduces to
$$d\sigma_w={C_{low}\over Ae}dV_{gate},$$
with $C_{low}$ given in this case by $A\epsilon/(x_g-x_w)$. This is
exactly what would be expected for an infinite DOS capacitor.

If confident of a model for $\eta$, and knowing the geometry of the
sample to obtain $C_{geom}$, Eqs.\ \equn{lever} \& \equn{dos}
give the desired quantities; but modeling of the sample typically will
{\it not} give sufficient accuracy in determination of unmeasured
parameters to allow for confidence in the DOS results.

Our method requires no such sample modeling. Two normalization conditions
are available to determine $C_{geom}$ and $\eta$
through knowledge of the experimentally measured
quantities, $C_{low}$ and $C_{high}$.
For the time being, we treat the parameters $C_{geom}$ and
$\eta$ as constants over the range of gate bias applied to the sample;
their values are expected to change by less than a few percent over
the range of our measurements.

The degeneracy of a Landau level, when
the sample is placed in magnetic field perpendicular to the electron
gas in the quantum well is the number of flux quanta threading the
sample per unit area times the spin degeneracy of two, or $2Be/h$.
Assuming a fixed value of $C_{geom}$ (for nonconstant $C_{geom}$
this procedure yields an averaged value of
$C_{geom}$ over the range of gate biases that comprises one Landau
level) as the density in the quantum well changes, we obtain
$${2Be\over h}={C_{geom}\over Ae}\int\limits_{L. level}({C_{low}\over
C_{high}}-1)dV_{gate}.\eqno{(\equn{dosint})}$$ 
This equation determines $C_{geom}$.
The minima in the DOS between Landau levels are easily
identified; they correspond to minima in the measured quantity,
$C_{low}/C_{high}$.

It is also known that the lever-arm, when integrated in
$V_{gate}$ over a Landau level, must give $\hbar\omega_c/e$, where
$\omega_c$ is the cyclotron frequency. $\eta$,
is determined through satisfying
the equation
$$\hbar\omega_c/e=\int\limits_{L. level}{C_{low}\over
C_{geom}}-({C_{low}\over C_{high}}-1)(1-{\eta C_{geom}\over
Ae^2})dV_{gate}.\eqno{(\equn{leverint})}$$
With $C_{geom}$ and $\eta$ determined by these conditions, Eq.\
\equn{dos} gives the DOS directly from the data and knowledge of the
sample area.
Further, through Eq.~\equn{lever} this technique yields the DOS as a
function of Fermi energy in the well, in contrast with existing work.

\v\par\penalty-1000
{\noindent{\sl {\bf 3.4.2} Further Contributions to the Analysis Procedure}}
\nobreak\v\nobreak
There are several other additions to the analysis procedure used in
this chapter that are not explicitly
included in the equations derived above. One of these is the motion of
the mean positions of the charge densities in the gate, well, and
substrate as the gate bias is varied. Of chief concern is the
variation in the spacing between the mean positions of the charge
densities in the well and in the substrate. Variation of this spacing
as the well is filled will cause $C_{geom}$ to vary also.
Lebens\cite{JOHN} estimates a shift of the mean position of the 
substrate charge of 10 $\AA$
over the range of gate biases used in sample {\bf A} using the
Thomas-Fermi approach of Baraff and Appelbaum.\pcite{APP-accum}
The same calculations carried out on samples {\bf B} and {\bf C} give
similar results.
A perturbation theory calculation on the electronic wavefunction in
the well and self consistent computer calculations\cite{STER-pri}
indicate that the mean position of electrons in the quantum well
also changes by only about $10 \AA$ as the well density is varied
through its range in each of our samples. Moreover, these two charge
densities move somewhat in tandem, both the charge in the substrate
and the well move closer to the top gate as the gate bias is made more
positive. A 10 $\AA$ change in the well charge to substrate charge
separation over the range of gate biases used would indicate a 3\%
change in $C_{geom}$ over the same range.
The parameter $\eta$ is also subject to small variations
as the gate bias is varied. The assumption that $\eta$ is a constant
is equivalent to the statement that there is little change in the
shape of the ground state wave function of the well as the well is
filled with different densities of electrons. Perturbation
theory arguments, calculating the change in the shape of the charge
density as the gate bias induces a change in the shape of the bottom
of the well, give an expected
variation in $\eta$ of about 3\%.

Simulations of our data have shown
show that the addition of a small term, linear in gate bias, to $C_{geom}$
can successfully take the these small variations into account in
calculating the DOS. This linear addition to $C_{geom}$ is adjusted so
that the analysis procedure results in a DOS which is constant in the
absence of magnetic field. This technique is described in detail in
appendix B.

There is also the question of the effects of the shift with gate
voltage of the mean positions of charges in the substrate, well, and
gate contact regions, and the contribution of these shifts to the
device capacitance.  Motion of these charges makes a contribution to
the measured device capacitance. Consider a parallel plate capacitor
with plate separation $x$. If the plate separation changes with a
rate $dx/dV$ then added to the simple geometric device capacitance
is
$$C_{motion}={dQ\over dx}{dx\over dV}=CV\biggl({-1\over
x}\biggr){dx\over dV}.\eqno{(\equn{motion})}$$
It is immediately clear that this capacitance due to the motion of
the charge plates grows linearly with the charge on the capacitor
plates. Using the numbers given above for the distances between
charged regions in our samples and for the motion of the well charge
as the gate bias is varied, we arrive at a rough estimate for this
effect in our samples. $C_{motion}$, due to the motion of charges in
the well, increases from zero to at most 1\% of the device capacitance
when the well density is $6\times10^{11}$ cm$^{-2}$.
This adds to both the low and the high frequency capacitances
measured.
The high frequency capacitance is also effected by
motion of the charge in the substrate.
These are small effects whose changes to
the DOS determination can be nulled by appropriate choice of
$C_{geom}$ using the technique described above.

\v\par\penalty-1000
{\noindent{\sl {\bf 3.4.3} Application of Analysis to Capacitance Data}}
\nobreak\v\nobreak
Before describing our DOS results in detail,
we give another brief description,
this time considering the data of Fig.~\fign{wfield}, of the
procedure which gives the DOS results from the capacitance data.
Referring to Fig.~\fign{wfield}, it is clear from Eq.~\equn{clow}
for the low frequency capacitance that
minima in $C_{low}$ correspond to minima in $\partial n/\partial\mu$
in the well.
We integrate the ratio $C_{low}/C_{high}$ over gate bias
between these minima, as indicated by Eq.~\equn{dosint}, in
order to determine $C_{geom}$. The two vertical lines shown
in Fig.~\fign{wfield}
demarcate the limits of integration. Integration is carried out here
over two Landau
levels in order to minimize error due to uncertainty in the position
of the minima. We estimate
errors in determining the exact
positions of these minima, which limit the accuracy of our measurement of
$C_{geom}$, to be about 3\%.
Errors in $\partial n/\partial\mu$ due to error this
in the value $C_{geom}$ used in the DOS calculation are of nearly
the same value. The value $C_{geom}$ obtained this way (corresponding
to a well charge to substrate charge distance of about 325
$\AA$ in sample {\bf B}) is in good
agreement with simple ideas about where charge is positioned in the
quantum well and substrate. In appendix B, we present a method for
obtaining the value of $C_{geom}$ to even greater accuracy.

\figinsert{\vfil\vskip6.5truein\includegraphics{2.0TH.ps}}{2.0TH}

Using this value of
$C_{geom}$ and starting with a value of $\eta$ determined from
perturbation theory,\pcite{JOHN} we determine the
lever-arm from Eq.~\equn{lever} and $\partial n/\partial\mu$
from Eq.~\equn{dos} as
functions of gate bias. Integrating the lever-arm, we plot the
DOS as a function of well energy. The value of $\eta$ is then adjusted
so that the peaks lie $\hbar\omega_c$ (using $m^*=.067m_0$)
apart in energy as in Fig.~\fign{2.0TH}.
We thus determine the only two unknowns in the DOS
determination. The only sample parameter relied upon
for this calculation is the area of the mesa on which the measurements
were taken.
This is known to better than 1\%.
Using two different values of the area in the analysis procedure,
differing from one another by 1\%, yields indistinguishable results
on the scale of the graphs presented in this chapter.

The analysis procedure discussed above assumes $C_{geom}$ and $\eta$
to be independent of gate voltage. The data of Fig.~\fign{nofield}
display a slight slope to both $C_{low}$ and $C_{high}$ in the region
of gate biases where the well contains electrons. This slope arises
largely from the fact that charges in the well and in the substrate
move closer to the gate as the gate bias is increased (made more
positive). These smaller distances of charges
to the gate at higher gate biases result in larger capacitances. 
Also changing may be the spacing between charges in the well and in
the substrate.
Empirically, we find that we need to
supplement the analysis described above by including a small variation
in $C_{geom}$, linear in gate voltage, amounting to 3\% of $C_{geom}$
over the full range of gate voltage. This term was included in the
analysis leading to the results of Fig.~\fign{2.0TH}.
The magnitude of the linear
correction is chosen to assure that the density of states deduced at
zero magnetic field is independent of filling. This term also serves
to correct the DOS results for
the small variation in $\eta$ as the well is filled; the DOS
calculation is much less sensitive to this variation than the
$C_{geom}$ variation. It also corrects for the very slight effects of
the contribution of the motion of charge planes to the
capacitance. This linear variation in $C_{geom}$ is the {\it only}
variation in a parameter with gate bias in our analysis. Data
simulated by computer and analyzed using our procedure demonstrates
the validity of correcting for these other variations by adjusting the
linear term in $C_{geom}$. In essence, there are three first
order variations which are corrected by the linear term in $C_{geom}$,
chosen empirically to give a constant DOS in zero magnetic field.

Note that the zero on the
abscissa in Fig.~\fign{2.0TH}
is chosen arbitrarily. The first oscillation in the graph is
the lowest Landau level in the well, at an energy of 0.5$\hbar\omega_c$
above the energy of the ground state of the well.
We set the second and third peaks seen in the figure
$\hbar\omega_c$ apart by adjusting the parameter $\eta$.
As is clear in the figure, the third and fourth peaks
also fall $\hbar\omega_c$ apart. This serves as one of several checks
of the validity of our method. Notice also that the spacing between
the first and second peaks is not $\hbar\omega_c$ as it should be. We
believe that this incorrect spacing occurs because of the breakdown of the
assumption, implicit in our model, of uniform filling in the well in
this region of well energy. As the gate bias is lowered into this
region, the well becomes depleted of electrons in a nonuniform
fashion,\pcite{DAVIES-pud} and this ``puddling'' makes interpretation
of the data more difficult in this region of energy. One can no longer
think of electric field lines between the well and the substrate all
pointing perpendicular to the plane of the 2d electron gas. Further,
portions of the 2d electron gas are depleted of electrons; it is
difficult to determine whether a decrease
in the low frequency capacitance may be due to an increase in depleted
area or a decrease in the 2d DOS.

\figinsert{\vfil\vskip6.5truein\includegraphics{intlev.ps}}{intlev}

The integrated lever-arm used to convert the DOS results from 
data of Fig.~\fign{wfield} from gate bias to well energy is shown in
Fig.~\fign{intlev}. Most interesting here are the oscillations
superimposed on a linearly increasing background.
These can be thought of
qualitatively in a very simple way. When the DOS is large, electrons
can be added to the well with little change in Fermi energy.
This explains the
regions of the figure that are more ``flat'' in gate bias. The steeper
regions occur where the DOS is low, and there is a larger change in the
Fermi energy in the well for each electron added to the well.

\figinsert{\vfil\vskip6.5truein\includegraphics{zerofield.ps}}{zerofield}

Figure~\fign{zerofield} displays $\partial n/\partial\mu$ in
zero field determined using
the same values for the parameters $C_{geom}$ and $\eta$ used in the
determination of the 2 Tesla DOS above. The horizontal line drawn in
the figure is the expected value of the 2d DOS, $2.8\times10^{13}
eV^{-1}cm^{-2}$. As stated earlier a term linear in gate bias is added
to $C_{geom}$ to make the DOS constant over a wide range of well
energy. Without this linear term, the curve obtained for
$\partial n/\partial\mu$ appears to have a slight
slope. Our model for this system works well in the range of energy
where $\partial n/\partial\mu$ shown here is flat.

The zero on the horizontal scales in Figs.\ \fign{2.0TH} \&
\fign{zerofield} have been adjusted to correspond to the same Fermi
energy with respect to the bound state in the well. Our method for
determining that they are the same is explained in appendix C. Our
analysis procedure should be valid over the region where the curve in
Fig.~\fign{zerofield} is flat (at energies above $-$2~meV). Hence,
it is believed that the DOS presented in Fig.~\fign{2.0TH} is also
valid at energies above $-$2~meV in that figure.

\v\par\penalty-1000
{\noindent{\sl {\bf 3.4.4} Effects of Thermal Broadening}}
\nobreak\v\nobreak
Recall again that our method results in determination of $\partial
n/\partial\mu$, the thermally broadened DOS.
Before attempting to describe systematically the lineshapes we
observe experimentally, it is important to understand the effects of
temperature on the lineshapes.
This section describes the observed effects of nonzero temperature on
the DOS results in magnetic fields weak enough so that effects of the
exchange enhanced spin splitting are not observed.

The bulk of the $\partial n/\partial\mu$ results to be shown in this chapter
are for 
temperatures around 2.0~K or $k_BT$ of approximately 0.2~meV.
This is much
smaller than the linewidths (full width half maximum)
of about 2~meV of the Landau levels shown so far.
The scattering
effects which cause the broadening of the DOS from ideal delta
functions are not expected to be temperature dependent in this
temperature range; any broadening due to increased temperature here is
likely to be simple thermal broadening.
In order to see just
how much change should be expected in the lineshapes, we have
experimented, on the computer, with adding thermal broadening
to various presumed underlying shapes
for the DOS. The thermally broadened DOS is given by:
$${\partial n\over \partial\mu}={\int\limits_{0}^{\infty}}g_0(E){\partial
f(E-\mu)\over\partial\mu}dE.\eqno{(\equn{fold})}$$
Here $g_0(E)$ is the zero temperature DOS.

We have looked at Lorentzian lineshapes of width 0.9~meV. These, when
convolved with the derivative of the Fermi function as in Eq.\
\equn{fold}, show a 5\% reduction in peak height with very little
change in the rest of the curve. Moreover, the qualitative shape of
the curve does not change with this small amount of thermal broadening.
Thermally broadened Gaussian lineshapes show even less
change owing to their more rounded maxima. We conclude that for the
widths of our lineshapes observed at 2~K, the lineshapes
are a reasonable approximation to the zero temperature DOS and can be
used for the purpose of determining lineshapes. A deconvolution of the
exact zero temperature DOS from our data has so far proved to be too
sensitive to error to be useful. We have taken data at much lower
temperatures (down to 90~mK)
on sample {\bf A}\cite{YOPRL}; though these data are
significantly noisier than those presented here, the peaks
sharpen only slightly at lower temperatures.

\figinsert{\vfil\vskip6.5truein\includegraphics{fold.ps}}{fold}

We now present a method for checking the
applicability of the DOS determination techniques described so far.
The systematic effects of inhomogeneity in the well
which could cause error in the DOS obtained using our
technique can be ruled out, and the general validity of our analysis
procedure confirmed, if the temperature
dependence of $\partial n/\partial\mu$ is in accord with the thermal
broadening given in Eq.~\equn{fold}.
Fig.~\fign{fold} shows the results of treating the 4~T data of sample
{\bf A} at 1.9~K as though it is the zero temperature DOS, and using
Eq.~\equn{fold} to 
simulate a thermally broadened DOS at 7.0~K.
The simulation is compared with data actually taken at 7.0~K. The two
curves are in generally very good agreement. At half maximum, the
7.0~K data is slightly narrower than the convolved 1.9~K data. This
is, at least in part, due to the convolution having been done on 1.9~K
and not zero temperature results. Simulations of 0.9~meV wide
Lorentzian levels broadened first to 1.9~K and then convolved again to
7.0~K show similar results when compared with 0.9~meV wide Lorentzians
broadened directly to 7.0~K. This effect is also partially 
responsible for some of the difference in peak heights seen in Fig.\
\fign{fold}. Also, the 7.0~K data is much noisier than the 1.9~K data,
and the DOS results at the peaks tend to be biased towards larger DOS
by noise. In all, the deviations between the two results
are slight, and serves as a confirmation of the sample model.
Both the quantitative values of the DOS measured and the qualitative 
``shape'' of the Landau peaks observed are meaningful
for comparison to models.

\v\par\penalty-6000
{\noindent{\bf 3.5 Low Field Landau Level Fitting}}
\nobreak\v\nobreak
The ``shape'' of Landau level peaks in the 2D electron gas as a
function of Fermi energy has been of great interest in the last
several years. Several experiments done with a fixed 2d electron gas
density\cite{EIS-DvA,STM-spc-ht,TSUI-heat} or Fermi
energy,\pcite{KVK-mcap} often cited as giving weight to a particular
shape of the DOS, determine the DOS at fixed density or fixed Fermi energy
as a function of magnetic field, DOS(B).
These groups have assumed a form for the DOS as a function
of energy, DOS(E), and determined what the DOS(B) at constant density
{\it would} look like given the assumed form. This deduced DOS(B) is
then compared to the data.
The novelty of the experiment presented here lies in the
direct observation of the DOS as a function of Fermi energy. We can
now test ideas for the DOS directly on our results.

In this section, we examine several key issues involving the Landau
level DOS. In particular, we test to see if the DOS peaks are fit
better by Lorentzian or Gaussian lineshapes. Also investigated are the
widths of the Landau level DOS peaks and the dependence of this width
on both the electronic density of the 2d electron gas and the strength
of the magnetic field. Finally, we carry out this survey on samples
{\bf A}, {\bf B}, and {\bf C} to gain information on the sample
dependence of our results.

Many different ideas exist for what the shapes of the levels should
be. We concentrate here on the DOS in magnetic fields low enough
that the exchange enhanced spin splitting does not make an important
contribution to the width of the observed Landau level, saving the
high field results for a later section.
The short range scattering model of Ando and
Uemura\cite{ANDO-ellip,ANDO-F-S} predicts elliptical lineshapes for
the Landau level peaks. In this model, the width of density of states
peaks can be determined directly from scattering times obtained from
the zero field sample mobility. Others\cite{STM-spc-ht}
have shown that these fit density of states data rather poorly. A more
complete theory including self consistently screened long range
Coulomb scattering and inter-Landau level coupling effects has been
developed by Das Sarma and Xie\cite{DSAR-dos}. For heterostructure
samples with doping in AlGaAs farther than a few 10's of angstroms
from the 2D electron gas the Das Sarma and Xie
theory predicts significantly broader Landau level peaks than
indicated by scattering times obtained from the sample mobility. The
qualitative shape of the Landau level peaks is predicted to depend
greatly on sample parameters other than the zero field mobility.
This more complete theory requires substantial computer calculation and
will not be a source of comparison here.

Several authors\cite{TREY-dos,STM-spc-ht,EIS-DvA,GOLD-exch}
have used a
Gaussian broadened density of states as a function of energy to fit
data of density of states as a function of field. Gornik et.\
al.\cite{STM-spc-ht} find for their specific heat results from their
semiconductor superlattices at magnetic fields up to 8 T that a
Gaussian density of states with an added field independent
background gives a superior fit to a
Lorentzian density of states. Recent theoretical
work\cite{GERH-between} has attempted to explain the necessity for
this background which was initially used for fitting data in an ad hoc
fashion.\pcite{STM-spc-ht}
We fit the DOS results
to the following Gaussian function including a background,
$$D(E)=\beta+({2Be\over h}-\beta\hbar\omega_c){\sum\limits_{i=1}^6}{1\over\sqrt{2\pi}\Gamma_i}exp[-{1\over2}{(E-(i+1/2)\hbar\omega_c)^2\over\Gamma_i^2}].\eqno{(\equn{gauss})}$$
Where $\beta$ is the background density of states, and $1.18\Gamma_i$
is the half width at half maximum (HWHM) for a particular level.
The factor $2Be/h$ in front of the summation arises because we
consider these peaks to be spin-degenerate Landau maxima.

\figinsert{\vfil\vskip6.5truein\includegraphics{fit4ta.ps}}{fit4ta}

For our fits, we first turn to the DOS results in sample {\bf
A} at 4~T and 1.9~K. $\partial n/\partial\mu$ results
are shown as the circles in Fig.~\fign{fit4ta}.
Note again that the lower half of the
lowest level is believed to be
out of the range of validity of our model, as discussed above.
The larger field here has increased the degeneracy of the levels
(the area underneath the peaks), the
cyclotron energy, and the peak to valley ratio of the DOS peaks
compared to the data of Fig.~\fign{2.0TH}, making it easier to
discriminate between the DOS from one peak and that from another.
Note also the change of scale on the abscissa in Fig.~\fign{fit4ta} 
compared to Fig.~\fign{2.0TH},
needed to display the wider range of densities available to sample {\bf A}.
$\beta,\Gamma_1,\Gamma_2,\Gamma_3$ are the four
fitting parameters used in the Gaussian fit of Fig.~\fign{fit4ta}.
The widths
of higher index Landau levels, $\Gamma_4, \Gamma_5,$ and $\Gamma_6$, were
constrained to be equal to $\Gamma_3$. The results of this fit give
$\beta=1.52\times10^{13}$eV$^{-1}$cm$^{-2}$, $\Gamma_1$=1.08~meV,
$\Gamma_2$=.689~meV, and $\Gamma_3$=.597~meV. We note that the
background required for this fit is more than half of the zero field
DOS.

In Fig.~\fign{fit4ta} we also fit the same data to a set of Lorentzians.
The form for this fit is
$$D(E)={2Be\over h}{\sum\limits_{i=1}^6}{1\over\pi}{\Gamma_i\over(E-(i+1/2)\hbar\omega_c)^2+\Gamma_i^2},\eqno{(\equn{lors})}$$
with {\it no} free background parameter.
In this case, there are three free fitting parameters,
$\Gamma_1,\Gamma_2,\Gamma_3$. Again,
$\Gamma_4, \Gamma_5,$ and $\Gamma_6$ were
set equal to $\Gamma_3$ in the fits. In this case the $\Gamma_i$ are the 
HWHM of the levels. The results of this fit give
$\Gamma_1$=1.34~meV, $\Gamma_2$=.969~meV, and $\Gamma_3$=.847~meV.

Both the Gaussians with the large added background and the Lorentzians
give reasonable characterizations of the data. The remarkable feature
of the Lorentzians here is that they fit the data of Fig.\
\fign{fit4ta} so well with only three
free parameters. Also, the Gaussians tend to bow out more than the
Lorentzians near the maximum of the DOS peak and broaden less than the
Lorentzians at the base of the Landau levels. Both the narrowness of
the DOS results near the peak and the breadth of the DOS peaks near
the base of a level favor the Lorentzian shape over a Gaussian shape
even after the ad hoc background is added to the Gaussians. With no
background, the Gaussians do much less well at fitting (fit not shown)
to the data of Fig.~\fign{fit4ta}. The main difficulty with no
background is that they cannot fit to the large interlevel DOS. This
contrasts with Lorentzian shapes, which more naturally fit the 
interlevel DOS as a consequence of their slower fall off 
compared to Gaussian shapes at energies well away from the peak DOS.

\figinsert{\vfil\vskip6.5truein\includegraphics{fit3ta.ps}}{fit3ta}

We continue fitting to results from sample {\bf A}. Figure
\fign{fit3ta} shows the 3 T results from sample {\bf A}, now displayed
with both Gaussian with background and Lorentzian fits. For the
Gaussians, there are now five free parameters, the widths of the first
four levels (again higher index levels are constrained to have
the same width as level 4)
and the background. In this case, the nonlinear least squares
fitting routine that we used to fit the data chooses a negligibly small
background. It does this because the interlevel DOS between different
levels is a strong function of filling, and a constant background does not
improve the fitting in this region. The Gaussian widths
widths from fits are made larger than necessary to fit the main part of the
peaks in an attempt (by the nonlinear least squares fit) to fit the
interlevel DOS. The Gaussian widths are a compromise between fitting
the Landau level peaks and the between level DOS. There is no such
conflict for the Lorentzian fits where no distortion of the peak width
occurs.
The data are fit in the range from
-7~meV to 10~meV where the DOS results are believed to be valid. The
four free parameters in the Lorentzians fits are the widths of the
first four levels.
The Lorentzian shapes in Fig.~\fign{fit3ta} again seem to capture
the essential shape of the levels better than the Gaussians.
Only in the low density regime is there significant deviation of
the Lorentzian fits from the data.
Aside from a very slight deviation between
Landau levels, the fits are almost identical to the data at higher
densities. 
The fitting parameters obtained are listed in the figure caption.

\figinsert{\vfil\vskip6.5truein\includegraphics{fit2tb.ps}}{fit2tb}

We now fit to the data of sample {\bf B} which shows much higher DOS
contrast at low fields than sample {\bf A}. in Fig.~\fign{fit2tb} we
fit the 2.0~T 2.1~K data of sample {\bf B}. This time, only Lorentzian
fits are plotted as they very obviously fit the DOS results better.
The widths of the Lorentzians are again described in the figure caption.
The quality of fit in this case is excellent.
At the second level and beyond, there is almost no deviation of the
fits from the data anywhere along the curves, even between Landau levels.
3~T results from the same sample give similar agreement. At 4~T,
the exchange enhanced spin becomes so large in this sample as to cause
great difficulty in determining the underlying lineshape.

\figinsert{\vfil\vskip6.5truein\includegraphics{fit2ta.ps}}{fit2ta}

For purposes of comparison, we move back to sample {\bf A}. Fig.\
\fign{fit2ta} displays $\partial n/\partial\mu$ in this sample at
2.0~T and 1.9~K. There are actually six levels plotted here. Five are
visible as large DOS oscillations, and the other is a very broad level
in the low density regime. In comparison with Fig.~\fign{fit2tb} for
sample {\bf B}, there are several features here that are particularly
noticeable. One is that the amplitude of the oscillations in sample
{\bf A} is much smaller than those of sample {\bf B}. Depending on the
Landau level index, the amplitude of the oscillation is between 30\%
to 50\% smaller. This, we attribute to lower mobility in sample {\bf
A} probably mostly due to the presence of a doped
layer in the AlGaAs blocking barrier. Note also that there is a larger
dependence of the amplitude of oscillation on Landau index in sample
{\bf A} than for sample {\bf B}.
Finally, and most significantly,
notice that despite these differences Lorentzians fit both data sets
well.

\figinsert{\vfil\vskip6.5truein\includegraphics{fit4tc.ps}}{fit4tc}

Finally, Fig.~\fign{4tc} displays $\partial n/\partial\mu$
in the well in sample {\bf C}
at 4.2~K and 4.0~T. Although the temperature is 4~K, the thermal
contribution to the width is small compared to the observed full
width of around 2.5~meV of these levels.
Notice here that the amplitude of oscillation is much
smaller than in either samples {\bf A} or {\bf B}, and the levels are
about 20-30\% broader than those in sample {\bf A} at the same field. We
think that this breadth arises from the very high doping density in the
AlGaAs blocking barrier in sample {\bf C}.
The solid line in this figure is a fit to a
Lorentzian lineshape.
The fit is again quite good with only a small discrepancy
between the fit and the data between levels. The levels here are much
broader, with $\Gamma_1$=1.70~meV and $\Gamma_2$=1.14~meV. For
the purposes of fitting, higher index Landau levels were forced to have
the same width as the second level.

\v\par\penalty-1000
{\noindent{\sl {\bf 3.5.1} Picture Implied By Fits}}
\nobreak\v\nobreak
The DOS shape fitting implies a picture which is surprisingly
different from that obtained by other experimentalists and many
theorists to this date. The results indicate that for fields below 4~T,
regardless of the doping configuration in the sample, Landau levels
are essentially Lorentzian in lineshape. All three samples exhibit
qualitatively similar lineshapes despite level widths that differ by
more than 50\% between sample {\bf B} and sample {\bf C}.

\figinsert{\vfil\vskip6.5truein\includegraphics{widths.ps}}{widths}

Another striking feature of the data is the magnetic 
field independence of widths of the Landau levels. It is clear in
the DOS results given above that the width of the
levels is dependent on the Landau level index. The level width always
narrows monotonically as the index increases. One notion is
that the level width is independent of field and depends only on
the electronic density in the quantum well. In order to test this
idea we plot, in Fig.~\fign{widths}, the half width $\Gamma$ from 
Lorentzian fits on sample {\bf A} vs.\ the 2d density at which the
Landau level peaks occur.
The results strongly support the idea of
an underlying universal curve for widths vs.\ density with level width
independent of magnetic field strength. Data from sample {\bf B} from
2 and 3 Tesla support the same conclusion. We have not taken
sufficient data on sample {\bf C} to verify this result in that sample.

The solid curve in Fig.~\fign{widths} is a power law fit using the
form, $$y=Cx^{\gamma}.$$
Here $C$ is a constant and $\gamma$ is a constant exponent. The
exponent used in the plot is $\gamma=-0.28$. The power law 
gives a reasonable characterization of the data with only two free
parameters. Experimental
work by Hirakawa and Sakaki\cite{SAK-mob} has demonstrated
a power law dependence of the low temperature
mobility of 2d electron gas systems on
the density in GaAs/AlGaAs heterostructures. They obtain exponents of
between 1.1 and 1.7 on plots of mobility vs.\ carrier concentration.
The value of the exponent in their experiment depends most strongly on
the distance from the 2d gas to the ionized impurities in the AlGaAs
layer adjacent to the well. If we consider the elastic scattering time
in our samples to be inversely proportional to the width of the DOS
peaks, and considering the mobility to be proportional to this
scattering time, then our results imply an exponent of 0.28 on a
mobility vs.\ carrier concentration plot. There is an obvious
discrepancy between these results and those of Hirakawa and Sakaki.

It has long been thought that the scattering times associated with
the widths of Landau level DOS peaks and the scattering times derived
from mobility determinations are not the same. Das Sarma and
Stern\cite{STER-mob} point out the differences, arising
from sensitivity of the conductivity to the angle of deflection of
an electrons in scattering events, between the single
particle scattering time and the scattering time that appears in the
Drude formula for the conductivity.
These two scattering times may differ considerably. The position of ionized
impurities which cause elastic scattering in the sample
is thought to be important in determining the relation between these two
times. Assuming that the widths of Landau levels in our samples are
more closely related to the single particle scattering time may
explain the difference between ours and the Hirakawa results.
We cannot measure the mobility of our samples directly;
however, if we assume, as in the Hirakawa experiment, that we would
obtain an exponent of between 1.1 and 1.7 from a mobility vs.\ carrier
concentration plot, then our experiment indicates that the ratio
between these two times varies as the carrier concentration is varied.

The thrust of our results can be summarized in three simple statements.
{\bf 1)}~Landau levels are characterized very well by Lorentzian lineshapes
for field 4~T and below. {\bf 2)} The shape of the DOS peaks is {\it
not} influenced by the details of the doping configuration in our
samples. Only the width varies.
{\bf 3)}~The width of Landau levels is independent of magnetic field
and varies only weakly with electron concentration.

\v\par\penalty-1000
{\noindent{\sl {\bf 3.5.2} Comparison Between These and Previous Results}}
\nobreak\v\nobreak
We first review published widths of
Landau level peaks. Gornik et.\ al.\cite{STM-spc-ht} determine
widths (HWHM=1.18$\Gamma_{Gaussian}$)
of around 0.9 and 1.5~meV, independent of $B$, for two
samples with mobilities of 80,000 cm$^{2}$/V$\cdot$s and
40,000 cm$^{2}$/V$\cdot$s respectively. Fits in that paper considered
the width to be independent of field. Eisenstein et.\
al.\cite{EIS-DvA} obtain a half width of 1.2~meV/T$^{1/2}\sqrt B$,
giving 2.5~meV at 5~T. The
capacitance measurements, at fields up to 2~T
of Smith et.\ al.\cite{TREY-dos} give half
widths of around 1.1~meV. Finally, Wang et.\ al.\cite{TSUI-heat},
obtain a half width of around 3~meV at 4.8~T. The range of half widths
from these groups runs from 0.9 to 3~meV.
Our half widths fall toward the lower end of this range. At densities
between $3\times10^{11}$cm$^{-2}$ and $4\times10^{11}$cm$^{-2}$, we
obtain widths ranging from 0.76~meV for sample {\bf B} to around 1.3
meV for sample {\bf C}.

A clear distinction between our results and those given previously is
the shape of the DOS fits. Three of the above
groups\cite{STM-spc-ht,EIS-DvA,TREY-dos} use Gaussian fits
for comparison to their results. In fact, Gornik et.\
al.\cite{STM-spc-ht} determine that a Gaussian with a constant
background gives a fit to their data which is superior to that given
by Lorentzians.
Wang et.\ al.\cite{TSUI-heat}, in specific
heat measurements, use a Gaussian with a level
width which both is a function of
magnetic field and oscillates with Landau index.
This form, for constant field
and varying filling fraction, produces, as a function of Fermi energy,
peaks which have more the appearance of Lorentzians than
Gaussians but bases which are more like Gaussians. This
is in somewhat better agreement with our results. However, Wang's
results indicate level widths which are more than twice those measured
here. It is important to note that these DOS measurements
as well as several others\cite{STM-spc-ht,EIS-DvA} were done on multiple
layer heterostructures. This raises the possibility that layer to
layer inhomogeneity in samples in these experiments could account for
the discrepancy in the observed lineshapes with those in our
experiment.

Our results indicating a field independence for the Landau level width
are in sharp contrast with the magnetization
results of Eisenstein et.\ al.\cite{EIS-DvA} who suggest a $\sqrt B$
dependence of the width of levels on magnetic field. The level width
of Wang et.\ al.\ is also field dependent. Below 4~T though, they
obtain a field dependence of the width which 
is much smaller than that of Eisenstein. There are also others who
either do not remark on a field dependence of the level
width\cite{STM-spc-ht} or suggest that it is masked by inhomogeneity
in the 2d gas.\pcite{TREY-dos}

\v\par\penalty-6000
{\noindent{\bf 3.6 High Fields}}
\nobreak\v\nobreak
The shape of the DOS peaks in fields above around 4~T in our samples
differs considerably from that in lower fields. This difference
arises from the exchange enhancement of the electronic $g$ factor in
the 2d electron gas in GaAs which causes a large splitting between the
spin subbands of a Landau level at high fields.

This exchange enhancement can be
understood in a simple heuristic fashion. If a
Landau level is half full, then the energy savings to
the system in placing one electron in the lower energy spin
subband instead of the higher energy subband is larger
than that indicated by the Zeeman energy,
$g_{L}\mu_BB$.  Here $g_L$ is the
Land\'e $g$ value for GaAs which has a value of around -0.44 due to
spin-orbit interactions.\pcite{HERM-g}
The increased spin splitting arises from the electron-electron repulsion.
In order to reduce the
electron-electron Coulomb energy, electrons should be spaced as far
apart as possible. If all of the electrons are in the same spin
subband, then by the Pauli principle the spatial wavefunction must be
antisymmetric. The antisymmetric spatial wavefunction here
automatically places electrons farther apart than does a symmetric
one. Energy can be saved by the system by placing all of the
electrons in the same spin subband. 
This increases the energy splitting of the spin subbands.
The augmented splitting can be characterized by an enhanced $g$
value, $g^*$.

Ando and Uemura\cite{ANDO-osc} first determined that the 
enhanced $g$ factor
should be an oscillatory function of filling as the Fermi energy
passes through Landau levels. Recent 
measurements of this exchange enhanced
$g$ value in GaAs have been made using level-coincidence (developed
by Fang and Stiles\cite{STIL-tilt}) of Shubnikov-de Haas oscillations
in a tilted magnetic field and by measuring low temperature
activation of $\sigma_{xx}$ when the Fermi energy is in the middle of
a spin split Landau level.\pcite{KVK-exch}
Goldberg, Heimann, and Pinczuk used a laser absorption probe to
measure thermal occupation of the spin subbands of the lowest
Landau level in order to determine $g^*$. In this section, we use
capacitance measurements to obtain the DOS at large magnetic fields to
which we fit the model of Ando and Uemura and determine an effective
$g$ and make comparisons to previous measurements.

\v\par\penalty-1000
{\noindent{\sl {\bf 3.6.1} DOS analysis in high fields}}
\nobreak\v\nobreak
We have made measurements only on sample {\bf A} at fields above 4~T.
Because they show the spin splitting effects best, we concentrate
here only on the results from 8.5~T, the highest magnetic field strength
used. The DOS determination proceeds the same way as the low field
DOS determination but with one complication. The mean position of the
excess charge density in the substrate
may be different at high magnetic fields than at low magnetic fields.
This can arise because $k_{Fermi}$ in the substrate,
in the direction perpendicular to the plane of the 2d electron gas,
decreases when the magnetic field is increased due to the large
degeneracy of states in the direction perpendicular to the magnetic
field. This means that $C_{geom}$ must be determined separately for
high fields. Eq.~\equn{dosint} is difficult to apply in this case
because there is only one full Landau level observed, and the low
density tail of this level falls into the region of the device
operation where the 2d gas empties in a nonuniform fashion. This
region is outside of the validity of our analysis, so we instead use
the normalization condition, Eq.~\equn{leverint}, to determine
$C_{geom}$, using the same value of $\eta$ as obtained in low fields.
The difficulty is that at 8.5~T the Landau level degeneracy is so
large that the second Landau level, comprised of the $\nu=3$ and
$\nu=4$ spin split levels, is at the high filling edge of our device.
Fig.~\fign{8.5all} illustrates this situation. This figure shows the
DOS at 8.5~T and 1.85~K in sample {\bf A}. The two vertical
lines drawn are $\hbar\omega_c$ apart in energy. It is difficult to
say exactly where the right hand vertical line should be placed due
to incomplete observation of the second Landau level.
We estimate that this situation introduces an error of about 5\% in
the observed spacing between the peaks of the lowest Landau level and
the same error in the effective $g$ value observed. This is still
relatively high precision for a measurement of the enhanced $g$
value.

\figinsert{\vfil\vskip6.5truein\includegraphics{8.5all.ps}}{8.5all}

\v\par\penalty-1000
{\noindent{\sl {\bf 3.6.2} Curve fitting to the spin split level}}
\nobreak\v\nobreak
With this caveat, we proceed with a detailed analysis of the lowest
Landau level. Figs.~\fign{8.5lor}~\&~\fign{8.5gauss}
zoom in on the lowest level from
Fig.~\fign{8.5all}. We now attempt to fit to the detailed shape
of this curve as well as the spin splitting.

\figinsert{\vfil\vskip6.5truein\includegraphics{8.5lor.ps}}{8.5lor}

\figinsert{\vfil\vskip6.5truein\includegraphics{8.5gauss.ps}}{8.5gauss}

The theory of Ando and Uemura\cite{ANDO-osc} indicates that the spin
splitting in a Landau level in the case of only one level occupied (or
for negligible overlap of different Landau level DOS peaks with
several filled levels)
should be given by the following formula:
$$g^*\mu_BB=g_L\mu_BB+E_{ex}(n_{+}-n_{-}).\eqno{(\equn{g})}$$
Here, $E_{ex}$ is known as the exchange energy and $n_{+}$ and
$n_{-}$ are the fractional fillings of each of the spin
split Landau levels.
Let the energy (mean energy) of the two spin subbands of a Landau level
with the Zeeman and
exchange interactions turned off be called $E_{\alpha}$. Turning these
interactions back on, the energies of the two subbands are given by
$$E_{\pm}=E_{\alpha}\pm{1\over2}[g_L\mu_BB+E_{ex}(n_{+}-n_{-})].\eqno(\equn{up})$$
For a given Fermi energy, the fractional occupations of $n_{+}$ and
$n_{-}$ are given by
$$n_{\pm}={h\over{eB}}\int\limits_{-\infty}^{E_F}f(E-E_F;T)g_{\pm}\bigl(E+{1\over2}[g_L\mu_BB+E_{ex}(n_{+}-n_{-})]\bigr)dE.\eqno(\equn{npm})$$
The term in front of the integral is the inverse of the Landau level
degeneracy which serves to normalize the integration, $f(E-E_F;T)$
is the Fermi distribution function, and $g_{+}(E)$ and $g_{-}(E)$
are the subband DOS for each spin direction. (The reader is cautioned
not to confuse the DOS, $g_{\pm}$, with the Land\'e and enhanced $g$
values.)

To make fits to
our data, we start with two DOS peaks, $g_{+}(E)$ and $g_{-}(E)$
(of either Lorentzian or
Gaussian shape), each initially centered at energy $E_{\alpha}$.
The two peaks typically have different linewidths because of the
variation in scattering times with electron concentration in our samples.
Equation~\equn{npm} suggests an iterative solution. We explain first
the most obvious method to obtain a DOS fit and then describe a
more computationally efficient technique that we use.
For a particular value of $E_F$, $n_{+}$ and $n_{-}$ can be calculated
from Eq.~\equn{npm}
starting from the assumption that $n_{+}=n_{-}$. The Zeeman term makes
the values of $n_{+}$ and $n_{-}$ slightly different from each other.
Then these values for $n_{+}$ and $n_{-}$ are placed on the right hand
side Eq.~\equn{npm} once more and new values for $n_{+}$ and $n_{-}$
are calculated. These steps can be continued until $n_{+}$ and $n_{-}$
cease to change between cycles. This process determines $n_{+}$ and
$n_{-}$ for a particular value of $E_F$.

The actual procedure used in computation is slightly different but
leads to the same results.
Starting with no spin splitting, we calculate the functions $n_{+}^0(E_F)$ and
$n_{-}^0(E_F)$ using the equation
$$n_{\pm}^0(E_F)={h\over{eB}}\int\limits_{-\infty}^{E_F}f(E-E_F;T)g_{\pm}(E)dE.$$
We then iterate the following equation, starting with values
$n_{\pm}^0(E_F)$ on the right hand side:
$$n_{\pm}=n_{\pm}^0\bigl(E_F\pm{1\over2}[g_L\mu_BB{\pm}E_{ex}(n_{+}-n_{-})]\bigr).$$
We do this as a function of $E_F$ to obtain $n_+(E_F)$ and
$n_{-}(E_F)$.
Finally, to obtain the total density of states, we use the
following equation:
$${\partial{n}\over\partial\mu}={eB\over
h}{d\over{dE_F}}[n_{+}(E_F)+n_{-}(E_F)].$$
This result is compared to the data.

Band nonparabolicity, even though it has
only a very small effect on the effective mass at these energies may
influence the Land\'e $g$ nonnegligibly.\pcite{HERM-g} We use a value
of $g_L=-0.40$ as do Nicholas et.\ al.\cite{KVK-exch} determined from
photoconductivity\cite{KVK-photo} measurements on their sample. We
recognize that $g_L$ in our samples may actually vary somewhat as the
electronic density is varied, but we do not attempt to correct for
this in fitting.

In Fig.~\fign{8.5lor} and Fig.~\fign{8.5gauss}
we fit the 8.5~T data to Lorentzian and
Gaussian spin split levels respectively
Because the process described above 
for generating the fits is rather complicated, we tune each of the
various parameters by hand, observe the results, and retune. The fits
shown in Figs.~\fign{8.5lor} and \fign{8.5gauss}
are subjectively the best fits for the
two different forms. For both the Lorentzians and the Gaussians,
there are 3 fitting parameters. These are $\Gamma_{+}$ and
$\Gamma_{-}$ the widths of each of the spin split levels and
the exchange energy, $E_{ex}$. The widths used are stated in the
figure caption. The exchange energy determined clearly depends on the
underlying lineshape used. For Lorentzians and Gaussians,
we determine exchange energies of 3.8~meV and 3.3~meV respectively.

Although the model curves in Figs.~\fign{8.5lor} and \fign{8.5gauss}
can be made to fit the experimental
peaks of the spin split DOS reasonably well, it is clear that neither
fit matches the data with nearly the same precision as did the
Lorentzian fits in low fields. With the Lorentzians one problem is
that the fits require a large $E_{ex}$; but when $E_{ex}$ is
increased to fit the peaks well, $\partial n/\partial\mu$ drops too low
in the density of states trough between the two levels. Both the
Lorentzians and the Gaussians have difficulty fitting to the very
steep sidewalls of the observed Landau level DOS. The Gaussians do
less well at fitting these sidewalls, and the Gaussian fit drops to
zero DOS between Landau levels where the data show substantial
interlevel DOS. As with the fits made to data at lower magnetic field
values, the Gaussian gives an inadequate DOS in the tails of the peak.
To compensate for this, the model calculation has a larger width than
otherwise necessary to fit the central portion of the DOS peak.

The value for $E_{ex}$ obtained here, 3.3~meV (from Gaussian fits), is
the same as that determined by Nicholas et.\
al;\cite{KVK-exch} Goldberg, Heiman, and Pinczuk obtain 2.8~meV.
Both of these groups use Gaussian lineshapes to fit their data. We
point out that it is important to consider the effect of the assumed
lineshape on the determination of $E_{ex}$. Our experience indicates
that the value of $E_{ex}$ can vary by 20\% or more depending on the
lineshape used.

One can easily determine the maximum value of $g^*$ from our fits by
equating the maximum value of $E_{+}-E_{-}$, obtained
with the Fermi energy at the Landau level center, to $g^*\mu_BB$. In
doing this, we obtain a value of $g*$ of 4.6 for Lorentzian fits
(maximum splitting of 2.3~mV) and 5.4 for Gaussian fits (maximum
splitting of 2.7~mV) for 8.5~T and 1.85~K. We have taken data at
temperatures down to 90~mK for our tunneling experiments\cite{YOPRL}
which seem to indicate an increased value of $g^*$ at lower
temperatures. This is consistent with Eq.~\equn{g} given that the
lowering of the temperature effectively narrows the width of the spin
states. However these data are not of sufficient quality for detailed
fits to be meaningful.

To conclude this section, we note that
the difficulty in fitting to the data in
Figs.~\fign{8.5lor} and \fign{8.5gauss} calls
into question either the underlying Gaussian or Lorentzian lineshapes
used or the validity of Eq.~\equn{g}. The sidewalls of the Landau
level DOS observed are so steep that no Gaussian or Lorentzian
lineshape, when entered into our fitting protocol, can
fit them and the rest of the DOS curve at the same time. Using
the better fitting
Lorentzian lineshapes, we determine a value for the exchange energy of
$3.8\pm.2$~meV.

\v\par\penalty-1000
{\noindent{\bf 3.7 Observations Regarding the Spatial Extent of States}}
\nobreak\v\nobreak
Since just after the discovery of the quantum Hall effect, the notion
has been widely discussed that two kinds of electronic states, with
very different character, exist in the 2d electron gas in the presence
of a magnetic field. These are localized states, thought to exist
between Landau levels, and extended states found at the ``core'' of a
Landau level. Most simply, Hall plateaus occur when the Fermi energy
is in a region of localized states which do not contribute to the
Hall voltage. Only when the Fermi level passes through the core of
extended states in a Landau level does the Hall voltage make a
transition from one plateau to the next.  Very low temperature
experiments\cite{PAAL-qhelow} demonstrating sharp transitions between
hall plateaus indicate a very narrow range of extended states.  One
often sees\cite{QHE} sketches of the 2d magnetic field DOS drawn with
lines separating regions of localized and extended states.  In
analogy to the metal-insulator transition, this line is referred to
as the mobility edge. In gauge arguments for the existence of the
quantum Hall effect,\pcite{LAUGH-quant,HALP-edge} the presence of
mobility gaps and a mobility edge, separating in energy localized and
extended states, is used to account for the Hall plateaus.

The transition region between localized and delocalized states has
been of great interest in recent years. The localization length
$\xi$ and its dependence on the position of the Fermi
level in the Landau level have been at the center of interest.
Specific attention has been placed on the region in energy near the
core of a Landau level. Here, $\xi$ is thought to diverge. The self
consistent theory of Ono\cite{ONO-diverge} and scaling
theory\cite{ANDO-loc} each make different predictions on the form of
the $\xi$ in the vicinity of the divergence. The region of the
divergence has been probed experimentally in transport
experiments.\pcite{TSUI-deloc} The transition from delocalized to
localized states, the effective mobility edge, is thought to occur
where the Fermi energy traverses a position in the Landau level
structure at which $\xi$ decreases below a characteristic length,
typically taken to be an inelastic scattering length.
Effectively, the behavior
of the localization length near the divergence can be probed in a
Hall experiment through measurements of $d\rho_{xy}/dB$ and
$d\rho_{xx}/dB$ as a function of temperature.\pcite{TSUI-deloc}
However, it is difficult using these means to achieve an absolute
measure of the size of localized states.  As we will show, our
measurements are instead more sensitive to the region between Landau
levels where the states are well localized, probing a different
regime than previous experiments. Further the method used here, with
appropriately designed samples, may offer a way of measuring the
absolute extent of localized states.

\v\par\penalty-1000
{\noindent{\sl {\bf 3.7.1} Sensitivity in our experiment to the
localization length}}
\nobreak\v\nobreak
Our low and high frequency capacitance measurements determine the 2d DOS, an
equilibrium property of the system not dependent on the localized or
extended character of the electronic states. We show below that if
tunneling rates vary for different regions of the 2d electron gas, then
the frequency
dependence of the capacitance in our experiment is sensitive to the in-plane
conductivity of the 2d electron gas. Specifically, when the in-plane
conductivity of the sample is very low, the 2d gas can be thought of
as made of isolated domains (each with slightly different tunneling
conductivities) which do not transfer appreciable charge to one
another during an RC time of the tunnel barrier. The simple circuit
model of Fig.~\fign{samp}b breaks down. Although we cannot make direct
contact to the 2d gas to measure the transport, we exploit this
sensitivity to the in-plane conductivity to explore the nature of
the states in the region of the Fermi level. Further, our
probe is sensitive to the bulk of the 2d gas and not particularly to
the edges of the sample. The theory of edge states in the quantum Hall
effect\cite{STRED-edge-exa} indicates that in transport experiments
when the Fermi energy is in
the region of localized states between Landau levels
currents used to probe the sample pass mainly
the edges of the sample and not the bulk.
Our experiment, microwave experiments\cite{KUCH-mic}, cyclotron
resonance experiments\cite{HEIT-res-width}, and surface acoustic-wave
attenuation experiments\cite{KOTT-saw} are instead
sensitive to the bulk regardless of the Fermi level position within
the Landau level structure.

\v\par\penalty-1000
{\noindent{\sl {\bf 3.7.2} Sample Model}}
\nobreak\v\nobreak
At low temperatures and when the Fermi energy is between Landau levels
in the 2d electron gas the longitudinal conductance, $\sigma_{xx}$,
drops to very low values. Physically, the only states in the bulk
at the Fermi level are highly localized states which do not conduct.
From the temperature dependence of the longitudinal resistance in
Hall samples,\pcite{STM-zero} it is thought that variable range
hopping between localized states forms the main conduction mechanism
in the bulk in high magnetic field
at temperatures below 4~K. At higher temperatures,
thermal excitation of carriers to bands of extended
states\cite{TSUI-temp} causes increased conductance as the
temperature is raised.

\figinsert{\vfil\vskip6.5truein\includegraphics{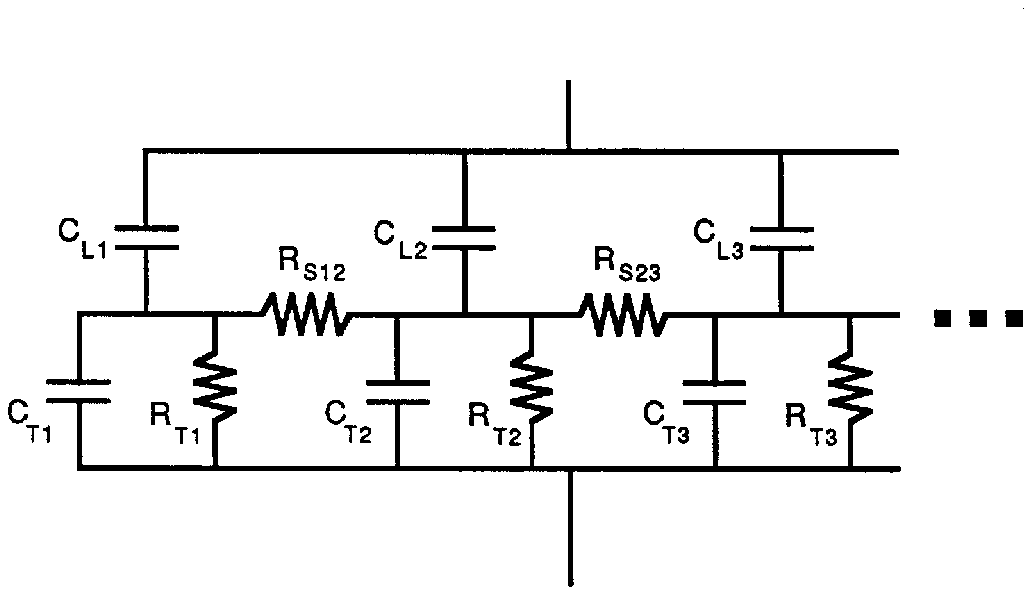}}{samp2}

In cases where the in-plane conductance ``freezes-out'' in our
samples, the circuit model of Fig.~\fign{samp}b can no longer be
used, and it should be replaced with a more sophisticated model such
as the one in Fig.~\fign{samp2}. This model is similar to those
used to describe the frequency dependence of the capacitance and loss
tangent of Si MOSFETs with electronic traps in the oxide
layer with different tunneling times from the different traps to the
electronic inversion layer of the MOSFET.\pcite{NICO-broad}
In the model of Fig.~\fign{samp2} for our device,
the 2d gas is considered to be made up of many different
domains,
linked by resistors $R_S$. Due to inhomogeneity (in the tunnel barrier
thickness or in other factors influencing tunneling) the $RC$ times
describing charge flow between the domains and the substrate
may vary among isolated domains.
The capacitance from the top gate
to each domain is modeled through the capacitors $C_L$. The
capacitance and tunneling resistance from each domain to the substrate
of the sample are modeled by capacitors $C_T$ and resistors $R_T$.
When the in-plane conductivity of the sample has frozen out the
variation in $RC$ times for the different domains (now electrically isolated
from one another) produces
loss tangent and capacitance vs.\ frequency curves for our
sample which are broadened as shown in Fig.~\fign{broad}.
This is in contrast with the curves of Fig.~\fign{freq} which correspond
to a single $RC$ time for the whole system. Figure~\fign{broad}
plots the loss tangent and capacitance as a function of
frequency for sample {\bf A} with a gate bias applied so that the
Fermi level is between Landau levels at a magnetic field of 8.5~T and
at a temperature of 875~mK. The solid and dotted 
lines are fits which will be described below.

\figinsert{\vfil\vskip6.5truein\includegraphics{broad.ps}}{broad}

We now describe the limiting cases of the circuit model of
Fig.~\fign{samp2}. In the case where the resistors $R_S$ have zero
resistance (i.e. when the 2d gas is perfectly conducting), then all of
the resistors $R_T$ can be seen as being in parallel, and the sets of
capacitors $C_L$ and $C_T$ can also be seen as adding in parallel. In
this case, we simply recover the circuit model of Fig.~\fign{samp}b
with a Debye loss tangent shape. In the other limit, the resistors
$R_S$ have infinite resistance. Then the sample can be seen
essentially as being made up of many small samples, each encompassing
one domain, all in parallel. These domains may each have different
$RC$ times for the tunnel capacitor-resistor combination. This means
that the different domains each have different frequencies $f_{peak}$
at which the loss tangent moves through a peak (or equivalently, the
domains have different frequencies at which the capacitance decreases
from $C_{low}$ to $C_{high}$). These ``mini-devices'' are all
effectively in parallel. Assuming that the capacitances per
unit area from the gate and substrate to the 2d gas remain the same,
both the capacitance at very low frequencies (compared to $1/2{\pi}RC$
for the slowest domains) and at high frequencies (compared to
$1/2{\pi}RC$ for the fastest domains) are unaffected by the breakup of
the 2d gas into electrically isolated domains. The differences in
$RC$ times for the tunnel barrier in the different domains only affect
the frequency dependence of the loss tangent and the capacitance
between low and high frequencies.

There obviously exists a crossover regime with the resistances $R_S$
between the limits of zero and infinity described above.
Very roughly, loss tangent
and capacitance curves begin to show broadening when the
characteristic time for one domain to transfer charge to another
becomes on order of the $RC$ time for a domain to transfer charge to
the substrate. This occurs when the
resistances $R_S$ in Fig.~\fign{samp2} approach the tunneling
resistances $R_T$. The exact details of the crossover are
complicated, depending on typical domain size, variation in domain
sizes, the number of closely neighboring domains,
and variation in the resistances $R_S$ shunting the domains.

In the presence of the broadening, we no longer fit
the capacitance vs.\ frequency curves with Eq.~\equn{cfit}.
We instead use the following form:
$$C(f)=.2C(f;f_{peak})+.4C(f;{\chi}f_{peak})+.4C(f;f_{peak}/\chi).\eqno{(\equn{cfitbroad})}$$
This equation can be read as regarding 20\% of the area of the sample
contributing to the capacitance with a peak frequency of $f_{peak}$,
40\% contributing with a peak frequency of ${\chi}f_{peak}$, and 40\%
contributing with a peak peak frequency of $f_{peak}/{\chi}$.
$f_{peak}$ is taken to be a ``mean peak frequency'' for the domains,
here taken to have individual peak frequencies distributed
symmetrically about this mean. We have introduced the parameter
$\chi$ which we call the ``broadening parameter''. The parameter
$\chi$ determined from fits is a measure in the variation of peak
frequencies from the different domains.  The weighting, 20\%, 40\%,
40\%, chosen in Eq.~\equn{cfitbroad} is somewhat arbitrary.
Experimentally, we find that this weighting tends to give a good fit
to data, such as that presented in Fig.~\fign{broad} regardless of
the level of broadening. Other weightings, for example 30\%, 35\%,
35\%, do nearly as well. Using this weighting increases the
deduced value for $\chi$ by up to around 20\% from that obtained
with the 20\%, 40\%, 40\% weighting. With a particular weighting
scheme chosen, $\chi$ is useful as a measure of {\it relative}
variations in the breadth of capacitance vs.\ frequency curves as the
sample temperature, gate bias, and magnetic field are varied. It is
also a rough measure of the absolute variation in peak frequencies,
$f_{peak}$, throughout the different domains. With more data points
and very high precision it may be reasonable to obtain a distribution
function of the peak frequencies from the capacitance.

\figinsert{\vfil\vskip6.5truein\includegraphics{broadtemp.ps}}{broadtemp}

We now briefly explore the effects of temperature on the broadening
parameter, $\chi$, obtained from fits. Fig.~\fign{broadtemp}
displays the broadening parameter as a function of temperature in
sample {\bf A} at a magnetic field of 8.5~T, with the gate bias
adjusted at each temperature so that the the Fermi level is at the
same position between the first and second Landau levels ($\nu=2$;
i.e. 2 spin subbands filled) at
each temperature plotted. Above 4~K, the value of $\chi$
parameter is one. This means that the fits detect no broadening of
the capacitance vs.\ frequency curves. The sample behaves as though
it is in the limit where the resistors $R_S$ of Fig.~\fign{samp2}
perfectly shunt the domains. There is no broadening, and the model of
Fig.~\fign{samp}b is appropriate. Below 4~K, the $\chi$ begins to
increase as the temperature decreases. A crossover region exists
between 0.5~K and 4~K, and below this $\chi$ levels off to a saturated
value $\chi_{sat}$. Curves such as that plotted in Fig.\
\fign{broadtemp} depend both on the magnetic field strength and the
position of the Fermi level with respect to the Landau level
structure.  Experimentally we find that the temperature range for the
crossover displays very little dependence on the position of the
Fermi energy within the Landau level structure, but this temperature
range does depend on the magnetic field strength.

\v\par\penalty-1000
{\noindent{\sl {\bf 3.7.3} Saturation of the Broadening Parameter
at Low Temperatures}}
\nobreak\v\nobreak
We briefly concentrate on the ``saturated'' value of the broadening
parameter at low temperatures.
The value of
$\chi_{sat}$ depends on both the Fermi
energy and the magnetic field strength. In general,
the value of $\chi_{sat}$ is the largest when the Fermi level is nearly
midway between two Landau levels.
We find in sample {\bf A} (the only
sample for which we have data below 1~K), that at the lowest temperatures
measured, we observe complete
saturation of the broadening parameter only at
the highest fields measured, 8.5~T and 6.5~T.
At 8.5~T with the Fermi energy midway between
Landau levels the onset of broadening typically occurs
at 4~K and saturation occurs by 0.5~K. 
From plots of $\chi$ vs.\ temperature it appears that
$\chi$ is close to saturation at 90~mK for a field of 4~T.  For both
fields of 8.5~T and 6.5~T, the largest saturated values of $\chi$
are 4.5 and 3.5 respectively. At 4~T, $\chi$ never
exceeds 2.5, and no broadening ($\chi=1$ for all positions of the
Fermi energy) is apparent for any gate bias in fields of 2~T and 1~T
in temperatures down to 90~mK.

What is the physical significance of $\chi_{sat}$? The following
arguments suggest an interpretation
of saturated broadening parameter as measure of the roughness
of the tunnel barrier. Recognizing that for domain sizes much larger
than $350\AA$, the separation between the 2d gas and the substrate
charge, the capacitance per unit area of the domain to substrate
capacitance is independent of domain size, we expect the variation in peak
frequencies to reflect mostly the variation in tunneling conductances
from the different domains to the substrate.
We identify the localization length $\xi$ as
the size scale of the domains here. 
The magnetic length acts as a lower bound on the localization length.
At fields of 8.5~T and below, it is thus certain that domains are
larger than $\approx100\AA$.
Depending on the
magnetic field strength and the position of the Fermi level within
the Landau level structure, the localization length
varies.
As the magnetic field increases and as
the Fermi level is moved towards the middle of the band of localized
states, this length decreases and so does the typical domain size.
Smaller domains probe smaller regions of the tunnel barrier and hence
result in larger variation in tunneling time from domain to domain.
If the tunnel barrier roughness were known as a function of lateral
distance along the tunnel barrier, the broadening of capacitance and
loss tangent curves vs.\ frequency could be translated into
information on the distribution of localization lengths for a
particular value of the Fermi energy.

\figinsert{\vfil\vskip6.5truein\includegraphics{8.5broad.ps}}{8.5broad}

Fig.~\fign{8.5broad} illustrates some of the ideas presented above.
Plotted along with  $\partial n/\partial\mu$ results taken at 8.5~T
for 1.85~K is the broadening parameter vs.\ Fermi energy for temperatures
ranging from 200~mK to 3~K. In the high DOS regions, the broadening
parameter $\chi$ is identically 1, regardless of temperature.
Between Landau levels $\chi$ increases with decreasing
temperature and, for most
regions of Fermi energy, $\chi$ follows a curve of saturated values
below 500~mK. The only exception is between the two spin subbands of
the lowest Landau level. Here $\chi$ is still increasing as the
temperature is lowered to 200~mK. This continued temperature
dependence of $\chi$ arises because the exchange interaction
splits the spin subbands still farther apart and diminishes the DOS
between them as the temperature is lowered, decreasing both the
in-plane conductivity and the localization length of states between the
subbands. The behavior of $\chi$ as a function of temperature for
these energies is complicated because it reflects both the increasing
spin splitting and the freezeout of the in-plane conductance as the
temperature is lowered. Concentrating instead on
the region between Landau levels, we see an interesting
universal behavior. The temperature range over which the $\chi$
increases from 1 to $\chi_{sat}$ is independent of the position of
the Fermi energy between Landau levels. We will return to this
interesting phenomena below.

The curve of $\chi_{sat}$ in Fig.~\fign{8.5broad} has a very
interesting detailed structure. A peak exists about midway
between Landau levels. Moving down on either side of this peak, there
are plateaus and then inflections and a sharp drop to $\chi=1$. The
origin of this structure, according to our model, may arise from
either a ``non-white'' distribution function for the tunnel barrier
roughness or, in the case of a white noise distribution function for
the roughness, an interesting behavior of the localization length.

\figinsert{\vfil\vskip6.5truein\includegraphics{4broad.ps}}{4broad}

Fig.~\fign{4broad} plots $\chi$ obtained from capacitance data
taken at 4~T and 140~mK on sample {\bf A}. Also plotted is $\partial
n/\partial\mu$ at 4~T and 1.9~K. This is plotted in place of the
noisier 140~mK DOS results. Again, the value of $\chi$ here appears,
from $\chi$ vs.\ $T$ plots, to be close to full saturation, and we
will refer to the results as reflecting $\chi_{sat}$. Upon
comparison with the 8.5~T data, we see an interesting
difference.
The peak value of $\chi_{sat}$ at 4~T is much
smaller than that obtained at 8.5~T. In fact, the peak value of
$\chi_{sat}$ at 4~T is about the same as $\chi_{sat}$ at 8.5~T midway
from the Landau level peak DOS position and the position of the peak
in $\chi_{sat}$. Considering the same value of $\chi_{sat}$ as
indicating the same typical localization length allows a
comparison of localization lengths in different fields.

Although we cannot measure the Hall resistance, we believe the
position in Figs.\ \fign{8.5broad} and \fign{4broad} at which $\chi$
begins to deviate from $\chi=1$ is different from the position
at which the Hall resistance enters a plateau.
Indeed, this position shows little temperature dependence; whereas the
breadth of Hall plateaus\cite{PAAL-qhelow,BRIG-qhelow} show much
temperature dependence. $\chi$ begins to deviate from 1 when $\xi$
becomes shorter than some temperature independent length fixed by the
tunnel barrier roughness; whereas a Hall plateau commences when $\xi$
becomes shorter than an inelastic mean free path or some other
temperature dependent length. 

One rough interpretation of the maximum value 4 for the
$\chi_{sat}$ is that it indicates a factor of about 4 variation in
tunneling conductance to the substrate from the different domains.
This is a reasonable amount of variation
given that monolayer (5.6$\AA$) thickness fluctuations on both sides of the
tunnel barrier, for the height in energy of our tunnel barrier, would
be expected to give rise to nearly the same size of tunneling
conductance fluctuation. This comparison assumes that the value of the
maximum value of $\chi_{sat}$ would not increase much more if the
magnetic field strength were increased beyond 8.5~T.

If a method could be devised to controllably place lateral
inhomogeneity in the tunnel barrier, broadening
parameter information in such a system
might yield a quantitative measurement of the
localization length. If a grid composed of thinner and thicker regions
of the tunnel barrier could be produced, large changes in the
broadening parameter would occur as the localization length ranges
through sizes smaller and larger than the period of the grid.

Although we have not made such a sample,
we have in our lab produced arrays of ``quantum
dots'',\cite{CHAP5} laterally confining electrons in the 2d gas into
small pockets less than 100~nm wide, and observed broadening in the
loss tangent and capacitance vs.\ frequency curves.
This dot sample was produced from the same MBE grown wafer from
which sample {\bf C} was processed. At first thought, one would think
that the broadening in
such a sample, where the size scale of the domains is known, should be
useful for comparison to results for the 2d gas in magnetic field.
Unfortunately, tunneling from
small dots is complicated by single electron charging
effects\cite{CHAP6} which also cause broadening, making comparison
to the broadening caused in a magnetic field by loss of in-plane
conductivity in the 2d electron gas difficult.

Additionally, we speculate that RHEED oscillations monitored during MBE
growth of the sample\cite{RHEED} can give information on the nature of
the nonuniformity of the tunnel barrier. Such knowledge be
useful in interpreting $\chi_{sat}$ data to find typical localization
lengths of states.

\v\par\penalty-1000
{\noindent{\sl {\bf 3.7.4} Connection to Magnetic Field Tunneling Suppression}}
\nobreak\v\nobreak
We have previously described\cite{YOPRL,CHAP4} a novel temperature
dependent suppression of the tunneling rate of electrons from the 2d
gas to the substrate in our samples that occurs only in the presence
of a magnetic field perpendicular to the plane of the 2d gas. This
suppression has been observed in samples {\bf A}, {\bf B}, and {\bf
C}. A key
feature of this tunneling suppression is that it occurs uniformly,
independent of the position of the Fermi energy within the Landau
level structure. The mechanism for this suppression is unknown,
although we have been able to fit the tunneling data to a model which
places an energy gap at the Fermi energy in the 2d electron gas. The
size of this energy gap varies nearly linearly with magnetic field
strength. The tunneling suppression is seen to commence at a
particular temperature as the temperature is lowered and saturate at
low temperatures.

A particularly intriguing behavior of the broadening parameter is that
the temperature range for $\chi$ to move from a value
of 1 to a value of $\chi_{sat}$ is independent of
the position of the Fermi energy
within the Landau level structure. 
Also, the temperature over which $\chi$ moves from 1 to
$\chi_{sat}$ is roughly {\it the same} as the temperature range over
which the tunneling suppression commences and saturates. As with the
tunneling suppression, the temperatures over which $\chi$ moves from
1 to $\chi_{sat}$ decrease as the magnetic field strength decreases.

The model presented earlier in this section predicts that $\chi$
approaches $\chi_{sat}$ as the resistances between the domains become
much larger than the tunneling resistances of the domains to the
substrate. The temperature dependence of $\chi$ arises from the
temperature dependence of the in-plane conductivity. Since the
tunneling suppression and $\chi$ have the same temperature dependence
this suggests that there is a connection between the processes which
causes freezeout of the in-plane conductivity and the tunneling
suppression. Strangely, the tunneling suppression occurs uniformly
throughout the Landau level structure, whereas the in-plane
conductivity depends heavily on the position of the Fermi energy
within the Landau level structure.

\v\par\penalty-1000
{\noindent{\bf 3.8 Summary}}
\nobreak\v\nobreak
In conclusion, we have developed a new technique which has allowed us
to make a systematic quantitative study of 2d Landau level DOS as a
function of Fermi energy in the 2d electron gas. This study was made
on three different samples with different doping configurations.
Analysis of the results in magnetic fields small enough so that the
exchange enhanced spin splitting small compared to the Landau level
width has lead to three principle conclusions.
Landau levels are described well by Lorentzian lineshapes; this
lineshape is independent of the doping profile in the sample; and,
the widths of Landau levels, while dependent on sample doping profile
and 2d electron gas density, are independent of magnetic field
strength. In high fields, the exchange enhanced spin splitting is
observed, and fits are made to the lineshapes using the model of
Ando and Uemura.\pcite{ANDO-osc} These fits determine a value for
the exchange energy that is in good agreement with
past determinations in 2d systems in
GaAs.\pcite{KVK-exch,GOLD-exch} Lastly, we have shown that the
frequency dependence of the capacitance and loss tangent in systems
such as our own can be used as a gauge of the localization length,
$\xi$, in the 2d electron gas.

\v\par\penalty-1000
{\noindent{\bf Appendix A - Explanation of $\eta$ term}}
\nobreak\v\nobreak
Here we give a brief explanation of the term $\eta$ which is a
correction for the nonzero extent of charge in the vertical ($x$)
dimension in the well. Recall that $\eta\sigma_w$ given in
Eq.~\equn{eta2} serves to correct the energy of the bound state in the
well as given by the sheet charge model ($U_w+E_0$ from
Eq.~\equn{usheet}) for the nonzero
width of distribution of charge in the well.
$\eta\sigma$ is an energy which is subtracted from the energy of the bound
state determined by the sheet charge model and which grows linearly
with charge density. This appendix describes the two contributions to
$\eta$ as well as giving some justification for why the sheet charge
model with the correction $\eta$ can be used in place of a distributed
charge model for the well.

\figinsert{\vfil\vskip6.5truein\includegraphics{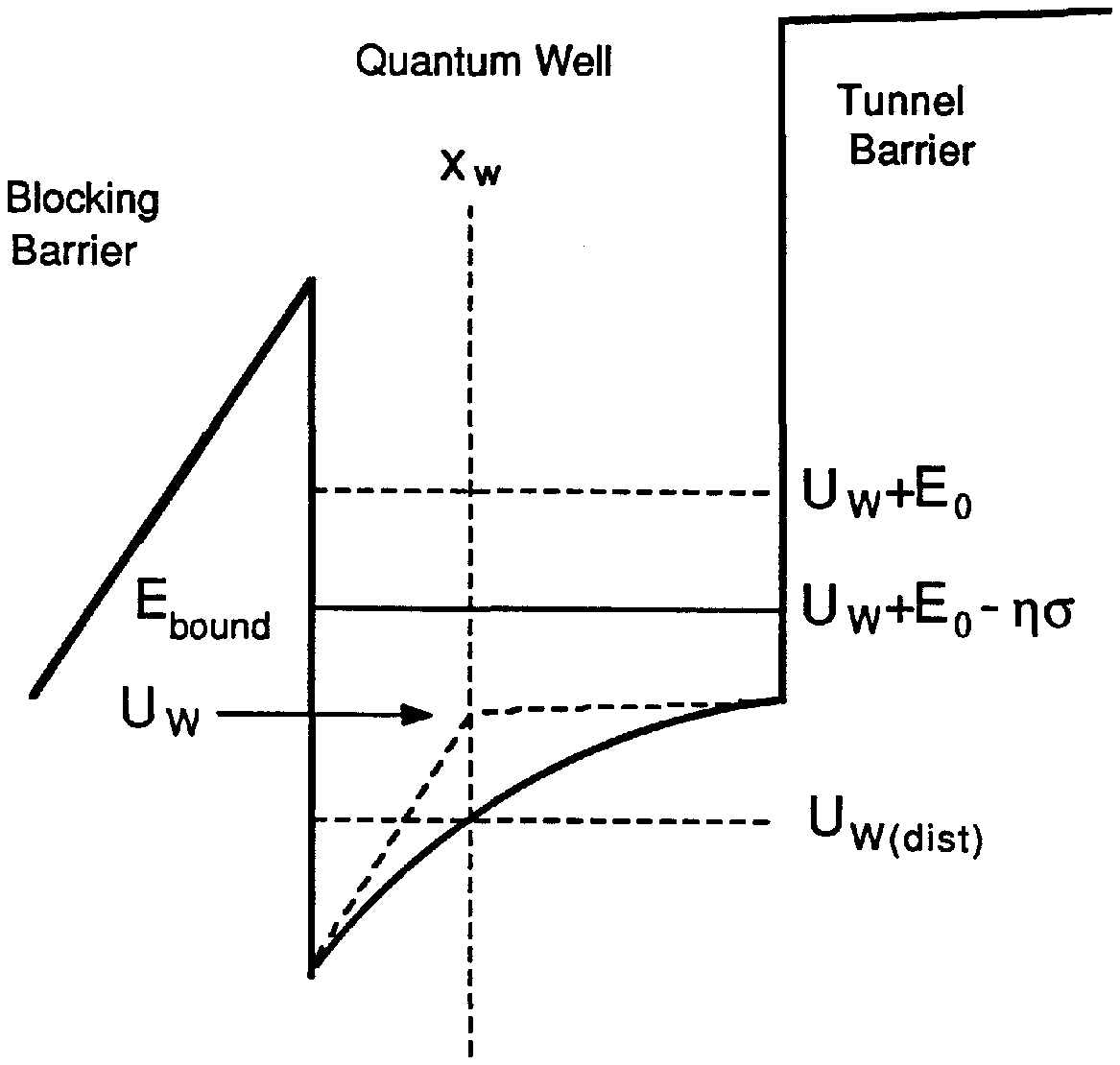}}{eta}

It should be noted that the capacitance, $C_{geom},$
deduced from the data through Eq.~\equn{dosint} specifies a position $x_w$
for the charge in the well. The significance of this position is
described pictorially in Fig.~\fign{eta}.
For the distributed charge model, it can be shown that the linear
extrapolations of $V(x)$ (the electrostatic potential due to the charge
densities $\sigma_g, \sigma_w,$ and $\sigma_s$) from charge free
barriers into the quantum well intersect at the position:
$$x_w={\int{x\rho(x)dx}}.$$   
Here $\rho(x)$ is the charge density distribution
(normalized to unity)
in the well in the $x$ (vertical) dimension.
Eq.~\equn{U_{gate}}, though derived in the sheet charge model, is
still correct in the distributed charge model provided
that the position $x_w$ is taken as the mean position of the charge in
the well.
In the series of equations that are solved in the sheet charge model to
obtain expressions for the capacitances, and hence the expressions for
the DOS using measured capacitances, only the expressions for $U_w$
(Eq.~\equn{U_w1}) and $U_{bound}$ (Eq.~\equn{usheet}) are different in
the distributed charge model.

First, we concentrate on Eq.~\equn{U_w1}. Figure~\fign{eta} depicts the
potential energy at the band edge at position $x_w$, $U_w$ in the sheet
charge model being greater than that given by the distributed
charge model. Eq.~\equn{U_w1}, modified in order to take into account
the nonzero $x$ extent of the charge in the well, becomes
$$U_{w (dist)}=-{e^2\over\epsilon}\sigma_sx_w - \Delta U_w,$$
where $\Delta U_w$ is given by,
$$\Delta
U_w={e^2\sigma_w\over\epsilon}\int\limits_{0}^{x_w}\rho(x)(x_w-x)dx.\eqno(\equn{U_w2})$$
Thus when the charge in the well is spread out, the potential energy
of electrons at the band edge at position $x_w$ is reduced.  If the
charge distribution function $\rho(x)$ retains the same shape and as
the well is filled, the reduction term $\Delta U_w$ will increase by
an amount proportional to $\sigma_w.$

Our sheet charge model considers the energy of the bound state to be
given by $U_{bound}=U_w+E_0$ (Eq.~\equn{usheet}).
The energy of the bound state is thus 
taken to be a fixed energy $E_0$ above the energy of electrons at the
band edge at position $x_w$. We define $E_0$ as being the energy added
to $U_w$ to give the actual bound state energy when the gate
bias is moved to the point where the first electron enters the well
(i.e.\ before there is any curvature to the well bottom).
In a na\"{\i}ve model, the energy of the bound state is $U_{w
(dist)}+E_0$.
However, as the well fills, the curvature of the band edge potential
energy at the
well bottom increases, and the quantum mechanical
bound state energy decreases with respect
to the potential energy at $x_w$.
This happens because electrons in the well sense the potential at
positions other than $x_w$, and the particular shape of the potential
in our samples, leads to a decrease in the bound state energy as the
well is filled.
In first order perturbation theory,
the energy difference between the bound state energy and $U_{w (dist)}$
decreases linearly with increasing charge density. 

With the assumptions that the shape of the charge distribution does
not change and that first order perturbation theory is correct in
calculating the shift of the bound state energy as the well fills, the
difference between the actual bound state energy in the well and that
indicated by the sheet charge model grows linearly. This linearity 
allows us to approximate the behavior in the distributed case by
inclusion the term $\eta\sigma_w$ in Eq.~\equn{eta2} in our analysis.

\v\par\penalty-1000
{\noindent{\bf Appendix B - Details of the DOS Calculation}}
\nobreak\v\nobreak
We have outlined a protocol earlier in this chapter that can be used
to determine the
values of $C_{geom}$ and $\eta$ from the low and high frequency
capacitance data. This appendix addresses some of the subtleties of
the DOS determination not immediately apparent in the model.

\v\par\penalty-1000
{\noindent{\sl Limits to our knowledge of the values of $C_{geom}$ and $\eta$}}
\nobreak\v\nobreak
We start first with the limitations of our knowledge of the
parameters $C_{geom}$ and $\eta$ and the effects of this uncertainty.
Some basic questions are:
1) How large is the error in the determination of the parameters
$C_{geom}$ and $\eta$? 2) How do the parameters $C_{geom}$ and $\eta$
vary as the gate bias is varied? and 3)
How do these effects influence the DOS determination?

We estimate the error in determining $C_{geom}$ from Eq.~\equn{dosint}
to be near two or three percent. These errors arise from several
factors. There is a breadth to the Landau minima in the capacitance
which prevents precise determination of the limits of integration in
Eq.~\equn{dosint}. Moreover, there may be some overlap of the DOS from
different levels, which for the case of adjacent levels with different
shapes, can change the expected value of the left hand side of
Eq.~\equn{dosint} from $2Be/h$ by a few percent. Given a value for
$C_{geom}$, determination of $\eta$ can be made with better precision.
Tips of the Landau level DOS peaks
plotted as a function of well energy are well defined;
the breadths of the Landau levels are typically a meV or less with
sharp tips. Also, the peak spacing or cyclotron energy $\hbar\omega_c$
is well known from the GaAs effective mass of 0.067m$_0$.

What is the effect of the error in the value of $C_{geom}$ used on
the DOS determination? Increasing the value of $C_{geom}$ used in the
DOS determination 
(for typical values of parameters for the samples measured here)
decreases the gate to well energy lever-arm, $dU_{bound}/d(eV_{gate})$,
and hence tends to bring closer together
peaks of Landau levels plotted as a function of Fermi energy in the
well. Also, as can be seen in Eq.~\equn{lever}, decreasing the value
of $\eta$ used in the DOS determination again decreases the lever-arm
and brings the positions of the observed Landau level peaks closer
together. If the error in $C_{geom}$ leads the value of $C_{geom}$
used in the DOS
determination to be larger than the actual value, then the value of
$\eta$ which positions the Landau level peaks $\hbar\omega_c$ apart
will be larger than its actual value. In a plot, sketched in
Fig.~\fign{imag}, of
$C_{geom}$ vs.\ $\eta$ there is a curve of values for these two
quantities upon which the DOS determination will give the correct
Landau level spacing. The important question is then: For the range of
values given by the error in the $C_{geom}$ determination, what is the
variation in the shape of observed Landau levels for levels that are
spaced $\hbar\omega_c$ apart? Upon trying different values ($\pm$3\%)
of $C_{geom}$ in the DOS calculation and using the corresponding
values of $\eta$ from Eq.~\equn{leverint}, we observe no visible
change in the shapes of levels. Using a value for $C_{geom}$ that is
3\% less than the actual value will, by Eq.~\equn{dosint}, lead to a
3\% error in the normalization of the total degeneracy of a Landau
peak. On the scale of the resolution of the plots shown in this chapter,
this error is invisible.

\figinsert{\vfil\vskip6.5truein\includegraphics{imag.ps}}{imag}

\v\par\penalty-1000
{\noindent{\sl Method for more precise determination of $C_{geom}$ and $\eta$}}
\nobreak\v\nobreak
Is it possible then, using our measurements, to ``pin down'' the values
of $C_{geom}$ and $\eta$ even though, with the constraint that Landau
level peaks have the correct separation, these values have no
observable effect on the shapes of the DOS peaks? We have developed a
method to do this. The zero field DOS is given by
$g_0=m^*/\pi\hbar^2$ or $2.8\times10^{13}$ cm$^{-2}$eV$^{-1}$. Referring back
to Fig.~\fign{imag},
there is only one set of values of $C_{geom}$ and $\eta$
which, in the absence of magnetic field, yield $g_0$ in the DOS
determination.

The protocol then becomes slightly more complicated. We start with
capacitance data and using Eq.~\equn{dosint} make the first rough
determination of $C_{geom}$. Then $\eta$ is determined by making
certain that Landau peaks have the correct spacing. This determines
one point on the imaginary curve. The zero field DOS is then
deduced using these parameter values. Depending on the value of the
zero field DOS obtained, we increase the value of $C_{geom}$
(increases observed zero field
DOS) or decrease it (decreases observed zero field DOS) and
redetermine $\eta$ by looking again at the data with field. We then
redetermine the zero field data using this new value of $\eta$. The correct
value of $C_{geom}$ can be interpolated from a graph of $g_0$
vs.\ $C_{geom}$ (where for each value of $C_{geom}$, $\eta$ has been
adjusted to satisfy Eq.~\equn{leverint}) to give the correct value of
$g_0$.

The values for $C_{geom}$ and $\eta$ obtained this way are in
accord with simple estimates. For
example, in sample {\bf B}, $C_{geom}$ is about
102.5~pF (which translates
into a distance $325 \AA$ for $x_w$ using a dielectric of 12 for
the medium) and $\eta=3.2\times10^{-14}$eVcm$^2$. These numbers are very
plausible. In sample {\bf B}, charge in the substrate is
expected to be about a Thomas-Fermi screening length of $100\AA$
away from the substrate-tunnel barrier
interface.\pcite{JOHN,APP-accum} Note, there is an undoped GaAs spacer
layer at the substrate, and the potential configuration here is quite
complicated. The tunnel barrier is $133 \AA$ thick, and the mean
position of the charge in the well is about $85 \AA$ from the
well-tunnel barrier interface.\pcite{STER-pri}
The sum of these numbers, 318$\AA$, is close to
the experimentally determined value for $x_w$. Considering the
electronic wavefunction in the well to be given by a sine wave and use of
first order perturbation theory\cite{JOHN} gives
$\eta=2.7\times10^{-14}$eVcm$^2$, again in close agreement with the
result determined using our protocol. However, the error
in the experimentally determined value of $\eta$ is still very large.
The steepness of the slope of the curve in Fig.~\fign{imag} is meant
to illustrate this; a small error in $C_{geom}$ will lead to a large
error in $\eta$.
Our fit to $g_0$, due to statistical error in the data,
uncertainty in the value of the effective mass in the quantum well,
and small drifts in capacitance shunting the sample (typically 0.1~pF)
between runs can only be
trusted to about $\pm2\%$. This leads to a $\pm2\%$ uncertainty in the
value for $C_{geom}$. In this range of probable values for $C_{geom}$
indicated by this uncertainty, the value of $\eta$ needed to keep
Landau levels spaced by $\hbar\omega_c$ is about 50\% smaller for
$C_{geom}$ at 100.5~pF than that for $C_{geom}$ set at 104.5~pF.

It is clear then, that to obtain a meaningful measure of $\eta$ for
comparison to models, one needs greater precision in the experiment
than we have in the data presented here. For the purposes of the DOS
determination however, the precise value of $\eta$ appears to be
irrelevant. 

\v\par\penalty-1000
{\noindent{\sl Variation of $C_{geom}$ and $\eta$ as the well fills}}
\nobreak\v\nobreak
Another complication in our DOS determination is the variation of
$C_{geom}$ and $\eta$ as the gate bias is varied. Both the mean
positions of the charge densities in the well and the substrate are
expected to move as the gate bias is varied. They tend to move in the
same direction (as the gate bias is made more positive, these charge
densities shift towards the gate) but not necessarily at the same
rate. This causes a change in $x_w$ and thus in $C_{geom}$. Also, the
shape of the conduction band edge energy in the well and the
charge distribution in the well change as well filling is varied. Thus
$\eta$ will also have some gate bias dependence.
We now refer back to Fig.~\fign{zerofield}.
This figure shows
the results for the DOS in zero magnetic field both with no corrections
to either $C_{geom}$ or $\eta$ for variation with gate bias and with
the correction described below.
In the curve which includes the correction, $C_{geom}$ varies linearly
with gate bias. For sample {\bf B} the formula for $C_{geom}$ that we
use is
$$C_{geom}=101.0{\rm{ pF}} + (0.0060{\rm{ pF/mV}})V_{gate},$$
where the gate voltage is measured in mV. The factor 0.0060~pF/mV was
determined empirically as the number which causes the observed zero
magnetic field DOS to be constant. This represents a 3\% variation in
the value of $C_{geom}$, or an approximately $10\AA$ variation in $x_w$, over
the range of gate biases used in sample {\bf B}. The value of 102.5~pF
given for $C_{geom}$ above is an average value.

Suppose the variation which causes the DOS in the uncorrected curve
of Fig.~\fign{zerofield} to
vary with gate bias were actually a variation of $\eta$ with gate bias
and not a variation of $C_{geom}$. We have also succeeded in making
the observed zero magnetic field DOS flat by adding a term linear in
gate voltage to $\eta$ and keeping $C_{geom}$ constant.
Whichever method was used to make the observed zero field DOS
flat, the DOS results with field were identical. This is not surprising
given our remarks above about the insensitivity of the DOS results on
the precise values of $C_{geom}$ and $\eta$ as long as these
parameters are adjusted so that the observed Landau level peaks lie
$\hbar\omega_c$ apart.

One concern is that the correction made, through observation of the
zero field DOS results, for variation in the parameters $C_{geom}$ and
$\eta$ using a term in $C_{geom}$ linear in gate bias might not be
appropriate when the magnetic field is applied.
For a given sample, the position of the excess charge in the substrate 
is solely a function of the electric field at the tunnel barrier. That
is to say, for a given electric field at the tunnel barrier the
self-consistent problem of the position of charges in the substrate
can be solved independently of variations in the positions of charges in
other parts of the sample. For a constant separation between the
excess charge in the substrate and the mean position of the charge in
the well, this electric field is proportional (neglecting small nonlinearity
arising from distributed charge in the well) to the bound state
energy in the well as measured with respect to the Fermi energy in the
substrate. 
One might expect then that lowest order variation in
$C_{geom}$ should be taken as a linear term in well energy and not
gate bias.

We have experimented with translating, using a lever-arm, the
variation in $C_{geom}$ from gate bias to well energy. In zero
magnetic field, the lever arm is nearly constant (about 1:42 for
Sample {\bf B}), and this translation is trivial. Instead of
0.0060~pf/mV
gate voltage variation, the variation is 0.254~pF/meV at the
well. We use
this same value in the well in the case with field and translate back
to gate voltage using the lever-arm determined by using the zero field
variation of $C_{geom}$ in gate bias. We then use the variation in
$C_{geom}$ obtained this way in the DOS and lever-arm determination.
Now an improved lever-arm is obtained and it can be used again to
determine the variation of $C_{geom}$ with gate bias. This procedure
can then be iterated until there is no further change in $C_{geom}$ as
a function of gate bias.

We find that the size of the variation in $C_{geom}$ is too small to
cause any observable change in the the Landau level DOS that we
obtain. In fact, in the iterative procedure outlined above we see only
a minute change (much too small to be observed in the DOS results)
to the lever-arm with field after the first
determination of $C_{geom}$ as a function of gate bias. Further
iteration yields no more change. In short, characterizing the effect, which
causes the deduced zero field DOS to have a slight slope in
Fig.~\fign{zerofield}, in terms of either
gate bias or well energy yields the same Landau level DOS.

\v\par\penalty-1000
{\noindent{\sl Robustness of the DOS calculation}}
\nobreak\v\nobreak
The procedure used to obtain the Landau level DOS given in this
chapter is very robust. Experimentation with the
protocol shows that the effects of adding a constant capacitance
shift as large as 2\% of the measured value to $C_{low}$, provided
that parameters $C_{geom}$ and $\eta$ are adjusted to insure that the
observe Landau levels are $\hbar\omega_c$ apart, are almost
unobservable in the DOS results. Similarly, a capacitance shift of
several percent can be added to both $C_{low}$ and $C_{high}$ with no
change in the DOS results. This indicates that even if there were a
small unobserved capacitance of several picofarads
shunting our device (we believe that any shunt capacitance
is less than 0.5~pF), the effects of this on the final DOS
determination would be inconsequential.

\v\par\penalty-1000
{\noindent{\sl Computer Simulations}}
\nobreak\v\nobreak
We have used computer simulations of data 
to understand better the observed robustness of the DOS results
and their insensitivity to the means used to correct them for the
variation in $C_{geom}$ and $\eta$. It is a simple
matter, starting with assumed values for $C_{geom}$, $\eta$, and $g$
(the DOS) to obtain the $C_{low}$ and $C_{high}$ as a function of gate
bias by reversing Eqs.\ \equn{lever} \& \equn{dos}. We can then run
the capacitance values obtained through the analysis procedure.

Typically, we start with Lorentzian or Gaussian lineshapes, convert
these into capacitance values after adding some deviation to the simple model
(such as a variation in $\eta$ with gate bias) and convert the
resulting values of $C_{low}$ and $C_{high}$ through the analysis procedure.
We can in this way isolate the effects of the
variability of $C_{geom}$ and $\eta$ as well as the effects of shunt
capacitance and its drift in the capacitance measurements. Our
correction of the variation of $C_{geom}$ using a term linear in gate
bias, even though it may be more appropriate to consider the variation
in $C_{geom}$ as linear in well Fermi energy, is justified by the
computer model, as well as is our method for correcting any variation
in $\eta$ by adjusting the correction term in $C_{geom}$. Also, these
simulations justify the correction of the effects of $C_{motion}$
(which increases linearly in charge density; see Eq.~\equn{motion})
described in section 4.3 by adjustment of the linear term in $C_{geom}$.
In each case, for the expected values for variation of these
parameters, the computer simulations show, on the scale of the graphs
in this chapter, errors incurred by using a linear variation in gate
voltage of $C_{geom}$ to correct all of the variations are at or below
the threshold of being visible.
The robustness of the method to shunt capacitance as large as several
pF is also verified by these calculations. 

\v\par\penalty-1000
{\noindent{\bf Appendix C - Comparison of Abscissas on DOS Plots}}
\nobreak\v\nobreak
In section 3.4.3 we stated that the abscissas on
Figs.~\fign{2.0TH} and \fign{zerofield} corresponded to the same Fermi
energies in the well with respect to the well bound state.
To explain how we know this to be true, we start by examining Eq.\
\equn{levx}. For constant parameter values of $x_w$, $x_s$, and
$\eta$, it is easy to show that
$$\Delta V_{gate}=A\Delta U_{bound} + B\Delta\sigma_w,$$
where $A$ and $B$ are constants. Because the DOS, averaged over a
Landau level is the same, independent of magnetic field, this equation
indicates that the lever-arm, averaged over a Landau level, is also
independent of field.
As stated earlier, our analysis procedure breaks down for low
electronic densities in the well. If it were valid in this regime,
we could state that the gate
bias which places the Fermi energy in a Landau minimum (in the case of
Landau levels symmetric in energy, the Landau maxima also) would give the
same Fermi energy with respect to the bound state energy in the well
in the absence of field. The collapse of our analysis in the low
density regime prohibits us from making this statement directly based
on our model. We observe experimentally however, that the gate voltage
which places the Fermi energy at Landau minima at 4~T
also corresponds to the same Fermi energy at 2~T.

We take this to mean that
the Fermi energy with respect to the
bound state energy in the well is the indeed the
same in the absence of magnetic
field for the same gate bias which places the Fermi energy in a Landau
minimum (or Landau maximum in the case of symmetric levels) in the
presence of field. In Figs.~\fign{2.0TH} and \fign{zerofield} we have
chosen the zero of energy on the abscissa to correspond to the same
gate voltage, a gate voltage which places the Fermi energy at a Landau
maxima of a nearly symmetric Landau level in Fig.~\fign{2.0TH}. We
thus believe that the horizontal scales in the two figures correspond
to the same Fermi energies above the bound state in the well.

\v
\vfill\supereject
\centerline{\bf References}
\v
\listrefs
\vfill\eject

\newchapter{4.}
\figitem{sampprl}{(a) shows the essential structure of the samples.
Tunneling from the GaAs quantum well to the substrate across the
AlGaAs tunnel barrier is observed by means of capacitive coupling
through the thick nonconducting barrier. (b)~displays a simplified
model of the sample.}

\figitem{relax}{The figure shows the band diagram of the quantum well,
tunnel barrier, and the substrate with the electronic density in the
well slightly out of equilibrium with the electronic density in the
substrate. The ``quasi-Fermi level'' in the well $E_{Fw}$ in time 
relaxes so that it becomes equal to the Fermi energy in the substrate,
$E_{Fs}$; as it changes, $U_{bound}$ also changes.}

\figitem{4tes}{Tunneling conductivity in sample {\bf A} (log scale)
vs.\ electron number density in the quantum well for zero field (bold
solid curve) and for a variety of temperatures with 4.0 T magnetic
field applied perpendicular to the plane of the electron gas in the
quantum well.  The smooth curves joining the points are guides to the
eye.}

\figitem{2tes}{Tunneling conductivity in sample {\bf A}
vs.\ electron number density in the quantum well at 2~T magnetic
field applied perpendicular to the plane of the electron gas in the
quantum well. The oscillations here correspond to Landau levels, the
first peak shown being the lowest level. Note that the oscillations
here continue to develop increased contrast as the temperature is
lowered below 1~K, even though the thermodynamic DOS is nearly fully
developed at 1~K.}

\figitem{8.5tes}{Tunneling conductivity in sample {\bf A}
vs.\ electron number density in the quantum well at 8.5~T magnetic
field applied perpendicular to the plane of the electron gas in the
quantum well. The first two peaks shown as the density is increased
are the spin split levels of the lowest Landau level. Note that the
vertical scale is different here than in Figs.~\fign{4tes} and
\fign{2tes}.} 

\figitem{4tesB}{Tunneling conductivity vs.\ number density in sample
{\bf B} at a magnetic field of 4.0~T. The suppression effect is
clearly visible in this sample. Also apparent is the spin splitting of
the Landau levels which becomes increasingly prominent as the
temperature is reduced.}

\figitem{summ}{Plotted as symbols are the averaged conductances
relative to the high temperature limit of the conductance plotted
against temperature (log scale). The smooth curves are fits described
the text. Shown in the inset is $\Delta$ in Kelvins plotted against
the applied magnetic field. The line is given by
$\Delta=0.047\hbar\omega_c/k_B$.}

\figitem{nofieldall}{Tunneling conductivities (linear scale)
in zero magnetic field vs.\ gate voltage
in sample {\bf A} at temperatures ranging from 95~mK to 16.1~K. Note
that at a gate voltage of -800~mV, the electron density is about
$1\times10^{11}$ cm$^{-2}$ and rises linearly with gate voltage to a
value of around $6\times10^{11}$ cm$^{-2}$ at a gate bias of 400~mV.
Below -800~mV gate bias, the quantum well begins to depopulate
nonuniformly, leaving
unoccupied regions of the well. In this region,
the electron density is no longer
linear in gate bias, and the rate of electron density change with gate
bias is decreased. The conductance is obtained from the capacitance
data using the ``puddling'' model described in the text.}

\figitem{sampnew}{This figure displays the circuit model used to fit
capacitance vs.\ frequency curves taken in the low density region of
the device operation. The capacitance on the right, $(1-\alpha)C_{high}$ is
the capacitance measured from unoccupied regions of the quantum well.
The two capacitor model on the left corresponds to the regions in the
quantum well which are occupied by electrons.}

\figitem{areafull}{Displayed is the fractional areal occupation of the
quantum well, $\alpha$, as determined from capacitance vs.\ frequency
curves fitted by Eq.~\equn{cfit2} and the model of
Fig.~\fign{sampnew}. The solid curve is a guide to the eye.}

\figitem{condlow}{The figure displays in detail the tunneling
conductivity (conductance per unit occupied area)
determined using Eqs.~\equn{cfit2} and \equn{garea} for a
variety of temperatures between 95~mK and 16.1~K. Two features of the
plot stand out. Firstly, there appear to be low and high temperature
limiting values to the conductivity.
The transition between
these two values always occurs at between 0.5 and 3~K, independent of
gate bias. Secondly, an unexplained plateau develops at low
temperatures in the conductivity at biases below -1050~mV.}

\figitem{lowtemp}{Plotted as boxes is the conductivity from
Fig.~\fign{condlow} at -1050~mV plotted as a function of temperature
(log scale). The solid curve is a fit using Eq.~\equn{lambda} and the
same ``linear gap'' of Eq.~\equn{gap} that was used to fit the
magnetic field induced suppression. The fit has $\Delta$ adjusted to a
value of 4.7~K.}

\figitem{lowmag}{Tunneling conductivity as a function of gate bias for
different magnetic field values at 290~mK. Notice that the curves
taken at higher magnetic fields are significantly suppressed below the
curve at zero field. Curves taken at 1 and 2~T display an interesting
merging with the curve taken in absence of field as the gate bias is reduced.}

{\centerline{\twelvebf Chapter IV}}
\v
\v
{\centerline{\twelvebf Equilibrium Tunneling from the}}
{\centerline{\twelvebf Two Dimensional Electron Gas}}
\v
\v
\v
\v
\v
\v
{\noindent{\bf 4.1 Introduction}}
\v
Study of the two-dimensional (2d) electron
gas formed at the interface of a
semiconductor and an insulator has revealed much
important physics, the most dramatic being the quantized Hall
effects.
The work presented in this chapter grew from a study intended to probe the
modification of the density of states (DOS) in the 2d electron
gas produced by a magnetic field perpendicular to the plane of the
gas, a measurement of ``the Landau-level DOS at the
Fermi level.'' The technique used provided two independent measures
of the DOS, one a thermodynamic DOS explored in chapter~3, the
other a tunneling or single-particle DOS described here.
The experiment extends
earlier capacitance spectroscopy of
the Landau level DOS\cite{TREY-dos} to lower temperatures and,
most importantly, also measures the
tunneling conductance between the 2d gas and an n$^{+}$ substrate.

It differs as well from other tunneling 
measurements\cite{KVK-magtun,EAVE-par} which determine an {\it I-V}
characteristic.
In our experiment, the Fermi energy in a quantum well is varied in a
controlled fashion by application of a gate voltage, and the Fermi energies
on both sides of the tunnel barrier are kept within $k_BT$ of one another.
We measure the equilibrium tunneling conductance as a function of
the Fermi energy in the well, not the more usual differential conductance
as a function of the difference in Fermi energies across the barrier.

The experiment shows, in addition to the structure expected from the
development of Landau levels, an unexpected suppression of the electron
tunneling which is greater than an order of magnitude at a field of 8~T
and a temperature of 100 mK.
We interpret these data as evidence 
for the development of
a new magnetic field induced
energy gap forming at the Fermi energy of the 2d gas.

This chapter is divided into two parts. First we explore the temperature
dependence of tunneling data taken in
the presence of magnetic field perpendicular to the plane of the 2d
electron gas. We then turn to data taken in the absence of magnetic
field. Unlike the data taken in the presence of magnetic field, these
data show no temperature dependence except at low electronic
densities in the well ($<1\times10^{11}$ cm$^{-2}$), where the data
again reveal a tunneling suppression as the temperature is lowered.
Strangely, the two different tunneling suppression effects have
similar temperature dependences. This similarity may indicate that
the same physical mechanism is responsible for the tunneling
suppression in both cases. At the end of the chapter, we offer a possible
explanation for the suppression effects seen at low density as well as
speculations regarding the suppression effects induced by a magnetic
field.

\figinsert{\vfil\vskip6.5truein\includegraphics{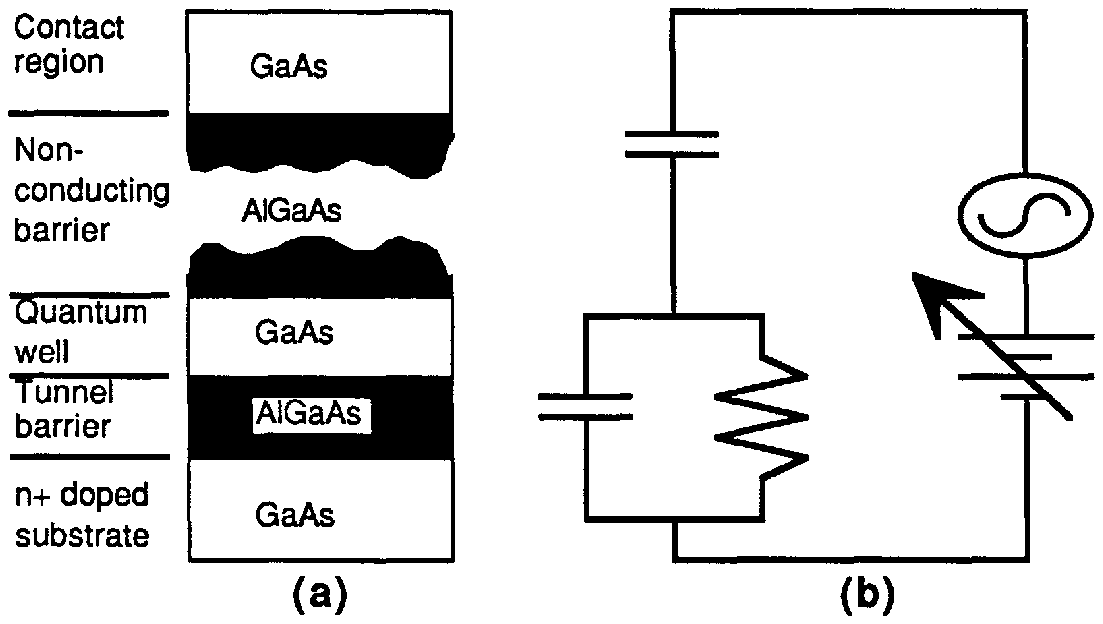}}{sampprl}

\v\par\penalty-1000
{\noindent{\bf 4.2 Samples and Method}}
\nobreak\v\nobreak
Mesas etched from three wafers grown using molecular beam
epitaxy have been studied. The essential
structure of the wafers is shown in Fig.~\fign{sampprl}a. The three
wafers, {\bf A}, {\bf B}, and {\bf C}, have been described in detail
in chapter~3. Also, tunneling measurements on sample {\bf A} have been
described in previous publications.\pcite{LSWa,LSWp,YOPRL,JOHN}
Each wafer consists of
a degenerately $n$ doped substrate in GaAs, an AlGaAs tunnel barrier,
a GaAs quantum well, a thick nonconducting AlGaAs barrier, and a
degenerately doped GaAs surface contact region. 
In all wafers,
only the lowest electronic subband of the well is occupied. The
electron density in the quantum well can be varied by the
application of a gate bias across the sample.

Tunnel barriers in our samples can be regarded as capacitors
shunted by a tunneling conductance. These were designed to have
$RC$ times which lie within the range of our measurements. The
capacitance and the loss tangent of patterned mesas were measured
at 20 frequencies between 15~Hz and 30~kHz.
Low-pass filtering\cite{CHAP2}
was employed on sample leads to reduce any noise that
might cause spurious voltage excitation across the tunnel
barrier. The loss tangent displays a Debye lineshape which peaks, and 
concurrently the measured capacitance decreases, as the
measuring frequency is swept through $1/2{\pi}RC$.

Fig.~\fign{sampprl}b shows the model which we use to characterize our
capacitance and loss tangent vs.\ frequency data set. The data can be
fit with suitable choices for the three circuit elements shown. If the
thermodynamic DOS in the well were infinite, the value of the
conductance (resistor shown in the figure) obtained by these fits
would be the tunneling conductance. With a finite DOS, this
correspondence no longer holds. The analysis of chapter 3, however,
allows deduction of the thermodynamic DOS from the data set, and a
simple extension of that analysis, given below, gives the tunneling
conductance as well.
In actuality, the capacitances shown in the model
depend both on sample dimensions and
the thermodynamic DOS\cite{LEE-mob-edge} in the quantum well.\pcite{CHAP3}
Discussed here are the results of a series of experiments in which the
the tunneling conductance is measured as
a function of the electron density in the quantum well, the temperature of the
sample, and the magnetic field strength (applied perpendicular to
the plane of the 2d electron gas).

\v\par\penalty-1000
{\noindent{\bf 4.3 Extraction of Tunneling Conductance from
Capacitance Data}}
\nobreak\v\nobreak
{\noindent{\sl Equilibration of 2d and 3d Charge Densities}}
\nobreak\v\nobreak
In this section we describe how fits of capacitance vs.\ frequency
and loss tangent vs.\ frequency curves are used to determine the
tunneling conductance between the 2d electron gas and the substrate.
To understand this problem, we first need to describe how the electron
densities in the well and the substrate, starting slightly out of
equilibrium, transfer electrons between one another through tunneling
to bring the two charge densities into equilibrium. Once the time
constants for approaching equilibrium are known, it is a simple matter
to determine how the overall device impedance behaves as a function of
frequency. The device operation is more complicated than is suggested
by the circuit model of Fig~\fign{sampprl}b due to the finite density
of states in the quantum well. In fact, both the
single particle (tunneling) DOS, $g_s$, and the
thermodynamic DOS, $g_d$, must be taken into account to correctly
deduce the tunneling conductance from the capacitance data. We note that
in chapter~3 we considered only the problem of the thermodynamic DOS.

\figinsert{\vfil\vskip6.5truein\includegraphics{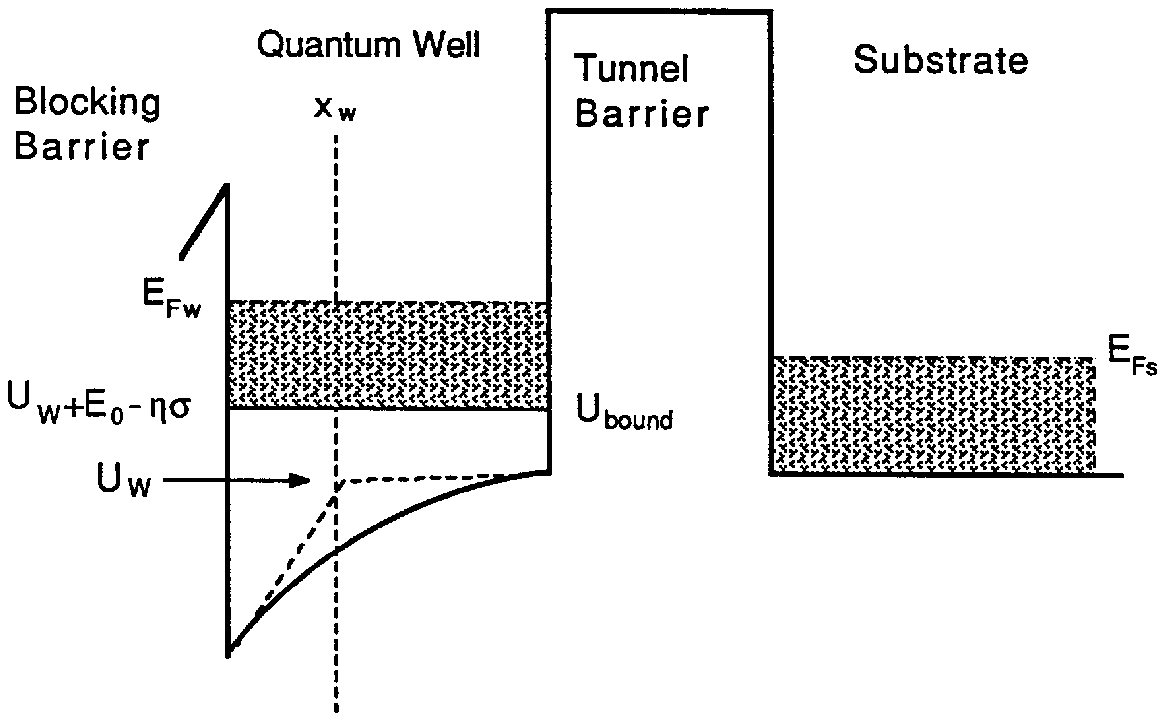}}{relax}

Figure~\fign{relax} shows the situation when the ``quasi-Fermi level''
in the quantum well $E_{Fw}$ is greater than the Fermi level in the
substrate, $E_{Fs}$. We are interested in the quantity
$E_{Fw}-E_{Fs}$, as it describes the degree to which the electron
gases are out of equilibrium. $E_{Fw}$ is determined by the sum of
two components.
One is the energy of the bound state in the well with respect to the
conduction band edge deep within the substrate.
The other is the width of the band of filled 2d electronic states
within the well well which arises from
the noninfinite thermodynamic DOS in the well. As the electronic 
charge in the well changes both of these energies change. The Fermi
energy in the substrate (all energies here are
measured with respect to the band edge deep
within the substrate) does not vary as charge in the well is changed.
Using the terminology of chapter~3, 
we can write
$${d(E_{Fw}-E_{Fs})\over dt}={dU_{bound}\over dt}+{1\over
g_d}{d\sigma_w\over dt},\eqno(\equn{relaxdev})$$
where, as in chapter~3, $U_{bound}$ is the bound state energy,
$\sigma_w$ is the number density of electronic charge in the well,
and it is understood that $g_d$ corresponds to the thermodynamic DOS
at the Fermi energy.
The inverse of the {\it thermodynamic} DOS multiplies $d\sigma_w/dt$
in Eq.~\equn{relaxdev} because the equilibration of Fermi energies
in the quantum well and the substrate occurs on long time scales (audio
frequencies).

In chapter~3, we described the term, $\eta$, which is a
correction to the energy of the bound state in the well determined using a
sheet charge model, given by $U_w+E_0$ where $E_0$ is a constant
energy, for the nonzero width of the wavefunction in the well.
The bound state energy including the $\eta$ correction term is given by
$$U_{bound}=U_w+E_0-\eta\sigma_w.$$
Here, the last term accounts for two effects. One is
the difference in the
electrostatic energy of the bound state associated with charge being
distributed in the well rather than in a sheet. The other is the quantum
mechanical change in the bound state energy due to the change of the
shape of the well bottom in the presence of charge.
We can then write
$${dU_{bound}\over dt}={dU_w\over dt}-\eta{d\sigma_w\over
dt}.\eqno(\equn{boundchange})$$

The tunneling current may be expressed in terms of a single particle DOS in
the well $g_s$, at the Fermi energy, and a mean tunneling rate per electron
of $1/\tau_{tun}$:\cite{WOLF-tunchap2}
$$I_{tun}=Ae(E_{Fw}-E_{Fs}){g_s\over\tau_{tun}},\eqno(\equn{tuncur})$$
where $e$ is the magnitude of the electronic charge and $A$ is the
sample area. Note that this expression is correct only for
for temperatures and applied biases small enough so that only electrons
in a narrow range over which $g_s$ is constant can tunnel. As will be
discussed in a later section, at higher
temperatures, $g_s$ must be replaced with its thermal average.
In the ``equilibrium tunneling'' measurements presented here,
$|E_{Fw}-E_{Fs}|$ is always kept less than $k_BT$ by suitable choice
of measuring voltage.
Given the tunneling current, it is simple to determine $d\sigma_w/dt$
and $dU_{w}/dt$. These are given by
$${d\sigma_w\over
dt}={-I_{tun}\over{Ae}}=-(E_{Fw}-E_{Fs}){g_s\over\tau_{tun}}\eqno(\equn{dsigdt})$$
and
$${dU_{w}\over dt}={-I_{tun}\over C_w}={{-Ae\over
C_w}(E_{Fw}-E_{Fs}){g_s\over\tau_{tun}}}.\eqno(\equn{Uwelldt})$$
Here, $C_w$ is the capacitance in the sheet charge model, of the quantum
well to the surroundings (substrate and top gate). It is given by
$$C_w=C_{geom}{x_g\over x_g-x_w},\eqno(\equn{C_w})$$
where, as in chapter~3, $x_w$ and $x_g$ are the distances from the
substrate charge to the charge in the quantum well and top gate respectively,
and $C_{geom}$ is the ``geometric capacitance'', $A\epsilon/x_w$, of
the quantum well sheet charge to the substrate. See chapter~3 for a
careful discussion of these terms.

Rewriting Eq.~\equn{relaxdev} again using
Eqs.~\equn{boundchange,tuncur,dsigdt,Uwelldt} and \equn{C_w} gives
$${d(E_{Fw}-E_{Fs})\over
dt}=-{(E_{Fw}-E_{Fs})\over\tau_{tun}}{\biggl[}{Ae^2g_s\over
C_{geom}}{\bigl(}1-{x_w\over x_g}{\bigr)}-\eta{g_s}+{g_s\over g_d}{\biggr]}.$$
The solution of this first order differential equation is of course
given by an exponential decrease of $E_{Fw}-E_{Fs}$ with time. The
relaxation rate is
$${1\over\tau_{r}}={1\over\tau_{tun}}{\biggl[}{Ae^2g_s\over
C_{geom}}{{\bigl(}1-{x_w\over x_g}{\bigr)}}-\eta{g_s}+{g_s\over
g_d}{\biggr]}.\eqno(\equn{relaxtime})$$
In our samples, the sum of the first two terms in the brackets is, in
zero magnetic field, of order 10, whereas the third is unity or
less. This means that the relaxation rates in our samples are
typically faster than the tunneling rates.

This difference between the relaxation rate and the tunneling rate can
be explained heuristically as follows. Consider one electron tunneling
from or to the quantum well.
This single electron, because of the electrostatic energy it carries
with it, brings the electron gases in the well and substrate much
closer to equilibrium than would an ``uncharged electron''
equilibrating the two regions simply by changing chemical potentials.
This ``speeds up'' the equilibration of the two electron gases. At
zero magnetic field in our samples, the average quantum level spacing
in the 2d electron gas, $1/Ae^2g_s$, is about 10\% of the electrostatic
energy for adding one charge to the well, $e^2/C_w$. However, if the
single particle density of states becomes small enough, the relaxation rate
simplifies to:
$${1\over\tau_{r}}={1\over\tau_{tun}}{\biggl[}{g_s\over g_d}{\biggr]}.$$
In the case where the thermodynamic and single particle densities of
states are equivalent, the relaxation time and the tunneling time are
the same in this low DOS limit.

\v\par\penalty-1000
{\noindent{\sl Fitting to Capacitance and Loss Tangent Curves}}
\nobreak\v\nobreak
Using the expressions above we can calculate the capacitance
and the loss tangent for the device as a function of frequency. We
start by calculating the current through the device after (at time
$t=0$) a voltage $\delta V$ is suddenly applied across between the top
gate and the substrate. Only the fraction of $(x_w/x_g)\delta V$ of
the applied voltage appears between the charge density in the well and
substrate. Further, only the fraction $(x_w/x_g)I_{tun}$ of the
current which moves between the
quantum well and the substrate should be counted in the device current
because the current only traverses part-way through the device.
The total device current is then
$$I_{dev}(t)=(\delta
V)Ae^2{g_s\over\tau_{tun}}{\biggl(}{x_w\over{x_g}}{\biggr)}^2e^{-t/\tau_{r}}
+(\delta V){C_{high}\over\tau_{fast}}e^{-t/\tau_{fast}}.\eqno(\equn{devcur})$$
We define $C_{high}$ (high frequency capacitance) as the capacitance
of the device with no current traversing the tunnel barrier. In terms
of device parameters, its value is $A\epsilon/x_g$.
$\tau_{fast}$ is the charging time of the capacitance $C_{high}$ due
to any resistance in series with our device. We consider
$\tau_{fast}$ to be much shorter than the
period of the measuring signals in our experiment.

The A.C.\ admittance of the device is given by $j\omega$ times the
Fourier transform of this ``step response''.
In the limit where $\tau_{fast}$ goes to zero,
the A.C.\ current through the sample is
$$I=Ae^2{g_s\over\tau_{tun}}{\biggl(}{x_w\over
x_g}{\biggr)}^2{\omega^2\tau_{r}^2+j\omega\tau_{r}\over1+\omega^2\tau_{r}^2}V+j\omega{C_{high}V},\eqno(\equn{accur})$$
where V is the amplitude of the measuring voltage $Ve^{j\omega{t}}$.
The tunneling conductance is given by
$$G_{tun}=Ae^2{g_s\over\tau_{tun}}.\eqno(\equn{gtun})$$
Rewriting Eq.~\equn{accur} again, dividing by the voltage, we find
that the device admittance is
$$Y(\omega)=G_{tun}{\biggl(}{x_w\over
x_g}{\biggr)}^2{\omega^2\tau_{r}^2\over
1+\omega^2\tau_{r}^2}+j\omega{\biggl[}G_{tun}{\biggl(}{x_w\over
x_g}{\biggr)}^2{\tau_{r}\over1+\omega^2\tau_{r}^2}+C_{high}{\biggr]}.\eqno(\equn{admit})$$
The term in the brackets of this equation can immediately be
identified as the device capacitance. At low frequencies, the device
capacitance from Eq.~\equn{admit} here reduces to the same form as
given in chapter~3 (using $g_d$ for $g$ in the expression for
$C_{low}$). The loss tangent for the sample is given by the real part
of the admittance divided by the imaginary part.

We can now write down the device capacitance and loss tangent as
functions of frequency. From Eq.~\equn{admit} for the sample admittance
above and using the expression for $C_{low}$ from chapter~3, we have
for the capacitance
$$C(f)={C_{high}C_{low}[1+(f/f_{peak})^2]\over
C_{high}+C_{low}(f/f_{peak})^2},\eqno(\equn{cfreq})$$
and for the loss tangent
$$D(f)={2D_{peak}}{f/f_{peak}\over1+(f/f_{peak})^2}.\eqno(\equn{dfreq})$$
Here, the loss tangent peak height $D_{peak}$ is given by
$$D_{peak}={\sqrt{C_{high}\over C_{low}}}{\biggl(}{C_{low}\over
C_{high}}-1{\biggr)},\eqno(\equn{dpeak})$$
and the frequency at which the loss tangent goes through a peak
$f_{peak}$ is related to the tunneling conductance in the following
fashion:
$$G_{tun}=2{\pi}f_{peak}C_{geom}{C_{geom}\over{\sqrt{C_{low}C_{high}}}}{\biggl(}{C_{low}\over
C_{high}}-1{\biggr)}.\eqno(\equn{gtunfit})$$
Note that to arrive at Eq.~\equn{gtunfit} we have made substantial use of
the formulas relating $g_d$ and $\eta$ to $C_{low}$, $C_{high}$, and
$C_{geom}$ developed in chapter~3. The value of $C_{geom}$ used here
is known to within 2\% using the methods of the previous chapter. (Note, as
discussed in detail the previous chapter, in the presence of low in-plane
conductivity of the 2d electron gas, Eqs.~\equn{cfreq} and \equn{dfreq}
must be modified somewhat in order to properly fit the data.) The fits
given by Eqs.~\equn{cfreq} and \equn{dfreq} are identical to the
functional forms for the capacitance and the loss tangent indicated by
the circuit model of Fig.~\fign{sampprl}b, but the interpretation
of the fitting parameters is different.

The present chapter concerns itself
with the value of $G_{tun}$ as determined from fits to the capacitance
and loss tangent. $G_{tun}$ is proportional to the single particle
DOS $g_s$, whereas the focus of the previous chapter was on the
thermodynamic DOS.

\v\par\penalty-1000
{\noindent{\bf 4.4 Sample Characterization}}
\nobreak\v\nobreak
The widths of thermodynamic DOS peaks attributed to Landau levels in
each of the samples studied here were discussed in chapter~3.
Elastic scattering times in the well can be estimated in the three wafers
from the widths of the DOS peaks of Landau levels in a magnetic
field.\pcite{EIS-DvA} According to the theory of Ando and
Uemura\cite{ANDO-ellip} the half-widths of Landau level DOS peaks are
related to the elastic scattering time by the relation
$$\Gamma=\hbar\biggl({2\over\pi}{\omega_c\over\tau_s}\biggr)^{1/2}.$$
Here $\omega_c$ is the cyclotron frequency and $\tau_s$ is the elastic
scattering time. In chapter~3, we showed that widths of Landau level
DOS peaks were nearly independent of magnetic field, in disagreement with
this relation. Because the Landau levels in chapter~3 fit well to
Lorentzian lineshapes, we can estimate elastic scattering times, and
hence sample mobilities from the relation for Lorentzian
lineshapes,
$$\tau_s={\hbar\over\Gamma}.$$
$\Gamma$ is the half-width of the Lorentzian lineshape.
Using this formula,
wafer {\bf A} would have a nominal 2.0~K mobility in the well
of approximately $\rm21,000\ cm^{2}V^{-1}sec^{-1}$
as defined by this single-particle scattering time;
wafer {\bf B} would have a mobility of $\rm25,000\
cm^{2}V^{-1}sec^{-1}$; and sample {\bf C} would have a mobility of 
$\rm15,000\ cm^{2}V^{-1}sec^{-1}.$ 
Actual transport mobilities are expected to be substantially
higher.\pcite{STER-mob}
The magnetic field 
suppression effect discussed here has been observed in all three samples.
More data have been taken on wafer {\bf A}, both with and without
magnetic field, than on either samples {\bf B} or {\bf C}. We will
concentrate mostly on the results from sample {\bf A}; over 200,000
capacitance measurements were taken
on one $400\mu$~m diameter mesa produced on this wafer.

\figinsert{\vfil\vskip 6.5true in \includegraphics{4tes.ps}}{4tes}

\v\par\penalty-1000
{\noindent{\bf 4.5 Tunneling in the Presence of Magnetic Field}}
\nobreak\v\nobreak
Figure \fign{4tes} displays the logarithm of the tunneling conductance
of sample {\bf A} at 4~T for different temperatures. The 
highest temperature curves oscillate about the zero field curve,\pcite{LSWa} 
indicating the development of Landau level structure in the DOS in the
2d gas. At lower temperatures, the tunneling conductance is strongly
suppressed. In contrast with this behavior in magnetic field, the zero field
tunneling conductance shows no substantial variation with temperature
over the range 90~mK to 10~K except for electron densities
near full depletion ($\rm<1\times10^{11}~cm^{-2}$).
The temperature dependent suppression occurs only in the presence of the
magnetic field applied perpendicular to the plane of the 2d electron
gas; a magnetic field applied parallel to
the plane\cite{LSWp} does not induce a temperature dependent
suppression. We note that
the doping levels in the substrate are high enough ($\rm10^{17}\ cm^{-3}$)
to discount magnetic freezeout\cite{HICK-fqhe}
as the cause of the effect. 
Finally, the thermodynamic DOS in the well,\pcite{CHAP3}
as determined from the capacitance
values distinct from conductance results, shows no unexpected behavior
reflecting the tunneling suppression.

The temperature dependence suggests
that the suppression is due to a tunneling anomaly
restricted to energies near the Fermi energy.
To explore this idea, a high frequency signal
(period shorter than the $RC$ relaxation time of the tunnel barrier)
was injected across the sample during capacitance measurements.
This signal provides an oscillating
Fermi level offset between the 2d gas and the
substrate allowing tunneling to occur from a band of states of width
given by double the amplitude of the injected signal.
We find that as the excitation voltage that appears
across the barrier due to the injected signal is made larger than
$k_BT/e$, the suppression effects induced by the low temperature of the
sample recede. At very low temperatures, where the suppression
appears to be saturated, the effect of excitation, of rms amplitude $V_e$,
on conductance is roughly the same as increasing the temperature in absence
of excitation to a value $eV_e/k_B$.
This implies that the suppression would be observed in a conventional
{\it I-V} characteristic as a zero bias anomaly, not as a general bias
independent suppression.

   The data shown in Fig.~\fign{4tes} are striking in that
the suppression is nearly independent of the Landau level filling number
in the well. Also, at low temperatures, where the contrast associated with
the Landau levels has developed, the suppression has roughly the same
strength when the Fermi level is between Landau levels as when it is at
a Landau maximum.

   In Fig.~\fign{4tes}, tunneling at densities higher 
than $\rm3.9\times10^{11}\ cm^{-2}$
should be forbidden in the absence of scattering.
At these densities, the Fermi level is in the third Landau level in the
well, whereas in the substrate the third Landau level lies outside of
the Fermi surface. 
Indeed, the tunneling conductance is markedly lower in this
range of density. Interestingly, the strength of
suppression due to the magnetic field is approximately the same in the
forbidden region as it is in the range of densities for which
tunneling is allowed. This observation suggests that the suppression
is not some consequence of the change in spatial character of the
one electron wave functions associated with the development of Landau
states.

\figinsert{\vfil\vskip 6.5true in \includegraphics{2tes.ps}}{2tes}
\figinsert{\vfil\vskip 6.5true in \includegraphics{8.5tes.ps}}{8.5tes}

For completeness here, we show tunneling conductivity vs.\ number
density curves in the same sample for different magnetic field
strengths. Figure~\fign{2tes} displays the tunneling conductivity at
various temperatures for a 2.0~T applied magnetic field. The tunneling
suppression here appears to be weaker and sets in at lower
temperatures than for the data in Fig.~\fign{4tes}. Just the opposite
is true of the data taken at 8.5~T shown in Fig.~\fign{8.5tes}. Strong
suppression commences at higher temperatures than for the data of
Fig.~\fign{4tes}. In Fig.~\fign{8.5tes}, the first two peaks that
appear in the tunneling conductivity correspond to the spin split
bands of the lowest Landau level. At 8.5~T, the second Landau level in
the substrate is outside of the Fermi surface, and tunneling at
electron densities above $4.1\times10^{11}$~cm$^{-2}$ should be
forbidden. Once again, despite the sharp decrease in the tunneling
conductance as the electron density is increased into this forbidden
region, the temperature dependent tunneling suppression still persists
in this region.

\figinsert{\vfil\vskip 6.5true in \includegraphics{4tesB.ps}}{4tesB}

Figure~\fign{4tesB} displays the tunneling conductivity
of sample {\bf B} at 4~T and at 1.9 and
4.2~K. The same suppression effect occurs in this sample as in sample
{\bf A}. Notice that while there is, as with sample {\bf A} at these
temperatures, some increased definition
of the Landau level as the temperature is
decreased, the suppression strength is again roughly 
independent of the 2d electron gas density. As with the data of sample
{\bf A}, on average the 1.9~K conductance data is again about
20-30\% less than the conductance at 4.2~K.
We have data in sample {\bf C} only at temperatures of 4~K and near
6~K. The suppression effect is observed in this sample
as well although it is
somewhat less transparent because, as with sample {\bf A} in this
temperature range at this magnetic field strength, the strength
of the suppression is small and there is at the same time
considerable temperature dependence of the Landau level widths.

\v\par\penalty-1000
{\noindent{\bf 4.6 Average Conductance}}
\nobreak\v\nobreak
In this section, to summarize the data from sample {\bf A} for
different fields, we perform an average of the conductance data over a
range of well fillings (over a half integral or integral number of
Landau levels) for each value of the magnetic field and for each
temperature measured. The averaging is needed to isolate changes in
the tunneling conductance arising from the tunneling suppression from
changes in the conductance due to varying detail in the Landau level
structure, such as the increased contrast of levels as the temperature
is reduced or spin splittings which are resolved only at low temperature.
We then relate this average conductance to the single
particle DOS in the well.

There is a subtlety involved in determining the average value of the
tunneling conductance.
When $k_BT$ is of order the Landau level width, the
Landau level structure changes markedly. The average of the
conductance must be taken with respect to the Fermi energy in the
well, not with respect to well density, to assure that changes in
contrast of the Landau level structure do not contribute to an
apparent but unreal change in the average value of the tunneling
conductance. The following paragraphs formalize these statements.

Considering the tunneling rate $1/\tau_{tun}$ to
be a constant over an energy of order $k_BT$, basic tunneling
theory\cite{WOLF-tunchap2} implies that $g_s$ in Eq.~\equn{gtun} should be
replaced by a thermal average so that the tunneling conductivity,
$G(T;B;E_{F})$, becomes
$$G(T;B;E_{F})=-{e^2\over\tau_{tun}(E_{F})}\int\limits_0^{\infty}
{g_{s}(E;B)}{\partial f(E;T)\over\partial E}dE,\eqno(\equn{gform})$$
where $f(E;T)$ is the Fermi distribution function, $E$ is the kinetic
energy of electrons in the 2d electron gas, and $E_F$ is the
Fermi energy in the well measured with respect to the bound state
energy in the well, $U_{bound}$.

We describe here why in Eq.~\equn{gform} $\tau_{tun}$ is taken to be a
function only of the Fermi energy in the well and not of the magnetic
field strength. First we discuss the situation in zero field.
$\tau_{tun}$ depends on the tunneling matrix element coupling states
in the well to those in the substrate as well as the density of states in
the substrate. 
These two elements depend on the ``longitudinal
momentum'' ${\hbar}k_z$,
the momentum of electrons in the substrate perpendicular
to the plane of electrons in the quantum well, but not on the
``transverse momentum''.
Because transverse momentum conservation specifies the
transverse wavenumbers, $k_x, k_y$,
of an electron in the substrate which has tunneled
from the quantum well, the appropriate substrate DOS to use in the
tunneling calculation is the one dimensional DOS\cite{LSWa} which
depends only on the value of $k_z$. Recalling again that $U_{bound}$
is the energy of the bound state in the well with respect to the
conduction band edge deep within the substrate, transverse momentum
conservation, along with the approximation of equal effective masses
in the 2d and 3d electron gases, dictates that
$$k_z={\sqrt{2m^*U_{bound}\over\hbar^2}},\eqno(\equn{kz})$$
where $m^*$ is the electron effective mass.
Because the Fermi energy in the well is fixed, equal to that in the
substrate, specifying $E_F$, the energy difference between the Fermi
energy and $U_{bound}$, also specifies $U_{bound}$. For these reasons,
we can write $\tau_{tun}(k_z)=\tau_{tun}(k_z(E_F))$.

In the case of magnetic field applied perpendicular to the plane of
the 2d electron gas, the magnetic field is pointed along the direction
tunneling and hence is not expected to alter matrix elements for tunneling
in this direction. Also, the Schr\"odinger equation is still separable
into terms involving $k_z$ and quantum numbers involving degrees of
freedom perpendicular to the field direction. These quantum numbers
are again conserved and Eq.~\equn{kz} still holds. $\tau_{tun}(E_{F})$
is thus independent of magnetic field strength.

In the absence of a magnetic field,
$g_s(E)$ is constant in energy, and Eq.~\equn{gform}
takes the form:
$$G_{B=0}(T;E_{F})={e^2\over\tau_{tun}(E_F)}g_0$$
where $g_0$ is the (single-particle) zero field DOS. In this model,
the ratio of the tunneling conductance in magnetic field to the
tunneling conductance in the absence of field,
for the same values of the Fermi energy in the well,
is equal to the ratio of
the thermally averaged DOS in magnetic field to $g_0$.

In a picture where the electrons are noninteracting free particles,
the ratio of these densities of states, averaged over a Landau level
is, of course, equal to one.
The fact that in some of our low temperature measurements the
tunneling conductance falls everywhere below the zero field
conductance (such as for the lowest temperature data plotted in
Figs.~\fign{4tes}, \fign{2tes} and \fign{8.5tes}) suggests that the
single particle DOS is lower, at energies near the Fermi energy,
in magnetic field than in the absence of magnetic field.
As discussed below,
the temperature dependence of the tunneling suppression
suggests that the single particle DOS is only decreased for energies
near the Fermi energy.

\figinsert{\vfil\vskip 6.5true in \includegraphics{summ.ps}}{summ}

The capacitance data provide sufficient information to allow a conversion
from filling to energy, yielding
the conductance as a function of Fermi energy in the
well.\pcite{CHAP3}
For all but the 8.5 T data, the average of the conductance divided by
the zero field conductance at the same value of the Fermi energy (and
hence the same value of $U_{bound}$),
$\Lambda(T;B)$, at high temperatures
approaches a value of one
which is taken to be the high temperature limit.
In Fig.~\fign{summ} the ratio $\Lambda(T;B)$ 
compared to its high temperature limiting value
is plotted as a function of temperature. 
Ambiguities of interpretation of the 8.5 T data at high temperature
imply a 10\% uncertainty in the normalization of the 8.5 T data.\par
We hypothesize that the tunneling suppression is a consequence of a
gap in the single particle DOS $g_s(E)$, centered at $E_F$.
In this case, the appropriate
form for $\Lambda(T;B)$ would be 
$$\Lambda(T;B)=-\int\limits_0^{\infty}
{g_s(E;B)\over g_0}{\partial f(E;T)\over\partial E}dE.\eqno(\equn{lambda})$$
For reasons described below, we
fit with the following gap:
$$\eqalign{g_s(E)&=Sg_0+{(1-S)g_0|E-E_f|\over\Delta}\cr g_s(E)&=g_0\cr}
\qquad\eqalign{ (E &\le 2\Delta)\cr  (E &>2\Delta).\cr}\eqno(\equn{gap})$$
The fits shown in Fig.~\fign{summ} are the
result using this ``linear'' gap. 

We call attention to a few principal features of the summary data in
Fig.~\fign{summ} and of the fits.
(a) For low fields, 1 T and 2 T, the width parameter
$\Delta$ depends little on B, but the depth of the gap $(1-S)$
increases with increasing field. (b) For high fields, 6.5 T and 8.5 T,
the gap is nearly fully developed in depth and the width is increasing
with field, with some indication of saturation at
high fields. (c) The data consistently show more temperature
dependence in the low temperature limit than do models in which the
gap is nonsingular at the Fermi energy, e.g.\ square or smooth bottomed
gap functions. The linear singularity is not unique in giving good
fits to the data; a variety (e.g.\ a gap where $g\propto{\sqrt{\vert
E-E_{F}\vert}}$) of other singular behaviors will do as
well. (d)
At fields 2 T and higher, the gap width $\Delta$
has a value of
about 5\% of the cyclotron energy, $\hbar\omega_c$. (e) The errors
generated in the averaging procedure at high temperatures (where the
Landau level structure washes out) by the
Landau level shape changes leave the results ambiguous as to whether
or not there is temperature dependence of the gap in this temperature
regime. 

We interpret the results in terms
of a gap, rather than of an influence of the field
on the single particle tunneling dynamics, because the temperature
and excitation dependences indicate that the tunneling is suppressed
only for states near the Fermi energy. Before discussing possible
candidates for such a gap, or other proposals, we turn first to
another tunneling suppression that we have observed.

\v\par\penalty-1000
{\noindent{\bf 4.7 Low Density Tunneling Suppression in the Absence
of Magnetic Field}}
\nobreak\v\nobreak
As noted above, in the case of zero magnetic field or magnetic field
parallel to the plane of the 2d electron gas, no temperature
dependent tunneling suppression is observed for most of the range of
densities of the 2d electron gas. Temperature dependence, other than
that associated with thermal broadening, is, however,
observed at densities below $\approx1\times10^{11}$~cm$^{-2}$.
Fig.~\fign{nofieldall} displays the tunneling conductance (obtained in
a model described in the next section) in sample
{\bf A} as a function
of device gate bias for sample temperatures ranging from 95~mK to 16~K.
The shape of this curve at densities above
$1.2\times10^{11}$~cm$^{-2}$ has been
explained previously in terms of momentum conservation rules in
tunneling.\cite{LSWa,JOHN} For gate voltages above about -800~mV
(about $1\times10^{11}$~cm$^{-2}$ mean well density as determined from
magnetic field measurements), low frequency
capacitance measurements indicate that the quantum
well area is fully occupied. At voltages below -800~mV a
significant temperature dependence of the conductance appears.
As in the case
of perpendicular applied magnetic field, the tunneling conductance is
suppressed as the temperature is decreased. A suppression at low
densities has been observed as well in samples {\bf B} and {\bf C}.
The technique used to plot the tunneling conductance here differs
at low densities from that used in previous figures. The next
few paragraphs describe the differences as well as why a different
method is needed in the low density limit.

\figinsert{\vfil\vskip 6.5true in \includegraphics{nofieldall.ps}}{nofieldall}

\v\par\penalty-1000
{\noindent{\sl Capacitance vs.\ Frequency Curve Fitting in the Low
Density Regime}}\nobreak
\nobreak\v\nobreak
Equation~\equn{gtunfit} above yielded the tunneling conductance of
samples in terms of the loss peak frequency and the low and high
frequency capacitances of the device.
This relation, as well as the framework used in developing it,
breaks down at low electron densities in the well if there
are portions of the well unoccupied by electrons. Experimentally, a
decrease in the low frequency capacitance of the device is observed as
the gate bias is lowered in the low density region. We explore here
reasons for this decrease and develop interpretations of the
capacitance and loss data in the low density regime.

\figinsert{\vfil\vskip6.5truein\includegraphics{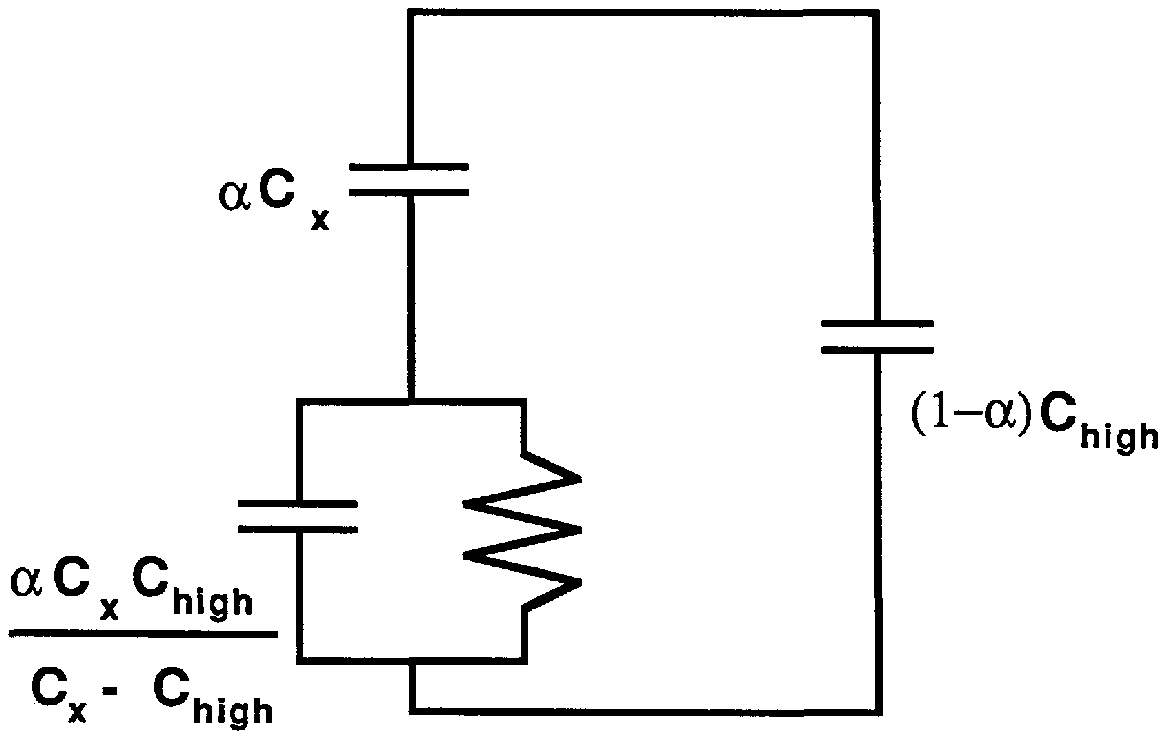}}{sampnew}

We proceed to model the sample with low densities in the well
using the the assumption that occupied
and unoccupied regions of the sample are larger than a few hundred
angstroms in size so that fringing fields can be neglected.
Because there are unoccupied regions of the well, a
capacitance shunting the top gate to the substrate must be added to
the circuit model of Fig.~\fign{sampprl}a.
The shunting capacitance in the model has a
value of $C_{high}(1-\alpha)$ where $\alpha$ is the fractional area of the 2d
electron gas which is occupied by electrons. This sample model is
shown in Fig.~\fign{sampnew}. The division the device this way into two
regions, one depleted and the other filled with electrons, gives the form
$$C(f)={{{\alpha}C_{high}C_{x}[1+(f/f_{peak})^2]\over
C_{high}+C_{x}(f/f_{peak})^2}}+(1-\alpha)C_{high}\eqno{(\equn{cfit2})}$$
for fits to capacitance vs.\ frequency curves. The
parameter $C_{x}$ is the capacitance that {\it would} be measured at
low frequencies if the well had no unoccupied regions. The measured
low frequency capacitance, $C_{low}$, is given by
$$C_{low}={\alpha}C_x+(1-\alpha)C_{high}.$$
At this point, fits can be made to the capacitance data using
$C_{low}$, $C_{high}$, $\alpha$, and $f_{peak}$ as fitting parameters.

We have made such fits and found that there is a
large error in the deduced values of $\alpha$ and $C_x$ (obtained from
$\alpha$ and other fitting parameters).
This occurs because, in the fitting, $\alpha$ is sensitive to subtle
changes in the shape of the capacitance vs.\ frequency curves,
such as broadening of the of the type discussed in chapter 3,
and to small random errors.

In order to obtain reasonable fits, we use Eq.~\equn{cfit2} and fix
the value of $C_x$. Equation~\equn{cfit2} can be written in
terms of the ratio $C_x/C_{high}$, $C_{high}$, $f_{peak}$, and $\alpha$.
We get results which fit the capacitance data well by setting the ratio
$C_x/C_{high}$ to the value of $C_{low}/C_{high}$ when the well is
fully occupied and using $C_{high}$, $f_{peak}$, and $\alpha$ as the
fitting parameters. The value of $C_{low}/C_{high}$, in the case where 
the density in the well is high enough so that it is undoubtedly occupied
everywhere (where $C_{low}$ ceases to be a strong function of gate
bias in zero magnetic field), depends, except for slight changes due to
variation in $\eta$ and $C_{geom}$ with gate bias discussed in chapter~3,
only on the thermodynamic density of states in the well.\pcite{CHAP3}
Setting $C_x/C_{high}$ equal to the value of $C_{low}/C_{high}$
when the well is full and letting $\alpha$ vary in the fits defines a
model which we call the ``puddling model''.

In the puddling model,
there are filled and unfilled regions of the well; in the filled
regions, the thermodynamic DOS is equal the constant 2d DOS given by
$m^*/\pi\hbar^2$. This contrasts with the ``filled well'' model, which
considers electrons to inhabit every portion of the well and
accounts for the decrease observed in the low frequency capacitance at
low gate biases through a gradual decrease in the thermodynamic DOS as
electrons are removed from the well. We proceed with the puddling
model discussing our reasons for using it instead of the filled well
model below. Models similar to this puddling model are used in
chapters 5 and 6 to understand quantum dots.

The tunneling conductance is easily evaluated in the puddling model.
Its value is given by:
$$G_{tun}=2{\pi}f_{peak}{\alpha}C_{geom}{C_{geom}\over{\sqrt{C_{x}C_{high}}}}{\biggl(}{C_{x}\over
C_{high}}-1{\biggr)}.\eqno(\equn{garea})$$
Of course, what is interesting physically is not the total
conductance, but the conductivity per unit filled area, which is this
formula divided by the quantity ${\alpha}A$, where $A$ again is the
area of the mesa.

We note that the puddling model considers all of the decrease in $C_{low}$, as
the electron density is lowered, to arise from depleted area.
It neglects DOS anomalies that can arise from a nonuniform potential
in the well; i.e. the potential in the well is not a constant
everywhere which can lead to the DOS being something other than a
constant as it is in 2d.
The results shown in Fig.~\fign{nofieldall} were determined using the
puddling model; the ``filled well'' model also displays a temperature
dependent tunneling conductivity.
Using the filled well model (forcing $\alpha$ to be equal
to one and allowing $C_x$ to vary, in which case $C_x\equiv{C_{low}}$
and the model becomes identical to that used at high densities)
leads to different conductance vs.\ gate bias curves in which
the tunneling conductivity falls off very rapidly at low densities,
much faster (for all temperatures) than does the device capacitance at
low densities. This is implausible in a framework where the tunneling
conductivity depends only on the DOS in the well but, as interaction
effects may influence the tunneling, this does not necessarily rule
out the filled well model as applicable to the device. The results from
attempts at fitting capacitance vs.\
frequency curves with both $\alpha$ and $C_{low}$ as free parameters,
while ambiguous, favor the puddling
model interpretation. Finally, theoretical models\cite{DAVIES-pud}
typically describe the 2d electron gas in heterostructures at densities
near full depletion in terms of filled and unfilled regions.

The low frequency capacitance may decrease at
low gate biases due both to regions of the well becoming unoccupied
and the occupied regions having a decreased DOS;
the puddling model and the filled well model are two limiting
interpretations. The rest of this
section and later sections use the puddling model to describe the
sample with the caveat that the real behavior of the system lies
between the two limits.

Finally, one salient feature of the puddling model is
the simplicity in determining the tunneling conductivity.
Because $C_{high}$ and $C_{geom}$
change very little over the range of gate biases
for the device, and $C_x/C_{high}$ is fixed to its value at high
densities, Eq.~\equn{garea} gives that the tunneling conductance is
simply proportional to the loss peak frequency.

\figinsert{\vfil\vskip 6.5true in\includegraphics{areafull.ps}}{areafull}

\v\par\penalty-1000
{\noindent{\sl Detailed Observation of Low Density Tunneling Conductance}}
\nobreak\v\nobreak
We use the puddling model framework to
plot in Fig.~\fign{areafull}, $\alpha$, the
fraction of the area of the quantum well which is occupied as a
function of gate bias. The curve was obtained from data taken
in zero magnetic field and at at 1.85~K.
We have fitted curves of capacitance vs.\ frequency
from a variety of temperatures from 95~mK to 16~K and
see little change in this curve.
Within the resolution of these results, we observe no
thermal smearing of this curve occurs up to the highest
temperatures. In the following discussion we treat the occupied area
and electron density
as fixed at a particular gate bias, independent of sample temperature.

Figure~\fign{condlow} ``zooms in'' on the tunneling conductivity,
determined in the puddling model, in low density region of
sample {\bf A} as a function of gate bias for temperatures from 95~mK
to 16~K. The temperature dependence is immediately striking. For the
whole region of gate biases plotted, above around 7~K the conductivity
saturates to a high temperature limiting value; below about 500~mK
the tunneling conductance saturates to a low temperature value. For
the entire curve, most of the shift of the conductance with
temperature happens between 1~K and 3~K. The other surprising feature
in Fig.~\fign{condlow} is the plateau which develops in the
conductivity at low temperatures. This reason for this
plateau, as well as the rest of the shape of the tunneling
conductivity plot as a function gate bias, is not understood
presently.

\figinsert{\vfil\vskip 6.5true in \includegraphics{condlow.ps}}{condlow}

\v\par\penalty-1000
{\noindent{\sl Low Density Conductivity as a Function of Temperature}}
\nobreak\v\nobreak
At this point, we note the striking similarity of
the temperature dependence of the tunneling conductivity at low
densities to the tunneling suppression induced by a magnetic field
illustrated in Fig.~\fign{summ}.
In the case of the low density data, we
simply take the conductivity at a particular value of gate bias and
plot it as a function of temperature. Figure~\fign{lowtemp} displays
the conductivity relative to its high temperature value plotted as a
function of temperature for a gate bias fixed at -1050~mV. The
resemblance the shape of this curve to the curves of
Fig.~\fign{summ} is readily apparent. The solid line is a fit to
the data obtained
using Eq.~\equn{lambda} and Eq.~\equn{gap}, the {\it same} gap used
to fit the magnetic field data. In Fig.~\fign{lowtemp}, $\Delta$ has
a value of about 4.7~K, and the gap depth parameter $S$ has a value of
0.4. By adjusting the values of $\Delta$ (which undergoes only slight
variations with gate bias in the low density regime)
and $S$ the gap model of Eq.~\equn{gap} can be made
to fit the data well for any gate bias in the low density region.
We will return below to this similarity in the temperature dependences
of the two suppression effects.

\figinsert{\vfil\vskip 6.5true in \includegraphics{lowtemp.ps}}{lowtemp}

\v\par\penalty-1000
{\noindent{\bf 4.8 Low Density Region with Magnetic Field Applied}}
\nobreak\v\nobreak
Now that we have examined in detail both the suppression of tunneling
by magnetic field at high densities and the suppression at low
densities in zero magnetic field, it is interesting to ask:
what happens at low density in the presence of a magnetic
field? We again use the puddling model of the last
section to fit the capacitance data and extract the tunneling
conductivity.

Before proceeding, we note that at high fields ($>$6.5~T) curves of the
occupied area obtained using the puddling model in the well vs.\ gate
bias are slightly shifted towards positive values of gate bias compared
to the results from zero field shown in Fig.~\fign{areafull} above.
This is largely due to  effects of the zero point energy in the well.
At lower fields (4~T and below), these effects are not visible on the scale
of Fig.~\fign{areafull}, and it is reasonable to consider, for these
low field data, that equivalent gate voltages for the sample for
different magnetic field strengths correspond to the same area
fraction of the well filled.

Figure~\fign{lowmag} displays the conductivity of sample {\bf A} at a
temperature of 290~mK at magnetic fields ranging from 0 to 8.5~T. At
the highest fields, the magnetic field conductivity suppression is
still very strong at these low electron gas densities, and the 6.5~T
and 8.5~T curves remain well below the 0~T curve over the full range
of gate biases shown. The data from lower fields show an intriguing
behavior. At higher gate bias values in Fig.~\fign{lowmag}, the 1 and
2~T curves fall below the zero field curve due to the magnetic field
suppression. As the gate bias (and thus the average electronic
density) is reduced, the low field
curves merge at a certain gate bias
with the zero field curve and then follow along the zero field curve
as the gate bias bias is reduced further. This same behavior persists
at a variety of different temperatures. Further, the curve at 2~T
always merges with the zero field curve at gate biases between
-950~mV and -975~mV, independently of the temperature.

\figinsert{\vfil\vskip 6.5true in \includegraphics{lowmag.ps}}{lowmag}

In effect, the magnetic field conductivity suppression at low
densities has an electronic density dependence, unlike the
suppression effect at higher densities. This density dependence
provides a strong indication that the source of the field induced
conductivity suppression lies in the 2d electron gas and not in the
substrate. Further, the merging behavior of the
curves in Fig.~\fign{lowmag} is suggestive that three regimes exist,
a high density regime where the magnetic field suppression dominates,
a low density regime where the low density suppression dominates,
and a crossover regime which is where the curves merge together.
Strangely, the magnetic field does not enhance the suppression in the
low density regime.

\v\par\penalty-1000
{\noindent{\bf 4.9 Speculations
Regarding Possible Causes for the Suppression Effects}}
\nobreak\v\nobreak
We briefly review the main characteristics of
suppression effects in the tunneling conductivity described above.
A magnetic field, only when perpendicular to the plane of the
2d electron gas, produces a tunneling suppression that can be
characterized by an energy gap at the Fermi energy in the 2d electron
gas. We focus on the 2d electron gas as the source of the suppression
for two reasons.
First, anomalies in the 3d gas should be present for arbitrary orientation
of the magnetic field, but the suppression is observed only with the field
perpendicular to the 2d gas. Second, density dependence of the suppression
is observed near depletion of the 2d gas, as described in the previous
section, suggesting the properties of the 2d gas as the source of the
suppression effect. In the case of zero applied
magnetic field, tunneling suppression is again observed at low
densities of the 2d electron gas and can be characterized by the same
``linear gap'' that fit the data well in the case of the magnetic
field conductivity suppression.

\v\par\penalty-1000
{\noindent{\sl Tunneling Conductivity Suppression at Low Densities}}
\nobreak\v\nobreak
We first focus on the low density zero field suppression because this
low density region of the 2d electron gas has shown a number of
interesting properties previously, some of which are understood
theoretically. Far infrared absorption
experiments\cite{ALLEN-freqdep} on the 2d electron gas in silicon MOSFETs
at low densities ($<4\times10^{11}$~cm$^{-2}$)
have shown a frequency dependence of
the electronic conductivity at photon energies below $\approx$0.5~mV.
These have been explained by arguments invoking localization in a
disordered system.\cite{GOLD-freqdep} Also, in low density inversion
layers in silicon MOSFETs, Bishop, Dynes, and Tsui\cite{BISHOP-SiPrl}
have observed a logarithmic temperature dependence in the low
temperature in plane conductivity of the 2d electron gas which they
interpret in terms of localization models.

Altshuler, Aronov, and Lee\cite{ALTSH-gap} have
suggested that localization effects should cause
an energy gap to form at the Fermi energy in the single particle
(tunneling) DOS in disordered systems. In the simplest heuristic
model, this gap arises for the following reasons. In a disordered
material with a background random potential, electrons eventually
find a configuration in which they minimize both their energy with
respect to the random potential and also the energy associated with
the electron-electron repulsion. Suddenly adding or subtracting
another electron forces the system to rearrange and requires it
overcome energy hurdles to find a new stable state.
In two dimensions, Altshuler et.\ al.\
predict that near the Fermi energy, this gap should have the form
$${{\delta}g(E)\over g(E)}=-{\hbar\over
2{\pi}E_{F}\tau_s}\ln\biggl({(E-E_{F})\tau_s\over\hbar}\biggr).\eqno(\equn{altgap})$$
Tunneling experiments have been done in both thin indium oxide
films\cite{IMRY-tun} and in thin tin films\cite{DYNE-2d-dos}. In both
of these experiments, the thicknesses of the films were varied to test
certain aspects of the theory. These measurements, in cases where the films
were thin enough to be considered two dimensional,
have shown the logarithmic behavior in the tunneling conductance as a
function of the voltage across the tunneling barrier (at voltages
$<$20~mV) expected from the Altshuler theory.

Due to difficulties in creating structures to observe tunneling with
small applied biases from the 2d electron gas, the capacitive
technique that we use has allowed the only detailed tunneling study of
the 2d electron gas in semiconductors samples at low voltages across
the tunnel barrier. Novel structures\cite{GORN-doubwell} may
allow study with more conventional I-V measurements.

We note that the gap used in Fig.~\fign{lowtemp} is, like the
Altshuler gap, singular (the derivative diverges) at the origin.
Indeed, this singularity was necessary to account for the low
temperature behavior of the tunneling conductance.
Again, there is not sufficient sensitivity in these fits to discern
the precise shape of an energy gap. Given the results of
previous work done in tunneling from disordered systems, it seems
reasonable to conclude that the Altshuler gap is the source of the
tunneling suppression at low densities in our samples. One intriguing
feature of the data of Fig.~\fign{condlow} is the independence of the
temperature range over which the tunneling conductance undergoes
large variation. This is always in the range of 1 and 3~K. With the
``linear gap'' used to fit the data, this yields a gap
parameter $\Delta$ which is independent of gate voltage. This is odd
given that Eq.~\equn{altgap} predicts that the gap should deepen and
widen as the Fermi energy is decreased. Lastly, we note that as the
gate bias is reduced, the system may at some bias move to the
insulating side of the metal-insulator transition. In this case, electrons
are fully localized, and it is more appropriate to speak of a Coulomb
gap.\cite{EFROS-coul}

\v\par\penalty-1000
{\noindent{\sl Tunneling Conductivity Suppression by a Magnetic Field}}
\nobreak\v\nobreak
At present, we have no clear understanding of the mechanism which
produces the apparent energy gap in the single particle DOS in a
magnetic field.
The similarities between the temperature dependence of the low
density suppression results in Fig.~\fign{lowtemp} and the magnetic
field results in Fig.~\fign{summ} lead us to speculate that these two
effects might be different incarnations of the same underlying
physics. The main problem with interpreting the magnetic field
results this way, however, is that
one would expect the strength of the gap to depend
upon the thermodynamic DOS which varies by more than a factor of 10
(at the highest fields used here) as the Fermi energy moves from
the center of a Landau level to between Landau levels. Our magnetic
field results instead indicate that the strength of the gap is independent
of the position of the Fermi energy within the Landau level structure.

Aside from an energy gap, another
proposal for the cause of the tunneling
suppression in magnetic fields
is the enhancement, by some mechanism involving the
magnetic field, of the coupling of the tunneling transition to other
excitations of the system, e.g. phonons or plasmons.
The limiting value of the suppression at low temperatures would,
as observed, decrease with increasing field
as more tunneling oscillator strength
is transferred from the elastic to the inelastic channels. There is
no obvious candidate for the coupled excitation which would need to have a
characteristic energy of order 0.5~meV in order to explain, even
crudely, the temperature dependence of the suppression.

Finally we mention two characteristics in the magnetic field data
that may relate to the cause of the tunneling suppression
in magnetic field. These involve the shape of the capacitance vs.\
frequency curves for the device and the contrast in the tunneling
conductance at the DOS maxima compared to that at the DOS minima.

As noted in chapter~3, when the Fermi energy is
positioned between Landau levels at low temperatures, loss tangent and
capacitance vs.\ frequency deviate from the simple 
curves given by the circuit model of Fig.~\fign{sampprl}a
due to freezeout of the in-plane conductance of the 2d electron gas.
Intriguingly, this freezeout has a similar temperature dependence to
the magnetic field tunneling suppression.\cite{CHAP3}

Another interesting feature of the data, described
previously,\pcite{JOHN} is that the amplitude of the oscillations due
to the Landau level structure are much smaller
in the tunneling conductance than in the thermodynamic DOS. This is
true for all three samples. At present, we have no explanation for
this effect. Qualitatively, it appears
that the curves taken at lower magnetic field values (such as
at 2~T), at the lowest temperatures (well below 1~K), do achieve
nearly the same 2:1 peak to valley ratios that are seen in the
thermodynamic DOS measurement. Strangely, tunneling
conductance results from 2~T  develop this contrast
only at temperatures well below 1~K even though the 
thermodynamic DOS Landau level peaks are nearly fully developed at 1~K.
At higher magnetic field strengths the 2:1 peak
to valley ratio is not surpassed in the tunneling conductance even
though the thermodynamic DOS results have much larger ($\approx$10:1)
peak to valley ratios.

We find it intriguing that thermodynamic DOS
results of chapter~3 and tunneling results show the same amount of
contrast at high temperatures where the Landau level structure is
nearly washed out due to thermal broadening. As the temperature is
decreased both the thermodynamic DOS and the tunneling conductance
results develop contrast equally. However, at some temperature (at
around 7~K for the data at 4~T shown in Fig.~\fign{4tes}), the
tunneling data ceases to develop contrast (or develops it
only slowly) while the thermodynamic DOS continues to show more
contrast as the temperature is reduced. Interestingly, this
bifurcation takes place at temperature around $2\Delta/k_B$, where
$\Delta$ is the gap parameter. We speculate that the development of
the gap may wash out features in the single particle DOS of energy
width on order of the gap energy.

\v\par\penalty-1000
{\noindent{\bf 4.10 Summary}}
\nobreak\v\nobreak
In summary, we observe
two novel tunneling suppression effects in the tunneling of electrons
between a 2d electron gas and a 3d substrate.
We have completed a detailed study of the temperature dependence of
these effects.
One of these, a tunneling
suppression that occurs only for low densities of the 2d electron
gas, may be related to the energy gap seen previously in tunneling from
thin metal film systems and thought to arise from electron-electron
interaction effects in the presence of disorder. The other tunneling
suppression effect occurs over a wide range of densities of the 2d
electron gas in the presence of a magnetic field perpendicular to the
plane of the 2d electron gas. This suppression can be characterized by
a field induced gap in the single particle DOS tied to the Fermi
energy of width roughly 5\% of $\hbar\omega_c$. 
\vfill\supereject
\centerline{\bf References}
\v
\listrefs
\vfill\eject

\newchapter{5.}
\figitem{wafer}{Layer structure of wafer used in quantum dot
experiments.}

\figitem{dotsamp}{A schmatic cut-away of a completed quantum dot
sample. The entire top surface of the sample is covered with chrome,
creating a Schottky barrier everywhere along this surface.}

\figitem{underdot}{The band structure underneath a pillar etched on
the surface as determined by computer calculation. The figure is shown
at zero applied bias, and the dotted line is the Fermi energy. The
``$+$'' symbols represent unfilled energy levels of donors dispersed
in the AlGaAs blocking barrier.}

\figitem{betweendot}{The band structure midway between two pillars
obtained by computer calculation.
Note that the cap layer has been etched away. The figure is shown
at zero applied bias, and the dotted line is the Fermi energy. The
``$+$'' symbols represent unfilled energy levels of donors dispersed
in the AlGaAs blocking barrier.}

\figitem{hexdots}{A scanning electron micrograph of and array of AuCr
hexagonal patterns after metallization and liftoff.}

\figitem{etchdots}{An SEM photo of 200~nm wide metallized patterns
after the freon etch described in the text.}

\figitem{mesaetch}{An SEM photo of a finished sample. Shown is the
edge of a sample mesa after the CAIBE etch described in the text. The
pillars shown in the lower part of the photograph are not used in the
experiment. These have been etched in the final CAIBE step that
defines the mesa.}

\figitem{large}{Low (solid) and high (dashed) frequency
capacitance results from a 200~$\mu$m diameter
broad area mesa with no patterning to
produce quantum dots. The sharp rise in the low frequency capacitance
occurs as the gate bias is increased beyond the threshold which allows
electrons to enter the quantum well. The rise in both the low and high
frequency capacitance at positive gate voltage
occurs as electrons start to tunnel from the
well to donors sites in the AlGaAs blocking barrier.
These results were obtained from fits to data of the device
capacitance vs.\ frequency.}

\figitem{hexcap}{Low (solid) and high (dashed)
frequency capacitances from a 200~$\mu$m wide square mesa of hexagonal
patterns at 95~Hz
and 25~kHz, frequencies low and high enough repectively 
so that the capacitance can be considered to be the low and high
frequency limiting values. Notice the shift in the position of the
bifurcation of the low and high frequency capacitances as compared to
Fig.~\fign{large}. The low frequency capacitance rises much more
slowly at gate biases above this position in this sample than it does
in the broad area mesa.} 

\figitem{dotmodel}{The figure shows the model of the quantum dot
samples used to extract the area of electron packets from capacitance
measurements. Note that the figure is not to scale. In actuality, the width
of the pillars on top of the sample is several times 
larger than the distance from the metallic cap to the substrate, and
the pillar is much wider than it is tall. Further, over much of
operating range, the dot size is larger than its spacing to either the
top metallic cap or the substrate.}

\figitem{areaall}{Areas of dots
determined using Eq.~\equn{occup} for various
shapes and sizes of pillars on the sample surface. The top dashed line
is from a sample with
400~nm wide square pillars spaced with centers 720~nm apart;
the dashed-dotted line is from 300~nm square pillars with centers
600~nm apart; the solid line is from hexagonal shaped pillars with
parallel sides 260~nm apart and with centers spaced 520~nm apart; the
lower dashed curve is from 200~nm square pillars with centers spaced
450~nm apart.
The actual pillar size may be slightly smaller due to undercutting
during the freon RIE etching which produced the pillars. The sharp
rise in the area of dots at positive gate biases should be ignored as
this is an artifact caused by electrons tunneling to donor sites in
the AlGaAs blocking barrier.}

\figitem{compare}{Area of dots underlying hexagonal pillars from
capacitance data (solid) compared
to computer calculation (boxes). The scatter in the calculation's
results arises from digitization error. Note the horizontal and
vertical scales are different here than in Fig.~\fign{areaall}.} 

\centerline{\twelvebf Chapter V}
\v
\v
\centerline{\twelvebf Creation of Quantum Dots}
\v
\v
\v
\v
\v
\v
\v
\v
{\noindent{\bf 5.1 Introduction}}
\v
Additional confinement of an initially two-dimensional
system has been a very popular subject of research in recent years.
In this chapter, we discuss a fabrication technique which allows us to
move from studying electrons free in two dimensions (2d) in a quantum well
to electrons trapped in all dimensions. 
The confinement technique described here produces
``quantum dots'', packets of from one to several hundred electrons.
The number of electrons in these packets can be varied by means of a gate
bias. Because ours is a capacitance experiment, we need on the order of
10$^5$ dots each having a capacitance of around 0.1~fF, to obtain
sufficient signal for our measurements.
For the average of capacitance signals from the
individual dots to be meaningful, our experiment requires that there
be high uniformity of the size and shape of the many electron packets.
Hence there must as also be uniformity in the potential which produces
the confinement from dot to dot. Lastly and most importantly, we seek
to minimize the number of process steps required to produce the dots
in order to enhance the chances for successful processing.

Before describing the confinement technique that we use, we briefly
motivate it by reviewing some different confinement schemes that we
have experimented with in our lab.
The most obvious way to
achieve this kind of lateral confinement has been to etch wires or
pillars out of the 2d electron gas. This technique of a ``fabrication
induced confining potential'' has been used successfully in
photoluminescence and resonant tunneling experiments.
Photoluminescence experiments\cite{KASH-small} done on arrays of small
pillars show that even though there is a large surface state density
in GaAs, the surfaces produced by etching do not serve as efficient
nonradiative recombination sites. However, in tunneling experiments
using this type of confinement surface states are a major
consideration.  Because surface states pin the Fermi energy at the
surface to a position of $\approx$0.9~V beneath the conduction band edge in n
doped GaAs,\cite{SPICER-schottky, KAIS-schottky} in $n^+$ pillar there
results a carrier
depletion layer at the edges of the sample.  Obviously,
in order to do a tunneling experiment the pillar must not be fully
depleted of electrons. Reed et.\ al.\cite{REED-dot} have made
measurements on single pillars made on a resonant tunnel structure.
From observation of the current in a pillar in comparison to that of
broad area samples patterned on the same wafer, they conclude that the
width of the conducting channel in a 1000~$\AA$ diameter pillar is
only 130$\AA$.
A previous attempt in our lab\cite{JOHN} to produce dots for
capacitance measurements by simply etching through pillars
resulted in fully depleted dots.
It was decided to explore a different confinement technique.

Gating techniques have been used to produce small wires. Thornton et.\
al.\cite{PEPP-wire} first used gating techniques to produce wires small
enough so that effects of the quantum level structure on the
wires\cite{PEPP-magwire} could be observed. Topological considerations
make such a technique to
produce dots considerably more difficult than for producing wires.
In our lab,
Lebens\cite{JOHN} developed a dual gate technique to produce quantum
dots. This technique, had it been successfully realized, would have
given some degree of independent control of both the number of
electrons in a dot and the size of a dot. The procedure for producing
these dots entailed a large number of process steps, a few of them
being rather tricky. Two samples were realized; however, both of these
had leakage problems either between the dual gates or between one of
the gates and the substrate. Because of the complexity involved, we
decided to forsake this confinement scheme and to develop another with
fewer process steps. 
Only very recently, Alsmeier, Batke, and Kotthaus\cite{KOTT-vdots}
have achieved confinement of electrons into dots using a very similar
dual gate technique. Their dots are produced on Si, which allows them
to use SiO$_2$ barriers that avoid the leakage problems.
Another feature of their fabrication method is that it requires no
metal-semiconductor ohmic contacts unlike our dual gate method.

Our next design (when I took over the fabrication),
was a single gate technique to reduce the number of process steps.
In this sample, we attempted to
make good ohmic contact to short $n^+$
GaAs pillars produced on the surface
of the sample etched down a few hundred angstroms to an AlGaAs
insulating barrier. This exposed edge of the
AlGaAs barrier was still about 800~$\AA$ away
from the quantum well (on the other side of the AlGaAs barrier)
containing the 2d electron gas. After producing
these pillars we evaporated Al metal to form a Schottky barrier
everywhere else. The basic idea behind this confinement scheme was to
use the electrostatic potential produced by the Schottky barrier in
the region surrounding a pillar and the lack of a Schottky barrier
potential under the $n^+$ pillar to produce an electrostatic potential
at the well which would confine electrons.
Two samples were made;
one produced on the same M.B.E. material that was used by
Lebens and the other from the M.B.E. material from which sample {\bf
B} of chapter 3 was fabricated. Capacitance measurements revealed no obvious
results indicative of confinement of electrons in small packets. At
the time these samples were produced,
we were uncertain of how reasonable it was to expect that GeAu ohmic
contact could be made to each of the pillars diffusing enough Ge into
the GaAs to counter the effects of surface depletion.
We gave up on this technique, deciding that one possible cause for the
lack of a small packet signature in the capacitance
was that ohmic contact was not being
made to all, or even a large fraction, of the dots. Recent work by
Blanc et.\ al.\cite{OHMIC} 
has shown that the average distance
between conducting grains in a GeNiAu contact similar to the type that
we used is about 4100~$\AA$, which is larger than the 2500~$\AA$
diameter of the pillars that we used.

Each of the three confinement techniques used up to this point had
required ohmic contact be made to the tops of small pillars.
We then sought out a technique to make quantum dots without having to
make these ohmic contacts. At about the same time, Smith's group at
IBM was also working on making quantum wires\cite{TREY-wire} and
quantum dots\cite{TREY-zeem} for capacitance measurements. They developed
a technique\cite{TREY-fab} which requires no ohmic contact and
requires only a single gate to produce confinement on samples grown on
conducting substrates. Although there are
differences between our experiment and theirs (we measure tunneling to
and from the dots as well as the capacitance and use a quantum well to
create the initially 2d electron gas, whereas they rely on a self
consistent potential well forming at a GaAs/AlGaAs interface),
we decided to adapt the some features of
their technique to our experiment. The rest of this chapter discusses
the confinement technique and its implementation on our samples.

\v\par\penalty-10000
{\noindent{\bf 5.2 Physics of the Confinement Technique}}
\nobreak\v\nobreak
The confinement technique used here requires only a 300~$\AA$
corrugation of the surface of the sample to confine electrons in the
quantum well which is about 600~$\AA$ below the surface of the sample.
This section describes how this surface corrugation produces the
confinement.

\figinsert{\vfil\vskip6.5truein\includegraphics{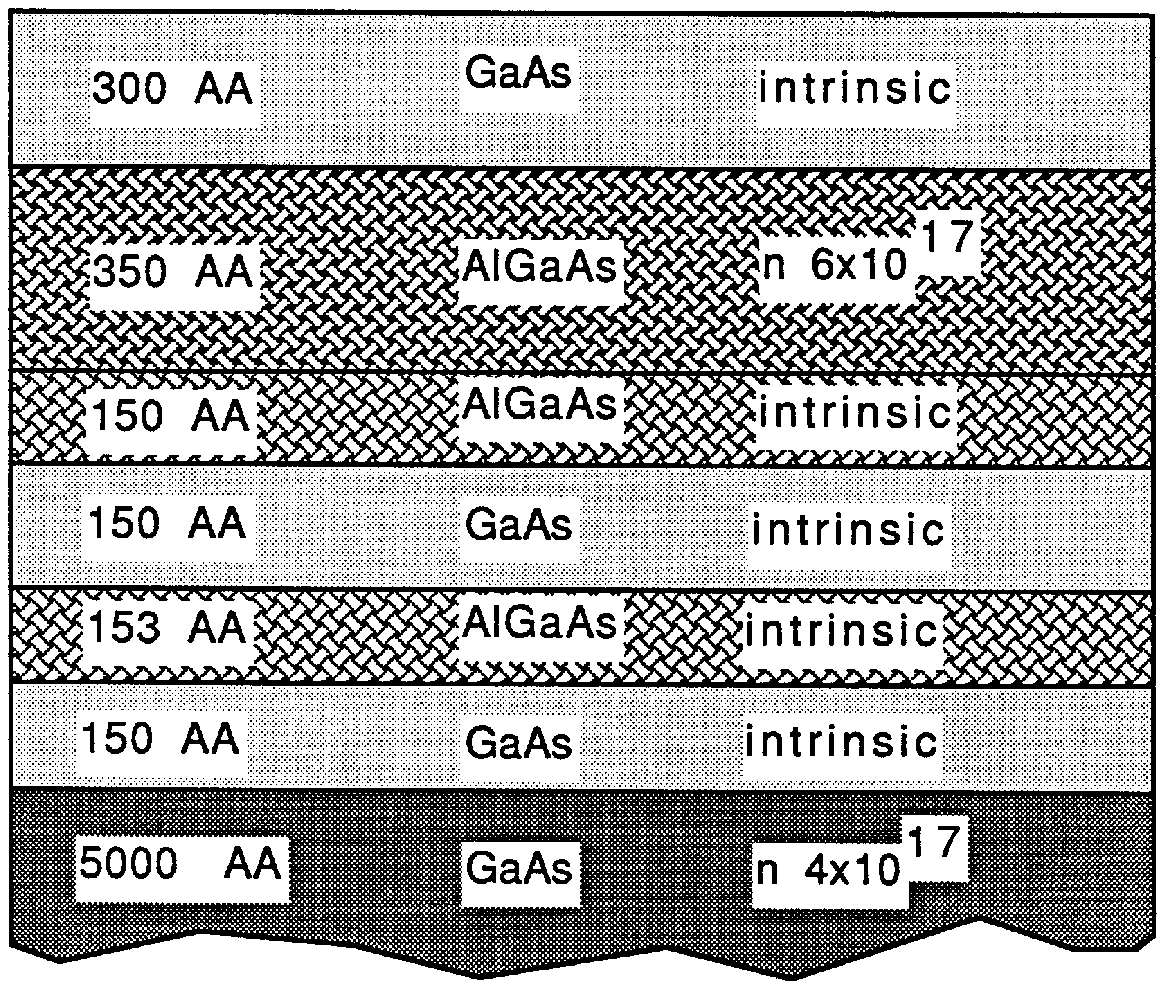}}{wafer}

Fig.~\fign{wafer} outlines the layer structure of the 
wafer that we use. The wafer was
grown using M.B.E.\ by L.\ Pfeiffer and K.\ West at AT\&T Bell
Laboratories. On top of a conducting substrate are grown: A 5000~$\AA$
GaAs $n^+$ layer, a 1500~$\AA$ $n^+$ ($4\times10^{17}$~cm$^{-3}$)
GaAs bottom electrode, a 150~$\AA$ intrinsic GaAs spacer layer, a
152~$\AA$ (54 monolayer)
Al$_{.3}$Ga$_{.7}$As tunnel barrier, a 150~$\AA$ quantum
well, a 150~$\AA$ GaAs quantum well, a 150~$\AA$ undoped
Al$_{.3}$Ga$_{.7}$As setback, a 350~$\AA$
n doped ($\approx6\times10^{17}$~cm$^{-3}$) Al$_{.3}$Ga$_{.7}$As
blocking barrier layer, and finally a 300~$\AA$ intrinsic GaAs cap
layer.

\figinsert{\vfil\vskip6.5truein\includegraphics{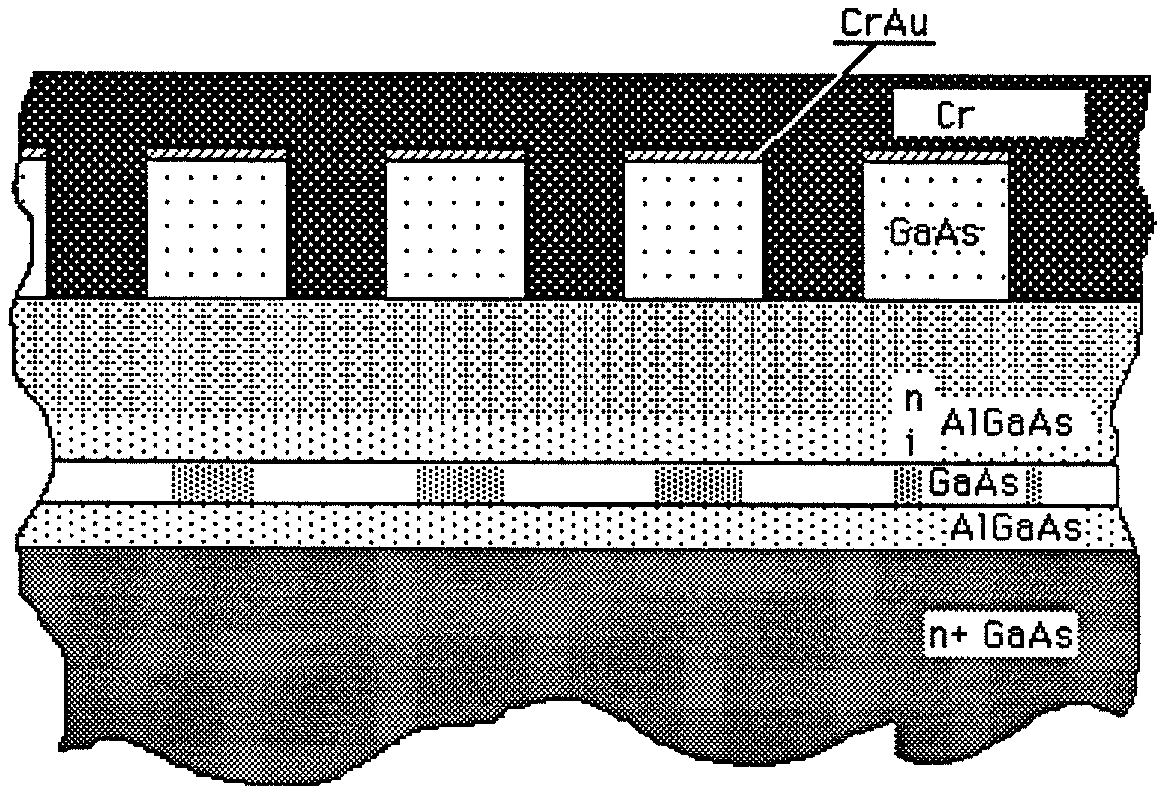}}{dotsamp}

Fig.~\fign{dotsamp} displays the sample after all processing has been
completed. We will describe the details of the processing later. The
finished sample consists of 300~$\AA$ tall pillars of GaAs, produced
by etching the GaAs cap layer of the wafer, embedded in Cr metal. Cr
metal contacts the surface of the semiconductor material surface
everywhere. Between the GaAs pillars, the semiconductor material
contacted by the Cr is Al$_{.3}$Ga$_{.7}$As. Everywhere along the
chrome-semiconductor interface, a Schottky barrier is formed.

\figinsert{\vfil\vskip6.5truein\includegraphics{underdot.ps}}{underdot}

This Schottky barrier is responsible for producing the confinement.
Fig.~\fign{underdot} is a band structure diagram of the sample
underneath the center of a 300~nm diameter (round)
pillar. The figure is shown with zero
applied gate bias. At the top surface, the Fermi energy is pinned to a
position of 0.8~eV below the conduction band edge, appropriate for a
surface exposed to air prior to Cr evaporation.\pcite{Cr-schottky}
The wafer was grown to have a large density of ionized donors in the
``blocking barrier''. Under a dot, the 0.8~V Schottky barrier is {\it far
enough} away from these donors that the electric field caused by
the donors multiplied by the distance to the Schottky barrier is
larger than 0.8~V. This means that the electric field from the Schottky
barrier is ``terminated'' at the donors, and the electric field just
at the edge of the quantum well is one which, due to the positive
donor charge, draws in electrons.

\figinsert{\vfil\vskip6.5truein\includegraphics{betweendot.ps}}{betweendot}

If the GaAs cap layer is etched away, the position of the Schottky barrier
will be {\it too close} to the donors for the Schottky electric field
to be terminated at the donors.
This is the situation in Fig.~\fign{betweendot}
which shows the electrostatic configuration calculated under the
region halfway between two 300~nm diameter pillars spaced with edges
340~nm apart and again at zero applied gate bias.
The electric field at the edge of the well repels electrons electrons
from the well, and the well remains empty at this position. Hence by
simply corrugating the surface, we can achieve lateral confinement of
electrons in the well. The relaxation method calculations for solving
Poisson's equation (classically) that produced Figs.\
\fign{underdot}~\&~\fign{betweendot} is described
elsewhere.\cite{JOHN} We will return to these calculations and compare
them with capacitance data from the a dot sample after a brief discussion of
the processing sequence which produces the dots.

\v\par\penalty-1000
{\noindent{\bf 5.3 Device Fabrication Sequence}}
\nobreak\v\nobreak
There are five basic steps in our device fabrication. These are:
1)~electron beam lithography and liftoff to produce CrAu metal pads on the
sample surface, 2)~photolithography to define mesas for capacitance
measurements (masks for liftoff of step \#4), 3)~CCl$_2$F$_2$ (freon)
etch to produce 300~$\AA$ tall GaAs pillars
on sample surface using the CrAu pads as an etch mask, 4)~evaporation
of Cr over the entire sample surface and liftoff of photoresist to
leave large (200~$\mu$m)
Cr pads covering $\approx10^5$ GaAs pillars, and 5) a mesa
isolation etch using the large Cr pads as a mask to produce separate
devices for capacitance measurements.

We describe these process steps serially, starting with the steps just
prior to electron beam lithography.
A small piece of the wafer, with AuGeNi ohmic contact already made to
the back side of the conducting substrate, is first cleansed
thoroughly in organic solvents and then subjected to a ``descumming''
oxygen plasma.
Extreme caution must be taken that the plasma potential is small
enough so that the bombarding oxygen ions do not damage the sample. We
use the Branson/IPC P2000 barrel etcher system at the National
Nanofabrication Facility (NNF). In this system, oxygen ions are
accelerated to $<10$~eV. Further, we use special shielding in the
etcher known as an ``etch tunnel'' which effectively neutralizes a
very large fraction of the oxygen ions.\cite{JERRY} The cleansing
effect of the oxygen plasma occurs largely as neutral oxygen radicals
diffuse to the surface and react with adsorbates.

We then spin on a bilayer resist. The top surface layer of the bilayer
is a resist with lower sensitivity to the electron beam than the
underlying layer. When this bilayer is exposed to the electron beam
and developed, the surface layer produces an opening which is undercut
in the underlying high sensitivity layer. This undercut is necessary
for easy liftoff of metallization.
We wet the sample surface
with 3\% solids copolymer (PMMA-MMA) solution and spin this at
5000~rpm. The sample is then baked for 1~hr at 170$^{\circ}$C. These
leaves about 700~$\AA$ of resist on the surface. Then a 2\% solids
mixture of PMMA is spun on at 6000~rpm. The sample is baked again for
1~hr, and this leaves a 1000~$\AA$ layer of PMMA. The sample is then
loaded into the JEOL JBX5D11 electron beam lithography system at the
NNF. 1~mm by 1~mm arrays, containing $\approx4\times10^6$ patterns
of either
hexagons or squares of particular sizes and spacings, are exposed.
Typically, 4 or more arrays are exposed, each containing different
pattern shapes or sizes. We use a beam current of 300~pA which
produces a spot size of around 20-30~nm\cite{JEOL} for high resolution of our
patterns. Exposures vary from between 160~$\mu$Coul/cm$^2$ to
240~$\mu$Coul/cm$^2$ depending on the size and spacing of the patterns
in the array. A hexagon, 300~nm across in its long dimension and
center separation from nearest neighbors of 520~nm in a hexagonal close
packed array consists of about 100 exposures per hexagon with a
current density of 200~$\mu$Coul/cm$^2$.

The resist is then developed in a 1:1 mixture of methyl isobutyl
ketone and isopropyl alcohol at 21$^{\circ}$C for 60~sec. This is
followed by an 8~sec.\ rinse with isopropyl alcohol to wash away any
excess resist still left in the patterns. As the alcohol causes 
some development of the underlying PMMA-MAA layer, care must be taken
to standardize this rinse time.

\figinsert{\vfil\vskip6.5truein\includegraphics{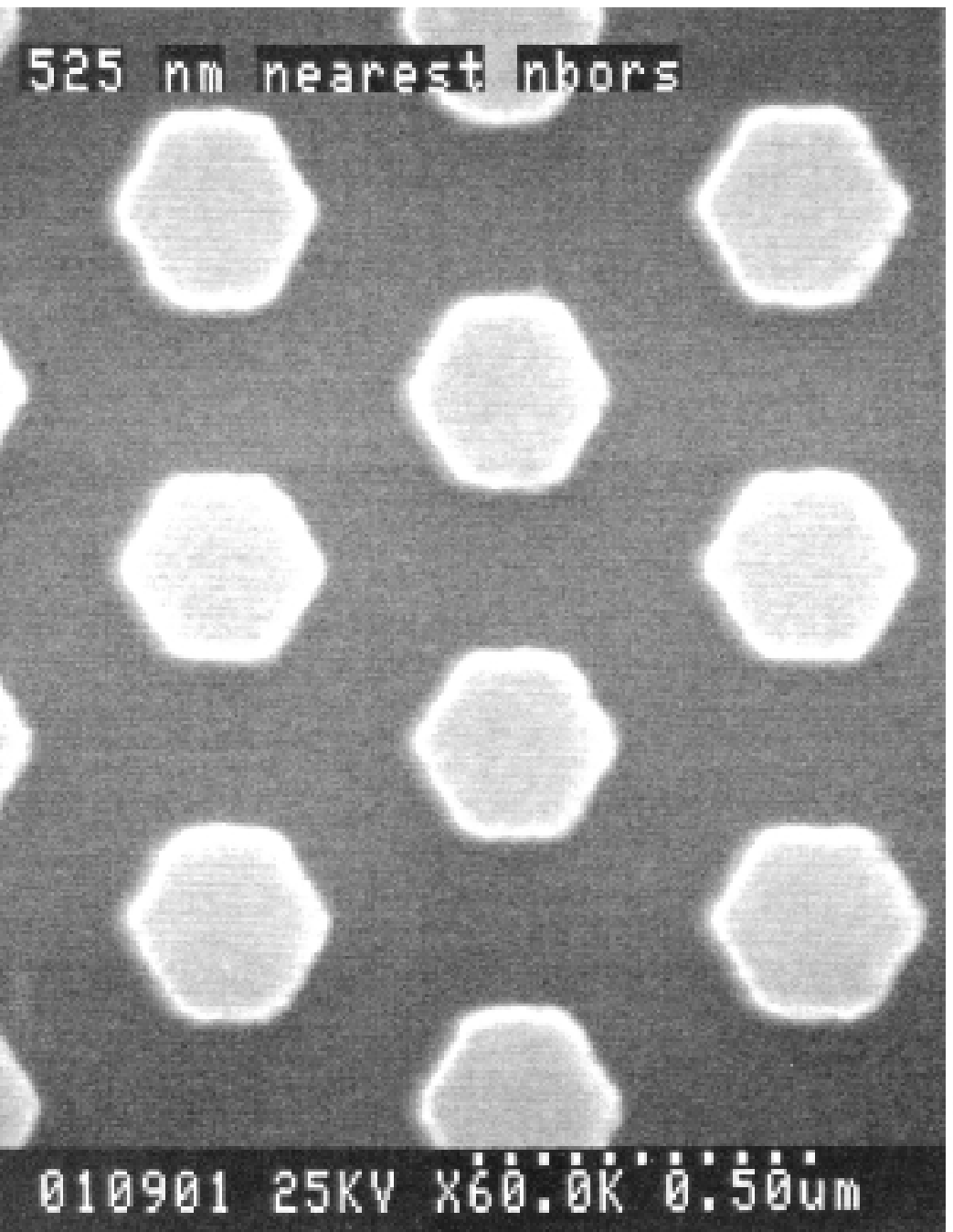}}{hexdots}

After development, the sample is prepared for evaporation to produce
small metal pads on the sample surface. The sample is again subjected
to a short oxygen descum to clear away any residual resist.
The sample is then dipped in a 1:5
dilution of hydrochloric acid to clear away any oxide that may have
formed on the sample surface. From the acids hood, the sample is rushed
to and quickly loaded in the electron beam evaporator at the NNF so
that the oxide does not have time to reform. We evaporate 200~$\AA$ of
chrome followed by 200~$\AA$ of gold. Liftoff is then performed in
methylene chloride and acetone. Fig.~\fign{hexdots} displays hexagonal
patterns on the surface of a sample after liftoff. The resolution
shown and the uniformity of the dots is typical for our samples.

At this point, the sample consists of several large arrays of metal
pads on the surface of a piece of the M.B.E.\ grown wafer. These
arrays are so large, that it is likely that there are several oval
defects contained in a large array. Because these oval defects can
create shorts in our devices, these must be avoided. To increase the
chances for finding a non-shorted device, we divide the array into 16,
200~nm on a side, square pieces using photolithography. Photoresist
(using no primer) is spun onto the sample and is exposed and developed
so that each large array has 16 box regions free from resist, each
separated by resist. Each box region contains between $0.8\times10^5$
and $2\times10^5$ patterns. Primer is not used before spinning on the
photoresist as it is thought to leave more residue after development
than photoresist when used alone.\pcite{WOODARD}

We now proceed to etch the metal patterns to produce pillars. Before
loading into the reactive ion etcher, we do another oxygen descum to
remove any remnant photoresist in the box regions and another
hydrochloric acid dip to remove any oxide that might impede
etching.\cite{JOHN}
The CCl$_2$F$_2$ etch\cite{CCL2F2} and the physical
processes\cite{CCL2F2-stop} which allow it to etch GaAs and not AlGaAs
have been well characterized by others. In order to reduce etch
damage from fast and light helium atoms,\cite{ROUK-unpub}
we chose not to mix in helium with the freon.
This has been done by others\cite{CCL2F2,JOHN} remove carbon contaminants from
the etch which induce poor surface morphology and to improve the
directionality of the etch. For the short, 300~$\AA$ etching distance
before the etch is stopped by an underlying AlGaAs layer in our
samples here,
carbon contaminant buildup and undercutting of the pillars occur
only insignificantly. There was no observable difference in sample
morphology with and without helium added to the etchant.
The freon
etch was run under conditions of a low D.C.\ self bias of 60~V (r.f.\
power was 0.2~W/cm$^2$) and a pressure of 24~mT. The low self bias was
again chosen to minimize etch damage. Etch times were typically around
10~sec., about one and a half to
twice as long as needed to etch through 300~$\AA$ of GaAs.

\figinsert{\vfil\vskip6.5truein\includegraphics{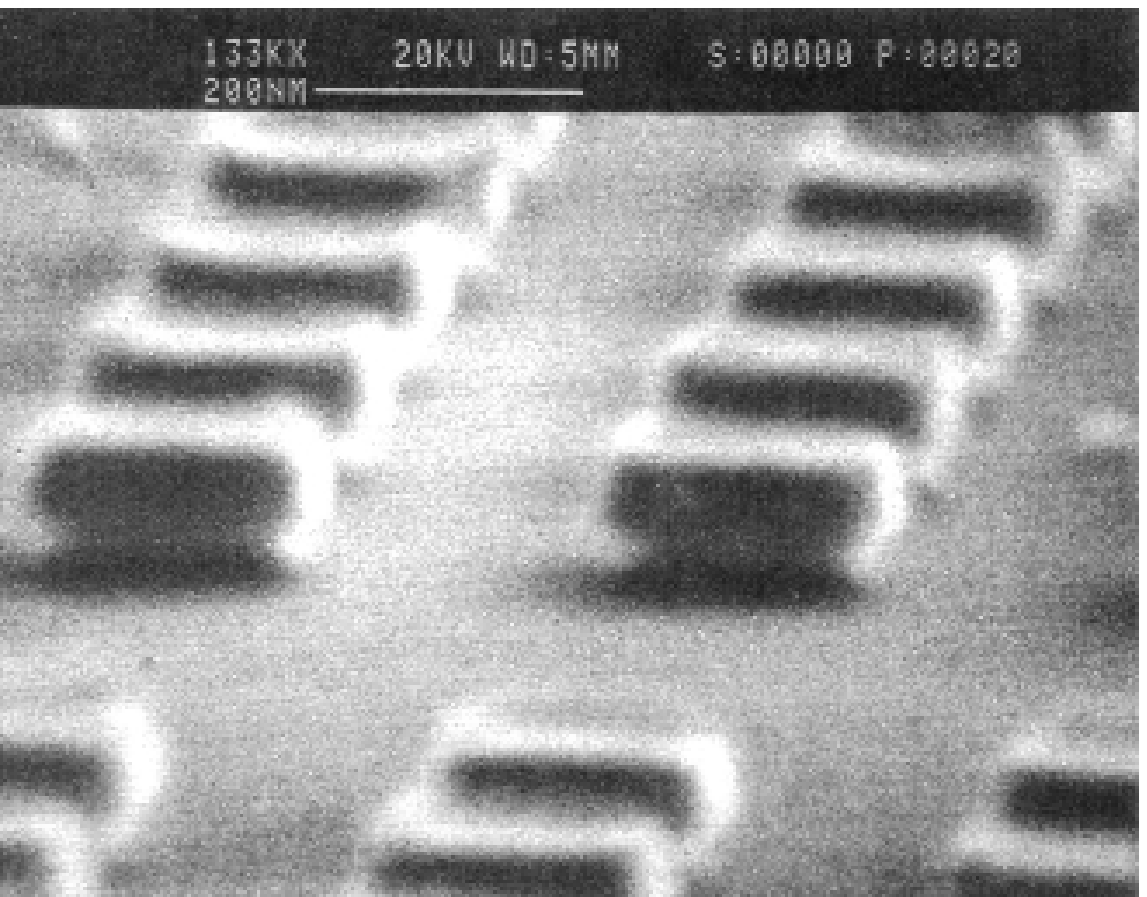}}{etchdots}

Fig.~\fign{etchdots} displays a set of 200~nm square dots after
etching. There is an obvious undercut, indicative that etching has
taken place, underneath the square CrAu metal pads which form ``hats''
on top of the pillars. The smooth surfaced material between the
pillars is AlGaAs.

Before doing the final chrome evaporation, we again descum the sample
with oxygen and do another hydrochloric acid dip. The sample is loaded
into the thermal evaporator and 2000~$\AA$ or more of Cr is evaporated
on the surface. We then liftoff the Cr in acetone leaving 16 boxes per
large array of pillars buried in chrome. 

\figinsert{\vfil\vskip6.5truein\includegraphics{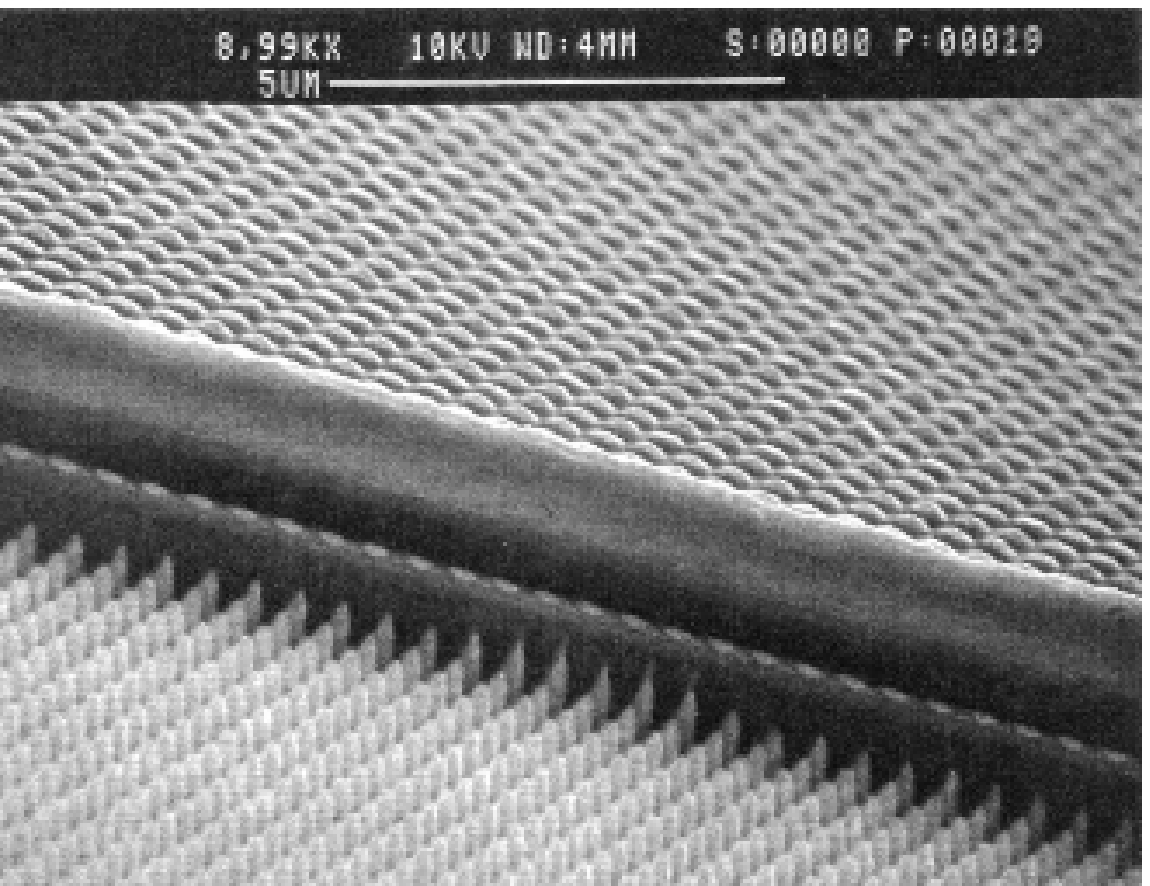}}{mesaetch}

Finally, we use the Cr as an etch mask in a CAIBE etch to produce
mesas for the device. This step was done with the assistance of George
Porkolab in the Technics Plasma GmbH R.I.B.-etch 160 system at the
NNF. A beam of 500~eV argon ions assists etching by a chlorine etch. The run
parameters used were: argon flow of 3.5~mL/min, chlorine flow
9.0~mL/min, and 0.1~mA/cm$^2$ beam current. The etch ran for 15
minutes and etched to a depth of 1~$\mu$m.
Fig.~\fign{mesaetch} is a picture of the edge of a mesa of
260~nm wide hexagons after this etch. The pillars off of the edge of the
mesa arise from metal pads previously covered with photoresist and
thus not protected by the large chrome boxes. The sample is now
mounted and wirebonds are made to the mesas.

\v\par\penalty-1000
{\noindent{\bf 5.4 Measurement of Dot Area}}
\nobreak\v\nobreak
In this section, we look at the capacitance vs.\ gate bias
characteristics of one of the device mesas. We interpret these results
in terms of the area in the quantum well which contains electrons.
This can then be compared with a computer model which predicts the
area of the sample as a function of gate bias.
The detailed results of high resolution capacitance measurements,
and our search for effects of the ``Coulomb Blockade''\cite{LAMB-coul}
will be discussed in Chapter~6.

As discussed in Chapter 3, the capacitance of our devices is a
function of frequency. The capacitance measured at low frequencies, $C_{low}$,
measures the capacitance of the device including charge transfer to
the quantum dots from the substrate.
The distinction between low and high frequency arises from the
$RC$ equilibration time the quantum dots and the substrate.
At high frequencies, there is insufficient time during one cycle of
the measuring voltage for appreciable charge to be transferred between
the electron density in the substrate and the quantum dots, and we
call this capacitance $C_{high}$.
The capacitance at low frequencies is thus larger than that measured at
high frequencies. As described below, knowledge of the value of
$C_{low}/C_{high}$ in comparison to its value in broad area samples
(no dots patterned) gives a measure of the size of the dots.

\figinsert{\vfil\vskip6.5truein\includegraphics{large.ps}}{large}

For the same wafer upon which the quantum dots were
fabricated, we have made measurements on broad area samples
of the low and high frequency capacitances. These are displayed in
Fig.~\fign{large}. At very low gate biases, $C_{low}$ and $C_{high}$
are the same; as the gate bias is increased beyond the
threshold at which electrons
begin to populate the well, the $C_{low}$ increases to a value larger
than $C_{high}$. In the broad
area case in the absence of magnetic field, the ratio of low to high frequency
capacitances is very nearly independent of the electronic density
(gate bias) in the well.
By measuring the ratio, $C_{low}/C_{high}$
in the quantum dot experiment and comparing this with the ratio
(about 1.36)
obtained in the broad area experiment, we can determine the fraction
of the area in the quantum well which is filled with electrons. Since
we know the number of quantum dots in a sample and the total area of
the sample, we can thus determine the area of a dot.

\figinsert{\vfil\vskip6.5truein\includegraphics{hexcap.ps}}{hexcap}

Figure~\fign{hexcap} displays the low and high frequency capacitances
from an array of dots
produced using hexagonal shaped pillars (260~nm wide from edge to
edge) in a hexagonal array (520~nm spacing between centers of nearest
neighbors). Notice two striking differences between these results and
the capacitances obtained for the broad area samples in
Fig.~\fign{large}. Firstly, the threshold gate voltage
at which electrons begin to enter the well is significantly increased
in the case of the quantum dot sample compared to the broad area
sample. This is expected as a result of the extra
repulsion of electrons away from the
quantum well by the surface Schottky potential in etched regions of
the sample surface.
Secondly the increase of the $C_{low}$ above $C_{high}$ is much
slower as the gate bias is increased compared to the broad area
sample. This is a consequence of a slow increase of the area of a
dot as the capacitance is increased.

\figinsert{\vfil\vskip6.5truein\includegraphics{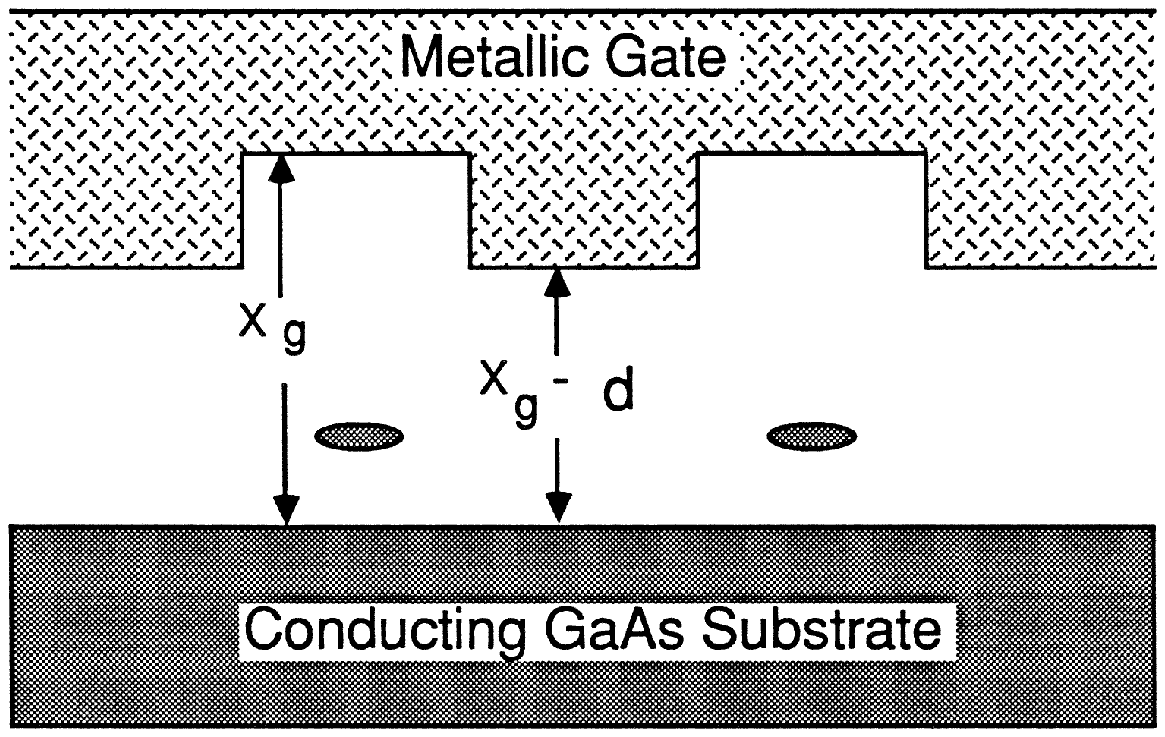}}{dotmodel}

Figure~\fign{dotmodel} graphically displays the model that we use to
determine the area of a dot. Note that this figure is not drawn to
scale. In reality the width of the pillars on top of the sample is
several times
larger than the distance from the metallic cap to the substrate.
Our smallest pillar widths are 200~$nm$, whereas the pillars are
$30$~nm high.
Also, the dot size is much larger than its spacing to the top
metallic cap or the substrate for most of the operating range of the
device. A main assumption in our model for the dot area determination
is that electric field lines point only vertically through the sample
structure. This assumption reasonable as long as the dot diameter is larger
than around 50~nm (.002 square micron dot area). 
Using the model of Fig.~\fign{dotmodel} and neglecting fringing
fields, it is easy to
to show that the fraction of the
area occupied by electrons in the well, $\alpha$ is given by:
$${\alpha}={\biggl[}{C_{low}\over
C_{high}}-1{\biggr]}{\biggl(}\Upsilon-{x_g-\delta\over
x_g-D\delta}{\biggr)}^{-1}.\eqno(\equn{occup})$$ 
Here we have denoted the
distance from the conducting substrate to the top
gate in unpatterned samples as $x_g$, the height of the pillars on the
sample surface by $\delta$, the fractional area of the surface
of the sample that is made up of pillars by $D$, and
$C_{low}/C_{high}$ for the broad area sample (in which $D$ is
effectively equal to one), with the
quantum well occupied everywhere, by $\Upsilon$.
Note that Eq.~\equn{occup} is only correct when the area of the dot is
smaller than the area of the pillar on the surface (true for all of
our samples).
We have assumed in the argument that led to
Eq.~\equn{occup} that the density of states per unit area and energy
is the same for the broad area 2d electron gas as it is for the
quantum dots. If this is indeed true, the density of states drops out
of the equation for determining the occupied area entirely, as in
Eq.~\equn{occup}. In any case level spacing in the well is an order of
magnitude smaller than the electrostatic energy so errors due to any
incorrectness of this assumption are likely to be very small.

\figinsert{\vfil\vskip6.5truein\includegraphics{areaall.ps}}{areaall}

The area of a dot is of course given by $\alpha$ times the area of the
entire sample divided by the number of pillars patterned onto the
sample surface. Figure~\fign{areaall} displays the area determined
from capacitance measurements on four samples with different size and
shape patterned pillars. One obvious feature here is that smaller
patterns produced on the sample surface create smaller sized dots
whose threshold voltage for filling increases as the pattern size
is made smaller. Results from above 0~mV gate
bias in the figure should be ignored; the sharp rise in the plotted
results here is an artifact caused by
electrons tunneling from the well into donors in the doped region of
the AlGaAs blocking barrier. Below 0~mV, it is clear that the size of
the dot remains smaller than the size off the patterned pillar.

\figinsert{\vfil\vskip6.5truein\includegraphics{compare.ps}}{compare}

We can now compare these results with the computer calculation
discussed earlier. Figure~\fign{compare} shows the area from
capacitance data for an array of hexagonal dots compared with the
calculated results. We make a few brief remarks on this
calculation.\pcite{JOHN}
The calculation assumes a Schottky barrier height of 0.8~V on the
sample surface. The doping density in the AlGaAs blocking barrier is
adjusted so that the calculation correctly
predicts the gate bias threshold for electrons to begin entering the
quantum well in a broad area sample. A threshold
in agreement with data is obtained if the doping density is adjusted
to a value of $5.75\times10^{17}$~cm$^{-3}$, very close to the value
estimated by the crystal grower\cite{LOREN} of $6\times10^{17}$~cm$^{-3}$.
Particularly interesting in Fig.~\fign{compare} is the fact that the
computer calculation correctly predicts the gate bias threshold for
filling the dots. This is true systematically for different size dots.
The calculation does less well at predicting the size of the dot as
it grows. The area from the calculation is, depending on gate bias,
up to 50\% larger than the area determined from capacitance data.

The disagreement in sample area between calculation and data
in Fig.~\fign{compare} may arise from several factors.
The computer calculation is two-dimensional. It assumes
cylindrical symmetry of potentials about the center of a dot.
In the actual sample, neither the square or hexagonal
pillars on the top surface or the positions of neighboring dots around a
central dot have cylindrical symmetry. In the computer calculation, we
simply adjust the radius of a cylindrical pillar so that it has the
same area as the true square or hexagonal pillars. We also adjust the
size of a ``unit cell'' in the calculation so that its area is the
same as that of a unit cell in the actual sample configuration.
A potentially important issue in the dot confinement
that of ``etch damage''. The
freon etch described above in the sample processing section which
produced the pillars on the surface of the sample may, despite the
very low bias voltage at which it was carried out, cause some etch
damage. Similar etch damage has successfully been used by Scherer and
Roukes to produce confinement in a heterostructure with the 2d gas
about 500~$\AA$ away from the sample surface
with beams of 200~eV Ne ions.\pcite{ROUK-resol} These damage effects
may confine our dots to a smaller area than would otherwise be
expected from an electrostatics calculation alone.

Another effect, this time ignored in the determination of the dot area
from the sample capacitance, is a contribution to the device
capacitance due to the rate of increase of the dot area as the gate
bias is increased.
Back of the envelope calculations indicate that the actual dot area
may, in a worse case picture of dots confined in lateral square
well potentials, be as much as 50\% smaller than the results
indicated by Eq.~\equn{occup}.
However, if the potential sidewalls of the dot are as gradual in slope as
expected from computer calculations, this effect would increase the
measured dot area by at most a few percent compared to the actual dot
area.
Correction for this effect would worsen agreement between the computer
calculation for the dot size and the experimental results.

With the quantum dot system now reasonably well characterized, we now
search in chapter~6 for new physics arising from the confinement of
electrons in small packets.

\vfill\supereject
\centerline{\bf References}
\v
\listrefs
\vfill\eject

\newchapter{6.}
\figitem{equilib}{(a) schematically describes the structure of our
devices. Small packets of electrons (dots) are embedded in a large
capacitor. Tunneling of electrons is permitted between the dots and
the bottom electrode of this capacitor; the dots are electrically
insulated from the top electrode.
Changing the voltage across the device (gate voltage) varies
the electron occupancy within the dots from
zero to as many as several hundred electrons. (b) depicts the
``equilibrium zone'' of width in energy $e^2/C$ on either side of the
substrate Fermi energy which develops as a
manifestation of the ``Coulomb blockade''. For simplicity, consider
quantum levels in the dots to be nondegenerate. Dark bars in the wells
shown to the left represent filled quantum levels. (i) depicts a dot
in equilibrium even though its chemical potential is greater than that
in the substrate. If this dot were to lose an electron, its chemical
potential, due to the change in its electrostatic energy of $e/C$
would fall below the lower edge of the equilibrium zone.
In (ii), a nonequilibrium situation is shown with an unfilled level
below the edge of the equilibrium zone. This level is filled in (iii)
and the electrostatic energy of the dot increases by $e/C$.}

\figitem{spect}{(a) portrays the electron addition spectrum for
circular dots of area .014~$\mu$m$^2$ after multiplication by the
inverse of the geometric lever arm (${1\over\kappa}\approx3.6$)
in this sample. (b) shows
this spectrum after the area correction described in the text.
The areas of the model before and after the varying area correction are the
same at a gate bias of 355~mV. At this gate bias, both before and
after the area correction the dots contain 34 electrons.}

\figitem{capcomp}{The solid curve is the device capacitance at low
frequencies minus the high frequency capacitance. This signal is the
capacitance resulting from electron transfer between the quantum dots
and the substrate. The dashed curve is the capacitance obtained
assuming a Gaussian distribution, of rms width 35~mV,
of threshold voltages for the different quantum dots in the array and
using the protocol for obtaining the device capacitance described in the
text. The dotted curve is the same as the dashed curve except that it
has been shifted by 10~mV.}

\figitem{ellipse}{The solid curve plots the ellipse function described
in the text which is convolved with the capacitance ideally measured
at infinitesimal measuring voltage to give the capacitance indicated
by the lock-in amplifier. The ellipse is for a 1.5~mV~rms sinusoidal
measuring voltage. The dashed curve shows the
convolution function imagined for
capacitance measurements made at higher frequencies where there is
insufficient time in a half cycle of the measuring voltage for
equilibration of the dots and substrate to occur. See section 6.5
for explanation.}

\figitem{noise}{The figure displays a typical capacitance fluctuation
spectrum for the addition of the $n^{\rm{th}}$ electron to each dot in
the quantum dot array. The spectrum was developed using the protocol
described in the text using a 2.44~mV~rms sinusoidal measuring voltage.
The results were obtained assuming a ${\overline{\rho}}(V)$
described by a Gaussian distribution of rms width 35~mV.}

\figitem{artspectl}{The figure displays a simulated capacitance fluctuation
spectrum developed from our model by
repeating the fluctuation shown in Fig.~\fign{noise} at every electron
addition (shown by the sticks in the bottom portion of the figure).
The smooth background has been subtracted.}

\figitem{artspecth}{The figure displays a capacitance fluctuation
spectrum developed from our model by
repeating the fluctuation shown in Fig.~\fign{noise} at every electron
addition (shown by the sticks in the bottom portion of the figure).
This is the result of the same
calculation used in Fig.~\fign{artspectl}. This
plot shows a different region of gate bias.}

\figitem{rndspect}{Displayed is the result of repeating the
capacitance fluctuation spectrum of Fig.~\fign{noise} at for a random
electron addition spectrum.}

\figitem{d13d}{The figure shows the sample capacitance (measured at
615~Hz with a 2.44~mV~rms measuring signal)
with the smooth background subtracted off. The quasi-periodic
peaks in the fluctuations in
the capacitance for gate voltages between $-250$~mV and $-50$~mV are
thought to occur with the nearly the
same frequency as the electron additions to the dots. For gate
voltages above $-50$~mV, where the spacing between successive electron
additions in our model is smaller than the peak to peak amplitude of
the measuring voltage, the fluctuations have a more random character.
Plotted on the abscissa are the electron additions from the model
spectrum.
(All experimental results in this and other figures in this chapter
are from the same array of
197,500 quantum dots made from 0.04~$\mu$m$^2$ square patterns.)}

\figitem{d13zh}{The figure shows the fluctuations in the device
conductance (at 1.7~kHz and with a 2~mV~rms measuring voltage).
It illustrates the difficulty in distinguishing fluctuations
associated with the electron addition spectrum and those, unrelated to
the addition spectrum, which occur with the measuring voltage much
smaller than the spacing between electron additions. Peaks in the 
fluctuations in
the lower gate bias region of the graph have a periodicity shorter
than the electron additions given by the model spectrum, whereas at
higher gate biases the periodicities are about the same. However
without the model spectrum as a guide, it is difficult to judge where
the fluctuations become ``regular'' and associated with the addition
spectrum.}

\figitem{d13zhcap}{The capacitance from the same run which produced
Fig.~\fign{d13zh} is displayed. While most all of the peaks in both
figures occur at the same gate bias values, the capacitance contains
additional fluctuations, such as those around -190~mV and -80~mV,
which are confined to a narrow enough range that they are not subtracted off
in the smooth background subtraction.}

\figitem{regspect}{The results of repeating the fluctuations from
Fig.~\fign{noise} (except that a 1~mV~rms measuring voltage is used in
the convolution of Eq.~\equn{clockcdev}) on a spectrum calculated with
all states considered to be completely degenerate. Note that the
removal of the quantum level splittings causes the
capacitance spectrum to contain much more regular oscillations than
those seen in Fig.~\fign{artspecth}}

\figitem{d13mbh}{The figure shows the fluctuations in the device
conductance (measured at 1.7~kHz) in a magnetic field of 4.3~T. Note
the long coherence of the oscillations.}

\figitem{d13mbl}{The figure displays conductance
fluctuations from the same data
run which was used to produce
Fig.~\fign{d13mbh} at a lower region of gate biases.}

\figitem{d13ag}{Shown is the frequency of the relaxation peak of a
quantum dot array,
determined from fits, plotted as a function of the rms amplitude of
the sinusoidal voltage used to make the measurements. The gate bias
voltage is set to zero. The solid curve is a guide to the eye.}

\figitem{capfreqdot}{The figure displays the capacitance and loss
tangent of the quantum dot sample as a function of frequency for
gate bias set to zero and using a 2.4~mV~rms measuring signal along
with fits to them as described in the text. The
parameters for the capacitance fit are as follows:
$f_{peak}=1.43$~kHz, $C_{high}=38.54$~pf, $\kappa=.156$, and
$\chi=4.05$.}

\figitem{broadamp}{Plotted is the broadening parameter $\chi$ obtained
from fits as a function of the amplitude of the measuring voltage.
The smooth curve is a guide to the eye.}

\figitem{caphigh}{The figure shows the high frequency capacitance
$C_{high}$ plotted as a function of the amplitude of the measuring
voltage. The smooth line is a guide to the eye.}

\figitem{d13agcomp}{The device capacitance is plotted as a function of
frequency for two different amplitudes of the measuring voltage. The
boxes are for a voltage of 0.39~mV~rms and the crosses are for
6.1~mV~rms. The capacitance at high frequencies for the two curves
appears to be approaching different limits.}

{\centerline{\twelvebf Chapter VI}}
\v
\v
{\centerline{\twelvebf Observation of the Single Electron Addition
Spectrum}}
{\centerline{\twelvebf of Quantum Dots}}
\v
\v
\v
\v
\v
\v
{\noindent{\bf 6.1 Introduction}}
\v
In the last 25 years, technological advances have allowed the
construction of ``particles'' so small that the addition or
subtraction of an individual electron to them leads to
experimentally observable effects. In a pioneering work, Giaver and
Zeller\cite{GIA-tun} demonstrated the effects of the Coulomb repulsion
of single electrons on the conductivity
of a film of very small tin particles embedded in an oxide layer.
The effects of this Coulomb repulsion of single electrons in small
particles can be described by a ``charging
energy'', $e^2/C$, where $C$ is the capacitance of the small particle
to its environment. Manifestations of charging energy have been
observed more recently by Fulton and Dolan\cite{FULT-charge} in
a sample in which tunneling currents pass between two electrodes
through a single small metal particle. The essential
effect that they see is a lack of electric conduction through their
device until the voltage across the device reaches a threshold beyond
which electrons can overcome the Coulomb repulsion associated with
tunneling to the particle. This effect has been dubbed the ``Coulomb
Blockade'' in tunneling. The Coulomb blockade can also be observed
in tunneling across small capacitors that do not contain a small
particle under certain conditions depending on the impedance of the
the leads which transport charge to the capacitor.\pcite{LIK-both}
Coulomb blockade effects have now been used to manipulate the motion
of single electrons in multi-capacitor devices,\cite{GEER-turnstile}
exposing the possibility that such devices may be used 
in ``single electron logic'' functions.

Aside from the Coulomb blockade, another more obvious effect in a small
particle is the nonzero energy splitting of the quantized states of
electrons in the small particle.
Reed et.\ al.\cite{REED-dot} believe to have observed the level
splitting, associated with quantum confinement in all three
dimensions, in tiny resonant tunneling diodes.
Further, in an experiment in which electrons are added or subtracted from a
small particle, these energies can be important as the Pauli principle
forces newly added electrons into higher lying energy states.

This chapter focuses on the effects of both single electron charging
energy and quantum level splitting in a large array of very small
particles (quantum dots) coupled by tunneling to a metallic substrate.
The dots were made by laterally confining electrons of a two
dimensional electron gas using techniques described
elsewhere.\pcite{CHAP5}
The dots have an occupancy of electrons 
which can be varied by means of a gate bias. In a simple model, a
single dot develops a spectrum of threshold gate voltages
at which electrons are
energetically allowed to tunnel into and out of the dot. The Coulomb
blockade regulates the entrance of electrons into a dot, allowing them
to be added only one at a time, with spacings between the electronic
additions at intervals in gate bias as large as 30~mV.

The sizes of our dots increase as more electrons are added to them. In
chapter 5, we determined the area of dots as a function of gate bias
using capacitance
measurements. Employing this, we develop a model which predicts a spectrum
of individual electronic additions. The spacing in gate voltage
between electron additions to the dots decreases as the dot size
increases.

We study the capacitance of these arrays to very high resolution as a
function of gate bias and amplitude of measuring signal. Intriguing
fluctuations, of amplitude on order 1~fF, are observed in the device
capacitance. Some of these fluctuations consists of nearly periodic
oscillations which can persist for many cycles. This
periodic structure can be identified as a reflection of an almost
regular spectrum of electron additions to the dots,
with one oscillation occurring
per electron addition to each of the dots in the array.

These experiments differ from previous experiments which studied
coarse features in the capacitance spectra of quantum
dots.\pcite{TREY-zeem}
In our model for that work,\cite{CHAP7} the capacitance 
fluctuations arise from
statistical anomalies in the spectrum of electron additions to a
single dot, on the scale of many electron additions.
Here, we observe features in the capacitance arising from
the addition of single electrons to the well.

Some of the work in this chapter has not been completed, and issues
brought forth in the data have not been entirely resolved.
A number of questions remain unanswered because further modeling of
the system is required.

\v\par\penalty-1000
{\noindent{\bf 6.2 Equilibrium State of Quantum Dots}}
\nobreak\v\nobreak
Figure~\fign{equilib}a shows the basic schematic
structure of our device. The
device structure and fabrication were described in detail in chapter~5. 
Electrons, laterally confined in a GaAs quantum well, embedded in the
dielectric of a semiconductor capacitor, can tunnel to an
$n^+$ GaAs conducting substrate (bottom capacitor plate)
via an AlGaAs tunnel barrier. A gate bias applied
between the two electrodes, a top metal gate and the conducting substrate,
of this capacitor varies the number of electrons in the dots. The
dots were produced using electron beam lithographic techniques; the
electron occupancy of those made
with the smallest lithographic patterns (patterns of
0.04~$\mu$m$^2$ area) can be varied from 0 to about 60 electrons per
dot by means of the gate bias.
There are capacitances $C_{top}$ and $C_{bott}$ coupling the dot to
the top gate and substrate electrodes respectively. We consider the
capacitance of the dot to the surroundings to be $C_{top}+C_{bott}$,
and refer to this sum as $C$.
Our measurements involve both a conductance measurement of this tunneling,
discussed later in this chapter, as well as measurements of the
bulk device capacitance.

\figinsert{\vfil\vskip5.5truein\includegraphics{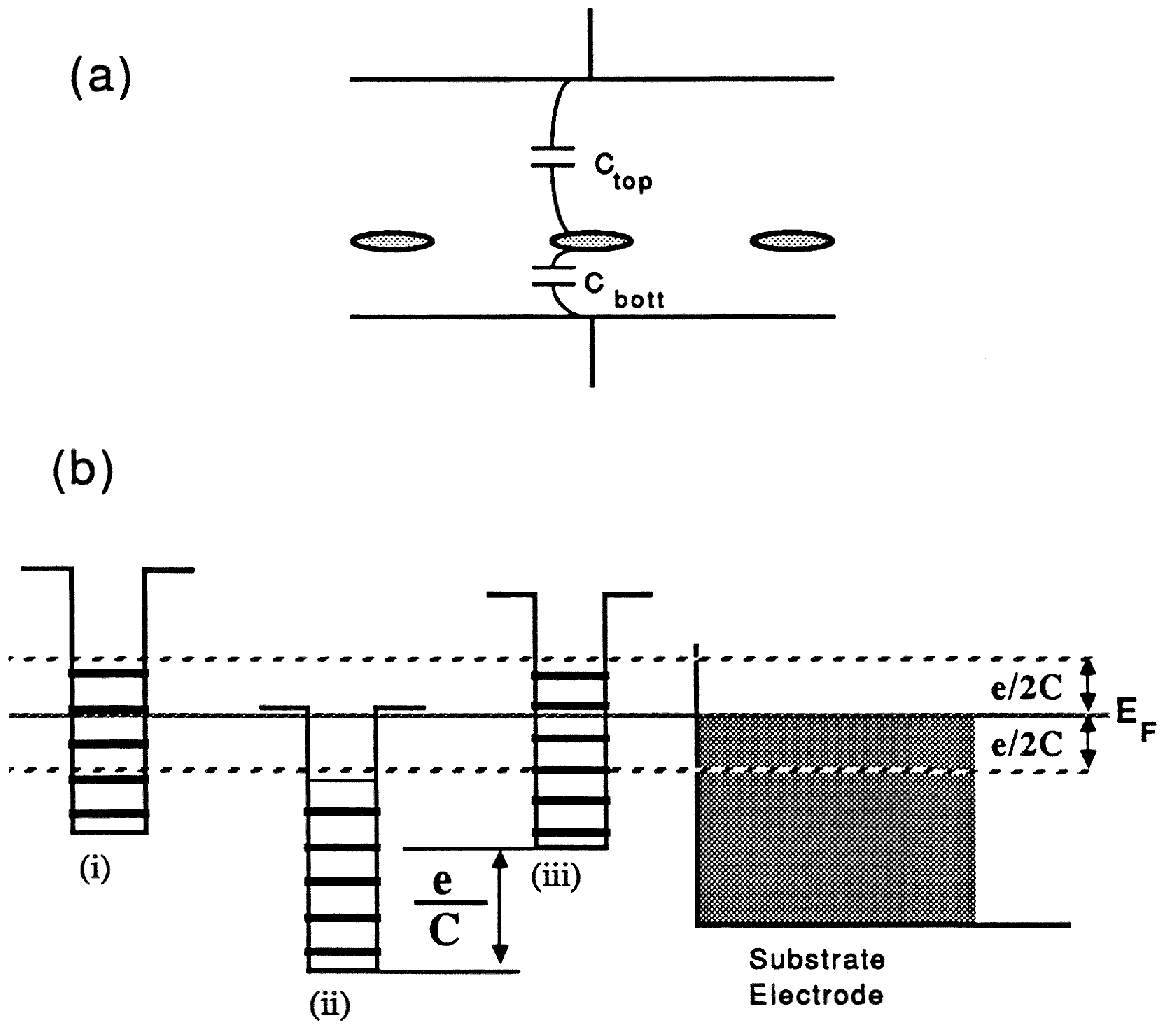}}{equilib}

We briefly explore the energetics of electron transfer between the
quantum dots and the substrate.
In order to minimize the Gibbs free energy of
the system, the Fermi energy in the well and the substrate tend to
equalize. The nonzero Coulomb charging energy
often makes exact alignment of Fermi energies impossible. Neglecting
charging effects, an empty
quantum level whose energy is less than the Fermi energy
in the substrate would ordinarily be filled by tunneling from the
substrate. Notice that, if charging effects are included, the
addition of this one electron from the substrate increases the
electrostatic potential of the dot. In fact, it may increase the dot
potential so much so that the energy of the quantum level just filled
may be greater than the Fermi energy in the substrate! The system is
in a strange ``pickle'', unable to equilibrate Fermi energies in the
dot and in the substrate.

The solution to this problem is, of course, that the dot-substrate
system minimizes
its Gibbs free energy as well as it can without actually equilibrating
the Fermi energies in the dot and in the substrate.
Lambe and Jacklevic\cite{LAMB-coul} first developed a model which was
further elaborated by Cavicchi and Silsbee\cite{SILS-coul} to
described tunneling of electrons between small metal particles,
embedded in an oxide capacitor, to a metal electrode. The same models
should be applicable to the GaAs quantum dot system, and we present
them here. Figure~\fign{equilib}b 
diagrams the energetics of transfer of electrons between the quantum
dot and the substrate. An ``equilibrium zone'' within $e^2/2C$ above
and below the Fermi energy of the substrate develops, and dots
with their Fermi energies within this zone do not gain or lose electrons.
To induce a dot to take on another electron, its Fermi energy
must first (by means of externally applied potentials) be made less
than that in the substrate by an amount $e^2/2C+\delta_n$, where
$\delta_n$ is the quantum level spacing between the highest energy
occupied state in the particle and that of the next one to be occupied when
another electron is added to the dot. Once the electron tunnels into
the dot, the Fermi energy of the dot rises by an amount $e^2/C+\delta_n$.

\v\par\penalty-1000
{\noindent{\bf 6.3 Spectrum of One Electron Additions to a Quantum Dot}}
\nobreak\v\nobreak
The discussion of the last paragraph suggests, if the quantum level
spacing in the dots is taken to be zero, that one electron will tunnel
into a dot every time the
electrostatic potential energy of a dot is decreased with respect to
the substrate, by an amount of $e^2/C$ (or $\Delta$). These single
electron additions
would occur periodically as a function of the electrostatic potential
energy of the dot if no changes were to take place in the capacitance 
of the dot to its surroundings as the electronic
occupation of the dot varies.
As discussed in chapter~5, the areas (lateral area in the quantum
well) occupied by the quantum dots in our samples vary as the
gate bias which controls the electron
occupation of the dots is varied. This causes the dot capacitance to
the surroundings and therefore the charging energy to also vary with
gate bias.

As alluded to above, in our samples the quantum level splitting is
actually a
substantial fraction of the charging energy further complicating the
spectrum of electron additions to the dot. The important parameter in
the case of capacitance fluctuations is the mean energy splitting
$\delta$ between quantum levels, {\it not} including in the average the
zero splitting of degenerate levels. We note
that as the size of the dots increases both the charging energy,
$\Delta$, and the mean quantum level splitting, $\delta$, decrease,
but their ratio remains fixed. These arguments are detailed in
chapter~7. In our samples, $\Delta/\delta\approx3$.

We now describe a model calculation which determines a spectrum of
electron additions to the quantum dots.
The results here are calculated assuming a circular
dot shape with infinite walls and with area equal to the area determined
using the capacitance given in chapter~5. The procedure consisted of
several steps. First the quantum level structure for a dot of
constant area $A_0$ appropriate to a given gate bias $V_0$
is calculated. The levels are spin degenerate and most are (doubly)
orbitally degenerate as well. For this model the mean level
splitting $\delta$ between sets of these degenerate levels is
about four times the inverse of the DOS, or
$\delta\approx4\pi\hbar^2/A_0m^*$.
The constant charging energy $\Delta_0$
corresponding to a dot of area $A_0$ is determined
using knowledge of the dielectric constant and distances from the
charge in the quantum well to the top gate and the substrate in our samples.
This energy is added for each {\it electron} (not each level) in the
quantum level structure. This results in a spectrum with initially
two and four fold spin and orbitally degenerate
levels spaced by an amount, $\Delta_0$, with spacings between electron
additions from different degenerate levels equal to
$\Delta_0+\delta_n$, where $\delta_n$ (the index $n$ is number of
electrons in the dot) is the quantum level splitting
between the sets of degenerate levels. Because $\Delta_0$ is typically
a factor of three larger than $\delta_n$, this spectrum is almost
periodic, but with slight irregularities in moving between the
degenerate levels.

This spectrum is now converted to the dot filling
spectrum one would expect as a function of gate bias for this fictitious
dot of constant area. This is accomplished by multiplying the spectrum
by the inverse of a
``geometric lever-arm'' $\kappa$
( this inverse value is approximately given by $[C_{bott}/C_{top}]+1$)
which describes the amount by which the
electrostatic potential at the dot varies for changes in the gate
bias. In our samples $\kappa$ is about 0.28, as determined from low
and high frequency capacitance measurements on unpatterned samples
discussed in chapter 5. The results of this
calculation for $A_0=0.014$~$\mu$m$^2$ are shown in Fig.~\fign{spect}a.

\figinsert{\vfil\vskip6.5truein\includegraphics{spect.ps}}{spect}

At this point, a correction
must be made for the variation of the dot area as the gate bias is
varied. As both Coulomb energies and quantum level spacings both
scale inversely with dot area,
this simply amounts to transforming the voltage scale through dividing
it by
the ratio of the dot area to the constant area of the dots assumed in
the calculations of the last paragraph. Conveniently, in chapter~5 we
determined the area of our dots
as a function of gate bias, $A(V_{b})$ from capacitance measurements.
There is a subtlety here which involves the fact that the
calculated spectrum is not yet in gate bias coordinates, $V_{b}$;
it is instead given in the coordinates of the model $V_{m}$ for a dot of
fixed area rather than for an area which depends on bias. Because all
energies in the problem scale with area, the two coordinate systems
are related by
$$V_{b}={A_0\over A(V_{b})}V_{m}.\eqno(\equn{conv})$$
We start with a spectrum known in terms of $V_m$ and must solve
this implicit equation to obtain the area in model coordinates
$A^*(V_m)$ so that the translation of the calculated
spectrum into device (gate
bias) coordinates can be accomplished. Once $A^*(V_m)$ is known the
mapping of the calculated results is made using
$$V_{b}={A_0\over A^*(V_{m})}V_{m}.\eqno(\equn{conv2})$$

We refer to the number of electrons in a dot as $n$.
Reversing the statements given above relating the spacing between electron
additions and the dot area given above, the following relation holds:
$$A(V_b)=K{dn\over dV_{b}}.$$
Here $K$ is a constant.
Using the chain rule, we now write the area of the dot as
$$A=K{dn\over dV_{m}}{dV_{m}\over dV_{b}}.\eqno(\equn{chain})$$
Obviously, since the model was calculated for a dot of constant size,
$dn/dV_{m}$ is just a constant.

The problem is made easily solvable using the empirical
fact that $A(V_{b})$ is fit well by a power law.
The voltage, $V_0$ at which the dot area
is equal to $A_0$ (by Eq.~\equn{conv}) is the same in both model
and gate bias coordinates, so we write
$$A(V_{b})=A_0{\biggl(}{V_{b}\over V_0}{\biggr)}^B.\eqno(\equn{avb})$$
Here $B$ is the
exponent in the power law determined from data of the area as a
function of gate bias. It is simple to show using
Eqs.~\equn{conv} and \equn{conv2} that if $A(V_b)$ is given by a power
law, so is $A^*(V_m)$. Hence we can write
$$A^*(V_{m})=A_0{\biggl(}{V_{m}\over V_0}{\biggr)}^M.\eqno(\equn{avm})$$
Placing this expression for $A^*(V_m)$ in Eq.~\equn{conv2}, it follows
that
$$V_{b}=V_0^MV_{m}^{1-M},$$
and
$${dV_{m}\over dV_{b}}={1\over1-M}\biggl({V_{m}\over V_0}\biggr)^M.$$
Equation~\equn{chain} then gives
$$A={K\over 1-M}{dn\over dV_{m}}\biggl({V_{m}\over V_0}\biggr)^M={K\over
1-M}{dn\over dV_{m}}\biggl({V_{b}\over V_m}\biggr)^{M\over 1-M}.$$
Because $dn/dV_m$ is a constant, we identify the exponent $M/(1-M)$ as
the exponent $B$ in Eq.~\equn{avb}. Thus the exponent $M$ is given by
$$M={B\over 1+B},\eqno(\equn{exp})$$
and $A^*(V_m)$ is now determined.

The model spectrum abscissa is thus simply converted to gate bias
using Eq.~\equn{conv2} with the dot area given by the power law in
Eq.~\equn{avm} with the exponent from Eq.~\equn{exp}.
The resulting electron addition
spectrum, after area correction using data of the area vs.\
gate bias from our smallest lithographically patterned dots,
is plotted in Fig.~\fign{spect}b.
These smallest dots were made using square patterns on the sample
surface of area .04~$\mu$m$^2$ as described in chapter~5.
The exponent $B$ in the power law fit to their area is 0.76.

\v\par\penalty-1000
{\noindent{\bf 6.4 Capacitance Fluctuations Expected in an Array of
Quantum Dots}}
\nobreak\v\nobreak
{\noindent\sl Capacitance Signal Due to Electron Transfer to Dots}
\nobreak\v\nobreak
At particular gate biases at which one electron can enter a dot, a
peak in the device capacitance occurs. This can be explained using the
following simple argument. When an electron leaves or enters a single
dot to or from the substrate, the charge on the entire device changes
by an amount $e\kappa$ because the electron traverses only part-way
through the device. Consider the
addition of the $n^{\rm{th}}$ electron to only this one dot,
which occurs at gate bias
$V_n$. At zero temperature, the charge on the device, due to this
transfer, as a time varying voltage $V(t)$ is applied across the device
is given by
$$Q=e{\kappa}\theta(V(t)-V_n),$$
where $\theta$ is the Heaviside function (zero for argument less than
zero, one for argument greater than zero). Taking the derivative with
respect to time, we find
$$I={dQ\over{dt}}=e{\kappa}\delta(V(t)-V_n){dV(t)\over{dt}}.$$
The contribution of the electron transfer to the
device capacitance (measured with infinitesimally small signals)
is the term multiplying  $dV(t)/dt$ in this equation.
At higher temperatures,
this $\delta$ function should be replaced by $df(V-V_n)/dV_n$, the derivative
of the Fermi distribution function (replacing the energy in $f$ with V,
the chemical potential with $V_n$, and the temperature $T$ with $T/e$).

Consider next the addition of the $n^{\rm{th}}$
electron to each of
the dots in a large array at zero temperature.
Due to inhomogeneities in the device, the gate voltage of this
transfer varies among the different dots. The transfers for the
$n^{\rm{th}}$ electron are
distributed according to some distribution function ${\rho}_n(V)$ which,
at zero temperature, is given by
$$\rho_n(V)=\sum\limits_{d=1}^{N}\delta(V-V_{n,d}),$$
where the $V_{n,d}$ are the voltages at which
transfer of the $n^{\rm{th}}$ electron from the substrate to the $d^{\rm{th}}$
dot takes place, and $N$ is the total number of dots. We define
${\rho}_n(V)$ at finite temperatures to be given by
$$\rho_n(V)=\sum\limits_{d=1}^{N}{\partial{f(V-V_{n,d})}\over\partial{V_{n,d}}}.$$
The device capacitance due to the addition of the $n^{\rm{th}}$
electron to each of the dots is given by $e\kappa\rho_n(V)$.

For the rest of this chapter, we consider 
the electron addition spectra of different
dots to be identical except for displacements of the origin
(this model is
justified somewhat by the results from measurements shown in section 6.6).
In this case, for each electron added (for each value of $n$),
$\rho_n(V)$ is identical, with only its origin shifted due to the
voltage spacing between electron additions. We then
define $V_d$ to be the voltage at which the $n^{\rm{th}}$
electron is transferred to the $d^{\rm{th}}$ dot,
measured from the mean voltage for the addition of the $n^{\rm{th}}$
electron ${\overline{V_n}}$ to each of the dots in the array. Since
the spectra of
different dots just have their origins shifted, $V_d$ is the same
for a particular
dot irrespective of the value of $n$. We define $\rho(V)$ by
$$\rho(V)=\sum\limits_{d=1}^{N}{\partial{f(V-V_{d})}\over\partial{V_{d}}}.$$
The device capacitance due to the transfer of the $n^{\rm{th}}$
electron to each of the dots in the
device is given by $e{\kappa}\rho(V-{\overline{V_n}})$ independent of $n$.
The total capacitance due to all electron transfers of any value of
$n$, $C_{dots}$ is given by
$$C_{dots}=e\kappa\sum\limits_{n=0}^{\infty}\rho(V-{\overline{V_n}}){\equiv}e\kappa\gamma(V).\eqno(\equn{cdevrho})$$
Here $\gamma(V)$ is just the sum of $\rho(V)$ repeated at the mean
voltage $\overline{V_n}$ for each electron addition.

\v\par\penalty-1000
{\noindent{\sl Rough Arguments For Capacitance Fluctuation Size}}
\nobreak\v\nobreak
At the temperatures of our
experiments ($\approx1.5~K$), $k_BT$ is large enough, and the
different $V_d$'s for the different dots closely spaced enough, that
$\rho$ is a smooth function of gate voltage.
There are however,
fluctuations in this shape due statistical variations in the
distribution of $V_d$'s. For simplicity, first consider the $V_d$ to
be randomly distributed with equal probability over a range $V_t$ and
with zero probability for all other voltages.
Then the mean value of $\rho$ is given by ${\overline{\rho}(V)}=N/V_t$
in this voltage range and zero for all other
voltages. We focus on $\rho$ over a voltage range of width
${\Delta}V$ within $V_t$. On average,
within the range spanned by ${\Delta}V$, $N{\Delta}V/V_t$
different dots in the sample will have allowed
electron transfers. Depending on the position in the span $V_t$ at
which we are
looking, the fluctuations in the number of electron transfers to
different dots over a range ${\Delta}V_d$ are
$\sqrt{N{\Delta}V/V_t}$.

Consider for this discussion the capacitance signal due to the
$n^{\rm{th}}$ electron addition to each of the dots in the array.
Given that $\rho$ consists of ${\overline{\rho}}+{\delta\rho}$,
where $\delta\rho$ is the fluctuations in $\rho$, fluctuations
will be seen in the device capacitance due to $\delta\rho$.
Now take
the voltage ${\Delta}V$, discussed in the last paragraph,
to be the amplitude of the measuring voltage, $V_{meas}$
used in capacitance measurements. Taking
${\Delta}V=V_{meas}$ to be much larger than $k_BT/e\kappa$, the average
capacitance signal
from an electron transfer to a single dot is equal to
$e{\kappa}/{\Delta}V$. The size of fluctuations in the device
capacitance due to fluctuations in $\rho$ is
$${\delta}C\approx{e\kappa\over\Delta{V}}{\sqrt{N\Delta{V}\over{V_t}}}=
{e\kappa\over{\sqrt{V_{meas}}}}{\sqrt{N\over{V_t}}}.\eqno(\equn{deltcvt})$$
From Eq.~\equn{deltcvt}, it is clear that the fluctuations in the
device capacitance decrease as the measuring voltage is increased.

In the simplest
picture in which the area of the dots is a constant and the
quantum level spacing in the dots is so small that it can be ignored,
the electron additions occur periodically in gate voltage for the
dots, with period $e/{\kappa}C$.
If $e/{\kappa}C$ were larger than the spread
in threshold voltages, $V_t$, then $C_{dots}(V)$ would consist of a
periodic reflection of $\rho(V)$, with capacitance fluctuations due to
$\delta\rho(V)$ also occurring periodically in gate voltage.
In another case, with $V_t$ larger than $e/{\kappa}C$, $C_{dots}$ is
aperiodic for the voltage range over the first few electron additions,
but for gate voltages larger than $V_t$, the spectrum again becomes
periodic. This behavior is due to the overlapping of $\rho(V-{\overline{V_n}})$
for different electron additions in Eq.~\equn{cdevrho}. The size of
the capacitance fluctuations is now different from that given in
Eq.~\equn{deltcvt}. At gate voltages larger than $V_t$, it is assured
that in any interval of gate voltage $e/{\kappa}C$ there will be
one electronic transfer to every dot in the system. In this case,
$V_t$ in Eq.~\equn{deltcvt} should be replaced by $e/{\kappa}C$. The
fluctuations in the capacitance are now given by
$${\delta}C_{dots}\approx{e\kappa\over{\sqrt{V_{meas}}}}{\sqrt{N\over{(e/{\kappa}C)}}}.\eqno(\equn{deltce/c})$$
For a sample with 200,000 dots, $\kappa$ of 0.28, and a 2.5~mV charging
energy (e/C), and measured with a 2.5~mV amplitude signal,
Eq.~\equn{deltce/c} indicates that the fluctuations observed in the
capacitance, due to statistical variation in the thresholds for
accumulating electrons among the different dots, will be of rms magnitude
4~fF. As shown later, the observed capacitance fluctuations in our devices
have nearly the same, but smaller, amplitude. The details of the
capacitance measurement technique must be described for better
comparison, as is done below.
Our samples are made more complicated to understand by
irregularities in the electron addition spectrum caused by the quantum
level structure in the samples as well as the changing area of the
quantum dots. Further below we discuss a model which includes these
irregularities.

\v\par\penalty-1000
{\noindent{\sl Evaluation of the breadth of the $V_d$ distribution}}
\nobreak\v\nobreak
The width of the distribution of thresholds can be
judged from the shape of the increase of the device capacitance as
the first electrons begin to enter the dots. We determine the
total capacitance due to electron transfer to the dots $C_{dots}$
using Eq.~\equn{cdevrho} and the spectrum of electron additions
shown in Fig.~\fign{spect}b.
We consider the smooth part of distribution of thresholds,
${\overline{\rho}(V)}$ to be 
given by a Gaussian shape and multiply this by $e\kappa$ to give the
capacitance signal (smooth part)
from each electron addition to the array of dots. We neglect
capacitance fluctuations caused by $\delta\rho$ because they are too
small to be of importance for our purposes here.
This Gaussian is normalized so that its
integral over gate voltage is equal to $N$, the number of dots in the
sample. It is repeated at every electron addition in
Fig.~\fign{spect}b as indicated by Eq.~\equn{cdevrho} to give the
total device capacitance due to charge transfer to the dots.

\figinsert{\vfil\vskip6.5truein\includegraphics{capcomp.ps}}{capcomp}

In Fig.~\fign{capcomp} shows the experimental
capacitance signal due to transfer of
electrons to dots in a sample of 200~nm wide square lithographic
patterns on the sample surface (the same
sample whose area as a function of gate bias was used to generate
Fig.~\fign{spect}b). The capacitance of our device is a function of
frequency. At frequencies low enough that equilibrating electron
transfer between the dots and the substrate occurs in one cycle of the
measuring voltage the capacitance $C_{low}$ is measured;
at high frequencies where no electron transfer occurs,
$C_{high}$ is measured. $C_{low}$ contains both the
background device capacitance and the capacitance due to
electron transfer to the dots; $C_{high}$ contains only the background
capacitance. $C_{low}-C_{high}$ is the capacitance signal due solely
to electron transfer to the dots, and this is the solid curve of
Fig.~\fign{capcomp}.
Shown also in Fig.~\fign{capcomp} is the device
capacitance modeled using the above protocol with an rms width of
${\overline{\rho}(V)}$ of 35~mV. Note that 355~mV has been subtracted from the
abscissa of the model curve to match the device threshold. -355~mV was
also used as the origin in the power law fit to the device area used
in the calculation of the electron addition spectrum.

We see the results from the model agree well with the measured
capacitance of the device. This is not altogether
surprising as the area of the dot
used in determining the electron
addition spectrum was initially obtained from these capacitance
data.\pcite{CHAP5} However, the agreement also arises from a
propitious choice of charging energy per unit area of the dot. In the
calculation of the electron addition spectrum,
we chose a value of $\Delta_0$ of 2.6~meV based on a simple parallel
plate capacitor argument for the capacitance of a 0.014~$\mu$m$^2$
area dot. Using a larger value for $\Delta_0$ would cause the
simulated results in Fig.~\fign{capcomp} to fall below the capacitance
data. The results confirm the validity of
the method of section 6.3 for the determination of the electron
addition spectrum. The region of importance for our purposes here is
the device capacitance at the gate bias threshold where electrons
begin to enter the dots. The
device capacitance displays a tail, and at gate voltages above this tail
the capacitance increases with a much steeper slope. This tailing
arises from inhomogeneity in the thresholds for adding the first
electron to each of the dots. We use this tail to estimate the
spread in dot thresholds.

The model does not fit the data well at gate biases
where electrons first begin to enter the dots. This is a difficult
region to understand because the dots are so small that fringing
fields are expected become important in determining the dot
capacitance and
the fact that there are only a few electrons in the
dot may demand a rigorous quantum mechanical solution to the problem.
We skirt these problems by assuming the breadth of the first electron
addition among the different dots is responsible for the observed
tail in the capacitance. We adjust the model results by shifting them
by 10~mV towards positive gate bias,
so that the capacitance at low bias for both the
model and the data have substantial overlap. Then the breadth of the
Gaussian distribution of thresholds is adjusted to give the best fit
of these two curves. The model curve adjusted this way is shown as
the dotted curve in Fig.~\fign{capcomp}.

Notice that the tail in the capacitance
appears slightly longer than that in the model. We have tried using
different widths of the threshold distribution function to fit this
curve better. We find that although wider (e.g. rms width of 45~mV)
distributions fit the tail of the capacitance better, they
match the region of high curvature (around -350~mV) less well. We find
that the 35~mV~rms wide Gaussian fits this region much better.
Perhaps forms other than a Gaussian distribution should be examined.
In any case,
we determine through examination of the threshold for accumulation
that the rms width of the distribution function is between 35 and
45~mV. Our models of the fluctuation spectrum of the capacitance are
not sensitive to the precise width; we proceed assuming a width of
35~mV.

This 35~mV width (83~mV full width at half maximum)
is larger than $e/{\kappa}C$ over almost the full range of gate
biases in the device. Further, this wide distribution is consistent
with our lack of observation of large capacitance features associated
with statistical anomalies ({\it not} in dot thresholds for accumulating
electrons in different dots as characterized by $\delta\rho$)
in the spectrum of single electron additions in a single
dot. In chapter~7, we show that a narrower threshold distribution
(11~mV for the electron addition spectrum shown in chapter~7) allows
observation of these fluctuations.

\v\par\penalty-1000
{\noindent\sl Sinusoidal Measuring Signals; More Formal Model}
\nobreak\v\nobreak
In our experiment, the capacitance of samples is measured using
sinusoidal signals. To understand better the capacitance spectra that
are observed, we briefly explore the implications of our capacitance
measurement technique for the observed capacitance fluctuations.

A sinusoidal measuring voltage, $V=\beta\sin(\omega_0t)$ is applied
across our samples, and the current through the device is observed by
means of a lock-in amplifier. The current due to transfer of electrons
to and from the dots is given by $I=C_{dots}dV/dt$ or
$$I=e{\kappa}\gamma\bigl(V_0+\beta\sin(\omega_0t)\bigr)\beta\omega_0\cos(\omega_0t).\eqno(\equn{current})$$
Here $V_0$ is the dc bias applied to the sample.
Notice that the time varying voltage forces the value of $\gamma$ to be
time varying as well.
The lock-in detects only the components of this signal at frequency
$\omega_0$ and not the harmonics generated by the time varying
amplitude in this equation. The output of the lock-in channel
out of phase with the reference signal $\beta\sin(\omega_0t)$ is given
by the coefficient of the
cosine component of a discrete Fourier transform of this signal
at frequency $\omega_0$.
This is given by
$$e{\kappa}{\omega_0\over\pi}\int\limits_0^{2\pi/\omega_0}\gamma\bigl(V_0+\beta\sin(\omega_0t)\bigr)\beta\omega_0\cos^2(\omega_0t)dt.$$
Using a change of variables, $U=V_0+\beta\sin(\omega_0t)$,
recognizing that this Fourier coefficient divided by $\beta\omega_0$
is the capacitance that the lock-in measures, $C_{lock}$, and
recalling that $C_{dots}=e\kappa\gamma$ gives
$$C_{lock}(V_0)={2\over\pi\beta}\int_{V_0-\beta}^{V_0+\beta}C_{dots}(U){\sqrt{1-\biggl({U-V_0\over\beta}\biggr)^2}}dU.\eqno(\equn{clockcdev})$$
In words, the capacitance spectrum measured by the lock-in is simply a
convolution of the device capacitance for infinitesimal measuring
voltage given in Eq.~\equn{cdevrho} by the ``ellipse function'' given
by the square root in Eq~\equn{clockcdev} and normalized by
multiplying the result by $2/\pi\beta$. This ellipse function is
plotted in Fig~\fign{ellipse}.

\figinsert{\vfil\vskip6.5truein\includegraphics{ellipse.ps}}{ellipse}

\v\par\penalty-1000
{\noindent\sl Simulated Capacitance Spectrum}
\nobreak\v\nobreak
We are now in a position to create a simulated capacitance spectrum
using the ideas given above and to compare it, and fluctuations in it,
with data that we have taken. A Gaussian form is assumed for
${\overline{\rho}}(V)$ is assumed
(the precise shape of ${\overline{\rho}}(V)$ is
irrelevant to our arguments). $\rho$ is given by
$$\rho(V)={N\over{\sqrt{2\pi}}\Gamma}exp\biggl[{-V^2\over2\Gamma^2}\biggr]+\delta\rho(V).\eqno(\equn{gaussdist})$$
Here $\Gamma$ is the rms width of the distribution, and
$\delta\rho(V)$ gives the fluctuations in $\rho$. Consider that over
an interval of gate bias
${\Delta}V$ the number of particles allowing tunneling is,
of course, $\rho{\Delta}V$. The rms size of the 
fluctuations in this number is given
by ${\sqrt{\rho{\Delta}V}}$. ${\delta}\rho$ is given the by fluctuation in
number divided by the interval over which the number was obtained. The
rms size of $\delta\rho$ when measured over the interval ${\Delta}V$
is equal to ${\sqrt{\rho/{\Delta}V}}$.

We now describe the method used to produce the simulated fluctuations
in the device capacitance. We commence by modeling the capacitance
fluctuations due to the $n^{\rm{th}}$ electron addition to each of the
dots in the array.
These fluctuations
are obtained in our computer models by starting with a noise signal
whose points are Gaussian distributed with an rms amplitude of one.
There is a discrete spacing between
points of ${\Delta}V_p$. This random noise signal is
multiplied with the square root of the smooth part of $\rho$ divided
by the point spacing. The spacing ${\Delta}V_p$ (typically 0.2~mV)
is chosen to be much
smaller than the amplitude of the measuring voltages used so as not to
influence the final outcome of the simulation. These fluctuations are
convolved with the derivative of the Fermi distribution function to
account for the effects of nonzero temperature. A temperature of
$(1/\kappa)T$ is used to account for the effects of the geometric
lever-arm. Given that most of our data was taken at 1.4~K, this
distribution function has a width of $(1/\kappa)k_BT$ of about 0.4~mV.
For modeling
fluctuations, a replica of $\delta\rho(V)$ is reproduced at every
electron addition in the spectrum of Fig.~\fign{spect}b. These repeating
fluctuations are then multiplied by $e\kappa$ and lastly convolved
with the ellipse function described above.

\figinsert{\vfil\vskip6.5truein\includegraphics{noise.ps}}{noise}

Shown in Fig.~\fign{noise} is a computer generated version of
$e\kappa\delta\rho(V)$, the capacitance fluctuations due to the
$n^{\rm{th}}$ electron addition to the dots, after convolution with
the ellipse function. We have used
$\Gamma$ in Eq.~\equn{gaussdist} equal to 35~mV, a measuring
voltage of 2.4~mV~rms ($\beta=3.4$~mV), and a temperature of 1.4~K.
Note that the effect of
the temperature, which is smaller than the measuring voltage in this
case, is mostly to smooth jagged edges in the fluctuations shown in
Fig.~\fign{noise}.

To produce the capacitance fluctuations expected for all electron
additions,
this  curve is repeated at every electron addition shown in
Fig~\fign{spect}b. (We note that this addition of convolved curves
gives the same result as adding unconvolved curves and convolving the
entire capacitance spectrum.)
The smooth background is subtracted, and the results are plotted in
Fig.~\fign{artspectl} and \fign{artspecth}. The two figures are for
different ranges of the gate bias.
At the base of these figures is replotted the
spectrum from Fig.~\fign{spect}b. 355~mV (the same voltage used as the
device threshold in power law fit to the dot area used in
the model spectrum calculation)
has been subtracted from the
abscissa of this curve so that the device threshold for the model
curves corresponds with the experimental thresholds on actual samples.

\figinsert{\vfil\vskip6.5truein\includegraphics{artspectl.ps}}{artspectl}

\figinsert{\vfil\vskip6.5truein\includegraphics{artspecth.ps}}{artspecth}

We mention several features of this artificial spectrum before
comparing it to our experimental results. Notice that the
breadth of peaks in Fig.~\fign{noise} is two or three times $\beta$.
In Fig.~\fign{artspectl},
where the spacing between electron additions
is several times this breadth, several peaks are seen in the
capacitance spectrum between electron additions. Conversely, beyond
about 150~mV (-220~mV on the plot)
above the device threshold where the spacing between
electron additions is less than 10~mV, typically only one peak per
electron addition is seen. In this region,
there is significant variation of the amplitude of peaks in
the capacitance spectrum
and the phase of the peaks varies somewhat with respect to the
electron addition spectrum.
Despite these observations,
the spacing between successive peaks in the
capacitance spectrum is nearly equal (to within about 30\%)
to the spacing of individual
electron additions in a single dot.
Put another way, the density of
peaks in the capacitance spectrum is very nearly the same as the
density of electron additions in the spectrum for one particle. 

We make a few statements based on observations of changes to
calculated capacitance spectra as the measuring voltage amplitude is varied 
using several different sets noise fluctuations, $\delta\rho$. Though
the different noise fluctuations give very different looking
capacitance spectra, some generalizations can be drawn.
All of these observations are made for $V_t$, the width of the distribution
of electron addition spectra among the different dots, larger than
$e/{\kappa}C$ anywhere in the electron addition spectrum. We
notice that the peak structure of the
fluctuations in the final capacitance spectrum
is rather insensitive to the shape of the $V_d$ distribution.
Using Gaussian or ``flat'' (boxcar) distributions with the same rms
width and starting with the same noise spectrum give nearly identical
capacitance fluctuation spectra. The rest of our observations detail
results from using a Gaussian distribution with rms width of 35~mV.

There are three limiting cases. {\bf 1)} The amplitude of the measuring
voltage, $\beta$, is less than about 25\% of $e/{\kappa}C$. In this
case, typically more than one peak in the capacitance fluctuation
spectrum is seen for each electron addition. This structure is a
consequence of the random peak structure such as that seen in
Fig.~\fign{noise}. Depending on the structure of the
fluctuations in the dot thresholds, the capacitance fluctuations may
or may not appear to have regions of quasi-periodic oscillations.
{\bf 2)} The next case roughly occurs when
$$0.25e/{\kappa}C<\beta<0.6e/{\kappa}C.$$
In this regime, the capacitance spectrum displays, over portions of the
spectrum, almost regular oscillations. Depending on the noise
fluctuations used, these may continue for many cycles or may have a
coherence of only a few cycles. The density of peaks in the
capacitance in these portions is nearly the same (it can be accurate
to better than 10\%
over a 100~mV interval) as the density of electronic additions.
In the other portions of the spectrum, the capacitance appears to be
random with few clear peaks defined, and the density of these peaks
bears no apparent relation to the density of electron additions. The
important point is that in this regime, where clear oscillations are
seen, our simulations indicate that their spacing {\it is} nearly
equal that of the underlying electron addition spectrum. {\bf 3)} The
measuring voltage amplitude approaches or becomes
larger than the spacing of electron additions or
$$0.6e/{\kappa}C<\beta.$$
In this case, fluctuations in the capacitance appear to be random,
with no obvious relation to the electron addition spectrum.

We remark briefly on the reasons for this behavior in the capacitance
spectrum. Consider two peaks in the capacitance fluctuations shown in
Fig.~\fign{noise} due to the addition of the $n^{\rm{th}}$ electron to
each of the dots. If these peaks are separated by an amount equal to
the spacing between successive electron additions in the dots, then
they reinforce one another in the capacitance fluctuation spectrum for
the sample. Peaks separated by voltages different the underlying
spacing between electron additions do not undergo this reinforcement.
In this way, the capacitance spectrum tends to display peaks separated
by the spacings of the electron addition spectrum.

\figinsert{\vfil\vskip6.5truein\includegraphics{rndspect.ps}}{rndspect}

For comparison, we also repeat the spectrum in Fig.~\fign{noise} at
each of the random electron additions shown in Fig.~\fign{rndspect}.
This random spectrum was developed by considering the spacing between
successive electron additions to be Gaussian distributed with the
mean spacing between additions given by
$(e/{\kappa}C)+(\delta_{mean}/\kappa)$, for a particular sized dot, where
$\delta_{mean}$ is the inverse of the electronic density of states in
the dot (not the $\delta$ defined earlier).
We then correct the spectrum for the variable area of the dot
using the method described in section 6.3 above. Clearly, the
repetition of the fluctuations of Fig.~\fign{noise} at random
intervals produces a qualitatively different final capacitance
fluctuation spectrum than repetition at the regular intervals of
Fig.~\fign{spect}b. We conclude that if regular oscillations are seen
in the capacitance fluctuation spectrum of our quantum dot array,
they are likely indicative of an underlying regular spectrum of electron
additions such as that in Fig.~\fign{spect}b.
We start off looking at capacitance data in the absence of magnetic
field first to gain some feel for the
problem and then turn to results taken in a magnetic field where the
connection between the capacitance fluctuations and the underlying
spectrum is much more clear.
 
\v\par\penalty-1000
{\noindent{\bf 6.5 Comparison of Simulation to Observed Capacitance Spectra}}
\nobreak\v\nobreak
We turn now to actual data. Most of our data has been taken on a
sample of dots made using 0.04~$\mu$m$^2$ square patterns. This sample
contains about 200,000 dots. We have also taken data on other
samples with larger pattern sizes and for hexagonal pattern shapes as
well. Although these samples show similar results, we concentrate here
on the sample with 0.04~$\mu$m$^2$ patterns because this sample
contains the most dots, giving the largest capacitance fluctuations,
and the dots are smaller than in other samples, leading the
energies of interest to be larger.

We measure the capacitance of our samples to a noise
resolution of $\approx$0.1~fF as the gate bias is varied. The smooth
variation in capacitance curves is removed by fitting the
capacitance to a polynomial (at most ninth order)
and subtracting off this fit. The order of this polynomial is always
kept small enough so that any fluctuations induced by the subtraction
of these fits would be of much longer period than the fluctuations of
interest. 

\figinsert{\vfil\vskip6.3truein\includegraphics{d13d.ps}}{d13d}

Fig.~\fign{d13d} shows capacitance results, obtained using the
procedure outlined above. These data were taken with an rms signal
amplitude of 2.4~mV~rms at a temperature of 1.4~K (smaller than
the signal amplitude as seen at the well) and a frequency of 615~Hz.
The results are in many
ways similar to those of Fig.~\fign{artspecth}. Firstly, both the
simulated and the actual show regions between -250~mV and between
-100~mV to -50~mV where there are almost periodic oscillations in the
device capacitance. These oscillations have nearly the same amplitude
and periodicity. Additionally, both spectra show some hint of a
``chirp'' in the oscillations with decreasing period as the
gate bias is increased. Secondly, the fluctuations in the device
capacitance become aperiodic for voltages greater than -50~mV and the
simulations of Fig.~\fign{artspecth} show a similar behavior at
voltages greater than -100~mV. 

We underscore that these fluctuations observed in the device
capacitance are reproducible; the statistical error bars (around
0.15~fF) from the signal averaging used in these measurements is minute on
the scale shown. These fluctuations change when the device is biased
so that leakage currents pass through the thick AlGaAs barrier
separating the top gate and the dots. They are also change
for warm up and subsequent cool down of the device. Both of these
occurrences are thought to change the occupation of donor sites in the
AlGaAs barrier, and thus change the distribution of thresholds for
adding electrons into the dots. 

It is not always obvious in observation of our data
where ``regular'' oscillations begin and end. Further, as discussed
above, even a random spectrum of electronic additions may produce
fluctuations such as those in Fig.~\fign{d13d} which appear regular.
We cannot yet conclude that the oscillations seen in the data of
Fig.~\fign{d13d} arise from the near periodicity of the electron
addition spectrum.

\v\par\penalty-1000
{\noindent\sl Measurements Involving the Real Part of the Device Admittance}
\nobreak\v\nobreak
Before proceeding, we make a few remarks on the methods used to take
the data. As with the broad area 2d electron gas samples discussed in
chapters 3 and 4, the capacitance and loss tangent of our samples (not
patterned to produce dots) display a relaxation behavior. The
capacitance decreases on moving from low to high frequencies over a
certain range in frequencies, governed by the characteristic charging
time of a dot, and the loss tangent has a peak in this range of
frequencies. It is a simple matter to show, using the equations of
chapter 4, for our 2d electron gas samples that any fluctuations in the
low frequency capacitance $C_{low}$ will lead to fluctuations of the
same size in the device conductance divided by the frequency (in
radians per second) when measured at the frequency of the relaxation
peak. Also, the size of fluctuations in the capacitance steadily
decrease as the frequency is increased. At the relaxation peak
frequency, the size in the fluctuations in the capacitance should be
half that measured in the device conductance divided by frequency.

The only electron transfer in our devices which occurs at the frequencies of
the observed relaxation peak is electron tunneling between the dots
and the substrate. Fluctuations in the conductance signal occur due to
changes in the height of the relaxation peak. The conductance
measurements are only sensitive to these fluctuations at frequencies
around the relaxation peak. This is not true of fluctuations in
the capacitance signal which depends on electron
transfer for all frequencies faster than the measuring frequency. Fast
transfer of electrons in other regions of the sample (perhaps between
donors in the AlGaAs barrier between the quantum well and the top
electrode) can contribute to this signal. In measurements at
frequencies around the relaxation peak, both the capacitance and the
conductance (divided by the measuring frequency) show similar
fluctuations, with peaks in the capacitance and the conductance
occurring at the same gate bias voltages.

Sometimes the capacitance
displays fluctuation which is not seen in the conductance. Measuring
the conductance allows us to reject spurious fluctuations.
The other feature of making measurements at the loss peak frequency,
for our particular device, is that it allows us to work at higher
frequencies (1.7~kHz) avoiding $1/f$ noise in low frequency
capacitance measurements (such as those in Fig.~\fign{d13d} taken at
615 Hz) which makes measurements more difficult. In all of this
``high'' frequency work, both the device capacitance and conductance
are measured simultaneously.

The convolution used to describe the effect of the measuring voltage 
discussed in section 6.4 must be changed at higher frequencies. This
is because Eq.~\equn{current} considers electron transfer to occur
quickly on the scale of the period of the measuring voltage. If the
relaxation time is not fast compared to the period of the measuring
voltage, a different convolving function must be used.
As shown in Fig.~\fign{ellipse} at measuring frequencies which are
not small compared to the relaxation peak frequency the convolving
function for capacitance fluctuations will to narrow. This shape is
different at higher frequencies because, as will be discussed below, the
tunneling rate from a dot depends on the position in gate bias
at which a dot's threshold (such as any particular stick in
Fig.~\fign{spect}b) lies for exchanging an electron with the
substrate. We expect that dots whose thresholds lie further from the
setting of the gate bias to have slower tunneling rates. At measuring
frequencies comparable to the tunneling rate, some of the dots do not
undergo a tunneling event in a half cycle of the measuring voltage
and hence do not contribute to the device capacitance. Because dots
with thresholds near the position of the gate voltage are more likely
have tunneling events in a half cycle
compared to those with thresholds far from this gate voltage
position, the convolving function narrows. This model is difficult to
solve for sinusoidal measuring signals and the dashed curve in
Fig.~\fign{ellipse} is an imagined form for the shape.

\figinsert{\vfil\vskip6.5truein\includegraphics{d13zh.ps}}{d13zh}

Now we refer to Fig.~\fign{d13zh} which shows the fluctuations in the
real part of the device admittance (conductance) of the same device
from which data of Fig.~\fign{d13d} was obtained.
(Note these data were taken at about two weeks after those in
Fig.~\fign{d13d}, and leakage currents had been passed through the
device, changing the fluctuation spectrum; so we do not expect these
two spectra to have features in common.) It is
difficult to distinguish where in this spectrum fluctuations which
have obvious correlation with the supposed spectrum of electron
additions commence. While above -150~mV there are some regular
oscillations that might be associated with the addition spectrum, at
lower gate voltages there are also regions where almost periodic
oscillations persist.

\figinsert{\vfil\vskip6.5truein\includegraphics{d13zhcap.ps}}{d13zhcap}

For comparison, we also show the capacitance signal from the same data
run in Fig.~\fign{d13zhcap}. Comparing this carefully with the results
of Fig.~\fign{d13zh}, it is clear that almost all of the peaks in the
two figures agree. However, the detailed shapes of peaks in the two
curves are different, with peaks having different amplitudes in some
cases, and the capacitance containing some large
deviations (the ones at -190 and -80~mV being most obvious)
which are not seen in the conductance. These extra fluctuations in the
capacitance are typical of our results.

\v\par\penalty-1000
{\noindent{\bf 6.6 Capacitance Spectra in Magnetic Field}}
\nobreak\v\nobreak
With the data shown so far, it is not entirely certain that the model
which we have developed is actually applicable.
Fortunately, the fluctuation spectrum with magnetic field shows features
which are far clearer.

At 4.3~T, the magnetic length is about 150$\AA$ and is much smaller
than the diameter of our dots for almost the full range of gate biases.
The dots are big enough that it is appropriate to think of the
electronic states in the dot in terms of
Landau levels.
For a magnetic field of 4.3~T, the Landau level degeneracy (including
spin degeneracy) in a dot of
area 0.01~$\mu$m$^2$ (the size of the dot sample produced with
$0.04~\mu$m$^2$ lithographic patterns
at -130~mV gate bias) is about 21 electrons. This is larger than the number of
electrons in the dot (about 16 from our calculated electron addition
spectrum) at the same gate bias.

We examine the
simplest picture with no broadening of the Landau levels. As the gate
voltage is scanned and electrons are added to a particular Landau level, the
change in gate voltage needed to added consecutive electrons is only
$e/\kappa{C}$ with no additional energy from quantum level splittings.
The electron addition spectrum is now much more regular than without
magnetic field; the variable energy level splittings in zero field
(which appeared in the addition spectrum after every second or forth
electron) have been removed.

For the only
sample on which we have taken capacitance fluctuation measurements in
magnetic field (the sample made from $0.04~\mu$m$^2$ lithographic
patterns), the model calculations indicate that only
the lowest Landau level is filled up to a gate voltage of -60~mV at
which there 25 electrons in the dots. Above this gate voltage, the
second level begins to fill as well as the first Landau level orbits as
the dot area increases.
In a picture where the dot area is not changing, electrons begin to
enter the second level
after a gap of $\hbar\omega_c/\kappa$ (plus the charging energy for
one electron) in gate bias after the first level is completely full.
In the real sample in which the area changes,
we find that one Landau orbit is added, due to increase in the area of
the dot, for a change in gate voltage (at the gate voltage at which
there are 25 electrons in a dot)
slightly larger than $\hbar\omega_c/\kappa$. We have not well
understood the complications arising from the changing area of the dot
as they concern these gaps. Some arguments dictate that a wide gap remains
between the Landau levels while others indicate that the gap may
contain electron additions from newly added Landau orbits in the dot.
For simplicity we proceed with simulations ignoring any gaps, which
after all appear relatively infrequently after many electron additions
to the dot, and adding electrons to the dot for every $e/\kappa{C}$
interval in gate bias.

\figinsert{\vfil\vskip6.5truein\includegraphics{regspect.ps}}{regspect}

Fig.~\fign{regspect} shows both the calculated electron addition
spectrum as well as a simulated fluctuation spectrum considering
electron additions to be spaced only by $e/\kappa{C}$. The same
``noise'' used to generate Fig.~\fign{noise} was used, this time
(for comparison to data taken at frequencies near the relaxation peak
frequency)
convolved with an ellipse function corresponding to a measuring
voltage of 1.0~mV. The simulation shows much more regular
oscillations than previous simulations.
The removal of the quantum level spacings has enhanced the periodic
nature of the fluctuations. We note even that when a gap of width
$\hbar\omega_c/\kappa$ (about 27~mV or about 4 times $e/\kappa{C}$ at
-60~mV) is placed in this electron addition spectrum at
a voltage of -60~mV, the simulation continues to show a strong
periodic behavior with oscillations occurring for each electron
addition (depending on the noise fluctuations $\delta\rho$ used the
oscillations may persist even through the region of the gap).

\figinsert{\vfil\vskip6.5truein\includegraphics{d13mbh.ps}}{d13mbh}

\figinsert{\vfil\vskip6.5truein\includegraphics{d13mbl.ps}}{d13mbl}

We compare this simulation (created without the $\hbar\omega_c$ gap)
to capacitance fluctuation data taken at
4.3~T and 1.8~K shown in Fig.~\fign{d13mbh}. This spectrum (taken again
at frequencies near the relaxation peak for the sample) shows
oscillations remarkably similar to those shown in
Fig.~\fign{regspect}. The ticks on the bottom of Fig.~\fign{d13mbh}
show the same electron addition spectrum as that used in
Fig.~\fign{regspect}. The oscillations in the data, like those in the
simulation, occur with nearly the same period as the separation
between electron additions in the model spectrum. The small
amplitude of the oscillations in the sample conductance relative to the
model spectrum may arise because the measuring frequency is not
exactly tuned to the relaxation peak frequency; also, the shape of
the relaxation peak, as discussed below, is somewhat broadened
which would decrease the amplitude of oscillations.
The data also more
clearly show the expected chirp in the oscillations arising from the
decrease in spacing between electron additions as the gate bias is
increased. For completeness, we also show in Fig.~\fign{d13mbl}
a different portion of the same gate bias sweep (lower
voltages). The oscillations again display good agreement with electron
addition spectrum.

We have a limited data set from measuring capacitance fluctuations.
We have taken only one other spectrum, at lower gate biases (from -350
to -100~mV) on the same sample, in magnetic field (also at 4.3~T).
It shows fluctuations which develop, as the gate bias is increased into
the range above -200~mV, into similar quality
oscillations as to those of Figs~\fign{d13mbh} and \fign{d13mbl}.
Though the quantity of the data is limited, we find broad agreement
with the model capacitance spectrum in spacing and chirp 
of oscillations. We believe this to be a convincing argument that the
observed fluctuations in the capacitance spectrum in magnetic field
do indeed occur at intervals given by the Coulomb charging energy of
the dots.

To close these sections on the capacitance fluctuations, we note
that the results in magnetic field also make a statement on the
uniformity of the dots. The obvious correlation seen in the
capacitance fluctuations over many cycles indicates that
electrons are added to the different dots with nearly the same
periodicity. We expect the periods of electron addition in different
dots to be the same if the dots are the same size (so that they have
the same charging energies). Electron beam lithography
produced very uniform patterns for creating the dots, but this may not
be the only factor influencing the uniformity of the size of the
electron packets.
Any quantitative statement to be derived from the data
on the uniformity of dot sizes requires further analysis.
We expect the autocorrelation function of the fluctuation spectra to
be useful in making these determinations. Our
model, of identical electron addition spectra in the dots but with
varying origins (thresholds), works well in simulating the fluctuation
data.

\v\par\penalty-1000
{\noindent{\bf 6.7 Tunneling Rate Measurements}}
\nobreak\v\nobreak
Tunneling of electrons from small metal particles to a metallic
electrode has been studied in detail in our lab previously. Cavicchi
and Silsbee\cite{DICK-dyn} developed a model which describes the
effects of the charging energy on tunneling. One of their essential
results, of importance to the work presented here, determines that the
average tunneling rate for an array small particles becomes dependent
on the amplitude of the signal used to make measurements on the sample. In
this section, we give a brief review of these ideas and examine their
applicability to tunneling measurements done on our quantum dot arrays.

We examine the tunneling in
our quantum dot sample at fixed gate bias, focusing on
a particular dot whose Fermi energy at this gate bias differs from
that in the sample substrate by an amount $eV_k$. Of course the
Fermi energy of this dot lies somewhere within the equilibrium zone
shown in Fig.~\fign{equilib}b. Now imagine that a square wave
measuring voltage, ${\beta}{\rm{sq}}({\omega}t)$ is applied to the
sample. Here ${\rm{sq}}(x)$ is a square wave of unit amplitude
which repeats every time its argument passes thorough $2\pi$. For
argument between $0$ and $\pi$, ${\rm{sq}}(x)$ is equal to one, and in
the other half cycle it is equal to negative one. The voltage which
appears between the quantum dot and the substrate due to this
measuring signal is given by $\kappa{\beta}{\rm{sq}}({\omega}t)$.
Thinking just of the tunneling response of this particular dot during the
first half cycle of the square wave, it is clear that no tunneling of
electrons from the particle to the substrate
occurs unless $V_k+{\kappa}{\beta}>e/2C$. This is the requirement that
the measuring voltage amplitude be large enough so as to move the
(quasi) Fermi energy of the dot out of the equilibrium zone.

Consider the square wave amplitude to be within the correct
range so that just one filled quantum level in the dot is exposed over
the top edge of the equilibrium zone during the first half cycle of
the square wave. 
We ignore degeneracies of levels for the present
discussion.
We refer to the tunneling rate (which can be determined in a WKB approximation)
from a single nondegenerate level as $1/\tau_0$. Now, if the
amplitude of the square wave is increased so that $n$ filled levels
are exposed outside of the equilibrium zone, the rate for tunneling
will be $n/\tau_0$.
If the spacing between nondegenerate levels is
equal to $\delta$, the tunneling rate, $1/\tau$ is given as follows:
$${1\over\tau}=\Biggl\lbrace\eqalign{&{V_k+{\kappa}{\beta}-e/2C\over\delta}{1\over\tau_0}\cr&0\cr}
\qquad\eqalign{\biggl(V_k+\kappa\beta&\ge{e\over2C}\biggr)\cr\biggl(V_k+\kappa\beta&<{e\over2C}\biggr).\cr}$$
In words, the tunneling rate for electrons moving from the dots to the
substrate is zero until measuring voltage is increased to some
threshold value. Beyond this threshold the rate increases linearly
with gate bias. These arguments are mirrored for the second half of
the square wave where the square wave amplitude threshold for
a dot to accept an electron (for moving the quasi-Fermi energy an amount
$\delta$ below the edge of the equilibrium zone so that an electron
can tunnel into the dot from the substrate) is given by
$$V_k-{\kappa}{\beta}<-{e\over{2C}}-\delta$$
Again, the tunneling rate from the particle increases linearly with
voltage beyond this threshold.

In our experiment, the tunneling current is measured. For a single
particle with square wave excitation, this current
is zero for the square wave
amplitude below the thresholds described above and rises linearly with
amplitude as these thresholds are surpassed. The tunneling current,
$I_{dot}$ for
a single dot, initially in equilibrium (square wave measuring voltage
just turned on) just after an upward step in the square wave voltage
with amplitude large enough to surpass the conductance threshold,
is
$$I_{dot}={V_k+{\kappa}{\beta}-e/2C\over\delta}{e\over\tau_0}.$$
Beyond threshold, the
differential conductance associated with electron tunneling is a
constant.

Consider now a uniform distribution of $V_k$'s for the array of dots.
As the measuring voltage amplitude, $\beta$ is increased between $0$
and $e/{\kappa}C$ there is a linear increase in the number dots
(initially in equilibrium for zero measuring voltage applied) from
which tunneling events can occur during the first half cycle. Recall
that beyond threshold, the differential conductance of
conductance of any one dot is a constant. Since the currents from the
different dots in the array are summed in a measurement of the sample
current, we expect a linear increase in the differential tunneling
conductance as the measuring voltage amplitude is increased.
Put in
other words, two factors are involved in the sample current. The
current from the dots which allow tunneling increases linearly with
measuring voltage beyond a threshold. Also, the number of dots which
allow tunneling increases linearly with the gate voltage. The
combination of these two factors causes a quadratic increase in the
sample current with voltage.

There are limits to this linear increase in sample conductance.
If the measuring voltage amplitude is increased so that
$\beta>e/\kappa{C}$ then tunneling is allowed (during both half cycles
of the square wave) from all of the dots, at which point the increase
in the sample current from adding more dots to the tunneling process
ceases. The conductance becomes a constant. The tunneling current also
becomes a constant for measuring voltages so small that
$\beta<\delta/\kappa$, where $\delta$ is the quantum level splitting.
In this case, only very few of the dots allow tunneling and the
tunneling rate for each of these dots is fixed at $1/\tau_0$ (for
nondegenerate levels) because at most only one level can contribute to
the tunneling current.
In the case of degenerate levels, we must consider $\delta$ to be the
energy spacing between sets of degenerate levels. The tunneling rate
at very low voltages, for those dots from which tunneling is allowed,
is the level of degeneracy of the exposed level times $1/\tau_0$. The
irregularity of the spacings between degenerate levels in our samples
should perhaps wash out this low amplitude cut-off.

As stated earlier, the ratio of the charging energy $\Delta$ to the
level splitting between degenerate levels is about 3 in our samples.
We thus expect the observed sample tunneling conductance to be level
at low measuring voltages, start a linear increase at a particular
measuring voltage and continue to rise for voltages to about 3 times
this voltage and then level off once again.

\figinsert{\vfil\vskip6.5truein\includegraphics{d13ag.ps}}{d13ag}

Figure~\fign{d13ag} displays the relaxation peak frequency, measured
using {\it sinusoidal} signals as a function of the rms amplitude of
measuring signals. The gate bias here is set to zero, corresponding to
a charging energy ($e/\kappa{C}$) of about 9~mV in our model.
The peak frequency, as discussed in chapter~4, is
proportional to the tunneling conductivity (which in turn is
proportional to the mean tunneling rate from the dots)
between the quantum well and the
substrate. The curve shows some of the expected features. There
appears to be a leveling off of the conductivity both at low and at
high amplitudes. Moreover, the high amplitude plateau occurs at around
the value of $e/\kappa{C}$ and the low amplitude plateau, much less
well defined, begins to set
in at a fraction of something near the expected 1/3
of the voltage of the high amplitude plateau.
The conductivity, however, does not increase nearly fast enough in the
region between the plateaus to agree with our model; the conductivity
in Fig.~\fign{d13ag} rises with an exponent much less than unity.

One explanation for this less than dramatic rise in the conductivity
is derived from the fact that sinusoids and not square waves were used
in the measurement. The calculation of the system response to
sinusoidal signals is complicated, and we have not
solved this problem. We suspect though, that the sinusoid should
blur the three sharply defined regions described in the square wave
model, rounding the transitions between them as a function of
measuring voltage.

We also know, from our work in chapter~4, that for low densities
of the 2d electron gas a temperature dependent
tunneling suppression arises, likely the result of collective
effects. These effects may influence the behavior of quantum dots.
The suppression seen in chapter~4, while occurring at a lower
temperature scale than the corresponding amplitude dependence of the
conductance in the dots, shows a similar range of increase (factor of
2 or 3) of the tunneling conductivity.
Further characterization of the dots is required to discern
the cause of the observed variation of the conductivity with
measuring voltage amplitude.
Specifically a series of curves
at different gate biases would be useful
to view the effects different charging
energies on the conductivity vs.\ measuring voltage amplitude curve.
Also, the use of square wave measuring signals may give clues to
whether collective or single particle effects cause the observed
behavior.

\v\par\penalty-1000
{\noindent{\bf 6.8 Shape of the Relaxation Peak}}
\nobreak\v\nobreak
It is clear from the discussion above that tunneling rates from the
different quantum dots in the array vary widely. Depending on the
measuring voltage amplitude, some dots do not allow tunneling at all
while those that allow tunneling do so with rates proportional to the
number of quantum levels exposed by the measuring voltage.

In this situation, we must combine models used in chapters~3 and 4 to
fit capacitance and loss tangent vs.\ frequency curves. In section
3.7.2, we described a model used for the samples composed of isolated
domains each with different tunneling times. This model gave
broadened loss peaks described by a ``broadening'' parameter $\chi$.
The model though, was for a quantum well whose area is fully
occupied with electrons and not
directly applicable to our results here for quantum dots with large
portions of the quantum well empty. We invoke the ``puddling''
model used in
section 4.7 to describe a well where a fraction $\alpha$ is filled.
We assume in this model that the density of states in the occupied
portions of the well is the same as the constant 2d DOS in the well
for a broad area sample.

We describe fits to the capacitance. For samples with the well
partially filled, the ``puddling'' model of section 4.7 gave the device
capacitance by
$$C(f)={\alpha}C_p(f)+(1-\alpha)C_{high}.$$
$C_p(f)$ is defined as
$$C_p(f)={C_{high}C_x[1+(f/f_{peak})^2]\over
C_{high}+C_x(f/f_{peak})^2}.$$
Here $C_{high}$ is the high frequency capacitance of the device, $C_x$
is the capacitance $C_{low}$ that {\it would} be measured at low
frequencies if the well were full and the DOS in the well were the
constant 2d DOS, and $f_{peak}$ is the relaxation peak frequency.
Combining this model with the model for the broadening parameter
given in section 3.7.2, we get
$$C(f)={\alpha}\bigl[.2C_p(f;f_{peak})+.4C_p(f;{\chi}f_{peak})+.4C_p(f;f_{peak}/\chi)\bigr]+(1-\alpha)C_{high}.$$
In fits, the free parameters are $f_{peak}$, $C_{high}$, $\alpha$,
and $\chi$. The value of $C_x$ is fixed (with respect to $C_{high}$),
and we set this value to
$1.36C_{high}$ from measurements of the ratio $C_{low}/C_{high}$ on
broad area samples from the same wafer on which the quantum dot
sample was produced.

\figinsert{\vfil\vskip6.5truein\includegraphics{capfreqdot.ps}}{capfreqdot}

Figure~\fign{capfreqdot} displays the capacitance and loss tangent of
the quantum dot sample (same device as in previous figures) as a
function of frequency. The gate bias is set to zero, and the
sinusoidal signal amplitude is 2.4~mV~rms
The ``broadened'' fits from the puddling model
were used to fit to the capacitance data in
Fig.~\fign{d13ag}.

\figinsert{\vfil\vskip6.5truein\includegraphics{broadamp.ps}}{broadamp}

We have measured the broadening parameter as a function of signal
amplitude for gate bias set to zero. This is plotted in
Fig.~\fign{broadamp}. Interestingly, the broadening parameter
increases as the signal amplitude is increased, reaching a maximum at
around 7~mV~rms, and then decreases again as the signal level is
further increased.

This behavior is consistent with the ideas about tunneling
from quantum dots given above. For simplicity, consider all quantum
levels in a dot to be nondegenerate with constant energy spacing between
levels given by $\delta_{mean}$. For a square wave measuring voltage of
amplitude $\beta$ (smaller than $e/\kappa{C}$),
for those dots from which tunneling is allowed,
tunneling rates range from $1/\tau_0$ to $\beta/\delta_{mean}\tau_0$.
As $\beta$ is increased, this range of rates is obviously increased.
The range reaches a maximum when $\beta=e/\kappa{C}$. Beyond this,
all dots allow tunneling and the slowest dots have a rate given by
$(\beta-(e/\kappa{C})/\delta_{mean}\tau_0)$ while the fastest dots have
a rate of $\beta/\delta_{mean}\tau_0$. Clearly, the ratio of the
fastest to slowest tunneling times decreases as $\beta$ is increased
in this regime. Again, issues are confused by our use of a sinusoidal
measuring voltage and the level degeneracies. However, the
qualitative behavior is still apparent in Fig.~\fign{broadamp}.

Finally, we do not necessarily expect $\chi$ to approach one even in
the extremes of high an low measuring voltage amplitude. Inhomogeneity
in the tunnel barrier or in the shapes of the dots may cause them to
have different tunneling rates even if the Coulomb blockade were not
present. In our model, at very low measuring voltages with amplitude
smaller than $\delta_{mean}/\kappa$ the tunneling rate from all
of the dots which allow tunneling is $1/\tau_0$. The fact that $\chi$
retains a value between two and three as the measuring voltage amplitude is
decreased to values (the lowest amplitudes plotted)
around $\delta_{mean}/\kappa$ may indicate some intrinsic broadening.

\figinsert{\vfil\vskip6.5truein\includegraphics{caphigh.ps}}{caphigh}

An intriguing behavior is seen in the high frequency capacitance
$C_{high}$ obtained from fits. This is plotted in
Fig.~\fign{caphigh}. $C_{high}$ decreases as the measuring voltage
amplitude is increased, reaches a minimum at around the same
amplitude at which the broadening parameter reaches a maximum, and
then starts to increase again.
Observation of capacitance vs.\ frequency curves at different
amplitude, such as those shown in Fig.~\fign{d13agcomp}, indicates
that these results are {\it not} an artifact of the fitting
procedure. In our model, the high frequency capacitance (the
capacitance measured at time scales so fast that no charge can be
transferred to the dots) should not depend on measuring voltage
amplitude. The results may possibly be explained by a few
dots which transfer electrons for very small measuring voltages and
do so at an enormous tunneling rate (compared to tunneling rates for
other dots) and thus contribute to the device capacitance even at the
highest measuring frequencies.
As at most one electron is transferred to these dots per half cycle of the
measuring voltage, the contribution of these dots to the device
capacitance decreases as the measuring voltage is increased (i.e.\ $e/V$
decreases as $V$ is increased).
This is a highly speculative picture, and we have no explanation why
fast transfer should exist only for a few dots which transfer
electrons at very low voltages.
This aspect of the high frequency capacitance data is at present a mystery.

Finally we note, as discussed elsewhere,\cite{DICK-thes} that the low
frequency capacitance in our model should not (and experimentally
does not) depend on the amplitude of the measuring voltage. While
many dots do not contribute to the tunneling at low amplitude, those
that do contribute transfer a disproportionately large amount of
charge (a whole electron) for the size of the measuring voltage used.

\figinsert{\vfil\vskip6.5truein\includegraphics{d13agcomp.ps}}{d13agcomp}

\v\par\penalty-1000
{\noindent{\bf 6.9 Summary}}
\nobreak\v\nobreak
We have made a series of capacitance measurements on quantum dot
arrays. High precision measurements have revealed reproducible
capacitance fluctuations which develop in the capacitance spectrum of
the dots as a function of gate bias. A model for the dots, involving
a spectrum of individual electron additions to the dot which occur as
the gate bias is increased produces simulated results which have
qualitative features of the data. In magnetic field, with highly
degenerate Landau levels, the fluctuations seen in the data appear as
regular oscillations with period thought to be given by the gate
voltage change necessary to add one electron onto a dot.

Tunneling measurements have also been made on the dot array. Some of
these results are in qualitative agreement with models while
others remain a unexplained. More data and analysis are required.

\vfill\supereject
\centerline{\bf References}
\v
\listrefs
\vfill\eject

\newchapter{7.}
%
%
%
%
\figitem{com}{(a) Each bar represents the gate voltage at which one electron
may enter a dot, from model described in text.
Brackets at low gate bias demarcate quantum levels. (b) DOS determined from
(a) after convolution with a Gaussian rms width 11 mV.  (c) DOS from
the data of Ref.1.  In both (b) and (c), the smooth variation has been
subtracted, and the results have been normalized to the DOS at .26 volts.}

{\centerline{\twelvebf Chapter VII}}
\v
\v
{\centerline{\twelvebf Computer Simulation of Quantum Dot Capacitance}}
\v
\v
\v
\v
\v
\v
\v
Hansen et al.\ have interpreted the magnetic field dependence of
capacitance spectra of quantum dot arrays as reflecting the evolution
of individual electron states.
They model the system as a two-dimensional quantum well with lateral
confinement on the scale, independent of bias, of 120~nm.\pcite{TREY-zeem}
We question whether the experiment has adequate resolution to reveal
structure associated with individual levels.

In a circular dot of diameter $D$, two energies are important, the
typical splitting $\delta\approx16\hbar^2/m^*D^2$
of the one electron energy
levels, usually both orbitally and spin degenerate, and a charging
energy
$\Delta\equiv e^2/C\approx16e^2d_e/\epsilon D^2$ ($cgs$ units).
$C$~is the capacitance of the dot to its surroundings,
$\epsilon$ is the dielectric constant of the medium, and $d_e$ is a
length of the order of the distance from the dot to the substrate or
gate structure. The ratio of the charging energy to the quantum
splitting is $\Delta/\delta\approx e^2d_em^*/\epsilon\hbar^2\approx5$ for
the configuration of the Hansen experiment and is independent of the
size of the dot.  

{\it If} the experiment can be carried out with adequate
resolution, each successive electron transfer will contribute a
distinct peak to the capacitance spectrum.\pcite{LAMB-coul}
With $\Delta>>\delta$, the spacing between peaks will be determined
principally by the charging energy $\Delta$, not by the level
splitting $\delta$.

Fig.\fign{com}a illustrates a model calculation for a dot with a
square well confining potential (infinite walls) with circular
symmetry. Individual energy levels were calculated for the dot whose
diameter was taken to
be proportional to the square root of the
dot capacitance deduced from Ref.1 by integration of
$dC/dV_{gate}$ . A maximum value of 270 nm was chosen for the dot
diameter, nearly equal to the
lithographically defined size of 300 nm.
The one electron spectrum for the dot was converted to a capacitance
spectrum by adding $\Delta$ for each electron added to the dot and
converting to gate voltage from well voltage using a geometric
lever-arm of 2 for this structure.
This scheme 
gives an areal density of electrons in the dot equal to that
implied by the interpretation of the strong maxima in the 2 T data
of Ref.1 as spin degenerate Landau maxima.
The increase in level density with increasing gate voltage
in fig.\fign{com}a results from
the increase in the diameter of the dot as it is filled
and the consequent decrease of both $\Delta$ and $\delta$.  

\figinsert{\vfil\vskip 6.5 true in\includegraphics{com.ps}}{com}

To simulate the experiment,
this spectrum is convolved with a Gaussian with rms width of 11 meV
at the gate (or $\sim$0.3 meV at the well)
to destroy resolution on the scale of individual energy
levels,
giving
the density of states (DOS) of
fig.\fign{com}b.  For comparison fig.\fign{com}c shows 
the normalized DOS,
obtained from the integration of $dC/dV_{gate}$ of the zero magnetic 
field data of Ref.1.
The two curves show the same qualitative behavior.
Comparison with the unbroadened spectrum shows that the first two or
three peaks may indeed be associated with filling of the first doublet
and the next one or two quartets of the single particle spectrum, but
structure at higher filling is the consequence of fluctuations in the
density of states. Each peak is associated, at the highest gate
voltages, with about 20-30 electrons.
The results of this simulation indicate:
over most of the range of
gate voltage, the experiment does not resolve individual quantum
levels, and changes in structure induced by a magnetic field
are
not simply interpretable
in terms of the behavior of the individual quantum levels.

\vfill\supereject
\centerline{\bf References}
\v
\listrefs
\vfill\eject

\bye